\documentclass[aps,amsmath,amssymb,pra,twocolumn,footinbib,superscriptaddress,floatfix,10pt]{revtex4-2}

\usepackage{amsthm}

\usepackage{soul}

\usepackage{xcolor}
\usepackage{color}
\usepackage{graphicx}
\usepackage{dcolumn}
\usepackage{amssymb} 
\usepackage{bm}
\usepackage{footmisc}
\usepackage{mathtools}
\usepackage{framed}
\usepackage{mathrsfs}
\usepackage{lipsum}
\usepackage{makecell}
\usepackage{physics}
\usepackage{braket}
\usepackage{diagbox}
\usepackage{hyperref}
\usepackage{multirow} 
\usepackage{tocvsec2}

\definecolor{darkblue}{rgb}{0,0,0.5}
\hypersetup{
colorlinks=true,
linkcolor=black,
filecolor=blue,
citecolor=darkblue,  
urlcolor=black,
}

\usepackage{booktabs}

\newcommand{\C}{\mathbb{C}}

\def\d{{\rm d}}

\def\>{\rangle}
\def\<{\langle}

\newcommand{\bs}[1]{\boldsymbol{#1}}

\newcommand{\map}[1]{\mathcal{#1}}

\usepackage{chngcntr}   

\newcommand{\beginsupplement}{
\setcounter{section}{0}  \renewcommand{\thesection}{Supplemental\ Note\ \Roman{section}}
\renewcommand{\appendixname}{} 
\counterwithout{equation}{section} 
\setcounter{equation}{0}   \renewcommand{\theequation}{S\arabic{equation}}
\counterwithout{theorem}{section}
  \setcounter{theorem}{0}
  \renewcommand{\thetheorem}{S\arabic{theorem}}
}

\usepackage{cellspace}
\newtheorem{theorem}{Theorem}
\renewcommand{\thetheorem}{\arabic{theorem}}

\newtheorem{corollary}[theorem]{Corollary}

\newtheorem{definition}[theorem]{Definition}

\newtheorem{lemma}[theorem]{Lemma}

\newtheorem{proposition}[theorem]{Proposition}
\newtheorem{remark}[theorem]{Remark}

\begin{document}
\title{Efficient learning of bosonic unitaries beyond the Gaussian class}
\date{\today}

\author{Xiaobin Zhao} \email{zhaoxb22@gmail.com} \affiliation{Ming Hsieh Department of Electrical and Computer Engineering, University of Southern California, Los Angeles, CA 90089, USA}

\author{Quntao Zhuang}
\email{qzhuang@usc.edu}
\affiliation{Ming Hsieh Department of Electrical and Computer Engineering,
University of Southern California, Los Angeles, CA 90089, USA}
\affiliation{Department of Physics and Astronomy, University of Southern California, Los Angeles, CA 90089, USA}

\nopagebreak

\begin{abstract}
Multimode quantum processes are generally difficult to learn, due to the large dimensionality and complex entanglement structure beyond the Gaussian class. Here, we show that the fundamental obstruction is not non-Gaussianity itself, but the buildup of irreducible multimode non-Gaussian correlations. We establish a tractability frontier for bosonic unitary learning: a general \(m\)-mode unitary with input energy at most \(E\) per mode requires at least \(\Omega(E^{2m})\) channel uses, whereas two broad non-Gaussian families---\(t\)-doped Gaussian unitaries and Gaussian-entanglable unitaries---can be learned with resources polynomial in \(m\). The latter can exhibit both extensive non-Gaussianity and strong multimode entanglement. Our forward-only protocols use coherent-state probes, Gaussian operations, local heterodyne detection, and classical post-processing to identify the global Gaussian mixing and reduce the remaining task to single- or few-mode learning. The analysis also yields a multimode quantum Darmois--Skitovich theorem showing that mode-spreading passive networks preserve product structure only for Gaussian input states, an almost-sure activation theorem for non-Gaussian processes showing that non-Gaussian unitaries yield non-Gaussian outputs for almost all coherent input states, and a method for learning unitaries from uncalibrated coherent probes. Our results identify that complexity of learning arises from irreducible mixing of non-Gaussianity and entanglement, rather than either resource alone.
\end{abstract}

\maketitle

\addtocontents{toc}{\protect\setcounter{tocdepth}{-1}}


Learning an unknown quantum process from experimental data is a fundamental task in quantum information science. It underlies the characterization, calibration, and verification of quantum devices \cite{eisert2020quantum,gebhart2023learning}, and provides a route to obtaining predictive models of their dynamics \cite{huang2023learning,caro2023out}.
A generic multipartite quantum operation requires a description whose size grows exponentially with the number of subsystems, and the same obstruction already appears in unrestricted tomography of finite-dimensional states and channels \cite{anshu2024survey,haah2016sample,haah2023query}. Continuous-variable systems add a further difficulty where each mode has an infinite-dimensional Hilbert space that creates additional challenge even under an energy constraint~\cite{mele2025learning}.

Structural assumptions can substantially change this conclusion. Gaussian states~\cite{fanizza2024efficient,mele2025learning,Bittel2025EnergyIndependent,bittel2025optimal,chen2026towards} and unitaries~\cite{fanizza2025efficient} form the canonical tractable class, as they are completely determined by a number of parameters polynomial in the number of modes. Recent results on bosonic state learning further show that efficient tomography can extend well beyond Gaussian states, including \(t\)-doped states with limited non-Gaussianity~\cite{mele2025learning} paralleling analogous results for finite-dimensional t-doped models~\cite{leone2024learning,leone2024learningU}, and even Gaussian-entanglable states~\cite{zhao2025complexity} whose non-Gaussianity and entanglement can be both extensive. These results suggest that Gaussianity itself is not the fundamental boundary of learnability. Whether an analogous tractability frontier exists beyond Gaussian unitaries for multi-mode quantum processes, however, remains largely unresolved.

In this work, we establish such a frontier for bosonic unitaries. First, we consolidate the general impression that learning general bosonic unitaries is hard---a general $m$-mode unitary with input-energy at most $E$ per mode takes at least $\Omega\left(E^{2m}\right)$ channel uses to learn. At the same time, we provide two connected classes of nontrivial non-Gaussian unitaries---the \(t\)-doped Gaussian unitaries and Gaussian-entanglable unitaries---that are learnable with $\textbf{poly}(m)$ channel uses. \(t\)-doped Gaussian unitaries are Gaussian unitaries interleaved with a constant number of local non-Gaussian gates, with limited non-Gaussianity; Gaussian-entanglable unitaries represent a surprising efficiently learnable case where both the entangling power and non-Gaussianity can be large. Together, these results point to a simple principle: learnability is controlled not by the amount of non-Gaussianity or entanglement alone, but by the complex mixing between the two resources.

Our protocol for learning these unitaries only requires coherent-state probes, Gaussian untiaries, heterodyne measurement and post-processing. We learn the entangling structure, reduce the unitary to single- or few-mode and then reconstruct the entire unitary. 
To establish efficient learning, we have developed several theorems that are of independent interest. First, we prove a multimode quantum version of the Darmois--Skitovich theorem \cite{ghurye1962characterization,kagan1973characterization,bryc1995normal,feldman2003characterization}---under nontrivial passive mixing, a product input can remain product only if the participating input states are Gaussian---that enables Gaussianity test and resource theory~\cite{hahn2025measuring}. Second, we prove an activation theorem for non-Gaussian processes---non-Gaussian unitary almost always output non-Gaussian output on input coherent states.
Third, we introduce a unitary learning protocol from uncalibrated coherent probes, allowing flexibility in unitary learning.

\begin{figure}[t]
\centering\includegraphics[width=0.47\textwidth,trim=15 0 0 0,clip,angle=0]{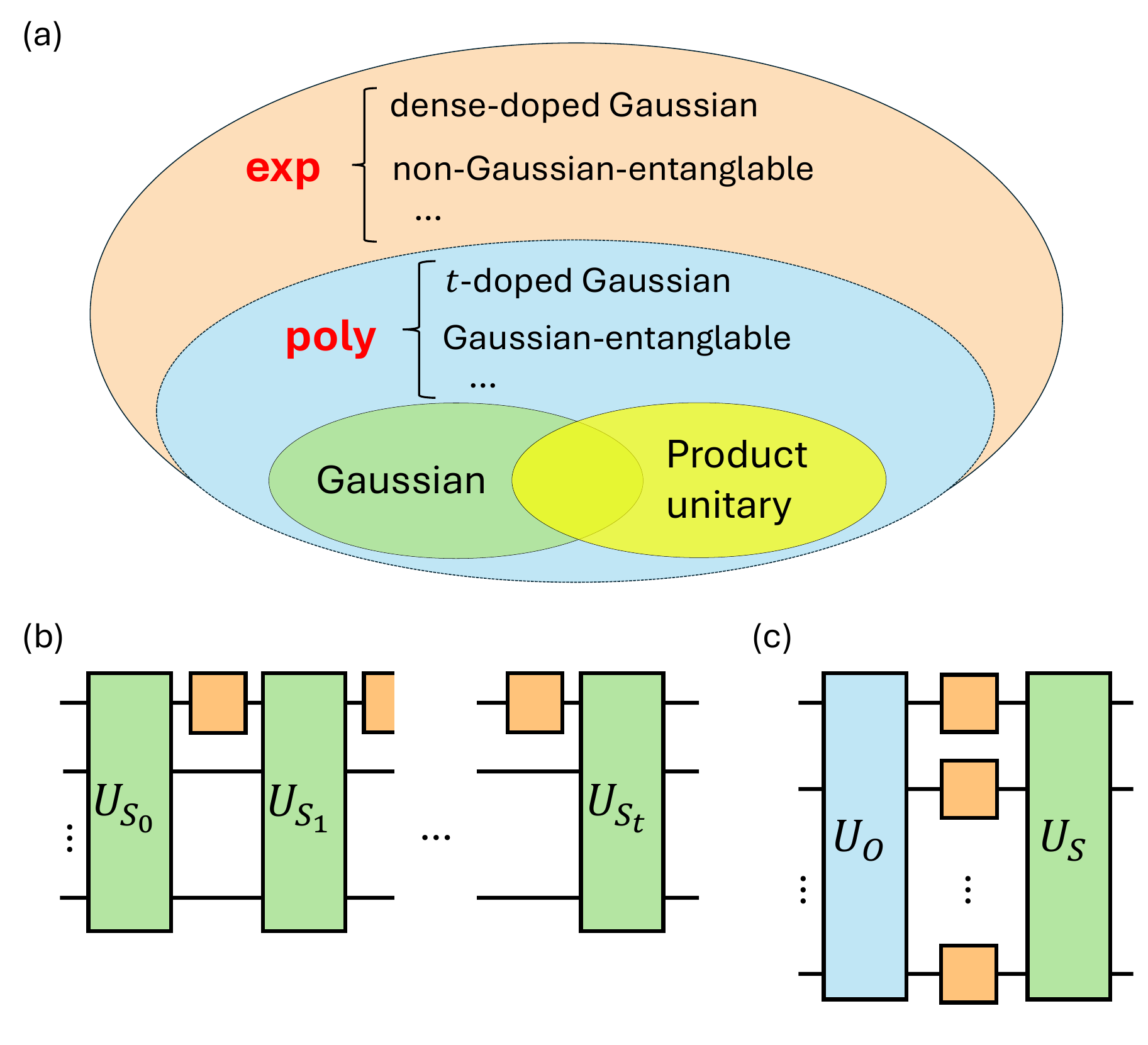}
\caption{
\textbf{Learning-complexity landscape of bosonic unitaries.} (a) Generic bosonic unitaries require a number of queries that grows
exponentially with the number of modes, whereas structural promises can reduce
the scaling to polynomial. Representative polynomially learnable families
include Gaussian unitaries and the two non-Gaussian families established here: \(t\)-doped Gaussian unitaries and Gaussian-entanglable unitaries. 
(b) A \(t\)-doped Gaussian circuit, in which local non-Gaussian unitaries, represented by brown boxes, are interleaved with multimode Gaussian layers $\{U_{S_0},\cdots U_{S_t}\}$. 
(c) A Gaussian-entanglable circuit, in which a product of arbitrary local unitaries is placed between a passive input Gaussian layer $U_O$ and
a general output Gaussian layer $U_S$. 
}
\label{fig:schematic}
\end{figure}


\section{Results}

We aim to learn an unknown unitary \(U\)---establishing an approximate classical description $\tilde{U}$, via accessing the unitary channel
\(\mathcal U(\rho)=U\rho U^\dagger\) with input probe and output measurement. We quantify the reconstruction error by the energy-constrained diamond norm deviation
\(\|\mathcal U-\widetilde{\mathcal U}\|_{\diamond}^{mE}\), where \(E\) is the per-mode mean physical-energy bound defining the
accuracy metric. We denote by \(E_{\rm probe}\ge E\) a per-mode physical-energy upper bound for every coherent
probe state $\rho_{\rm in}$ actually queried by the protocol, and assume
\begin{align}
\begin{cases}
&\Tr\!\left(\widehat H_m\rho_{\rm in}\right)
\leq mE_{\rm probe} ,\\
&\sqrt{
\Tr\!\left[
\widehat H_m^2\,
\mathcal U(\rho_{\rm in})
\right]
}
\leq mE_{\rm II}
\end{cases}
\label{eq:structured-energy-bounds}
\end{align}
for every coherent probe queried by the protocol, where $\widehat H_m
=
\sum_{j=1}^m a_j^\dagger a_j+\frac{m}{2}$ denotes the Hamiltonian, \(E_{\rm II}\) is a
uniform per-mode bound on the root second moment of the output energy. The local reconstruction stages additionally require the finite
dynamical-energy bounds specified in Methods. Within this setting, we first establish a benchmark for general bosonic unitaries. 
We then present our main result on the efficient learning of \(t\)-doped Gaussian unitaries at fixed doped width. Finally, we consider Gaussian-entanglable unitaries as a complementary family, in which non-Gaussian gates may act on every mode while the query complexity remains polynomial in the number of modes.

We emphasize that the efficient learning protocol only query the unknown unitary in the forward direction, using coherent-state inputs and local heterodyne detection, together with Gaussian unitaries and classical post-processing when required. This is in constract to other works that require accessing $U^\dagger$ or $U^T$~\cite{haah2023query,zhao2024learning,ma2025construct,tang2025conjugate}---which are often not available due to nonreciprocity~\cite{jalas2013and,sounas2017non,metelmann2015nonreciprocal}. 

\subsection{General bosonic unitaries}

Gaussian unitaries provide an analytically tractable benchmark. Their generators only involve second-order polynomial of quadrature operators and their action
on the quadrature operators is affine and is fully specified by a
\(2m\times2m\) symplectic matrix together with a displacement
\cite{serafini2023quantum}. An \(m\)-mode Gaussian unitary therefore contains only \(\mathcal O(m^2)\) real parameters and can be learned with resources polynomial in \(m\)~\cite{fanizza2025efficient}. However, Gaussian unitaries are a rather limited class of unitaries---they are not universal for quantum computation~\cite{lloyd1999quantum}.

At the opposite extreme lie unrestricted bosonic unitaries. The
difficulty is already visible at the level of state learning: reconstructing
a general \(m\)-mode state with average energy \(E\) per mode takes $\mathcal O(E^m)$ number of identical copies 
\cite{mele2025learning}. Taking a step back and considering finite-dimensional systems, learning a generic \(d\)-dimensional
unitary requires a number of channel uses that scales quadratically with
\(d\) \cite{haah2023query,mele2025optimal}. These observations suggest that
unrestricted bosonic process tomography is generically inefficient
\cite{lobino2008complete,rahimi2011quantum}. In Methods,  Theorem~\ref{thm:asymptotic-scaling}
makes this benchmark precise:

\begin{quote}
A general $m$-mode unitary with input-energy less than $E$ per mode takes at least $\Omega\left(E^{2m}\right)$
channel uses. 
\end{quote}

Taken together, Gaussian and unrestricted unitaries delimit two ends of the
learning benchmark. Our central goal is to extend the efficiently learnable
regime beyond the Gaussian class while retaining genuinely multimode,
non-Gaussian dynamics. An immediate, but limited, extension is a product of single-mode unitaries, whose learning cost grows only linearly with \(m\). Such circuits, however, contain no multimode entangling structure and therefore do not address the regime of interest. The nontrivial question is then:
\begin{quote}
Are there nontrivial classes of $m$-mode non-Gaussian unitary that is efficiently learnable with $\textbf{poly}(m)$ queries?
\end{quote}

The first nontrivial positive answer is provided by \(t\)-doped Gaussian
circuits, whose collective non-Gaussian action can be compressed to at most
\(\kappa t\) modes.

\subsection{\(t\)-doped Gaussian unitaries}

As shown in Fig.~\ref{fig:schematic}b, an \(m\)-mode \(t\)-doped Gaussian unitary consists of Gaussian layers
interleaved with \(t\) local non-Gaussian unitaries,
\begin{align}
U_{\rm doped}
=
U_{S_t} W_t U_{S_{t-1}}\cdots U_{S_1} W_1 U_{S_0} ,
\label{eq:t-doped-sequential-main}
\end{align}
where each \(U_{S_i}\) is an \(m\)-mode Gaussian unitary and each \(W_i\) acts on
at most \(\kappa\) modes. The supports of the \(W_i\)'s may overlap, and the intervening Gaussian
layers can delocalize their action across the full system.

A key structural fact is that this circuit admits a three-layer
decomposition \cite{mele2025learning}. Up to a relabelling of modes, it can
be written as
\begin{align}
U_{\rm doped}
=
U_S
\left(
U_X\otimes D_{\bar X}
\right)
U_O ,
\qquad
|X|=x\leq\kappa t ,
\label{eq:t-doped-compressed-main}
\end{align}
where \(U_S\) is a general Gaussian unitary, \(U_O\) is passive-Gaussian,
\(U_X\) is an arbitrary unitary on subsystem \(X\), and \(D_{\bar X}\) is a
displacement on its complement. Eq.~\eqref{eq:t-doped-compressed-main}
shows that all non-Gaussian dynamics can be confined to a block of at most
\(\kappa t\) modes. For process learning, however, this decomposition is not
directly accessible: \(U_S\), \(U_O\), the subsystem \(X\), and the coherent
amplitudes entering \(U_X\) are all unknown. The following theorem shows that
the overall channel can nevertheless be learned efficiently.

\begin{theorem}[Efficient learning of \(t\)-doped Gaussian unitaries]
\label{thm:t-doped-learning-main-text}
Consider an \(m\)-mode $t$-doped Gaussian unitary with \(\kappa t=\mathcal O(1)\), satisfying the energy bounds in
Eq.~\eqref{eq:structured-energy-bounds} and the finite block
dynamical-energy condition stated in
Proposition~\ref{prop:xmode-unknown-coherent-main} in Methods. There exists a forward-only
learning protocol using coherent-state probes, local heterodyne detection, Gaussian unitaries and classical post-processing that returns an estimate 
\(\widetilde U_{\rm doped}\) such that
\begin{align}
\left\|
\mathcal U_{\rm doped}
-
\widetilde{\mathcal U}_{\rm doped}
\right\|_{\diamond}^{mE}
\leq
\epsilon
\end{align}
with probability at least \(1-\delta\). 
At constant \(\kappa t\) and fixed energy, accuracy and stability parameters, its query complexity satisfies
\begin{align}
M
=
\textbf{\em poly}(m).
\end{align}
\end{theorem}

The proof of Theorem~\ref{thm:t-doped-learning-main-text} yields an explicit
learning protocol. Its central idea is to successively expose the fixed-width
non-Gaussian block, reconstruct it despite the unknown coherent amplitudes
induced by the input passive  mixing, and reduce the remaining task to the efficient
learning of a structured Gaussian unitary. It also draws on the block-identification, activation techniques, and the unitary-learning approach based on uncalibrated coherent inputs, all established in Methods. We summarize the protocol below. 

\begin{enumerate}

\item[(1)] \emph{Gaussian disentangling.}
Estimate the covariance matrix of the output generated by a random coherent
probe and use it to construct a Gaussian counter-rotation
\(U_{\widetilde S}^{\dagger}\). The same counter-rotation is physically applied to all
subsequent outputs, reducing the output family to generalized
passive-separable states.

\item[(2)] \emph{Learning the  passive-layer.}
Using a fresh coherent probe, apply the shadow-tomography protocol of
Ref.~\cite{zhao2025complexity} to identify the residual passive transformation
\(U_{\widetilde O_2}\), up to its intrinsic gauge. The
corresponding inverse is implemented at the data level by a linear rotation of
the heterodyne outcomes. In this passive frame, the output state factorizes into an
\(x\)-mode state with \(x\leq\kappa t\) and an \((m-x)\)-mode coherent state. The
block Darmois--Skitovich result 
(Proposition~\ref{prop:cross-block-coherent-preservation-main} in Methods) constrains the
residual passive ambiguity to preserve this noncoherent--coherent
decomposition. The block activation result 
(Proposition~\ref{prop:block-activation-main} in Methods) further shows that this
decomposition is revealed by a random coherent probe state almost surely.

\item[(3)] \emph{Learning the non-Gaussian unitary.}
The unknown input passive layer \(U_O\) leaves the coherent amplitudes entering
\(U_X\) uncalibrated. Proposition~\ref{prop:xmode-unknown-coherent-main} in Methods provides a learning protocol that reconstructs
\(\widetilde U_X\) directly from the corresponding output states, up to a
coherent-state-preserving Gaussian gauge that is absorbed into the residual
Gaussian unitary.

\item[(4)] \emph{Learning the remaining Gaussian unitary and assembly.}
Classically undo the $U_X$ by applying $U_X^\dag$ to the reconstructed output states. Reconstruct the remaining Gaussian unitary $\widetilde U_G$ from input-output relations. The final estimate is assembled as
\begin{align}
\widetilde U_{\rm doped}
=
U_{\widetilde S}\,
U_{\widetilde O_2}\,
\left(
\widetilde U_X\otimes I_{\bar X}
\right)
\widetilde U_{\rm G}.
\end{align}
\end{enumerate}


The finite-error
analysis and the complete query-complexity bound is given in \ref{supp:learning t doped unitary}.

Note that Eq.~\eqref{eq:t-doped-compressed-main} guarantees only the existence of
a small non-Gaussian block, whereas Theorem~\ref{thm:t-doped-learning-main-text}
makes this representation operational. It identifies an equivalent
factorization from forward coherent-state data without recovering the
original gate sequence. Here, the only potentially non-polynomial contribution to complexity comes from learning the \(x\)-mode block. Since we have \(x\leq\kappa t=\mathcal O(1)\), the overall protocol is efficient.

\subsection{Gaussian-entanglable unitaries}

The \(t\)-doped unitary-learning result exploits the fact that each such unitary admits an
equivalent representation in which the non-Gaussian component acts on at
most \(x\leq\kappa t\) modes. This bounded-width representation is sufficient,
but not necessary, for efficient learning. As shown in Fig.~\ref{fig:schematic}c, we now consider
Gaussian-entanglable unitaries, whose non-Gaussian layer may act on all
\(m\) modes while remaining factorized across modes:
\begin{align}
U_{\rm ge}
=
U_S
\left(
\bigotimes_{j=1}^m W_j
\right)
U_O ,
\label{three_layer_maintext}
\end{align}
where \(U_O\) is a passive-Gaussian unitary, \(U_S\) is a general Gaussian
unitary, and each \(W_j\) is an arbitrary single-mode unitary. The two Gaussian layers can generate multimode entanglement, whereas the middle layer contains no coupling between distinct modes. The present result is restricted to single-mode \(W_j\), where the associated  gauges reduce to mode routing, including permutations and local phase rotations, allowing a clearer presentation of protocol. Genuinely multimode cases would retain a blockwise mixing gauge that is not addressed here. In addition, this single-mode restriction  allows an extensive number of non-Gaussian local unitaries, scaling with \(m\), rather than confining the non-Gaussian
dynamics to a fixed-size subsystem.

For a coherent probe state 
\(\ket{\underline{\bs\alpha}}\), the passive input layer preserves the
coherent-state structure, as we have $U_O\ket{\underline{\bs\alpha}}
=
\ket{\underline{\bs\beta}}$ for some amplitude vector \(\bs\beta\). Hence, we have  $U_{\rm ge}\ket{\underline{\bs\alpha}}
=
U_S
\bigotimes_{j=1}^m
W_j\ket{\underline{\beta_j}}$, which is a Gaussian-entanglable state and is efficiently learnable using
local heterodyne detection \cite{zhao2025complexity}. This state-level
reduction does not yet determine the unitary $U_{\rm ge}$. Process learning additionally
requires a Gaussian counter-rotation that remains valid when the coherent
probe is changed, together with a reconstruction of the local non-Gaussian
unitaries from uncalibrated coherent inputs. We establish both ingredients in
Methods, leading to the following theorem.

\begin{theorem}[Efficient learning of Gaussian-entanglable unitaries]
\label{thm:learning-main-text}
Consider an \(m\)-mode Gaussian-entanglable unitary of the form
in Eq.~\eqref{three_layer_maintext}, satisfying the energy bounds in
Eq.~\eqref{eq:structured-energy-bounds} and the finite dynamical-energy
condition stated in
Theorem~\ref{thm:output-only-efficient-unknown-coherent-main} in Methods, and the finite uniform intermediate evaluation-energy condition
\(E^{**}<\infty\) defined in the Supplemental Material. 
For any \(\epsilon,\delta\in(0,1)\), there exists a forward-only
learning protocol using coherent-state probes, local heterodyne detection,
Gaussian unitaries and classical post-processing. The protocol returns an
estimate \(\widetilde U_{\rm ge}\) such that
\begin{align}
\left\|
\mathcal U_{\rm ge}
-
\widetilde{\mathcal U}_{\rm ge}
\right\|_{\diamond}^{mE}
\leq
\epsilon
\end{align}
with probability at least \(1-\delta\). Its query complexity satisfies
\begin{align}
M
=&
\textbf{\em poly}\!\left(
m,\,
E^\star,\,
\frac{1}{\epsilon},\,
\log\frac{1}{\delta}
\right),\nonumber \\
\,
E^\star=&\max\{E_{\rm probe},E_{\rm II},E^{**}\}.
\end{align}
\end{theorem}

The Gaussian-entanglable learning protocol retains the first, the second, and the final steps of the protocol for \(t\)-doped unitaries, while only the non-Gaussian reconstruction changes
substantially. In Step~(1), the Gaussian counter-rotation $U_{\widetilde S}^\dag$ inferred from one
coherent probe can be reused for all subsequent probes. In Step~(2),
passive-separable tomography identifies the residual passive layer $U_{\widetilde O_2}$ up to its
intrinsic gauge. In particular, the quantum
Darmois--Skitovich theorem
(Theorem~\ref{prop:DS-passive-main} in Methods)
restricts the residual passive gauge on modes carrying activated
non-Gaussian states to mode permutations and local phase rotations, while nontrivial passive mixing is confined to covariance-degenerate Gaussian
subsystems. In addition, Theorem~\ref{prop:random-coherent-witness-nongaussianity}  in Methods ensures that a
random coherent probe activates every non-Gaussian unitary \(W_j\) almost surely.
Any unresolved passive mixing is therefore confined to modes that output Gaussian states.
After the learned local unitaries are removed classically on reconstructed states, the residual transformation is
Gaussian and is reconstructed as in Step~(4).

The essential new ingredient is Step~(3). Whereas the \(t\)-doped protocol
reconstructs one joint \(x\)-mode block, the Gaussian-entanglable protocol
reconstructs the middle layer mode by mode. The coherent state incident on a
given \(W_j\) is not calibrated, because its amplitude has been transformed
by the unknown passive input layer \(U_O\). Standard coherent-state process
tomography therefore cannot be applied directly.
Theorem~\ref{thm:output-only-efficient-unknown-coherent-main} in Methods provides a protocol with polynomial sample complexity for precisely this
setting. In particular, it reconstructs an arbitrary single-mode unitary from uncalibrated
coherent inputs, up to the unavoidable phase-rotation gauge. Local tomography
and a purity test select a set \(X_{\rm pure}\) containing all activated
non-Gaussian modes, and the theorem is applied independently to obtain
\(\{\widetilde W_j\}_{j\in X_{\rm pure}}\). Residual phases, mode
relabelings and any Gaussian factors included in \(X_{\rm pure}\) are
absorbed into the final Gaussian layer $U_{\rm G}$. The resulting estimate is
\begin{align}
\widetilde U_{\rm ge}
=
U_{\widetilde S}\,
U_{\widetilde O_2}\,
\left[
\left(
\bigotimes_{j\in X_{\rm pure}}
\widetilde W_j
\right)
\otimes
I_{[m]\setminus X_{\rm pure}}
\right]
\widetilde U_{\rm G}.
\label{eq:ge-estimator-main}
\end{align}

The non-Gaussian unitary learning step thus consists of at most \(m\) sample-efficient
single-mode learning tasks, rather than a general \(m\)-mode process
tomography problem. Consequently, its cost accumulates polynomially with
\(m\), even when the number of non-Gaussian gates is extensive. The
finite-error analysis and the complete query-complexity bound are given in 
\ref{supp:Learning of Gaussian entanglable unitaries}.



\subsection{Structural results beyond query complexities}

Three results that we establish during the proof of our main theorems are of independent interest.

The first result determines when passive optical mixing can preserve a
product structure. 
\begin{theorem}[The multimode quantum Darmois--Skitovich theorem
(informal)]
If every input
mode is coupled to at least two outputs in a passive-Gaussian unitary, then a product input can remain
product only when all input states are Gaussian.    
\end{theorem}
A concrete version is shown in Theorem~\ref{prop:DS-passive-main} in Methods. This implies that when a non-Gaussian state remains product after a passive-Gaussian unitary, then the non-Gaussian component can only go through permutations and local phase rotations, rather than
genuine mixing; while nontrivial mixing can only happen among Gaussian
modes. The block extension
(Proposition~\ref{prop:cross-block-coherent-preservation-main}) gives the
corresponding statement for subsystems preserving the form of coherent states. If there are some other modes being passively mixed with that subsystem during the passive mixing, then these modes must output coherent states.

The second result establishes that coherent probe states reveal non-Gaussian
structure generically. 
\begin{theorem}[Almost-sure coherent-state witness of non-Gaussianity (informal)]
If \(W\) is not a Gaussian unitary, then for any random coherent amplitude
\(\bm\alpha\) drawn from a continuous probability distribution, the output state $W|\underline{\bm\alpha}\rangle$ is non-Gaussian with unit probability.
\end{theorem}
Although a special coherent input may produce a
Gaussian output from a non-Gaussian unitary,
we show that all such
inputs form a set of Lebesgue measure zero, as we detail in the full version in Theorem~\ref{prop:random-coherent-witness-nongaussianity} in Methods. A random coherent probe therefore
witnesses the non-Gaussianity of the process almost surely. The block
activation result (Proposition~\ref{prop:block-activation-main}) further shows that if one
output subsystem remains coherent for a positive-measure family of coherent
inputs, then the unitary must have the block-passive form. Persistent coherent output blocks
therefore certify an exact structural separation of the process, rather than
an accidental property of a particular probe.

The third result removes the need for a calibrated coherent source.
Theorem~\ref{thm:output-only-efficient-unknown-coherent-main} shows that an
arbitrary single-mode unitary can be reconstructed from its output states
with polynomial sample complexity even when the absolute input amplitude and
phase are unknown, up to the unavoidable input phase-rotation gauge.
Proposition~\ref{prop:xmode-unknown-coherent-main} extends this construction
to a fixed-width \(x\)-mode unitary, modulo the displacement-passive-Gaussian
gauge that preserves the coherent-state family. In contrast to standard
coherent-state process tomography, which assumes calibrated probe amplitudes
\cite{lobino2008complete,rahimi2011quantum,anis2012maximum}, the same output
data are used to infer the accessible probe geometry and reconstruct the
unknown unitary within precisely the gauge that coherent probes cannot
resolve.

\section{Discussion}

Our results extend efficient  learning of bosonic unitaries beyond the Gaussian class. The protocol facilitates efficient characterization and verification of structured photonic processors combining Gaussian unitaries with local nonlinear operations 
\cite{killoran2019cvqnn,yoshida2025sequential}. More broadly, our result add to a growing separation between learning and
sampling complexity. In particular, boson-sampling states are efficiently reconstructible,
and boson-sampling models can admit classically tractable training objectives,
while their photon-counting distributions may remain hard to sample
classically
\cite{aaronson2011computational,zhao2025complexity,
kurkin2026universality}. Here, we extend this perspective to the
forward-only reconstruction of structured bosonic processes.

The three structural statements sharpen this conclusion and suggest concrete
uses beyond the learning protocols. In the ideal setting, the multimode quantum
Darmois--Skitovich theorem  gives a diagnostic of hidden mode
mixing that complements direct transfer-matrix characterization of
linear-optical networks \cite{rahimi2013direct}. The stable two-input result of Ref.~\cite{cuesta2020stable} provides a
starting point for extending our exact theorem to approximately factorized
outputs. The activation theorem resolves the probe-selection
problem. Exceptional coherent inputs may hide non-Gaussianity, but a probe
drawn from a continuous distribution reveals a non-Gaussian unitary almost
surely. Finally, the learner for unknown inputs shows that
source calibration and process reconstruction need not be performed
separately. An arbitrary single-mode unitary can be learned efficiently when the absolute coherent amplitude and phase are
unknown, up to the unavoidable input phase gauge. This extends
coherent-state process tomography beyond fully calibrated probes
\cite{lobino2008complete,rahimi2011quantum,anis2012maximum}.

Below, we discuss some open directions. While our quantum Darmois--Skitovich theorem and activation theorem do not include error, a finite-error version of them can enable learning in the presence of loss, noise and mixed outputs. We identified two classes of non-Gaussian states to be sample-efficient learnable, potential extension to active Gaussian input layers in $U=U_{S_2} \left( \bigotimes_j W_j \right) U_{S_1}$ beyond our Gaussian-entanglable unitaries is plausible.

\

\

\ 

{\noindent\bf \Large Methods}\\[0.5em]

The Methods collect the structural statements used in the main text. We defer the full proofs and learning protocols to the Supplemental Material. Specifically, the proof for Gaussian-entanglable unitaries is in \ref{supp:Learning of Gaussian entanglable unitaries}, the proof for \(t\)-doped unitaries is in \ref{supp:learning t doped unitary}, and the general-unitary benchmark is in \ref{sec:coh-het-unitary}.\\[-0.5em]

\noindent{\bf Preliminary}\\[-0.5em]

We work in natural units \(\hbar=1\). For an \(m\)-mode bosonic system, let us denote the vector of the annihilation operators as $\bm a=(a_1,\ldots,a_m)^{\mathsf T}$ with the commutation relations $[a_j,a_k^\dagger]=\delta_{jk}$. For a subsystem \(A\subseteq[m]\), \(\bm a_A\) denotes the corresponding restriction. The displacement operator on \(A\) is $D_A(\bm\alpha)
=
\exp\!\left(
\bm\alpha^{\mathsf T}\bm a_A^\dagger
-\overline{\bm\alpha}^{\mathsf T}\bm a_A
\right)$, and the coherent state is $|\underline{\bm\alpha}\rangle_A
=
 D_A(\bm\alpha)|0\rangle_A$. 

A Gaussian unitary is generated by a Hamiltonian at most quadratic in the
canonical operators, including possible linear terms. We denote by \(\mathcal G_m\) the set of pure \(m\)-mode Gaussian states, which are generated by applying a Gaussian unitary to vacuum state. 
An $m$-mode passive-Gaussian unitary \(U_O\) is equivalently represented by a
mode-mixing matrix \(U\in\mathrm U(m)\), through
\(U_O^\dagger\bm a\,U_O=U\bm a\), and by the corresponding
symplectic-orthogonal matrix
\(O=\begin{psmallmatrix}\Re U&-\Im U\\ \Im U&\Re U\end{psmallmatrix}
\in\mathrm{Sp}(2m,\mathbb R)\cap\mathrm O(2m)\)
in the \((q_1,\ldots,q_m,p_1,\ldots,p_m)\) ordering.

Let us denote the total photon number operator $\hat n_m=\sum_{j=1}^m a_j^\dagger a_j$ and the Hamiltonian $\widehat H_m=\hat n_m+\frac m2$. For an \(m\)-mode linear map \(\mathcal V\) and a total physical-energy bound
\(m E\), define the energy-constrained diamond norm as
\begin{align}
\|\mathcal V\|_\diamond^{m E}
=
\sup_{\substack{\Psi_{AR}\ge0,\ \Tr\Psi_{AR}=1\\\operatorname{Tr}[(\widehat H_m\otimes I_R)\Psi_{AR}]\le m E}}
\left\|
(\mathcal V\otimes \map I_R)(\Psi_{AR})
\right\|_1 .
\end{align}\\[-0.5em]






\noindent{\bf Multi-mode quantum Darmois--Skitovich theo-
rem}\\[-0.5em]
\noindent 

Preservation of product states imposes a strong structural constraint on passive-Gaussian unitary mixing. The following theorem converts this constraint into a Gaussianity test by showing that, when a passive layer genuinely spreads every input mode, a product output can arise only from Gaussian input states. This observation allows the non-Gaussian state to be used as a mode-resolved signature of the passive layer. Beyond the learning problem, the statement gives a network-level criterion for when passive mixing can preserve statistical independence.

\begin{theorem}[The quantum Darmois--Skitovich theorem]
\label{prop:DS-passive-main}
Let $U_O$ be an $m$-mode passive-Gaussian unitary with mode-mixing matrix
$U\in\mathrm U(m)$.  Assume that every input mode is spread to at least two
distinct output modes, namely, for each $j\in\{1,\dots,m\}$ there exist
$k\neq \ell$ such that $U_{kj}\neq 0$ and $U_{\ell j}\neq 0$.

Let $\rho_{\mathrm{in}}=\bigotimes_{j=1}^m \rho_j$ be an arbitrary product input.
If the output remains a product across output modes, i.e., 
\begin{align}
U_O \rho_{\mathrm{in}} U_O^\dagger = \bigotimes_{k=1}^m \rho_k',
\end{align}
then every input state $\rho_j$ is Gaussian.
\end{theorem}

\noindent The concrete proof of  Theorem~\ref{prop:DS-passive-main} is shown in \ref{supp:Learning of the second passive-Gaussian layer}. The proof idea is as follows. Heterodyne detection converts the claim into a statement about independent Husimi \(Q\)-distributed variables. The passive layer acts as a linear transformation on these variables, so the vector Darmois--Skitovich theorem forces Gaussianity of any input mode that contributes to two independent output linear forms. Gaussianity of the Husimi \(Q\) function is then lifted back to Gaussianity of the quantum state. The support condition used here is compared with invariant-subset mixing in Remark~\ref{rem:mixing-comparison}.

Related quantum Darmois--Skitovich results were established for a nontrivial beam splitter between two bosonic subsystems, where product output implies Gaussian inputs with the same covariance matrix \cite{cuesta2020stable}. Our statement is tailored to passive-network learning. It allows an arbitrary multimode passive matrix \(U\in{\rm U}(m)\), including complex phases and nonuniform support, and uses the local column-spreading condition needed to identify routed non-Gaussian modes. The proof is also different. Rather than reducing Weyl characteristic functions along fixed phase-space directions, we use heterodyne \(Q\) distributions and the vector classical Darmois--Skitovich theorem. The theorem determines which input states must be Gaussian. The residual
passive freedom within covariance-degenerate Gaussian sectors is characterized separately in Theorem~\ref{thm:gauge-general-gaussian-probes} of the Supplemental Material.

For a \(t\)-doped circuit, the compressed non-Gaussian component occupies a subsystem \(A\) of size \(|A|=x\leq\kappa t\), rather than a product of single-mode states \cite{mele2025learning}. If the coherent output subsystem sees all directions of \(A\), coherent-state preservation forces the entire input block to be coherent after a passive change of basis. The protocol requires the rank-\(r\) refinement below, because only the remaining \(x-r\) directions can carry noncoherent dynamics.

\begin{proposition}[Coherent output subsystems force coherent input directions]
\label{prop:cross-block-coherent-preservation-main}
Let \(A\sqcup B=[m]\), with \(|A|=x\), and let \(U_O\) be a passive-Gaussian unitary with mode-mixing matrix \(U\). Suppose
\begin{align}
U_O\bigl(|\psi\rangle_A\otimes|\underline{\bm\alpha}\rangle_B\bigr)
=
|\phi\rangle_A\otimes|\underline{\bm\beta}\rangle_B .
\end{align}
Let us denote \(r=\operatorname{rank}(U_{B,A})\) where $U_{B,A}\in\mathbb C^{|B|\times |A|}$ denotes the submatrix of $U$ with rows indexed by $B$ and columns indexed by $A$. Then, after a passive change of basis on subsystem \(A\), the \(r\) input directions visible from \(B\) are coherent. Equivalently, there exist a passive-Gaussian unitary \(V_A\), an amplitude \(\bm\mu\in\mathbb C^r\), and a state \(|\chi\rangle\) on the remaining modes such that
\begin{align}
V_A|\psi\rangle_A
=
|\underline{\bm\mu}\rangle_{1,\ldots,r}
\otimes
|\chi\rangle_{r+1,\ldots,|A|}.
\end{align}
\end{proposition}

\noindent
The proof is given in Lemma~\ref{prop:cross-block-coherent-preservation} in \ref{supp:learning t doped unitary}. The submatrix \(U_{B,A}\) measures how many independent directions of subsystem \(A\) are visible from subsystem \(B\). Its singular-value decomposition converts coherent-state preservation on \(B\) into annihilation constraints on \(A\). In the full-rank case \(r=x\), this gives the clean DS-like statement that the whole subsystem \(A\) must be coherent after a passive change of basis. In the learning protocol, the rank-\(r\) form is essential, as it identifies exactly which directions are forced to be coherent and leaves the remaining \(x-r\) directions as the noncoherent subsystem to be learned.

Beyond its role in \(t\)-doped unitary learning,
Proposition~\ref{prop:cross-block-coherent-preservation-main} is a
block-level identifiability statement for passive optical networks with a
correlated input subsystem and coherent ancillary modes. The rank
\(r=\operatorname{rank}(U_{B,A})\) has a direct operational meaning: it is the
number of independent directions in the correlated input block that are
forced to be coherent by the observation of a coherent output subsystem.
This differs from the modewise column-spreading condition in the quantum
Darmois--Skitovich theorem and quantifies how many genuinely noncoherent
degrees of freedom may remain hidden from a complementary block.\\[-0.5em]

\noindent{\bf Non-Gaussian activation theorem}\\[-0.5em]
\noindent 

The passive-layer reconstruction above requires identifying which output subsystems are non-Gaussian. 
However, a coherent probe can accidentally hide non-Gaussianity. For example, the Kerr operator $e^{-it (a^\dag a)^2}$ leaves the vacuum unchanged. The following activation theorem shows that such failures are nongeneric.

\begin{theorem}[Almost-sure coherent-state witness of non-Gaussianity]
\label{prop:random-coherent-witness-nongaussianity}
Let \(\Omega\subset\mathbb C^m\) be a nonempty connected open set, and let
\(W\) be a unitary operator acting on an \(m\)-mode bosonic system. If \(W\) is not a Gaussian unitary, then for any random coherent amplitude
\(\bm\alpha\in\Omega\) drawn from a probability distribution that is absolutely
continuous with respect to Lebesgue measure, we have 
\begin{align}
\Pr\!\left[
W|\underline{\bm\alpha}\rangle\in\mathcal G_m
\right]=0,
\qquad
\Pr\!\left[
W|\underline{\bm\alpha}\rangle\notin\mathcal G_m
\right]=1.
\end{align}
\end{theorem}

\noindent
The proof follows from that of  Theorem~\ref{prop:measure-zero-bargmann-multimode} in \ref{supp:activation theorem}. The argument uses the Bargmann--Fock representation and a holomorphic differential Gaussianity witness. A nonzero holomorphic function of several complex variables has a
zero set of Lebesgue measure zero. Hence, if the witness vanishes for a
positive-measure family of coherent amplitudes, it vanishes identically.
Proposition~\ref{prop:all-coherent-to-gaussian-implies-gaussian} then implies
that the unitary is Gaussian. Therefore every non-Gaussian unitary is activated by a random coherent probe with probability one.

This result is used here as an activation theorem, rather than as a quantitative non-Gaussianity measure. It complements the resource-theoretic and witness-based approaches to non-Gaussian states and operations \cite{zhuang2018resource,takagi2018convex,walschaers2021non}. Those works ask how to quantify or certify non-Gaussianity once a state or process is probed. Here the question is different: for a non-Gaussian unitary, almost every coherent probe already produces a non-Gaussian output, so activation failure is a measure-zero event.

For \(t\)-doped learning, we need a block version of activation. Instead of asking whether the entire output is Gaussian, we ask whether a complementary output block remains in a coherent state. If this occurs for a positive-measure family of coherent inputs, the unitary must consist of an arbitrary operation on the complementary block, a displacement on the coherent block, and an input passive-Gaussian unitary.

\begin{proposition}[Almost-sure coherent-state witness of subsystem separation]\label{prop:block-activation-main}
Let \(W\) be a \(y\)-mode unitary and let the output modes be partitioned as \(A|B\), with \(|B|=r\). Let \(\Omega\subset\mathbb C^y\) be a nonempty connected open set. If, for a positive-measure set of coherent probes \(\bm\alpha\in\Omega\), the output factorizes as $W|\underline{\bm\alpha}\rangle
=
|\psi_{\bm\alpha}\rangle_A
\otimes
|\underline{\bm\eta_{\bm\alpha}}\rangle_B$, then, with probability one, we have 
\begin{align}
W=(U_A\otimes D_B(\bm\beta))U_{O_u},
\end{align}
where \(U_A\) is a unitary on \(A\), \(D_B(\bm\beta)\) is a displacement on \(B\), and \(U_{O_u}\) is passive-Gaussian. 
\end{proposition}

\noindent
This is the block activation statement used for \(t\)-doped unitary learning. Its proof is found in that of Corollary~\ref{cor:random-coherent-block-factorization} in \ref{supp:learning t doped unitary}. In particular, it is proved by showing that a positive-measure set of coherent probes can leave a coherent state subsystem only when the unitary has the displayed block-passive form. Otherwise the exceptional probes have measure zero.

Independently of the learning protocol, this proposition is a rigidity
statement for bosonic processes. Persistent production of a coherent output
subsystem over a positive-measure family of coherent inputs certifies an exact
block-passive factorization of the underlying unitary. It may therefore be
read as an architectural test for subsystem isolation, rather than merely as
a probabilistic tool for choosing successful probes.\\[-0.5em]

\noindent{\bf Efficient unitary learning from unknown coherent
probe states}\\[-0.5em]
\noindent 

After the learned Gaussian and passive frames have been removed, the
effective non-Gaussian unitary is driven by coherent states whose absolute
amplitudes are unknown. The next theorem shows that this lack of calibration
does not prevent process learning. In the single-mode case, a controlled
family of output states determines an arbitrary unitary up to an unavoidable
input-side phase-rotation gauge.

\begin{theorem}[Efficient unitary learning from unknown coherent probe states]
\label{thm:output-only-efficient-unknown-coherent-main} Let \(\mathcal W_\star(\rho)=W_\star\rho W_\star^\dagger\) be an unknown single-mode unitary channel. The learner can prepare, but not characterize, coherent probes $|\underline{s e^{i\vartheta}\alpha}\rangle$ with $\alpha=\sqrt{\mu}\,e^{i\theta_0}$, where \(s\ge0\) and \(\vartheta\in\mathbb R\) are chosen by the learner, while \(\mu\in[\mu_{\min},\mu_{\max}]\) and \(\theta_0\) are unknown. Fix a photon-number bound \(N\) for definition of diamond norm, set \(E=N+\frac12\), let \(\epsilon\in(0,1)\) be the target accuracy and $J=\lceil C_J N/\epsilon^2\rceil$. Assume that all queried output states have mean photon number at most \(N_{\rm out}\), and define $N_{\rm dyn}
\coloneqq
\sup_{\substack{
|\varphi\rangle\in\mathcal H_{\le J}\\
\|\varphi\|=1}}
\langle\varphi|W_\star^\dagger \hat n W_\star|\varphi\rangle
<\infty$, for a sufficiently large universal constant \(C_J\).

Then there exists an output-only learning protocol returning an estimator
\(\widehat{\mathcal W}\) such that
\begin{align}
\Pr\!\left[\inf_{\theta\in\mathbb R}
\bigl\|
\widehat{\mathcal W}-\mathcal W_\star\circ \mathcal C_{\theta}
\bigr\|_{\diamond}^{E}
\le
\epsilon
\right]
\ge
1-\delta ,
\end{align}
where $\map C_\theta(\cdot)=e^{-i\theta a^\dag a }\cdot e^{i\theta a^\dag a }$ denotes the phase-rotation channel. The total number of channel uses satisfies
\begin{align}
M
\le&
C(N_{\rm out}+1)
\frac{\Lambda_\epsilon^{7}}{\epsilon^4}
\left[
(1+S_{\rm cal})
\log\frac{12(1+S_{\rm cal})}{\delta}\nonumber\right. \\
&+\left.
(1+\Lambda_\epsilon^2)
\log\frac{12(1+\Lambda_\epsilon^2)}{\delta}
\right],
\end{align}
where \(C>0\) is a universal constant, 
$\Lambda_\epsilon
:=
1+\frac{N}{\epsilon^2}+\log\frac1\epsilon$, and $S_{\rm cal}
=
\mathcal O\!\left(
\log\frac{\mu_{\max}}{\mu_{\min}}
\right)$ is the number of calibration
scales. The average input photon number satisfies $N_{\rm probe}\le 2J.$
\end{theorem}

\noindent
This theorem is the single-mode process-learning step used after the second passive layer have been counter-rotated. The proof in \ref{supp:Learning the remaining part of the overall unitary} uses output-only intensity calibration, Fourier-ring synthesis of Fock columns, local phase synchronization, column recovery, and polar completion. The corresponding estimates are given in  Lemmas~\ref{lem:output-only-calibration}--\ref{lem:output-only-polar-completion}. The calibration window needed after passive mixing is controlled by Lemma~\ref{lem:selected-mode-calibration-window}.

The corresponding protocol is shown in \ref{supp:Single-mode unitary reconstruction from state tomography}. This should be distinguished from standard coherent-state process tomography, where the coherent probe amplitudes are calibrated and the process is reconstructed from their outputs \cite{lobino2008complete,rahimi2011quantum,anis2012maximum}. Our setting controls only relative scalings and phases, while the absolute coherent amplitude is unknown. The reconstruction therefore uses unitarity to self-calibrate the intensity, synthesize Fock columns through Fourier-ring probes, and recover the unitary modulo the coherent phase gauge.

The \(t\)-doped protocol uses the same idea at fixed width. Here, the finite non-Gaussian subsystem is learned from unknown random coherent probes by first reconstructing the probe geometry from output-state overlaps and then fitting the input-output relation, up to the coherent-state-preserving  gauge. For \(x\)-mode channels, we measure error modulo the coherent-state-preserving gauge. Before showing the result, let us define the following measure 
\begin{align}
&d_{\diamond,xE_x}^{\rm coh}(\mathcal W_1,\mathcal W_2)
:=
\inf_{\bs \gamma,R}
\left\|
\mathcal W_1
-
\mathcal W_2\circ\mathcal G_{\bs\gamma,R}^{\dagger}
\right\|_{\diamond}^{xE_x},\nonumber \\
&\mathcal G_{\bs \gamma,R}(\rho)=G_{\bs \gamma,R}\,\rho \,G_{\bs \gamma,R}^\dagger ,
\end{align}
where \(G_{\bs \gamma,R}=D(\bs \gamma)\Gamma(R)\) denotes a Gaussian unitary, with \(D(\bs \gamma),\,\bs \gamma\in\mathbb C^x\) being a displacement operator, and
\(\Gamma(R)\) being a passive-Gaussian unitary. Let \(W_\star^{\rm gf}\) be a gauge-fixed representative of the unknown unitary. For $q=\left\lceil C_x N/\epsilon^2\right\rceil$, define the dynamical photon-number bound $N_{\rm dyn}
\coloneqq
\sup_{\substack{
|\varphi\rangle\in\mathcal H_{\le q}^{(x)}\\
\|\varphi\|=1}}
\langle\varphi|
(W_\star^{\rm gf})^\dagger \hat n_x  W_\star^{\rm gf}
|\varphi\rangle
<\infty$, where \(C_x>0\) is a universal constant. Then, the following proposition is given: 

\begin{proposition}[Unitary learning with random coherent probe states]\label{prop:xmode-unknown-coherent-main} Consider a learning task of an $x$-mode unitary \(W_\star\) with unknown coherent probe states. Let $N:=x(E_x-\tfrac12)$ be an evaluation photon-number bound and $N'$ be the total photon-number bound of the output states. 
  
Then there exists a learning protocol that reconstructs $d$ output states with a trace distance error $\eta_{\rm stab}$ and probability at least \(1-\delta\). In the meantime, it returns an estimator $\widehat W$ such that 
\begin{align}
d_{\diamond,xE_x}^{\rm coh,x}
(\widehat{\mathcal W},\mathcal W_\star)
  \le
  \epsilon .
\end{align}
Total number of channel uses of \(W_\star\) is at most 
\begin{align}
M
  =&\,
  d
  \left\lceil
  2^{21}
  \left(
\frac{eN'}{x\eta_{\rm stab}^{2}}
+
2e
  \right)^x
  \eta_{\rm stab}^{-2}
  \log\frac{4d}{\delta}
  \right\rceil,
\end{align}
where $d=\binom{x+r}{x}$ denotes the number of distinct coherent probe states used in the protocol, $r\ge \left\lceil C_x \max\{N,N_{\rm dyn}\}/\epsilon^2\right\rceil$ is the cutoff in space truncation.  
\end{proposition}

\noindent
Note that this proposition indicates a sample complexity 
\begin{align}
M=&\textbf{ exp}\left[\textbf{ poly}\left(x,N,N_{\rm dyn},\frac1\epsilon,\log (N'+1)\right)\right]\nonumber \\
&\times \textbf{ poly}\left(\log\frac1\delta\right).
\end{align}
The corresponding protocol is useful when the number of doped unitary $\kappa t$ is $\mathcal O(1)$. In particular, it is based on the \(x\)-mode unknown-coherent reconstruction theorem, Theorem~\ref{thm:xmode-main}, together with the stability threshold \(\eta_{\rm stab}\). Since \(x\le \kappa t\) is fixed in the \(t\)-doped setting, the non-polynomial dependence is independent of the total mode number \(m\).

Beyond \(t\)-doped circuits, this result provides self-calibrating process
tomography for bosonic unitaries. The absolute coherent probe
amplitudes need not be known, and the only remaining ambiguity is the
displacement-passive gauge that preserves the coherent-state family. This
complements standard coherent-state process tomography, which assumes
calibrated probe amplitudes, by separating process identification from source
calibration.

Combining these statements gives the two structured-unitary learning protocols. For GE unitaries, the quantumm DS theorem identifies the passive layers up to the intrinsic  gauge, activation makes the non-Gaussian mode partition stable with probability one, and the learning protocol with unknown coherent probe states reconstructs the local non-Gaussian unitaries. The remaining Gaussian cleanup and query accounting are given in Theorem~\ref{thm:query-complexity-GE-final}. For \(t\)-doped unitaries, the same logic is applied blockwise: cross-block preservation gives the passive gauge, block activation stabilizes the \(x|(m-x)\) split, and the fixed-width unknown-coherent learner reconstructs the \(x\)-mode block.

The Supplemental Material gives the remaining stability and accounting details for the block protocol. The Gaussian counter-rotation is in \ref{supp:Learning of the Gaussian layer-t doped}. The second passive layer is reduced to a generalized passive-separable problem in \ref{supp:learning the second layer of t doped}. After counter-rotation, Corollary~\ref{cor:random-coherent-block-factorization} gives stable factorization into an \(x\)-mode block and coherent modes. The output mode locations are identified by covariance tests in \ref{supp:Identification of output mode types_doped}. The finite block is learned in \ref{supp:Learning of the local unitary that might not preserve coherent states_doped}. Energy concentration is controlled by Theorem~\ref{lem:selected-mode-energy-bound}. Together, these statements establish
Theorem~\ref{thm:t-doped-learning-main-text}. the complete query accounting
is given in
Theorem~\ref{thm:query-complexity-t-doped-final}.\\[-0.5em]

\noindent{\bf Benchmark for learning general unitaries
}\\[-0.5em]
\noindent

The preceding results exploit structure. Without a passive-product or fixed-width subsystem promise, coherent probes and heterodyne measurements still give a valid finite-cutoff learner, but the cutoff dimension is generically exponential in the number of modes. The theorem below gives the unstructured benchmark. It combines the energy-constrained diamond norm framework for infinite-dimensional channels \cite{shirokov2018energy} with the finite-dimensional \(d^2\)-scale obstruction for unitary learning \cite{haah2023query}.

\begin{theorem}[Query complexity of general unitaries]
\label{thm:asymptotic-scaling} 
Consider a learning task of a general \(m\)-mode unitary with per-mode physical input-energy bound \(E\ge \frac12\).
Let $d_E \coloneqq
\binom{\lfloor m(E-\frac12)\rfloor+m}{m}$ be the dimension of the energy-accessible Fock subspace, and let \(d_L:=\binom{L+m}{m}\) denote the effective cutoff dimension with 
$L:=\max\left\{
K,\,
\left\lceil
\frac{256E_U(K)}{\epsilon^2}
\right\rceil
\right\}$ with $K:=\left\lceil
\frac{256m(E-\frac12)}{\epsilon^2}
\right\rceil$ and $E_U(K)
:=
\left\|
\Pi_K U^\dagger \hat n_m U\Pi_K
\right\|_\infty<\infty$.

Then, any protocol that learns arbitrary \(m\)-mode bosonic unitaries under the error $\epsilon$ in energy-constrained diamond norm must involve  
\begin{align}
M\ge \Omega\!\left(\frac{d_E^2}{\epsilon}\right)
\end{align}
channel uses. 
Assume that the cutoff-\(L\) coherent-probe reconstruction is stable. Then
there exists a learning protocol using coherent-state probes and heterodyne
detection such that
\begin{align}
M\le \mathcal O\!\left(\frac{d_L^{C}}{\epsilon^{2}}\log\!\frac{d_L}{\delta}\right)
\end{align}
channel uses suffice to achieve error at most \(\epsilon\) with success probability at least \(1-\delta\), where \(C\ge 5\) is a protocol-dependent positive number.
\end{theorem}

\emph{Proof idea.} The lower bound embeds finite-dimensional unitary learning on the \(d_E\)-dimensional energy-accessible Fock subspace. The upper bound first
uses Theorem~\ref{thm:effective-dimension-corrected} to replace the bosonic
unitary by an effective cutoff-space unitary and then reconstructs its matrix
elements by a conditioned interpolation of coherent-probe heterodyne data.
The precise stability conditions are given in \ref{sec:coh-het-unitary}. \\

\begin{acknowledgements}
The project is supported by NSF (OMA-2326746, 2350153, CCF-2240641), AFOSR MURI FA9550-24-1-0349, ONR MURI N000142612102 and an unrestricted gift from Google.
    
\end{acknowledgements}

\bibliographystyle{apsrev4-1}
\bibliography{reference}

\newpage

\begin{widetext}

\addtocontents{toc}{\protect\setcounter{tocdepth}{2}}

\clearpage
\beginsupplement

\tableofcontents

\section*{Supplemental information}

\section{Preliminary}

\subsection{Continuous-variable quantum systems}\label{subsec:Continuous-variable quantum systems}

An $m$-mode bosonic quantum system is completely specified by the statistical moments of its canonical quadrature operator vector  ${\bs r} =[ q_1,\cdots,  q_m, p_1,\cdots, p_m]^{\rm T}$, which is defined by the relations $\left\{q_j=\frac{a_j+a_j^\dag}{\sqrt 2},p_j=\frac{a_j-a_j^\dag}{\sqrt 2 i}\right\}$ in natural units $\hbar=1$ with $\{a_j\}(\{a_j^\dag\})$ being the annihilation (creation) operator of the $j$-the mode, respectively. Its first and second moments are captured by the mean vector $\bs\xi$ and the covariance matrix $V$~\cite{serafini2023quantum}:
\begin{align}
\begin{cases}
\overline{\bs r}:=&\Tr[{\bs r} \rho ]\\
V:=&\Tr [\{({\bs r}-\overline{\bs r}),({\bs r}-\overline{\bs r})^{\rm T}\} \rho],
\end{cases}
\end{align}
where $\{A,B\}=AB+BA$ denotes the anticommutator. 

The covariance matrix must fulfill the quantum uncertainty condition $V+i\varpi\ge 0$ where $\varpi=\left(\begin{matrix}
0&1\\-1&0
\end{matrix}\right)\otimes I_m$ is the symplectic form. Note that the quantum uncertainty condition is an inequality constraint and therefore does not reduce this parameter count. To completely specify the covariance matrix of an $m$-mode continuous-variable quantum system, one must estimate at least $2m^2+m$ independent real parameters. 

An arbitrary $m$-mode Gaussian unitary can be written as:
\begin{align}
U= D(\bs\xi) U_S
\end{align}
where $D(\bs\xi)$ denotes an $m$-mode displacement operator associated with a displacement $\bs\xi$, $U_S$ is a Gaussian unitary associated with an $m\times m$ symplectic transformation in phase space. After the action of the Gaussian unitary, we have the following transformation in moments:
\begin{align}
\begin{cases}
\overline{\bs r}&\to S\overline{\bs r} + \bs\xi \\
V&\to S V S^{\mathsf T} 
\end{cases}.
\end{align}

\subsection{Learning theory}

Two basic lemmas in learning theory are given as follows: 

\begin{lemma}[Median-of-means under finite variance \cite{nemirovskij1983problem,jerrum1986random}]
\label{lem:Median-of-means}
Let $X$ be a real-valued random variable with variance $\sigma^2<\infty$, and let
$X_1,\dots,X_n$ be i.i.d.\ copies of $X$. Suppose that the total sample size is written as
$n=Kn'$ with integers $K\ge 1$ and $n'\ge 1$, and partition $\{1,\dots,n\}$ into $K$
disjoint blocks $B_1,\dots,B_K$ of equal size $n'$. For each block, define the empirical mean $m_j=\frac{1}{n'}\sum_{i\in B_j}X_i,$ for $j=1,\dots,K$. Let $m_{(1)}\le \cdots \le m_{(K)}$ denote the order statistics of $(m_1,\dots,m_K)$, and define the median-of-means estimator by $\hat\mu(n,K)\coloneqq m_{(\lceil K/2\rceil)}$. If $n'\ge \frac{34\sigma^2}{\epsilon^2}$, then for every $\epsilon>0$,
\begin{align}
\Pr\!\left(\bigl|\hat\mu(n,K)-\mathbb E[X]\bigr|\ge \epsilon\right)\le 2e^{-K/2}.
\end{align}
As a consequence, for any $\delta\in(0,1)$, choosing $K\ge 2\log\!\frac{2}{\delta}$ and therefore $n=Kn'\ge \frac{68\sigma^2}{\epsilon^2}\log\!\frac{2}{\delta}$ guarantees that
\begin{align}
\Pr\!\left(\bigl|\hat\mu(n,K)-\mathbb E[X]\bigr|\ge \epsilon\right)\le \delta.
\end{align}
\end{lemma}

\begin{lemma}[Union bound for simultaneous estimation]
\label{lem:union bound}
Let $\theta_1,\dots,\theta_n$ be deterministic quantities, and let $\widehat\theta_1,\dots,\widehat\theta_n$ be estimators such that $\Pr\!\left(|\widehat\theta_j-\theta_j|>\epsilon_j\right)\le \delta_j,$ for $j\in[n]$. Then, we have 
\begin{align}
\Pr\!\left(
|\widehat\theta_j-\theta_j|\le \epsilon_j
\ \text{for all } j\in[n]
\right)
\ge
1-\sum_{j=1}^n \delta_j.
\end{align}
\end{lemma}

\subsection{\texorpdfstring{\(\epsilon\)}{epsilon}-covering net of quantum states}

The following definition will be useful for characterizing the sample complexity of state learning and the query complexity of unitary learning up to a target error.

\begin{definition}[\(\epsilon\)-covering net for a set of quantum states in trace distance]
Let \(\mathcal S\subseteq \textbf{\em St}(\mathcal H)\) be a set of quantum states, where \(\textbf{\em St}(\mathcal H)\) denotes the set of density operators on a Hilbert space \(\mathcal H\). For \(\rho,\sigma\in \textbf{\em St}(\mathcal H)\), define the trace distance by
\begin{align}
d_{\rm tr}(\rho,\sigma)
\coloneqq
\frac{1}{2}\|\rho-\sigma\|_1 .
\end{align}
A subset \(\mathcal N_\epsilon\subseteq \mathcal S\) is called an \(\epsilon\)-covering net of \(\mathcal S\) with respect to trace distance if, for every \(\rho\in \mathcal S\), there exists some \(\sigma\in \mathcal N_\epsilon\) such that
\begin{align}
d_{\rm tr}(\rho,\sigma)\le \epsilon .
\end{align}
Equivalently, we have $\mathcal S \subseteq \bigcup_{\sigma\in \mathcal N_\epsilon}
\left\{
\rho\in \mathsf D(\mathcal H)\,\middle|\, d_{\rm tr}(\rho,\sigma)\le \epsilon
\right\}$. The minimum cardinality of such a net is called the \(\epsilon\)-covering number of \(\mathcal S\) in trace distance. 
\end{definition}

In addition to mapping all states of interest into $\epsilon$-balls, the following definition will also be useful for verifying whether a state is Gaussian: 
\begin{definition}[\(\epsilon\)-close Gaussian state]
Let
\(\mathsf G_1\subseteq \textbf{\em St}(\mathcal H)\) denote the set of all single-mode Gaussian states, where \(\mathcal H\) be the Hilbert space of a single bosonic mode. A single-mode quantum state \(\rho\) is \(\epsilon\)-close Gaussian if its trace-distance to the set of single-mode Gaussian states is at most \(\epsilon\), namely,
\begin{align}
\inf_{\sigma\in \mathsf G_1}\frac{1}{2}\|\rho-\sigma\|_1 \le \epsilon .
\end{align}
\end{definition}

\section{Learning of Gaussian-entanglable unitaries}\label{supp:Learning of Gaussian entanglable unitaries}

This Supplemental Note presents an efficient protocol for learning bosonic unitaries composed of an initial passive-Gaussian layer, followed by a layer of single-mode unitaries that may be non-Gaussian, and finally a general Gaussian unitary layer. The overall protocol only uses Gaussian unitary, coherent state probing, and local heterodyne measurement. In particular, we first show that it is possible to learn and counter-rotate the last Gaussian layer into a passive-Gaussian layer by covariance matrix estimation and physical action of Gaussian unitaries. In this way, the learning problem reduces to learning a passive-separable unitary. We then introduce a quantum Darmois–Skitovich theorem. Using this theorem, we show how to identify the passive-Gaussian layers from learned Gaussian-entangleable states generated by coherent-state probes. Furthermore, we provide a theorem analyzing the possibility of activation failure in the middle layer, namely when the observed output state does not certify the non-Gaussianity of that layer. After the classical counter-rotation of the learned second passive layer, we additionally provide a unitary-learning protocol with unknown coherent states. Finally, we implement a standard learning protocol for the left Gaussian unitary. At the end of this Supplemental Note, we summarize the protocol in detail and derive the corresponding query-complexity bounds.

\subsection{Learning task}\label{supp:learning task of ge unitary}

Let us first describe the set-up and formulate the learning task explicitly. Given the fact that any multimode Gaussian unitary transforms a layer of displacement operators to displacement operators, let us move all the displacement operation to the middle layer. Without loss of generality, we assume to be given repeated access to the following unitary
\begin{align}\label{def:U_three layer with S}
U_{\rm ge} = U_{S} \left(\bigotimes_{j=1}^m W_j\right) U_{O},
\end{align}
where $U_{S}$ is an $m$-mode  Gaussian unitary corresponding to symplectic matrix $S$, $U_{O}$ is an $m$-mode passive-Gaussian unitary corresponding to symplectic-orthogonal matrix $O$, and $\{W_j\}_{j=1}^m$ are unknown single-mode unitaries. To make the analysis clear, we first assume the following constraints on energy and its second moment
\begin{align}\label{supp:three energy constraints}
\begin{cases}
\Tr\left\{\left[ \left(\sum_{j=1}^m a_j^\dag a_j+\frac m 2\right)\right] |\Phi_{\rm in}\>\<\Phi_{\rm in}|\right\}\le & mE_{\rm probe}\\[0.7em]
\sqrt{\Tr\left\{\left[ \left(\sum_{j=1}^m a_j^\dag a_j+\frac m 2 \right)^2\right]|\Psi_{\rm out}\>\<\Psi_{\rm out}|\right\}}\le & mE''
\end{cases},
\end{align}
where $|\Psi_{\rm out}\>=U_{\rm ge}|\Phi_{\rm in}\>$ refers to the output state. We will later establish a query complexity theorem that relies only on these two conditions. In general, we may prepare arbitrary probe states $\{|\Phi_{\rm in}\>\}$ and perform general measurements on the output. In the next subsections, we demonstrate an efficient protocol that is based on coherent state probing and local heterodyne measurements. 

The learning task is to construct an estimate $\widetilde U_{\rm ge}$ such that  
\begin{align}
\|\mathcal U_{\rm ge} - \widetilde{\mathcal U}_{\rm ge}\|_\diamond^{mE} \le \epsilon,
\end{align}
where $\map U(\cdot)=U\cdot U^\dag$ denotes the associated channel of a unitary $U$, 
\begin{align}\label{def:diamond norm}
\|\mathcal V\|_{\diamond}^{mE}:=
\sup_{\Tr\{[ \left(\sum_{j=1}^m a_j^\dag a_j+\frac m 2 \right)\otimes I]\Psi\}\le mE}\ \big\|(\mathcal V\otimes I)(\Psi)\big\|_1
\end{align}
represents the energy constrained diamond norm, \(E\) is the per-mode evaluation-energy bound in the
energy-constrained diamond norm, whereas \(E_{\rm probe}\) bounds the physical
energy of the coherent probes queried by the protocol, $\epsilon$ is a small number representing the learning accuracy. In the meantime, we need to guarantee that the success probability is larger than $1-\delta$, where $\delta$ is a small constant.  

In the following discussion, we adopt a perturbative error-propagation assumption. Specifically, we neglect the coupling between estimation errors arising in different steps of the protocol and treat the error contributed by each layer independently. Equivalently, when analyzing the reconstruction error at a given step, we assume that all preceding steps have been carried out exactly. Under this first-order approximation, the total reconstruction error is taken to be the sum of the errors contributed by the individual steps.

\subsection{Learning the general Gaussian layer and physical counter-rotation}\label{supp:Learning of the general Gaussian layer and physical counter-rotation}

Let us first look at the learning protocol of the final layer $U_S$ in Eq. (\ref{def:U_three layer with S}). Without loss of generality, we consider coherent-state probes, since they are known to be informationally complete for quantum process tomography \cite{rahimi2011quantum}. Moreover, with a random coherent probe state 
$|\underline{\bs\alpha}\rangle \coloneqq \bigotimes_{j=1}^m |\underline{\alpha_j}\rangle$, the unitary $U_{\rm ge}$ produces a Gaussian-entanglable (GE) state \cite{zhao2025complexity}. Indeed, the first passive-Gaussian layer $U_O$ maps coherent states to coherent states and therefore preserves the product structure. The intermediate layer, consisting of single-mode unitaries, also preserves separability across modes. Consequently, after the final Gaussian layer $U_S$, the output state takes the form
\begin{align}\label{eqs9:ge state}
|\Psi_{\bs\alpha}\>&= U_S \bigotimes_{j=1}^m |\psi_{j,\bs \alpha}\>,
\end{align}
where $\{|\psi_{j,\bs \alpha}\>\}_{j=1}^m$ are single-mode states that depend on the input amplitude $\bs\alpha$.

Therefore, we can apply the Gaussian-disentangling protocol introduced in Ref.~\cite{zhao2025complexity} and reduce the learning problem to that of learning a simpler state.  In particular, it consists of the following steps:  
\begin{enumerate}
\item [(GD1)] Estimate the covariance matrix $V_{\bs\alpha}$ of the GE state $|\Psi_{\bs\alpha}\>$ via heterodyne measurements.
\item [(GD2)]
Once the covariance matrix $\widetilde V_{\bs\alpha}$ is reconstructed, it is possible to calculate the symplectic matrix $\widetilde S_{\bs\alpha}$ that fulfills the requirement $\widetilde V_{\bs\alpha}= \widetilde S_{\bs\alpha}\left(\widetilde \Lambda_{\bs\alpha}\otimes I^{(2)}\right)\widetilde S_{\bs\alpha}^T$ using eigendecomposition  \cite{serafini2023quantum}:  $i\Omega \widetilde V_{\bs\alpha}=\widetilde W_{\bs\alpha}\left[\widetilde \Lambda_{\bs\alpha} \otimes\left(\begin{matrix}
-1&0\\
0&1
\end{matrix}\right)\right] \widetilde W_{\bs\alpha}^{-1}$ and $\widetilde S_{\bs\alpha}=\Omega^T \widetilde W_{\bs\alpha} \left[I \otimes \frac{1}{\sqrt 2}\left(\begin{matrix}
1&-i\\
1&i
\end{matrix}\right)\right]\Omega$  where $\widetilde \Lambda_{\bs\alpha} =\text{diag}(\widetilde\lambda_1,\cdots,\widetilde\lambda_m)$ is a $m\times m$ diagonal matrix fulfilling the condition $\det (i\Omega\widetilde V_{\bs\alpha} \pm \widetilde \lambda_j I )=0$, $\widetilde W_{\bs\alpha}=[\bs x_{1,+},\bs x_{1,-}\cdots,\bs x_{m,+},\bs x_{m,-}]$ is the eigenvector matrix of $i\Omega \widetilde V_{\bs\alpha}$ from the solution of $(i\Omega \widetilde V_{\bs\alpha} \pm \widetilde \lambda_j I)\bs x_{j,\pm}=0$. On the other hand, the true value of the covariance matrix of the pure GE state $|\Psi_{\bs\alpha}\>$  is $V_{\bs\alpha}= S_{\bs\alpha}
\left(\Lambda_{\bs\alpha}\otimes  I^{(2)}\right) 
S^T_{\bs\alpha}$ where $ \Lambda_{\bs\alpha} =\text{diag}(\lambda_1,\cdots,\lambda_m)$ is a $m\times m$ diagonal matrix, 
\begin{align}
S_{\bs\alpha}=S \left(\bigoplus_j S_{j,{\bs\alpha}}\right)
\end{align}
is associated with the true symplectic matrix $S$, where $\{S_{j,{\bs\alpha}}\}$ denote single-mode symplectic matrices. Therefore, one can use the estimated symplectic matrix $\widetilde S_{\bs\alpha}^{-1}$ as an approximation of $S_{\bs\alpha}^{-1}$ up to an unknown symplectic-orthogonal operation due to symplectic eigenvalue degeneracy and permutation \cite{zhao2025complexity}.
\item [(GD3)] Physically apply the unitary $U_{\widetilde S_{\bs \alpha}}^\dag$. 
\end{enumerate}

Note that a displacement operator may be placed either before or after the Gaussian unitary, since for any \(\bs\gamma'\) there exists \(\bs\gamma\) such that
\(
U_S D(\bs\gamma') = D(\bs\gamma) U_S.
\) For the present purpose, this distinction is irrelevant. Indeed, the reconstruction of the counter-rotation \(U_{\widetilde S_{\bs\alpha}}^\dag\) is based only on the covariance matrix \(V_{\bs\alpha}\), which is invariant under displacements. Although the mean vector does depend on the displacement and is therefore associated with the particular representation of the state in which the displacement acts after \(U_S\), this affects only the first moments and not the Gaussian unitary extracted from \(V_{\bs\alpha}\). Therefore, there is no need to counter-rotate the displacement before applying \(U_{\widetilde S_{\bs\alpha}}^\dag\).

More importantly, local states $\{|\psi_{j,\bs\alpha}\>\}$ in Eq.~(\ref{eqs9:ge state}) do not in general determine a general unitary acting on them. It only probes the effective action of $U_S$ on the particular input product state, so part of the original multimode information may be inaccessible. Consequently, the reconstructed Gaussian unitary $U_{\widetilde S_{\bs\alpha}}^\dag$ may depend on the chosen probe $|\underline{\bs\alpha}\>$ and need not equal the original $U_S$. The real question is therefore not whether one recovers the same Gaussian unitary for every probe, but 
\begin{quote}
\emph{``whether the reconstructed unitary still serves as a valid disentangler when the probe state is changed from $|\underline{\bs\alpha}\>$ to another coherent product state $|\underline{\bs\alpha'}\>$'', }
\end{quote}
as it is required for the further tomography protocol for local unitaries $\{W_j\}$. Specifically, this change produces a different family of local states $\{|\psi_{j,\bs\alpha'}\>\}$ in Eq.~(\ref{eqs9:ge state}). The relevant issue is then whether steps (GD1)--(GD3) still work for this new probe, namely, whether the physically reconstructed Gaussian unitary $U_{\widetilde S_{\bs\alpha}}^\dag$ maps the corresponding output state $|\Psi_{\bs\alpha'}\>$ to a passive-separable state, as in Ref.~\cite{zhao2025complexity}. Otherwise, applying the Gaussian-disentangling protocol would be of no use.

Here, it is straightforward to have the following lemma: 

\begin{lemma}[Gaussian-disentangling for different input]\label{lem:Gaussian-disentangling for different input}
Assume precise estimation in the learning protocol (GD1)-(GD3), the reconstructed Gaussian unitary $U_{\widetilde S_{\bs\alpha}}^\dag$ is able to partially disentangle the effect of $U_S$ in the state $|\Psi_{\bs\alpha'}\>= U_S \bigotimes_{j=1}^m |\psi_{j,\bs \alpha'}\>$ with an alternative input $|\underline{\bs\alpha'}\>$, i.e., $U_{\widetilde S_{\bs\alpha}}^\dag U_S\simeq  U_{ O_{2,{\bs\alpha}}} \left(\bigotimes_{j=1}^m U_{ S_{j,{\bs\alpha}}}^\dag\right)$. After applying $U_{\widetilde S_{\bs \alpha}}^\dag$, the state $|\Psi_{\bs \alpha'}\>$ is always mapped to a passive-separable state 
\begin{align}
|\Psi_{\bs\alpha,\bs\alpha'}^{\rm ps}\>&= U_{\widetilde S_{\bs \alpha}}^\dag |\Psi_{\bs\alpha'}\>\simeq U_{ O_{2,{\bs\alpha}}} \left(\bigotimes_{j=1}^m U_{ S_{j,{\bs \alpha}}}^\dag \right)\bigotimes_{k=1}^m |\psi_{k,\bs \alpha'}\>,\label{eqs10}
\end{align}
where $U_{O_{2,{\bs\alpha}}}$ is an $m$-mode passive-Gaussian unitary linked to the degeneracy of the symplectic eigenvalues of local states $\{|\psi_{j,{\bs\alpha}}\>\}$ and a unknown permutation sorting the symplectic eigenvalues that depend on $\bs\alpha$, $\left\{U_{ S_{j,{\bs \alpha}}}\right\}$ are local Gaussian unitaries with $S_{j,{\bs \alpha}}^{-1}$ diagonalizing the covariance matrix of $\{|\psi_{j,{\bs\alpha}}\>\}$, $\bigotimes_{j=1}^m|\psi_{j,{\bs\alpha'}}\>= \left(\bigotimes_{j=1}^m W_j\right)U_O|\underline{\bs\alpha}\>$ is the output of the second layer of $U_{\rm ge}$. 
\end{lemma}

\begin{proof}
If we prepare a coherent state  $|\underline{\bs\alpha}\>$ to probe $U_{\rm ge}$ defined in Eq. (\ref{def:U_three layer with S}), the covariance matrix takes the form 
\begin{align}
V_{\bs\alpha}=S \left(\bigoplus_{j=1}^m \Gamma_{j,\bs\alpha}\right)S^T 
\end{align}
where $\bigoplus_{j=1}^m\Gamma_{j,\bs\alpha}=\bigoplus_{j=1}^m\lambda_j S_{j,\bs\alpha}S_{j,\bs\alpha}^T$ denotes the covariance matrix of the state $\left(\bigotimes_{j=1}^m W_j\right)U_O|\underline{\bs\alpha}\>$. By applying Step~(GD2) together with Theorem~8.11 and Proposition~8.12 of Ref.~\cite{de2006symplectic}, one can reconstruct a symplectic matrix \(\widetilde S_{\bs\alpha}\) such that
\begin{align}
V_{\bs\alpha}=&\widetilde S_{\bs\alpha}\left(\text{diag}(\widetilde \lambda_1,\cdots,\widetilde \lambda_m)\otimes I^{(2)}\right)\widetilde S_{\bs\alpha}^T\\
\widetilde S_{\bs\alpha}=& S\left(\bigoplus_{j=1}^m S_{j,{\bs\alpha}}\right) O_{2,\bs\alpha}^T,
\end{align}
where \(O_{2,\bs\alpha}\) denotes the symplectic-orthogonal matrix associated with the reordering of the symplectic eigenvalues, local phase rotations, and mode mixing among modes with degenerate symplectic eigenvalues. 

When we probe $U_{\rm ge}$ with another input state $|\underline{\bs\alpha'}\>$, we can apply $U_{\widetilde S_{\bs\alpha}}^\dag$ to have an output state
\begin{align}
U_{\widetilde S_{\bs\alpha}}^\dag U_{\rm ge}|\underline{\bs\alpha'}\>= U_{ O_{2,{\bs\alpha}}} \left(\bigotimes_{j=1}^m U_{ S_{j,{\bs \alpha}}}^\dag \right)\left(\bigotimes_{k=1}^m W_k\right)U_O |\underline{\bs\alpha'}\>.
\end{align}
Therefore, Eq.~(\ref{eqs10}) follows, which complets the proof of Lemma~\ref{lem:Gaussian-disentangling for different input}. \end{proof}

Next, a natural question is how many distinct coherent probes are needed to determine the complexity of the passive-Gaussian unitary $U_{O_{2,\bs\alpha}}$. Let us change the probe notation back to $|\underline{\bs\alpha}\>$. As discussed in Ref.~\cite{zhao2025complexity}, $U_{O_{2,\bs\alpha}}$ can be trivial, namely equal to a permutation, if the local states $\{|\psi_{j,\bs\alpha}\rangle\}$ generated by the probe $|\underline{\bs\alpha}\rangle$ have distinct symplectic eigenvalues. In that case, no nontrivial passive-Gaussian unitary remains, and the learning task reduces directly to the learning of local states, which has polynomial sample complexity in the number of modes.

Since the local states $\{|\psi_{j,\bs\alpha}\rangle\}$ in Eq.~(\ref{eqs9:ge state}) are obtained by applying the single-mode unitaries $\{W_j\}$ to coherent-state inputs, the following lemma will be useful. It will be useful to clarify how the structure of $U_{O_{2,\bs\alpha}}$ depends on the choice of the input $\bs\alpha$.

\begin{lemma}[A single random coherent state suffices to test the coincidence of the single-mode symplectic eigenvalue]
\label{prop:single-mode-symplectic-zero-measure-energy}
Let $\Omega\subset\mathbb C$ be a nonempty connected open set, and let $|\underline{\alpha}\>$ be the single-mode coherent state with amplitude $\alpha$. Given two single-mode unitaries $W_1$ and $W_2$, define for each $\alpha\in\Omega$ $\rho_j(\alpha)\coloneqq W_j\ket{\underline{\alpha}}\!\bra{\underline{\alpha}}W_j^\dagger, (j\in\{1,2\})$. For each $j$, let $V_j(\alpha)$ denote the covariance matrix of $\rho_j(\alpha)$, and let $\nu_j(\alpha)$ be its unique symplectic eigenvalue. Assume that the output energy is finite, i.e., $\Tr\!\bigl[\rho_j(\alpha)\,a^\dagger a\bigr]< E' ,$ $
\text{for all }\alpha\in\Omega, (j=1,2)$.

Define the set of input amtitudes as follows
\begin{align}
\mathcal E
\coloneqq
\bigl\{
\alpha\in\Omega \,\big|\, \nu_1(\alpha)=\nu_2(\alpha)
\bigr\}.
\end{align}
Then, exactly one of the following alternatives holds
\begin{enumerate}
\item $\mathcal E=\Omega$
\item $\mathcal E$ has two-dimensional Lebesgue measure zero in $\Omega$
\end{enumerate}
In particular, if $\nu_1(\alpha)\not\equiv \nu_2(\alpha)$ on $\Omega$, then $\mathcal E$ has Lebesgue measure zero. In physical terms, if the two output families do not have identically equal symplectic eigenvalue on $\Omega$, then the coherent input states that accidentally produce the same symplectic eigenvalue form a set of area zero in $\Omega$. Hence, if $\alpha$ is drawn from any probability distribution on $\Omega$ that is absolutely continuous with respect to Lebesgue measure, then
\begin{align}
\Pr_\alpha\!\bigl[\nu_1(\alpha)=\nu_2(\alpha)\bigr]=0,
\qquad
\Pr_\alpha\!\bigl[\nu_1(\alpha)\neq \nu_2(\alpha)\bigr]=1.
\end{align}
\end{lemma}

\begin{proof}
Let us define the unnormalized coherent state $\|\alpha\rangle \coloneqq e^{\alpha a^\dagger}\ket{0}
=
\sum_{n\ge 0}\frac{\alpha^n}{\sqrt{n!}}\ket{n}$, and the associated output state of $W_j$: 
$|\Phi_j(\alpha)\>\coloneqq W_j\|\alpha\rangle,$ for $j\in\{1,2\}$. The Fock expansion of $\|\alpha\rangle$ has infinite radius of convergence in the Hilbert-space norm, for example by the ratio test. So $\alpha\mapsto \|\alpha\rangle$ is entire. Since $W_j$ is bounded, the same is true for $\alpha\longmapsto |\Phi_j(\alpha)\>$ \cite{Dineen1999InfiniteDimHolomorphy}.

Next, the assumed output-energy bound implies $\|\sqrt{a^\dag a +I}|\Phi_j(\alpha)\>\|^2
=
\||\Phi_j(\alpha)\>\|^2+\|(a^\dag a )^{1/2}|\Phi_j(\alpha)\>\|^2
\le
(1+E')e^{|\alpha|^2}$. Hence $\sqrt{a^\dag a +I}|\Phi_j\>$ is locally bounded on $\Omega$. Since $\sqrt{a^\dag a+I}$ is self-adjoint, hence closed, and $|\Phi_j\>$ is holomorphic as a Hilbert-space-valued map, by a standard result on Banach-valued holomorphic maps and closed operators \cite{Dineen1999InfiniteDimHolomorphy}, $\alpha\longmapsto \sqrt{a^\dag a +I}|\Phi_j(\alpha)\>$ is also holomorphic on $\Omega$. Now the quadrature operators $q$ and $p$ are $\sqrt{a^\dag a +I}$-bounded. Indeed, for every $|\psi\>\in D(\sqrt{a^\dag a +I})$,
we have $\|q\psi\|^2+\|p\psi\|^2
=
\langle \psi|(q^2+p^2)|\psi\rangle
=
\langle \psi,(2a^\dag a+I)|\psi\>\rangle
\le
2\|\sqrt{a^\dag a +I}|\psi\>\|^2$. Therefore, $\alpha\longmapsto q\Phi_j(\alpha)$ and $\alpha\longmapsto p\Phi_j(\alpha)$
are holomorphic as Hilbert-space-valued maps on $\Omega$. Moreover, it follows that the scalar functions
\begin{align}
m_{q,j}(\alpha)
&\coloneqq
e^{-|\alpha|^2}\langle \Phi_j(\alpha),q\Phi_j(\alpha)\rangle,\\
m_{p,j}(\alpha)
&\coloneqq
e^{-|\alpha|^2}\langle \Phi_j(\alpha),p\Phi_j(\alpha)\rangle,\\
s_{qq,j}(\alpha)
&\coloneqq
e^{-|\alpha|^2}\langle q\Phi_j(\alpha),q\Phi_j(\alpha)\rangle,\\
s_{pp,j}(\alpha)
&\coloneqq
e^{-|\alpha|^2}\langle p\Phi_j(\alpha),p\Phi_j(\alpha)\rangle,\\
s_{qp,j}(\alpha)
&\coloneqq
e^{-|\alpha|^2}\Re\langle q\Phi_j(\alpha),p\Phi_j(\alpha)\rangle
\end{align}
are real analytic on $\Omega$.  Hence every entry of the covariance matrix $V_j(\alpha)$ is real analytic on $\Omega$. Therefore, the function $\det V_j(\alpha)$ is real analytic on $\Omega$ as well.

For a single-mode covariance matrix, the unique symplectic eigenvalue satisfies
\begin{align}
\nu_j(\alpha)^2=\det V_j(\alpha).
\end{align}
Since $\nu_j(\alpha)\ge 0$, we have
$\nu_1(\alpha)=\nu_2(\alpha)
\iff
\det V_1(\alpha)=\det V_2(\alpha)$. Thus, if we define
\begin{align}
F(\alpha)\coloneqq \det V_1(\alpha)-\det V_2(\alpha),
\end{align}
then $F$ is real analytic on $\Omega$ and
\begin{align}
\mathcal E=\{\alpha\in\Omega\mid F(\alpha)=0\}.
\end{align}

Since $\Omega\subset\mathbb C\simeq\mathbb R^2$ is connected and open, the standard zero-set theorem for real-analytic functions \cite{KrantzParks2002} implies that either $F\equiv 0$ on $\Omega$, or else its zero set has two-dimensional Lebesgue measure zero. If $F\equiv 0$, then $\nu_1(\alpha)=\nu_2(\alpha)$ for all $\alpha\in\Omega$, hence $\mathcal E=\Omega$. Otherwise, $\mathcal E$ has Lebesgue measure zero.

For the final statement, if $\nu_1(\alpha)\not\equiv \nu_2(\alpha)$ on $\Omega$, then $F\not\equiv 0$, so the zero set $\mathcal E$ is Lebesgue-null. Therefore, for any probability distribution on $\Omega$ that is absolutely continuous with respect to Lebesgue measure,
\begin{align}
\Pr_\alpha\!\bigl[\nu_1(\alpha)=\nu_2(\alpha)\bigr]
=
\Pr_\alpha[\alpha\in\mathcal E]
=
0,
\end{align}
and hence
\begin{align}
\Pr_\alpha\!\bigl[\nu_1(\alpha)\neq \nu_2(\alpha)\bigr]=1.
\end{align}
This proves the proposition.
\end{proof}

Following from  Lemma~\ref{prop:single-mode-symplectic-zero-measure-energy}, with probability one each pair $(W_j,W_k)$ defined in Eq. (\ref{def:U_three layer with S}) is either intrinsically indistinguishable at the level of the single-mode symplectic eigenvalue for all probes $|\underline{\bs \alpha}\>,(\bs\alpha\in\C^m)$, or distinguished by the sampled probe. Therefore, with probability one over the choice of the coherent probe $\bs\alpha$, the left passive-Gaussian unitary $U_{O_{2,\bs\alpha}}$ in Eq.~(\ref{eqs10}) acts on the same mode groups for different $\bs\alpha$, because the modes sharing the same symplectic eigenvalue are probe-independent almost surely. If the symplectic eigenvalues of $V_{\bs \alpha}$ are different from each other for a random input $\bs\alpha$, we have probability one that $U_{O_{2,\bs\alpha}}$ is just a permutation operation combined with phase rotations. 

Now, let us look at the reconstruction error of $U_{\widetilde S_{\bs\alpha}}$ in diamond norm. After sufficiently many rounds of measurement, the covariance matrix \(V_{\bs\alpha}\) can be estimated accurately. By Eq.~(S160) of \cite{zhao2025complexity}, one can then reconstruct \(U_{\widetilde S_{\bs\alpha}}\) such that the corresponding symplectic matrix \(\widetilde S_{\bs\alpha}\) satisfies
\begin{align}
\|S_{\bs\alpha}'-\widetilde S_{\bs\alpha}\|_F \le C,
\end{align}
where \(C=\mathcal O(m^{5/2},E'',\Delta^{-1},\sqrt{\|\widetilde V_{\bs\alpha}-V_{\bs\alpha}\|_\infty})\) is a constant independent of the gauge structure with $\Delta$ being the minimal non-zero eigengap, \({S_{\bs\alpha}'}\) is the unitary to which \({\widetilde S_{\bs\alpha}}\) converges asymptotically, and \(\|A\|_F \coloneqq \sqrt{\operatorname{Tr}(A^\dagger A)}\) denotes the Frobenius norm of \(A\), $\|\cdot\|_\infty$ denotes the operator norm. Here we use the fact that the output energy per mode is bounded by $E''$, due to the Cauchy-Schwarz inequality. Then, using the Gaussian Solovay-Kitaev theorem \cite{becker2021energy,zhao2025complexity}, the error of reconstructing the Gaussian unitary $U_{\widetilde S_{\bs\alpha}}$ as follows: 
\begin{align}
\left\|\map U_{ S'_{\bs\alpha}}-\map U_{\widetilde S_{\bs\alpha}}\right\|_\diamond^{mE''}&\le \textbf{poly}\left(\mathcal O\left(m^{2},E^{''5/4},\sqrt{\| S_{\bs\alpha}'-\widetilde S_{\bs\alpha}\|_F}\right)\right)\label{eqs29}\\
&=\epsilon_{\rm gd}.
\end{align} 
See Eqs.~(S121) and (S125) of \cite{zhao2025complexity} for the detailed form of Eq.~(\ref{eqs29}).

\subsection{Learning of the second passive-Gaussian layer}\label{supp:Learning of the second passive-Gaussian layer}

After the learning and physical counter-rotation steps (GD1)--(GD3), and using any other probe coherent states $|\underline{\bs\alpha'}\>$, the learning problem reduces to that of learning the following passive-Gaussian-entanglable unitary
\begin{align}\label{def:U_three layer_PGE0}
U_{\rm pge} = U_{ O_{2,{\bs \alpha}}} \left(\bigotimes_{j=1}^m U_{ S_{j,{\bs \alpha}}}^\dag \right)\left(\bigotimes_{j=1}^m W_j\right) U_{O},
\end{align}
where $U_{O_{2,{\bs \alpha}}}$ is defined in Eq. (\ref{eqs10}), $U_{ S_{j,{\bs \alpha}}}$ are introduced in step (GD2), $U_O$ is defined for $U_{\rm ge}$ in Eq. (\ref{def:U_three layer with S}). Within this subsection, for simplicity, let us temporarily fix $\bs\alpha$ and denote
\begin{align}\label{def:U_three layer_PGE}
U_{\rm pge} = U_{O_{2}} \left(\bigotimes_{j=1}^m \widetilde W_j\right) U_{O_1}.
\end{align}
We will use back the complete notation of Eq. (\ref{def:U_three layer_PGE}) in the summary of the overall protocol in \ref{supp:final protocol of GE layer}.

Given a coherent probe state \( |\underline{\bs\alpha'}\rangle \), the output passive-separable state $|\Psi_{\bs\alpha,\bs\alpha'}^{\rm ps}\>$ of $U_{\rm pge}$ can be efficiently reconstructed using local heterodyne measurements and polynomially many measurement rounds, following the shadow tomography protocol of Ref.~\cite{zhao2025complexity}. Nevertheless, this protocol generally returns multiple solutions for \(U_{O_2}\) that all produce the same overall passive-separable state $|\Psi_{\bs\alpha,\bs\alpha'}^{\rm ps}\>$. The problem of uniquely reconstructing the description of $U_{O_2}$ remains to be solved.

Before showing the algorithm that learns the unique representation of $U_{O_2}$ up to a gauge structure, let us first look at a useful proposition, which is the complete version of Theorem~\ref{prop:DS-passive-main} in the main text, that will play an important role in the subsequent discussion.

\begin{proposition}[the quantum Darmois--Skitovich theorem for multimode general passive layer]
\label{prop:DS-passive-general}
Let $U_O$ be an $m$-mode passive-Gaussian unitary associated with a symplectic-orthogonal matrix 
$O\in \mathrm O(2m)\cap \mathrm{Sp}(2m,\mathbb R)$.
Let $U\in \mathrm U(m)$ be the associated mode-mixing matrix on annihilation operators,
\begin{align}
U_O^\dagger \bm a\,U_O = U\,\bm a,
\qquad
\bm a=(a_1,\dots,a_m)^{\mathsf T},
\label{UO_U}
\end{align}
so that in quadratures one has
$O=\begin{psmallmatrix}\Re U&-\Im U\\ \Im U&\Re U\end{psmallmatrix}$.
Assume the following mixing condition
for every input index $j$ there exist two distinct output indices $k\neq \ell$ such that
$U_{k j}\neq 0$ and $U_{\ell j}\neq 0$.

Let $\rho_{\rm in}=\bigotimes_{j=1}^m \rho_j$ be an arbitrary product input and
$\rho_{\rm out}=U_O\rho_{\rm in}U_O^\dagger$.
If $\rho_{\rm out}$ is a product across the output modes,
$\rho_{\rm out}=\bigotimes_{k=1}^m \rho_k'$,
then each $\rho_j$ is a single-mode Gaussian state.
\end{proposition}

\begin{proof}
Let us first prove that heterodyne measurement gives a classical density and its independence for product inputs. For one mode case, the positive operator-valued measure (POVM) for the heterodyne measurement is $E(\alpha)\coloneqq \frac{1}{\pi}\,|\underline{\alpha}\rangle\langle\underline{\alpha}|,(\alpha\in\mathbb C)$, so that for any single-mode state $\rho$ the outcome density equals the Husimi-$Q$ function
\begin{align}
p_\rho(\alpha)=\mathrm{Tr}\!\bigl[\rho\,E(\alpha)\bigr]
=\frac{1}{\pi}\langle\underline{\alpha}|\rho|\underline{\alpha}\rangle
\eqqcolon Q_\rho(\alpha).
\end{align}
For $m$ modes, we use the product POVM $E(\bm\alpha)\coloneqq \bigotimes_{j=1}^m E(\alpha_j),$ with $\bm\alpha=(\alpha_1,\dots,\alpha_m)\in\mathbb C^m$, and the corresponding density is $p_\rho(\bm\alpha)=\mathrm{Tr}[\rho\,E(\bm\alpha)]\eqqcolon Q_\rho(\bm\alpha)$.

If the input is a product state $\rho_{\rm in}=\bigotimes_{j=1}^m \rho_j$, then for every $\bm\alpha\in\mathbb C^m$,
\begin{align}
Q_{\rho_{\rm in}}(\bm\alpha)
&=\mathrm{Tr}\!\left[\left(\bigotimes_{j=1}^m\rho_j\right)\left(\bigotimes_{j=1}^m E(\alpha_j)\right)\right] \nonumber\\
&=\prod_{j=1}^m \mathrm{Tr}\!\bigl[\rho_j E(\alpha_j)\bigr]
=\prod_{j=1}^m Q_{\rho_j}(\alpha_j).
\label{eq:Q-factor-in-proof}
\end{align}
Therefore, if we denote the heterodyne outcomes on the input modes by random variables
$A_1,\dots,A_m\in\mathbb C$ with joint density $Q_{\rho_{\rm in}}(\bm\alpha)$, then $A_1,\dots,A_m$ are independent
and each $A_j$ has density $Q_{\rho_j}$.

It is convenient to identify $\alpha=(x+ip)/\sqrt 2$ with $(x,p)\in\mathbb R^2$. We therefore define random vectors
\begin{align}
Z_j\in\mathbb R^2 \quad\text{as the heterodyne outcome on input mode $j$},
\end{align}
so $Z_1,\dots,Z_m$ are independent and have densities $Q_{\rho_j}$ under this identification.
Similarly, let
\begin{align}
Y_k\in\mathbb R^2 \quad\text{be the heterodyne outcome on output mode $k$},
\end{align}
so the joint density of $(Y_1,\dots,Y_m)$ is $Q_{\rho_{\rm out}}$.

Now, let us look at how passive linear optics acts linearly on heterodyne outcomes. Because $U_O$ is passive, there exists $U\in\mathrm U(m)$ such that
$U_O^\dagger \bm a\,U_O = U\,\bm a$ for $\bm a=(a_1,\dots,a_m)^{\mathsf T}$, and coherent states transform as
\begin{align}
U_O|\underline{\bm\alpha}\rangle = |\underline{U\bm\alpha}\rangle,
\qquad |\underline{\bm\alpha}\rangle=\bigotimes_{j=1}^m|\underline{\alpha_j}\rangle.
\end{align}
Therefore, for $\rho_{\rm out}=U_O\rho_{\rm in}U_O^\dagger$, we have: 
\begin{align}
Q_{\rho_{\rm out}}(\bm\alpha)
&=\frac{1}{\pi^m}\,\langle \underline{\bm\alpha}|U_O\rho_{\rm in}U_O^\dagger|\underline{\bm\alpha}\rangle \nonumber\\
&=\frac{1}{\pi^m}\,\langle \underline{U^\dagger\bm\alpha}|\rho_{\rm in}|\underline{U^\dagger\bm\alpha}\rangle
=Q_{\rho_{\rm in}}(U^\dagger\bm\alpha).
\label{eq:Q-covariance-proof}
\end{align}

Write the real embedding of $U$ as
\begin{align}
O=\begin{pmatrix}\Re U&-\Im U\\ \Im U&\Re U\end{pmatrix}\in \mathrm O(2m)\cap\mathrm{Sp}(2m,\mathbb R),
\end{align}
so that in real coordinates, stacking $Z=(Z_1,\dots,Z_m)\in\mathbb R^{2m}$ and
$Y=(Y_1,\dots,Y_m)\in\mathbb R^{2m}$, the identity \eqref{eq:Q-covariance-proof} is equivalently the following form
\begin{equation}\label{eq:Y=OZ-in-law}
Y = O\,Z .
\end{equation}

Next, we can show that the output product state implies independence of two linear forms. Assume $\rho_{\rm out}=\bigotimes_{k=1}^m \rho_k'$. Repeating \eqref{eq:Q-factor-in-proof} on the output gives
\begin{equation}\label{eq:Q-factor-out-proof}
Q_{\rho_{\rm out}}(\bm\alpha)=\prod_{k=1}^m Q_{\rho_k'}(\alpha_k),
\end{equation}
hence $Y_1,\dots,Y_m$ are independent random vectors in $\mathbb R^2$.

Fix two distinct outputs $k\neq \ell$. Since $Y_k$ and $Y_\ell$ are independent and
$(Y_1,\dots,Y_m)$ has the same distribution as $O(Z_1,\dots,Z_m)$ by \eqref{eq:Y=OZ-in-law},
we can regard $Y_k$ and $Y_\ell$ as two linear forms in the independent inputs $Z_1,\dots,Z_m$. To write these linear forms explicitly, use the complex form.
For $c=a+ib\in\mathbb C$, let
\begin{align}
M(c)\coloneqq
\begin{pmatrix}a&-b\\ b&a\end{pmatrix}\in\mathbb R^{2\times 2},
\qquad
\det M(c)=|c|^2.
\end{align}
Then multiplication by $c$ on $\mathbb C\cong\mathbb R^2$ corresponds to $M(c)$ on $\mathbb R^2$.
Writing $U_{k j}$ for the entries of $U$, the $k$th output heterodyne vector has the form
\begin{equation}\label{eq:linear-forms-proof}
Y_k=\sum_{j=1}^m M(U_{k j})\,Z_j,
\qquad
Y_\ell=\sum_{j=1}^m M(U_{\ell j})\,Z_j.
\end{equation}
Moreover, $M(U_{k j})$ is nonsingular if and only if $U_{k j}\neq 0$.

Then, we can apply the classical Skitovich--Darmois theorem for random vectors. We now invoke the random-vector Skitovich--Darmois theorem (for the original scalar definitions see Theorem~5.3.1 in~Ref. \cite{bryc1995normal} and Theorem~3.1.1 in~Ref. \cite{kagan1973characterization}, for vector version see the main Theorem in \cite{ghurye1962characterization} and Theorem 1 of Ref. \cite{feldman2003characterization})
for independent $\mathbb R^d$ valued random vectors.
It states that if $Z_1,\dots,Z_m$ are independent $\mathbb R^d$ valued random vectors and two linear forms
\begin{align}
L_1=\sum_{j=1}^m A_j Z_j, \qquad L_2=\sum_{j=1}^m B_j Z_j
\end{align}
are independent, where $A_j,B_j\in\mathbb R^{d\times d}$, then every $Z_j$ for which both $A_j$ and $B_j$
are nonsingular must be (possibly shifted) Gaussian in $\mathbb R^d$.
We apply it with $d=2$, $L_1=Y_k$, $L_2=Y_\ell$, and coefficients from \eqref{eq:linear-forms-proof}.
Since $Y_k$ and $Y_\ell$ are independent, we conclude that every $Z_j$ such that
$U_{k j}\neq 0$ and $U_{\ell j}\neq 0$ is a Gaussian random vector in $\mathbb R^2$.

Now fix an input index $j$. By the mixing assumption, there exist $k\neq \ell$ such that
$U_{k j}\neq 0$ and $U_{\ell j}\neq 0$. Hence $Z_j$ is Gaussian for every $j$.
Equivalently, each single-mode heterodyne density $Q_{\rho_j}$ is a Gaussian density on $\mathbb R^2$.

Finally, we lift Gaussianity of $Q$ back to Gaussianity of the input quantum states. Let $\chi_{\rho_j}(\eta)\coloneqq \mathrm{Tr}[\rho_j D(\eta)]$ be the Weyl characteristic function on $\mathbb R^2$.
Define the anti-normal ordered characteristic function
\begin{align}
\chi_{\rho_j}^{(Q)}(\eta)\coloneqq e^{-\|\eta\|_2^2/4}\,\chi_{\rho_j}(\eta).
\end{align}
A standard phase-space identity relates $Q_{\rho_j}$ and $\chi_{\rho_j}^{(Q)}$ by a symplectic Fourier transform
\begin{equation}\label{eq:Q-Fourier-proof}
Q_{\rho_j}(z)=\frac{1}{(2\pi)^2}\int_{\mathbb R^2} e^{i\eta^{\mathsf T}\omega z}\,\chi_{\rho_j}^{(Q)}(\eta)\,{\rm d}\eta,
\qquad z\in\mathbb R^2.
\end{equation}
Since $Q_{\rho_j}$ is a Gaussian density, its Fourier transform $\chi_{\rho_j}^{(Q)}$ is a Gaussian function of $\eta$.
Multiplying by the fixed Gaussian factor $e^{\|\eta\|_2^2/4}$ preserves Gaussianity, hence $\chi_{\rho_j}$ is Gaussian.
Therefore $\rho_j$ is a single-mode Gaussian state for every $j$. This proves the claim.
\end{proof}

We have proved the quantum Darmois–Skitovich theorem for passive-Gaussian unitaries that satisfy a mixing condition. A natural question is whether there are alternative way of defining mixing conditions. Here, we provide a remark comparing two of them.

\begin{remark}[Mixing for a general passive layer]\label{rem:mixing-comparison}
Let $U\in\mathrm U(m)$ be the mode mixing matrix associated with a passive-Gaussian unitary $U_O$.
There are several natural notions of mixing, and they are logically different.

\begin{enumerate}
\item[(1)] \textbf{Invariant subset mixing.} One may call the passive layer mixing if there is no nonempty proper subset
$\mathcal J\subset\{1,\dots,m\}$ such that the span of $\{a_j\}_{j\in\mathcal J}$ is invariant, meaning
\begin{align}
U\,\mathrm{span}\{e_j\}_{j\in\mathcal J}\subset \mathrm{span}\{e_j\}_{j\in\mathcal J}.
\end{align}
Equivalently, after a simultaneous relabelling of inputs and outputs, $U$ cannot be written in block diagonal
form. This rules out a decomposition of the circuit into independent mode groups.

\item[(2)]\textbf{The mixing condition in Proposition~\ref{prop:DS-passive-general}.} The assumption used there is local and support-based.
For every input column $j$, there exist two distinct outputs $k\neq \ell$ with
$U_{kj}\neq 0$ and $U_{\ell j}\neq 0$.
In graph terms, in the bipartite support graph with an edge $(k,j)$ whenever $U_{kj}\neq 0$,
each input vertex has degree at least two.
This excludes mere routing, where an input mode feeds a single output mode, and it is the minimal support
requirement that enables the Darmois--Skitovich argument based on two independent output linear forms.
\end{enumerate}

Note that (1) does not imply (2), for example a single cycle permutation
is invariant subset mixing but each column has exactly one nonzero entry.
Conversely, (2) does not imply (1), since $U$ may be block
diagonal with each block mixing internally so that every column still has at least two nonzero entries.
Thus, invariant subset mixing prevents decoupled mode groups, while the condition in
Proposition~\ref{prop:DS-passive-general} enforces per input spreading, which is the feature needed for
Gaussianity from the Darmois--Skitovich theorem.
\end{remark}

Another interesting question is which passive-Gaussian operations fall outside the mixing condition defined in Proposition \ref{prop:DS-passive-general}. Here, the following lemma addresses this question.

\begin{lemma}[Failure of the column spreading condition implies a routed mode]
\label{prop:fail-i-implies-swap-phase}
Let $U_O$ be an $m$-mode passive-Gaussian unitary. Equivalently, there exists a unitary matrix
$U=[U_{kj}]\in \mathrm U(m)$ such that in the Heisenberg picture $U_O^\dagger \bm a\,U_O = U\,\bm a,$ with $\bm a=(a_1,\dots,a_m)^{\mathsf T}$. Suppose that the mixing condition in Proposition~\ref{prop:DS-passive-general} fails, namely, there exists
an input index $j$ such that
\begin{align}
\#\{k\in\{1,\dots,m\}\mid U_{k j}\neq 0\}\le 1.
\end{align}
Then there exist an output index $k$, a phase $\theta\in\mathbb R$, a unitary matrix
$U'\in \mathrm U(m-1)$, and permutation matrices $P_{\rm in},P_{\rm out}\in\mathbb R^{m\times m}$ such that
\begin{align}
P_{\rm out}\,U\,P_{\rm in}^{\mathsf T} = e^{i\theta}\oplus U'.
\end{align}
Equivalently, up to relabeling of input and output modes, $U_O$ factorizes as
\begin{align}
U_O
=
U(P_{\rm out})^\dagger\,
\Bigl(R_1(\theta)\otimes U_{O'}\Bigr)\,
U(P_{\rm in}),
\end{align}
where $U(P)$ is the passive-Gaussian unitary implementing the mode permutation $P$,
$R_1(\theta)\coloneqq e^{-i\theta a_1^\dagger a_1}$ is a single-mode phase rotation on mode $1$,
and $U_{O'}$ is a passive-Gaussian unitary acting only on the remaining $m-1$ modes with mixing matrix $U'$.
In particular, the distinguished input mode $j$ is merely routed to a single output mode $k$ up to a local
phase shift, and it does not interfere with the other modes.
\end{lemma}

\begin{proof}
The assumption says that column $j$ of $U$ has at most one nonzero entry. Since $U$ is unitary,
the Euclidean norm of the $j$th column equals $1$, so the column cannot be identically zero.
Hence it has exactly one nonzero entry. Let that entry be $U_{k j}$ for some $k$.
Then $|U_{k j}|=1$, so there exists $\theta\in\mathbb R$ such that $U_{k j}=e^{i\theta}$, and
$U_{r j}=0$ for all $r\neq k$.

For any other column index $j'\neq j$, orthogonality of columns gives
\begin{align}
0 = \langle U_{\cdot j}, U_{\cdot j'}\rangle
= \sum_{r=1}^m \overline{U_{r j}}\,U_{r j'}
= \overline{U_{k j}}\,U_{k j'}
= e^{-i\theta}U_{k j'}.
\end{align}
Thus $U_{k j'}=0$ for all $j'\neq j$, so row $k$ also has exactly one nonzero entry, located at column $j$.

Let $P_{\rm in}$ be a permutation matrix that maps $e_1$ to $e_j$, and let $P_{\rm out}$ map $e_1$ to $e_k$.
Define $\widetilde U\coloneqq P_{\rm out}\,U\,P_{\rm in}^{\mathsf T}$.
By construction, $\widetilde U_{11}=U_{k j}=e^{i\theta}$, and the first column and first row of $\widetilde U$
have no other nonzero entries. Hence $\widetilde U$ has the block form
\begin{align}
\widetilde U=
\begin{pmatrix}
e^{i\theta} & 0\\
0 & U'
\end{pmatrix}
= e^{i\theta}\oplus U'
\end{align}
for some matrix $U'\in\mathbb C^{(m-1)\times(m-1)}$. Since $\widetilde U$ is unitary, the bottom right block
$U'$ is unitary, hence $U'\in\mathrm U(m-1)$.

Finally, mode relabeling on input and output is implemented by passive-Gaussian unitaries $U(P_{\rm in})$
and $U(P_{\rm out})$ satisfying $U(P)^\dagger \bm a\,U(P)=P\,\bm a$.
The block diagonal mixing matrix $e^{i\theta}\oplus U'$ corresponds to the tensor product
$R_1(\theta)\otimes U_{O'}$, where $R_1(\theta)$ acts on the distinguished mode and $U_{O'}$ acts on the
remaining $m-1$ modes with mixing matrix $U'$.
\end{proof}

Now, we are ready to prove the following theorem, which is useful for learning a the second passive-Gaussian layer of $U_{\rm pge}$ in Eq. \ref{def:U_three layer_PGE} from the Gaussian-entanglable state induced by coherent input states.

\begin{theorem}[Learned passive-Gaussian unitary from passive-separable states]
\label{thm:gauge-general-gaussian-probes}
Fix an $m$-mode passive-Gaussian unitary $U_O$ with mode-mixing matrix
$U\in \mathrm U(m)$ defined by $U_O^\dagger \bm a\,U_O = U\,\bm a$.
Let $|\Psi\rangle = U_O\bigl(\bigotimes_{j=1}^m |\psi_j\rangle\bigr)$ be a passive-separable pure state. Assume a tomography procedure returns a passive-Gaussian unitary $U_{\widetilde O}$ and single-mode states
$\{|\widetilde\psi_k\rangle\}_{k=1}^m$ such that
\begin{equation}
\label{eq:disentangle-assumption-general-passive-fixed}
U_O\left(\bigotimes_{j=1}^m |\psi_j\rangle\right)
=
U_{\widetilde O}\left(\bigotimes_{k=1}^m |\widetilde\psi_k\rangle\right).
\end{equation}
Define the passive disentangler $V \equiv U_{\widetilde O}^\dagger U_O,$ with $V^\dagger \bm a\,V = W\,\bm a,$ and $W\in \mathrm U(m)$. Define the non-Gaussian input index set
\begin{align}
\mathcal N = \{ j\in[m]\mid |\psi_j\rangle \text{ is non-Gaussian}\},
\qquad
\mathcal G = [m]\setminus \mathcal N.
\end{align}

Then there exist permutations $\pi_{\rm in},\pi_{\rm out}$ of $[m]$ such that, after relabeling by $\pi_{\rm in}$,
all indices in $\mathcal N$ become routed. More precisely, letting $\mathcal N^{\rm in} = \pi_{\rm in}(\mathcal N),$ and $\mathcal G^{\rm in} = \pi_{\rm in}(\mathcal G)$, one can write
\begin{equation}
\label{eq:W-global-decomp}
P_{\pi_{\rm out}}\,W\,P_{\pi_{\rm in}}^{\mathsf T}
=
\left(\bigoplus_{j\in \mathcal N^{\rm in}} e^{i\theta_j}\right)
\oplus
\left(\bigoplus_{r=1}^R W_r\right),
\end{equation}
for some phases $\{\theta_j\}_{j\in\mathcal N^{\rm in}}$ and some block unitaries
$W_r\in \mathrm U(d_r)$ acting on disjoint nonempty index blocks
$\mathcal G_1^{\rm in},\dots,\mathcal G_R^{\rm in}$ that partition $\mathcal G^{\rm in}$, with $d_r = |\mathcal G_r^{\rm in}|$.

Moreover, every input state $|\psi_j\rangle$ with $j\in \mathcal G_r^{\rm in}$ is Gaussian, and all these
single-mode Gaussian states share the same covariance matrix up to a local phase rotation.
Equivalently, for $j\in\mathcal G_r^{\rm in}$ define the second moments $n_j = \langle \Delta a_j^\dagger \Delta a_j\rangle,$ and $m_j = \langle \Delta a_j \Delta a_j\rangle,$ computed on $|\psi_j\rangle$ after removing the displacement.
Then, within each block $\mathcal G_r^{\rm in}$, the values $n_j$ and $|m_j|$ are constant in $j$.

More precisely, each Gaussian block falls into exactly one of the following two cases.

\begin{enumerate}
\item Phase insensitive block. The common value satisfies $|m_j|=0$ on $\mathcal G_r^{\rm in}$.
Then there is no further restriction on $W_r$ beyond unitarity, so $W_r$ can be any element of $\mathrm U(d_r)$.

\item Phase sensitive block. The common value satisfies $|m_j|>0$ on $\mathcal G_r^{\rm in}$.
Then there exist diagonal phase matrices $D_r^{(1)},D_r^{(2)}$ and a real orthogonal matrix $T_r\in \mathrm O(d_r)$ such that
\begin{equation}
\label{eq:Wr-real-orthogonal-normal-form}
W_r = D_r^{(1)}\,T_r\,D_r^{(2)}.
\end{equation}
\end{enumerate}

In quadratures ordered as $(x_1,\dots,x_m,p_1,\dots,p_m)$, let $O(X)$ denote the symplectic-orthogonal embedding
of a unitary mixing matrix $X$,
\begin{align}
O(X)=
\begin{pmatrix}
\Re X & -\Im X\\
\Im X & \Re X
\end{pmatrix}.
\end{align}
Then the symplectic matrix $\widetilde O$ differs from $O$ by right composition with a gauge element
of the same structural form as \eqref{eq:W-global-decomp} after taking transpose,
so there exist permutations $\pi,\pi'$ and block data as above such that
\begin{equation}
\label{eq:Otilde-gauge-structure}
\widetilde O
=
O\,
P_{\pi}
\left[
\left(\bigoplus_{j\in \mathcal N^\star} R(\vartheta_j)\right)
\oplus
\left(\bigoplus_{r=1}^R O(W_r^\star)\right)
\right]
P_{\pi'}.
\end{equation}
Here $\mathcal N^\star= \pi^{-1} (\mathcal N)$ is a relabeling of $\mathcal N$ induced by $\pi^{-1}$, $R(\vartheta)$ is the $2\times2$ phase-space rotation, and each $W_r^\star$ is either an arbitrary unitary in the phase
insensitive case or a real orthogonal matrix in the phase sensitive case, up to diagonal phases as in
\eqref{eq:Wr-real-orthogonal-normal-form}.
\end{theorem}

\begin{figure}[ht]
\centering\includegraphics[width=0.43\textwidth,trim=0 0 0 0,clip,angle=0]{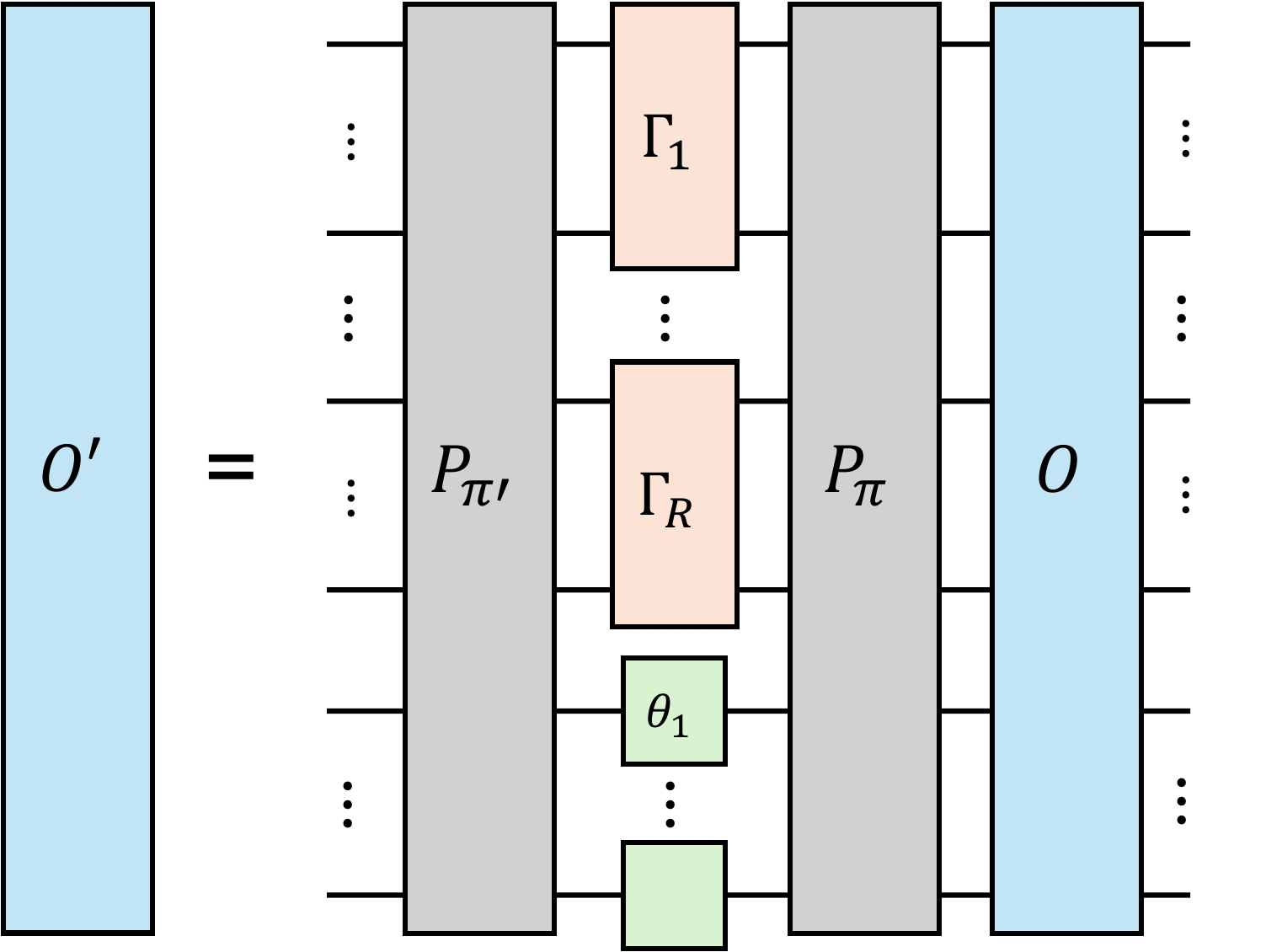}
\caption{\textbf{Residual passive gauge of a learned passive-separable
state.} After input and output relabellings, modes carrying non-Gaussian states are
routed individually and remain ambiguous only up to local phase rotations.
Residual passive mixing \(\Gamma_r\) is confined to covariance-degenerate
Gaussian blocks, with the allowed block transformations characterized in
Theorem~\ref{thm:gauge-general-gaussian-probes}.}
\label{fig:rebuilt_PG}
\end{figure}

\begin{proof}
From \eqref{eq:disentangle-assumption-general-passive-fixed}, we have $V\left(\bigotimes_{j=1}^m |\psi_j\rangle\right) = \bigotimes_{k=1}^m |\widetilde\psi_k\rangle$, so $V$ maps a product input to a product output and is passive.

Now, let us peel routed modes and fix the relabeling. If there exists an input index whose column of $W$ has at most one nonzero entry, then
Lemma~\ref{prop:fail-i-implies-swap-phase} applies to $V$ and yields, up to input and output permutations,
a factorization that isolates one mode as a single-mode phase shift and reduces the remaining transformation to
a passive unitary on $m-1$ modes.
Iterating this peeling procedure produces permutations $\pi_{\rm in},\pi_{\rm out}$ such that
\eqref{eq:W-global-decomp} holds with a routed sector indexed by $\mathcal N^{\rm in}$ and a core acting on
$\mathcal G^{\rm in}$.

Then, one can show that non-Gaussian inputs must be routed.
Assume for contradiction that some non-Gaussian input mode remains inside the core, meaning that after the relabeling
$\pi_{\rm in}$ there exists $j\in\mathcal N^{\rm in}$ on which the core block acts nontrivially.
By construction, every column of the core mixing matrix has at least two nonzero entries, so the mixing condition
in Proposition~\ref{prop:DS-passive-general} holds on the core subsystem.
The core passive unitary maps the restricted product input to a product output.
Proposition~\ref{prop:DS-passive-general} then implies that every input factor on the core subsystem is Gaussian,
contradicting that $|\psi_j\rangle$ is non-Gaussian.
Hence all indices in $\mathcal N^{\rm in}$ lie in the routed sector.

Furthermore, we have a block structure on the Gaussian sector.
Restrict to the Gaussian sector indexed by $\mathcal G^{\rm in}$ and remove displacements mode-wise.
For each Gaussian input factor define $n_j$ and $m_j$ as in the theorem.
Let $N=\mathrm{diag}(n_j)$ and $M=\mathrm{diag}(m_j)$ on this sector.
Under a passive mixing $W_{\mathcal G}$ one has
\begin{align}
N_{\rm out}= W_{\mathcal G} N W_{\mathcal G}^\dagger,
\qquad
M_{\rm out}= W_{\mathcal G} M W_{\mathcal G}^{\mathsf T}.
\end{align}
Since the output state is a product Gaussian state on this sector, both $N_{\rm out}$ and $M_{\rm out}$ are diagonal.
The condition that $N_{\rm out}$ is diagonal implies $W_{\mathcal G}$ cannot mix modes with different $n_j$ values,
so it is block diagonal up to permutation along the degeneracy classes of $n_j$.
Similarly, the condition that $M_{\rm out}$ is diagonal implies $W_{\mathcal G}$ cannot mix modes with different $|m_j|$
values.
Therefore $W_{\mathcal G}$ is block diagonal up to permutation along the joint degeneracy classes of the pair $(n_j,|m_j|)$.
This yields the partition $\mathcal G_1^{\rm in},\dots,\mathcal G_R^{\rm in}$ and the block decomposition
in \eqref{eq:W-global-decomp}, and shows that within each block $n_j$ and $|m_j|$ are constant.

Then, let us look at the classification inside one Gaussian block. Fix a block of size $d$ and write its mixing matrix as $W_r\in\mathrm U(d)$. If the common value satisfies $|m_j|=0$, then $M=0$ on the block and the constraint from $M_{\rm out}$ disappears.
Since $N=nI_d$ on the block, $N_{\rm out}=W_r (nI_d) W_r^\dagger = nI_d$ is always diagonal.
Thus any $W_r\in\mathrm U(d)$ preserves productness of the block, which is the phase insensitive case. Otherwise, if the common value satisfies $|m_j|>0$, apply local phase rotations on the inputs so that $M=m I_d$ with $m>0$.
This does not change productness and only conjugates $W_r$ by diagonal phase matrices.
The constraint that $M_{\rm out}=W_r (mI_d) W_r^{\mathsf T}$ is diagonal is equivalent to requiring
$W_r W_r^{\mathsf T}$ to be diagonal.
Let $S=W_r W_r^{\mathsf T}=\mathrm{diag}(e^{i\beta_1},\dots,e^{i\beta_d})$.
Set $D=\mathrm{diag}(e^{-i\beta_1/2},\dots,e^{-i\beta_d/2})$ and define $W_r' = D W_r$.
Then $W_r' (W_r')^{\mathsf T} = I_d$.
Since $W_r'$ is also unitary, we have $(W_r')^{-1}=(W_r')^\dagger$ and the condition
$W_r' (W_r')^{\mathsf T}=I_d$ implies $(W_r')^{\mathsf T}=(W_r')^\dagger$.
Taking complex conjugates gives $\overline{W_r'}=W_r'$, so $W_r'$ is real and orthogonal.
Writing $T_r=W_r'\in\mathrm O(d)$ and undoing the diagonal phases yields \eqref{eq:Wr-real-orthogonal-normal-form}.

The symplectic matrix $G$ of $V$ equals $O(W)$, and because $V=U_{\widetilde O}^\dagger U_O$ is passive, we have 
\begin{align}
G = \widetilde O^{-1} O = \widetilde O^{\mathsf T} O,
\qquad
\widetilde O = O\,G^{\mathsf T}.
\end{align}
All factors appearing in \eqref{eq:W-global-decomp} are unitary, so their embeddings are orthogonal, and taking transpose
simply inverts them inside the same family.
Thus $\widetilde O$ differs from $O$ by right composition with a gauge element of the same structural form,
which gives \eqref{eq:Otilde-gauge-structure} after renaming the permutation labels and parameters.
\end{proof}

Fig.~\ref{fig:rebuilt_PG} illustrates the decomposition of the reconstructed passive-Gaussian unitary. Finally, once \(U_{O_2}\) in Eq.~(\ref{def:U_three layer_PGE}) has been learned from shadow tomography, we can partially  counter-rotate \(U_{\rm pge}\) through postprocessing. Specifically, for each heterodyne outcome \(\bs\gamma\), we apply the classical transformation $\bs\gamma \;\mapsto\; W_{\widetilde O_2}\bs\gamma$, where \(W_{\widetilde O_2}\) is the complex mode-mixing matrix associated with \(\widetilde O_2\). As a result, the original learning problem reduces to learning the remaining passive-Gaussian layer \(U_{O_1}\) together with the local unitaries \(\{\widetilde W_j\}\) defined in Eq.~(\ref{def:U_three layer_PGE}), up to the passive mixing, permutation, and single-mode phase-rotation ambiguities characterized in Theorem~\ref{thm:gauge-general-gaussian-probes}.

Here, a crucial question is whether the effective counter rotation \(U_{\widetilde O_2}^\dagger\), obtained by reconstructing the state \( |\Psi^{\rm ps}_{\bs\alpha,\bs\alpha^{(\ell)}}\rangle \) in Eq.~(\ref{eqs10}) using the probe \( |\underline{\bs\alpha}\rangle \), also works for the passive separable state \( |\Psi^{\rm ps}_{\bs\alpha,\bs\alpha^{(h)}}\rangle \) obtained from another input $|\underline{\bs\alpha^{(h)}}\>$. In other words, the question is 
\begin{quote}
\emph{whether we can replace the unitary $U_{O_2,{\bs\alpha}}$ in Eq.~(\ref{eqs10}) by mixing on Gaussian states, permutation, and single-mode phase rotation via this counter-rotation process. }
\end{quote}

Following Eq.~(\ref{eq:Otilde-gauge-structure}) and the fact that, with probability one, a random coherent state activates the non-Gaussianity of the output whenever at least one \(W_j\) is non-Gaussian, as will be rigorously proved in Theorem~\ref{prop:measure-zero-bargmann-multimode} in the next subsection, we obtain the following lemma.
\begin{lemma}\label{lem:s9}
In the learning of the counter-rotated unitary $U_{\rm pge}$ defined in Eq. (\ref{def:U_three layer_PGE0}), a probe state $|\underline{\bs\alpha^{(h)}}\>$ leads to the passive-separable state \( |\Psi^{\rm ps}_{\bs\alpha,\bs\alpha^{(h)}}\rangle \) defined in Eq.~(\ref{eqs10}). Then, we can use shadow tomography to reconstruct $U_{\widetilde O_2,{\bs\alpha},\bs\alpha^{(h)}}$ defined in Eq. (\ref{def:U_three layer_PGE0}) up to a gauge structure shown in Theorem \ref{thm:gauge-general-gaussian-probes}. Then, the state $U_{\widetilde O_2,{\bs\alpha},\bs\alpha^{(h)}}^\dag |\Psi^{\rm ps}_{\bs\alpha,\bs\alpha^{(h)}}\rangle$ is a product of a multimode Gaussian state and single-mode non-Gaussian states with unit probability. 
\end{lemma}

Finally, by applying the shadow-tomography protocol to the passive-separable state
\(
\ket{\Psi^{\rm ps}_{\bs\alpha,\bs\alpha^{(h)}}},
\)
one can reconstruct a passive-Gaussian unitary
\(
U_{\widetilde O_{2,\bs\alpha,\bs\alpha^{(h)}}}^{\dagger}
\)
with arbitrarily small reconstruction error \cite{zhao2025complexity}. More precisely, by identifying the target state through fidelity witnesses over an \(\epsilon\)-covering net of passive-separable states with bounded second energy moment, one obtains an estimate of the associated symplectic-orthogonal matrix
\(
\widetilde O_{2,\bs\alpha,\bs\alpha^{(h)}}.
\)

The resulting error in the reconstructed matrix can then be converted into an error bound for the corresponding unitary channel. In particular, by the Gaussian Solovay--Kitaev theorem \cite{becker2021energy}, there exists an explicit polynomial function such that
\begin{align}
\left\|
\mathcal U_{O_{2,\bs\alpha,\bs\alpha^{(h)}}}
-
\mathcal U_{\widetilde O_{2,\bs\alpha,\bs\alpha^{(h)}}}
\right\|_{\diamond}^{mE''}
\le
\mathrm{poly}\!\left(
m,\sqrt{E''},
\sqrt{
\left\|
O_{2,\bs\alpha,\bs\alpha^{(h)}}
-
\widetilde O_{2,\bs\alpha,\bs\alpha^{(h)}}
\right\|_{F}
}
\right)
\coloneqq
\epsilon_{\rm ps}.\label{eqs67}
\end{align}
Therefore, the reconstruction of the second passive-Gaussian layer
\(
\mathcal U_{\widetilde O_{2,\bs\alpha,\bs\alpha^{(h)}}}
\)
achieves a controlled error in the energy-constrained diamond norm. A more explicit expression for
\(
\epsilon_{\rm ps}
\)
can be seen from Refs.~\cite{becker2021energy,zhao2025complexity}.

\subsection{Activation of non-Gaussian states}\label{supp:activation theorem}

In \ref{supp:Learning of the second passive-Gaussian layer}, we show that the reconstruction of the second passive-Gaussian layer depends on whether its input state is Gaussian or non-Gaussian. In general, we can choose a family of coherent probes $\{|\underline{\bs\alpha^{(\ell)}}\rangle\}_{\ell=1}^L$ and send them through $U_{\rm pge}$ defined in Eq. (\ref{def:U_three layer_PGE}). This generates a family of output Gaussian-entangleable states
\begin{align}
\left\{\,|\Psi^{(\ell)}\rangle
=
U_{O_2}\!\left(\bigotimes_{j=1}^m |\psi_j^{(\ell)}\rangle\right)\right\}_{\ell=1}^L,
\end{align}
each of which can be learned efficiently. In the ideal setting, the chosen probes are sufficiently rich to activate every non-Gaussian middle-layer gate whenever such activation is possible. We then define
\begin{align}
\map N
:=
\left\{\,j\in\{1,\cdots,m\}\ \middle|\ \exists\,\ell\ {\rm such\ that}\ |\psi_j^{(\ell)}\rangle\ {\rm is\ non\mbox{-}Gaussian}\right\},
\qquad
\map G
:=
\{1,\cdots,m\}\backslash \map N.
\end{align}
Under this ideal activation pattern, Theorem \ref{thm:gauge-general-gaussian-probes} implies that $U_{O_2}$ can be reconstructed from the resulting system of equations. The main difficulty, however, is that for a fixed finite family of coherent probes, it is not known in advance whether the chosen probes will indeed activate a given non-Gaussian middle-layer gate and produce a non-Gaussian output.

In this subsection, we will show that a single random coherent probe identifies an arbitrary multimode non-Gaussian unitary with almost probability one. Before proving the main theorem, we first present a useful lemma and a useful proposition.

\begin{lemma}[Positive-measure Gaussian outputs force Gaussian outputs on the whole open set]
\label{prop:positive-measure-implies-open-set-gaussian-multimode}
Let $m\ge 1$, let $\Omega\subset\mathbb C^m$ be a nonempty connected open set, and let $W$ be an $m$-mode unitary.
Define
\begin{align}
\mathcal E_W(\Omega)
\coloneqq
\bigl\{
\bm\alpha\in\Omega
\,\big|\,
W|\underline{\bm\alpha}\rangle \text{ is a pure Gaussian state}
\bigr\}.
\end{align}
If $\mathcal E_W(\Omega)$ has positive $2m$-dimensional Lebesgue measure in $\Omega$, then
\begin{align}
W|\underline{\bm\alpha}\rangle
\text{ is a pure Gaussian state for every }\bm\alpha\in\Omega.
\end{align}
\end{lemma}

\begin{proof}
We work in the $m$-mode Bargmann--Fock representation \cite{Bargmann1961,Folland1989}. Let $\mathcal F_m$ be the space of entire functions on $\mathbb C^m$ with inner product
\begin{align}
\langle f,g\rangle_{\mathcal F_m}
=
\int_{\mathbb C^m}\frac{d^{2m}\bm z}{\pi^m}\,
e^{-|\bm z|^2}\,
\overline{f(\bm z)}\,g(\bm z).
\end{align}
Let $\mathcal B$ be the Segal--Bargmann transform, normalized by
\begin{align}
(\mathcal B|\psi\rangle)(\bm z)=\langle \bm z\|\psi\rangle=\sum_{n_1,\cdots n_m\ge 0}\frac{\<n_1,\cdots,n_m|\psi\>}{\sqrt{\bs n!}}\bs z^{*\bs n} , 
\end{align}
where $\|{\bm\alpha}\rangle
\coloneqq
e^{\sum_{j=1}^m \alpha_j a_j^\dagger}|0\rangle
=
e^{|\bm\alpha|^2/2}|\underline{\bm\alpha}\rangle$ denotes the unnormalized coherent state, $\bm n\in\mathbb N^m$ is multi-indice, with
$\bm z^{\bm n}\coloneqq \prod_{j=1}^m z_j^{n_j}$, and
$\bm n!\coloneqq \prod_{j=1}^m n_j!$. For unnormalized coherent states, we have:
\begin{align}
\mathcal B\|{\bm\alpha}\rangle (\bs z)=  \<\bs z\|\bs \alpha\>=
k_{\bm\alpha}(\bs z)= e^{\bm\alpha^{\mathsf T}\bm z}.
\end{align}
We can also equivalently write the functional equality as follows 
\begin{align}
\mathcal B\|{\bm\alpha}\rangle
=
k_{\bm\alpha}.
\end{align}

Define the representation of $W$ as $T\coloneqq \mathcal B\,W\,\mathcal B^{-1}$ on $\map F_m$ and set $f_{\bm\alpha}(\bm z)\coloneqq (Tk_{\bm\alpha})(\bm z)\in \map F_m$ for $\bs z$. We first show that $(\bm\alpha,\bm z)\mapsto f_{\bm\alpha}(\bm z)$ is jointly entire on $\mathbb C^m\times\mathbb C^m$,  i.e. holomorphic on the whole complex plane \cite{Ahlfors1979ComplexAnalysis}. To see this, we can express
\begin{align}
f_{\bs \alpha}(\bs z)
=\mathcal B\,W\,\mathcal B^{-1}\mathcal B\|{\bs \alpha}\rangle (\bs z)=\mathcal B\,W\,\|\alpha\rangle (\bs z)=\langle \bs z\|W\|{\bs \alpha}\rangle=e^{\frac{|\bs \alpha|^2+|\bs z|^2}{2}}\braket{\underline{\bs z}|W|\underline{\bs \alpha}}<e^{\frac{|\bs \alpha|^2+|\bs z|^2}{2}},
\end{align}
which is finite for any unitary $W$. Also, we can write down the series form. 
\begin{align}
f_{\bm\alpha}(\bm z)
=
\sum_{\bm n,\bm p\in\mathbb N^m}
\frac{\langle \bm n|W|\bm p\rangle}{\sqrt{\bm n!\,\bm p!}}\,
\bm\alpha^{\bm p}\bm z^{\bm n}.
\end{align}
Since $|\langle \bm n|W|\bm p\rangle|\le 1$, on every closed polydisc
$\{|\alpha_j|\le R_j,\ |z_j|\le S_j\}_{j=1}^m$ we have
\begin{align}
\sum_{\bm n,\bm p}
\left|
\frac{\langle \bm n|W|\bm p\rangle}{\sqrt{\bm n!\,\bm p!}}\,
\bm\alpha^{\bm p}\bm z^{\bm n}
\right|
\le
\prod_{j=1}^m
\left(\sum_{p_j\ge 0}\frac{R_j^{p_j}}{\sqrt{p_j!}}\right)
\left(\sum_{n_j\ge 0}\frac{S_j^{n_j}}{\sqrt{n_j!}}\right),
\end{align}
and each one-variable series converges by the ratio test, since
\begin{align}
\frac{R_j^{\,p_j+1}/\sqrt{(p_j+1)!}}{R_j^{\,p_j}/\sqrt{p_j!}}
=
\frac{R_j}{\sqrt{p_j+1}}
\to 0
\qquad (p_j\to\infty),
\end{align}
and similarly for $S_j$. Hence the double series converges absolutely and uniformly on compact sets $\{|\alpha_j|\le R_j,\ |z_j|\le S_j\}_{j=1}^m$ , so $f_{\bm\alpha}(\bm z)$ is jointly entire in $(\bs \alpha,\bs z)$. It defines an entire function in both variables. In particular, for
each fixed $\bs z$, the map of higher-order derivatives $bs \alpha \to f_{\bs \alpha^{(j)}}(\bs z)$ is entire for every integer $j\ge 0$.

We next introduce the multimode differential \emph{Gaussianity witness}: an explicit functional that vanishes identically for (and
only for) pure Gaussian states in the Bargmann representation. For an entire function $f$ on $\mathbb C^m$, define a function for $1\le i,j,k\le m$
\begin{align}
\mathsf J_{ijk}[f]
\coloneqq\ &
f^2\,\partial_i\partial_j\partial_k f
-
f\Bigl(
(\partial_i\partial_j f)(\partial_k f)
+
(\partial_i\partial_k f)(\partial_j f)
+
(\partial_j\partial_k f)(\partial_i f)
\Bigr)
\nonumber\\
&\quad
+
2(\partial_i f)(\partial_j f)(\partial_k f),
\end{align}
where $\partial_i\equiv \partial/\partial z_i$ denotes the differential operator on the $i$-the mode. Note that, if we have 
\begin{align}
f(\bm z)=\exp\!\Bigl(\tfrac12 \bm z^{\mathsf T}A\bm z+\bm b^{\mathsf T}\bm z+c\Bigr),
\qquad
A=A^{\mathsf T},
\end{align}
then $q=\log f$ is quadratic, hence all third partial derivatives of $q$ vanish, and therefore
\begin{align}
\mathsf J_{ijk}[f]= f^3 \partial_i \partial_j \partial_k q=  0
\qquad
\text{for all }i,j,k.
\end{align}

Conversely, suppose $f\not\equiv 0$ is entire and
\begin{align}
\mathsf J_{ijk}[f]\equiv 0
\qquad
\text{for all }i,j,k.
\end{align}
We claim that $f$ must be the exponential of a quadratic polynomial.

Now, let us pick $\bm a\in\mathbb C^m$ with $f(\bm a)\neq 0$.
For any $\bm v\in\mathbb C^m$, define the one-variable entire function for $t$: 
\begin{align}
h_{\bm a,\bm v}(t)\coloneqq f(\bm a+t\bm v).
\end{align}
By the chain rule,
\begin{align}
h'''_{\bm a,\bm v}(t)\,h_{\bm a,\bm v}(t)^2
-3h''_{\bm a,\bm v}(t)\,h'_{\bm a,\bm v}(t)\,h_{\bm a,\bm v}(t)
+2h'_{\bm a,\bm v}(t)^3
=
\sum_{i,j,k=1}^m
v_i v_j v_k\,
\mathsf J_{ijk}[f](\bm a+t\bm v)
=0.
\end{align}
Hence the one-variable witness vanishes identically for every line restriction.
By the single-variable argument, each $h_{\bm a,\bm v}$ is the exponential of a quadratic polynomial in $t$.
In particular, $h_{\bm a,\bm v}(t)$ never vanishes.
Taking $\bm v=\bm z-\bm a$ and $t=1$, we conclude that $f(\bm z)\neq 0$ for every $\bm z\in\mathbb C^m$.
Thus $f$ is zero-free on $\mathbb C^m$.

Now define
\begin{align}
g_i(\bm z)\coloneqq \frac{\partial_i f(\bm z)}{f(\bm z)},
\qquad i=1,\dots,m.
\end{align}
Since $f$ is zero-free, each $g_i$ is entire.
Moreover,
\begin{align}
\partial_j\partial_k g_i
=
\partial_i\partial_j\partial_k(\log f)
=
\frac{\mathsf J_{ijk}[f]}{f^3}
=
0.
\end{align}
Hence each $g_i$ is affine linear in the following form: 
\begin{align}
g_i(\bm z)=b_i+\sum_{\ell=1}^m A_{i\ell}z_\ell.
\end{align}
Since we have 
\begin{align}
\partial_j g_i
=
\partial_j\partial_i(\log f)
=
\partial_i\partial_j(\log f)
=
\partial_i g_j,
\end{align}
the matrix $A$ is symmetric.
Therefore, we have 
\begin{align}
q(\bm z)\coloneqq \tfrac12 \bm z^{\mathsf T}A\bm z+\bm b^{\mathsf T}\bm z
\end{align}
satisfies $\partial_i q=g_i=\partial_i(\log f)$ for all $i$.
Thus $\log f-q$ is constant, and
\begin{align}
f(\bm z)=\exp\!\Bigl(\tfrac12 \bm z^{\mathsf T}A\bm z+\bm b^{\mathsf T}\bm z+c\Bigr).
\end{align}

So we have proved the following equivalence for nonzero entire $f$.
\begin{align}
\mathsf J_{ijk}[f]\equiv 0\ \forall\,i,j,k
\quad\Longleftrightarrow\quad
f(\bm z)=\exp\!\Bigl(\tfrac12 \bm z^{\mathsf T}A\bm z+\bm b^{\mathsf T}\bm z+c\Bigr)
\end{align}
with $A=A^{\mathsf T}$.
For vectors in $\mathcal F_m$, we have $I-A^\dag A>0$. This is exactly the Bargmann form of a pure multimode Gaussian state \cite{Bargmann1961}.

Now, let us prove that the positive area of Gaussian outputs forces Gaussian outputs everywhere on $\Omega$. Assume $\mathcal E_W(\Omega)$ has positive Lebesgue measure in $\Omega$.
Fix $\bm z\in\mathbb C^m$ and indices $i,j,k$.
Define
\begin{align}
F_{\bm z;i,j,k}(\bm\alpha)
\coloneqq
\mathsf J_{ijk}[f_{\bm\alpha}](\bm z).
\end{align}
Because $(\bm\alpha,\bm z)\mapsto f_{\bm\alpha}(\bm z)$ is jointly entire, each
$F_{\bm z;i,j,k}$ is holomorphic in $\bm\alpha$ on $\Omega$. Given that $F_{\bs z}$ is an entire function of $\bs \alpha$, it vanishes on $\map E_W(\Omega)$, which has positive area inside the open set $\Omega$. Additionally, $\map E_W(\Omega)$ has positive measure and therefore has an accumulation point within.

A holomorphic function on a domain in $\mathbb C^m$ that vanishes on a subset of positive $2m$-dimensional Lebesgue measure must vanish identically \cite{Ahlfors1979ComplexAnalysis}.
Hence, we have 
\begin{align}
F_{\bm z;i,j,k}(\bm\alpha)=0
\qquad
\text{for all }\bm\alpha\in\Omega.
\end{align}
Since this holds for every $\bm z$ and every $i,j,k$, we obtain
\begin{align}
\mathsf J_{ijk}[f_{\bm\alpha}]\equiv 0
\qquad
\text{for every }\bm\alpha\in\Omega,\ \text{for all }i,j,k.
\end{align}
By the already proved characterization, each $f_{\bm\alpha}$ is the Bargmann function of a pure multimode Gaussian state.
Equivalently,
\begin{align}
W|\underline{\bm\alpha}\rangle
\text{ is a pure Gaussian state for every }\bm\alpha\in\Omega.
\end{align}
This proves the claim.
\end{proof}

\begin{proposition}[A unitary mapping all coherent states to pure Gaussian states is Gaussian]
\label{prop:all-coherent-to-gaussian-implies-gaussian}
Let \(W\) be an \(m\)-mode unitary on the bosonic Fock space.
Assume that for every \(\bm\alpha\in\mathbb C^m\), the state \(W\ket{\underline{\bm\alpha}}\) is a pure Gaussian state.
Then \(W\) is a Gaussian unitary.
\end{proposition}

\begin{proof}
Since \(W\ket{\bm 0}\) is a pure Gaussian state, by definition there exists a Gaussian unitary \(G\) such that
\begin{align}
G\ket{\bm 0}=W\ket{\bm 0}.
\end{align}
Let us define $U\coloneqq G^\dagger W$. Then \(U\) is unitary such that 
\begin{align}
U\ket{\bm 0}=\ket{\bm 0}.
\end{align}
For every \(\bm\alpha\in\mathbb C^m\), the state \(U\ket{\underline{\bm\alpha}}\) is again a pure Gaussian state, since acting both sides a multimode displacement operator preserve pure Gaussianity. Then, for \(\mathbf z,\bm\alpha\in\mathbb C^m\), define the following function 
\begin{align}
F(\mathbf z,\bm\alpha)
\coloneqq
e^{(\|\mathbf z\|^2+\|\bm\alpha\|^2)/2}
\bra{\underline{\overline{\mathbf z}}}U\ket{\underline{\bm\alpha}}.
\end{align}
Using the multimode coherent-state expansions
\begin{align}
\ket{\underline{\bm\alpha}}
=
e^{-\|\bm\alpha\|^2/2}
\sum_{\mathbf m\in\mathbb N^m}
\frac{\bm\alpha^{\mathbf m}}{\sqrt{\mathbf m!}}\ket{\mathbf m},
\qquad
\bra{\underline{\overline{\mathbf z}}}
=
e^{-\|\mathbf z\|^2/2}
\sum_{\mathbf n\in\mathbb N^m}
\frac{\mathbf z^{\mathbf n}}{\sqrt{\mathbf n!}}\bra{\mathbf n},
\end{align}
we obtain
\begin{align}
F(\mathbf z,\bm\alpha)
=
\sum_{\mathbf n,\mathbf m\in\mathbb N^m}
\frac{\mathbf z^{\mathbf n}\bm\alpha^{\mathbf m}}{\sqrt{\mathbf n!\,\mathbf m!}}
\bra{\mathbf n}U\ket{\mathbf m}.
\end{align}
Since \(\bigl|\bra{\mathbf n}U\ket{\mathbf m}\bigr|\le 1\) and
\begin{align}
\sum_{\mathbf n\in\mathbb N^m}\frac{|\mathbf z^{\mathbf n}|}{\sqrt{\mathbf n!}}
=
\prod_{j=1}^m\sum_{n_j\ge 0}\frac{|z_j|^{n_j}}{\sqrt{n_j!}}
<\infty,
\qquad
\sum_{\mathbf m\in\mathbb N^m}\frac{|\bm\alpha^{\mathbf m}|}{\sqrt{\mathbf m!}}
=
\prod_{j=1}^m\sum_{m_j\ge 0}\frac{|\alpha_j|^{m_j}}{\sqrt{m_j!}}
<\infty.
\end{align}
For each fixed \(j\), the ratio test gives
\begin{align}
\frac{|z_j|^{n_j+1}/\sqrt{(n_j+1)!}}{|z_j|^{n_j}/\sqrt{n_j!}}
=
\frac{|z_j|}{\sqrt{n_j+1}}
\longrightarrow 0
\qquad
\text{as } n_j\to\infty.
\end{align}
Thus, we have
\begin{align}
\sum_{\mathbf n\in\mathbb N^m}\frac{|\mathbf z^{\mathbf n}|}{\sqrt{\mathbf n!}}
=
\prod_{j=1}^m\sum_{n_j\ge 0}\frac{|z_j|^{n_j}}{\sqrt{n_j!}}
<\infty,
\end{align}
and similarly for the \(\bm\alpha\)-series. Therefore the series for \(F(\mathbf z,\bm\alpha)\) converges absolutely and locally uniformly on \(\mathbb C^m\times\mathbb C^m\). Hence \(F\) is jointly entire in \((\mathbf z,\bm\alpha)\). Hence \(F\) is jointly entire in \((\mathbf z,\bm\alpha)\).

Now fix \(\bm\alpha\in\mathbb C^m\). Since \(U\ket{\underline{\bm\alpha}}\) is a pure Gaussian state, its Bargmann function has the standard form \cite{Bargmann1961,Folland1989}
\begin{align}
F(\mathbf z,\bm\alpha)
=
c(\bm\alpha)
\exp\!\Bigl(
\tfrac12 \mathbf z^{\mathsf T}A(\bm\alpha)\mathbf z
+
b(\bm\alpha)^{\mathsf T}\mathbf z
\Bigr),
\end{align}
where \(c(\bm\alpha)\neq 0\), \(b(\bm\alpha)\in\mathbb C^m\), and \(A(\bm\alpha)\in\mathbb C^{m\times m}\) is symmetric and satisfies $\|A(\bm\alpha)\|<1$. In particular, we have a relation  $F(\mathbf 0,\bm\alpha)=c(\bm\alpha)\neq 0
\,
\text{for all }\bm\alpha\in\mathbb C^m$.

For each \(j,k\in\{1,\dots,m\}\), define the following function 
\begin{align}
\Xi_{jk}(\bm\alpha)
\coloneqq
\frac{
F(\mathbf 0,\bm\alpha)\,\partial_{z_j}\partial_{z_k}F(\mathbf 0,\bm\alpha)
-
\partial_{z_j}F(\mathbf 0,\bm\alpha)\,\partial_{z_k}F(\mathbf 0,\bm\alpha)
}{
F(\mathbf 0,\bm\alpha)^2
}.
\end{align}
Substituting the Gaussian form of \(F\) gives
\begin{align}
\Xi_{jk}(\bm\alpha)=A_{jk}(\bm\alpha).
\end{align}
Since \(F\) is jointly entire and \(F(\mathbf 0,\bm\alpha)\neq 0\) for all \(\bm\alpha\), each \(\Xi_{jk}\), hence each \(A_{jk}\), is an entire function of \(\bm\alpha\). Moreover, we have 
\begin{align}
|A_{jk}(\bm\alpha)|
\le
\|A(\bm\alpha)\|_\infty
<1
\qquad
\text{for all }\bm\alpha\in\mathbb C^m.
\end{align}
Therefore each \(A_{jk}\) is a bounded entire function on \(\mathbb C^m\). By Liouville's theorem \cite{Ahlfors1979ComplexAnalysis}, every \(A_{jk}\) is constant. Thus \(A(\bm\alpha)\) is independent of \(\bm\alpha\).

Next we evaluate at \(\bm\alpha=\bm 0\). Since \(U\ket{\bm 0}=\ket{\bm 0}\), we have $F(\mathbf z,\bm 0)
=
e^{\|\mathbf z\|^2/2}\<\overline{\mathbf z}\|\cdot|\bm 0\rangle
=
1$. Hence, we have  $A(\bm 0)=0$. Because \(A(\bm\alpha)\) is constant in \(\bm\alpha\), we conclude
\begin{align}
A(\bm\alpha)\equiv 0
\qquad
\text{for all }\bm\alpha\in\mathbb C^m.
\end{align}
Therefore, we have 
\begin{align}
F(\mathbf z,\bm\alpha)
=
c(\bm\alpha)\exp\!\bigl(b(\bm\alpha)^{\mathsf T}\mathbf z\bigr).
\end{align}
Thus \(U\ket{\underline{\bm\alpha}}\) is a coherent state up to a phase for every \(\bm\alpha\).

Define a function
\begin{align}
\eta_j(\bm\alpha)
\coloneqq
\frac{\partial_{z_j}F(\mathbf 0,\bm\alpha)}{F(\mathbf 0,\bm\alpha)},
\qquad
j=1,\dots,m.
\end{align}
From the form of \(F\), we have $\eta_j(\bm\alpha)=b_j(\bm\alpha)$. So \(\bm\eta(\bm\alpha)\coloneqq(\eta_1(\bm\alpha),\dots,\eta_m(\bm\alpha))\) is exactly the coherent amplitude of the output state. Since \(F\) is jointly entire and \(F(\mathbf 0,\bm\alpha)\neq 0\), each \(\eta_j\) is entire on \(\mathbb C^m\). Hence \(\bm\eta\colon\mathbb C^m\to\mathbb C^m\) is holomorphic.

Now \(U\) is unitary and \(U\ket{\underline{\bm\alpha}}\), \(U\ket{\underline{\bm\beta}}\) are coherent states up to phases. Taking the modulus of coherent-state overlaps, we get
\begin{align}
e^{-\frac12\|\bm\alpha-\bm\beta\|^2}
&=
\bigl|\langle\underline{\bm\alpha}|\underline{\bm\beta}\rangle\bigr|
=
\bigl|\langle \underline{U\bm\alpha}|\underline{U\bm\beta}\rangle\bigr|
=
\bigl|\langle\underline{\bm\eta(\bm\alpha)}|\underline{\bm\eta(\bm\beta)}\rangle\bigr|
=
e^{-\frac12\|\bm\eta(\bm\alpha)-\bm\eta(\bm\beta)\|^2},
\end{align}
and
\begin{align}
\|\bm\eta(\bm\alpha)-\bm\eta(\bm\beta)\|
=
\|\bm\alpha-\bm\beta\|
\qquad
\text{for all }\bm\alpha,\bm\beta\in\mathbb C^m.
\end{align}

Fix \(\bm\alpha\in\mathbb C^m\) and \(\mathbf v\in\mathbb C^m\). For real \(t\neq 0\), we have 
\begin{align}
\left\|
\frac{\bm\eta(\bm\alpha+t\mathbf v)-\bm\eta(\bm\alpha)}{t}
\right\|
=
\|\mathbf v\|.
\end{align}
Since \(\bm\eta\) is holomorphic, letting \(t\to 0\) gives
\begin{align}
\|D\bm\eta(\bm\alpha)\mathbf v\|=\|\mathbf v\|
\qquad
\text{for all }\bm\alpha,\mathbf v\in\mathbb C^m,
\end{align}
where \(D\bm\eta(\bm\alpha)\) denotes the complex Jacobian matrix of \(\bm\eta\) at \(\bm\alpha\) \cite{HornJohnson2013}. Therefore
\begin{align}
D\bm\eta(\bm\alpha)^\dagger D\bm\eta(\bm\alpha)=I_m
\qquad
\text{for all }\bm\alpha\in\mathbb C^m.
\end{align}
So each \(D\bm\eta(\bm\alpha)\) is unitary. In particular, every matrix entry of \(D\bm\eta(\bm\alpha)\) is bounded by \(1\) in absolute value. Since each entry is entire in \(\bm\alpha\), Liouville's theorem \cite{Ahlfors1979ComplexAnalysis}  implies that every entry of \(D\bm\eta(\bm\alpha)\) is constant. Hence there exists a fixed matrix \(V\in\mathbb C^{m\times m}\) such that
\begin{align}
D\bm\eta(\bm\alpha)=V
\qquad
\text{for all }\bm\alpha\in\mathbb C^m.
\end{align}
Integrating, we obtain
\begin{align}
\bm\eta(\bm\alpha)=V\bm\alpha+\bm\gamma
\end{align}
for some \(\bm\gamma\in\mathbb C^m\). Since \(D\bm\eta(\bm\alpha)\) is unitary, \(V\in U(m)\). Let \(\Gamma(V)\) be the passive-Gaussian unitary determined by 
\begin{align}
\Gamma(V)^\dagger \bm a\,\Gamma(V)=V\bm a.
\end{align} 
Then, we have $\Gamma(V)\ket{\underline{\bm\alpha}}=\ket{\underline{V\bm\alpha}}$ and $D(\bm\gamma)\ket{\underline{\bm\beta}}
=
e^{i\phi(\bm\beta)}\ket{\underline{\bm\beta+\bm\gamma}}$ for a real phase \(\phi(\bm\beta)\). Therefore the Gaussian unitary
\begin{align}
U_0\coloneqq D(\bm\gamma)\Gamma(V)
\end{align}
satisfies
\begin{align}
U_0\ket{\underline{\bm\alpha}}
=
e^{i\chi(\bm\alpha)}\ket{\underline{\bm\eta(\bm\alpha)}}
\end{align}
for some real-valued function \(\chi(\bm\alpha)\).

Hence, the operator $X\coloneqq U_0^\dagger U$ satisfies
\begin{align}
X\ket{\underline{\bm\alpha}}=e^{i\lambda(\bm\alpha)}\ket{\underline{\bm\alpha}}
\qquad
\text{for all }\bm\alpha\in\mathbb C^m.
\end{align}
Taking the overlap with \(\ket{\underline{\bm\beta}}\), we find
\begin{align}
\langle\underline{\bm\beta}|X|\underline{\bm\alpha}\rangle
=
e^{i\lambda(\bm\alpha)}\langle\underline{\bm\beta}|\underline{\bm\alpha}\rangle.
\end{align}
Since \(X\) is unitary, we have
\begin{align}
\langle\underline{ \bm\alpha}|\underline{\bm\beta}\rangle
=
\langle \underline{X\bm\alpha}|\underline{X\bm\beta}\rangle
=
e^{i(\lambda(\bm\beta)-\lambda(\bm\alpha))}\langle\underline{ \bm\alpha}|\underline{\bm\beta}\rangle.
\end{align}
Since coherent-state overlaps are never zero, it follows that
\begin{align}
e^{i(\lambda(\bm\beta)-\lambda(\bm\alpha))}=1.
\end{align}
because \(\langle\underline{\bm\beta}|\underline{\bm\alpha}\rangle\neq 0\) for all \(\bm\alpha,\bm\beta\). Thus \(\lambda(\bm\alpha)\) is constant modulo \(2\pi\). Since coherent states form a total set, we conclude
\begin{align}
X=e^{i\varphi}I
\end{align}
for some \(\varphi\in\mathbb R\). Therefore
\begin{align}
U=e^{i\varphi}U_0
\end{align}
is Gaussian. Finally,
\begin{align}
W=GU
\end{align}
is a product of Gaussian unitaries, hence itself a Gaussian unitary.
\end{proof}

Now we are ready to have the main theorem on non-Gaussianity activation. The following theorem will also be useful in proving Corollary~\ref{cor:measure-zero-coherent-multimode} for the discussion of $t$-doped Gaussian unitary learning.

\begin{theorem}[A random multimode coherent state suffices to witness non-Gaussianity]
\label{prop:measure-zero-bargmann-multimode}
Let $\Omega\subset\mathbb C^m$ be a nonempty connected open set.
Let $W$ be an $m$-mode unitary, and define
\begin{align}
\mathcal E_W(\Omega)
\coloneqq
\bigl\{
\bm\alpha\in\Omega \,\big|\,
W|\underline{\bm\alpha}\rangle \text{ is a pure Gaussian state}
\bigr\}.
\end{align}
Then the following are equivalent.
\begin{enumerate}
\item $\mathcal E_W(\Omega)$ has positive $2m$-dimensional Lebesgue measure in $\Omega$.
\item $W$ is a Gaussian unitary.
\end{enumerate}
In particular, if $W$ is not Gaussian, then $\mathcal E_W(\Omega)$ has Lebesgue measure zero.

In physics terms, if $W$ is not Gaussian, then in any open phase-space region $\Omega$, the coherent input states that accidentally produce a pure Gaussian output state form a set of Lebesgue measure zero. Hence, if $\bm\alpha$ is drawn from any probability distribution on $\Omega$ that has a density with respect to Lebesgue measure, then
\begin{align}
\Pr\!\bigl[\bm\alpha\in\mathcal E_W(\Omega)\bigr]=0,
\qquad
\Pr\!\bigl[W|\underline{\bm\alpha}\rangle \text{ is non-Gaussian}\bigr]=1.
\end{align}
\end{theorem}

\begin{proof}
We first prove $(2)\Rightarrow(1)$.
If $W$ is a Gaussian unitary, then it maps every multimode coherent state to a pure Gaussian state.
Hence, we have 
\begin{align}
\mathcal E_W(\Omega)=\Omega.
\end{align}
So $\mathcal E_W(\Omega)$ has positive $2m$-dimensional Lebesgue measure in $\Omega$.

We now prove $(1)\Rightarrow(2)$.
Assume that $\mathcal E_W(\Omega)$ has positive $2m$-dimensional Lebesgue measure in $\Omega$.
By Lemma~\ref{prop:positive-measure-implies-open-set-gaussian-multimode}, we already know that
\begin{align}
W|\underline{\bm\alpha}\rangle
\text{ is a pure Gaussian state for every }\bm\alpha\in\Omega.
\label{eq:gaussian-on-Omega}
\end{align}

We next upgrade Eq. \eqref{eq:gaussian-on-Omega} from $\Omega$ to all of $\mathbb C^m$.
In the proof of Lemma~\ref{prop:positive-measure-implies-open-set-gaussian-multimode}, for each fixed $\bm z\in\mathbb C^m$ and each $i,j,k\in\{1,\dots,m\}$, the entire function
\begin{align}
F_{\bm z;i,j,k}(\bm\alpha)
\coloneqq
\mathsf J_{ijk}[f_{\bm\alpha}](\bm z)
\end{align}
was introduced, where $f_{\bm\alpha}$ is the Bargmann function associated with $W|\underline{\bm\alpha}\rangle$ and $\mathsf J_{ijk}$ is the multimode Gaussianity witness from that lemma.
Moreover, that proof established the equivalence
\begin{align}
W|\underline{\bm\alpha}\rangle \text{ is a pure Gaussian state}
\quad\Longleftrightarrow\quad
F_{\bm z;i,j,k}(\bm\alpha)=0
\ \text{for all }\bm z\in\mathbb C^m,\ 1\le i,j,k\le m.
\label{eq:witness-characterization}
\end{align}
By Eq. \eqref{eq:gaussian-on-Omega}, the functions $F_{\bm z;i,j,k}$ vanish on the nonempty open set $\Omega$.
Since each $F_{\bm z;i,j,k}$ is entire on the connected domain $\mathbb C^m$, the identity theorem \cite{Ahlfors1979ComplexAnalysis} in several complex variables implies
\begin{align}
F_{\bm z;i,j,k}(\bm\alpha)=0
\qquad
\text{for all }\bm\alpha\in\mathbb C^m,\ \bm z\in\mathbb C^m,\ 1\le i,j,k\le m.
\end{align}
Using Eq. \eqref{eq:witness-characterization} again, we conclude that
\begin{align}
W|\underline{\bm\alpha}\rangle
\text{ is a pure Gaussian state for every }\bm\alpha\in\mathbb C^m.
\end{align}
We may therefore apply Proposition~\ref{prop:all-coherent-to-gaussian-implies-gaussian}, which yields that $W$ is a Gaussian unitary.
This proves $(1)\Rightarrow(2)$.

We next prove the measure-zero conclusion.
Assume that $W$ is not Gaussian.
By contraposition of the equivalence already proved, $\mathcal E_W(\Omega)$ cannot have positive Lebesgue measure.
It therefore remains only to note that $\mathcal E_W(\Omega)$ is measurable. Using the same witness functions as above, for each $\bm q\in(\mathbb Q+i\mathbb Q)^m$ and each $i,j,k$, define
\begin{align}
F_{\bm q;i,j,k}(\bm\alpha)\coloneqq \mathsf J_{ijk}[f_{\bm\alpha}](\bm q).
\end{align}
Each $F_{\bm q;i,j,k}$ is continuous on $\Omega$, and by Eq. \eqref{eq:witness-characterization},
\begin{align}
\mathcal E_W(\Omega)
=
\bigcap_{i,j,k=1}^m
\ \bigcap_{\bm q\in(\mathbb Q+i\mathbb Q)^m}
\Bigl\{
\bm\alpha\in\Omega
\,\Big|\,
F_{\bm q;i,j,k}(\bm\alpha)=0
\Bigr\}.
\end{align}
Indeed, for fixed $\bm\alpha$, each function $\bm z\mapsto \mathsf J_{ijk}[f_{\bm\alpha}](\bm z)$ is entire, and an entire function vanishes identically if and only if it vanishes on the dense countable set $(\mathbb Q+i\mathbb Q)^m$.
Hence $\mathcal E_W(\Omega)$ is a countable intersection of closed subsets of $\Omega$, so it is Borel measurable \cite{Folland1999RealAnalysis}.
Since it is measurable and cannot have positive Lebesgue measure, it must satisfy 
\begin{align}
\mathrm{Leb}_{2m}\bigl(\mathcal E_W(\Omega)\bigr)=0.
\end{align}

Finally, let $\bm\alpha$ be a random variable on $\Omega$ whose law is absolutely continuous with respect to Lebesgue measure.
Then, we have 
\begin{align}
\Pr\!\bigl[\bm\alpha\in\mathcal E_W(\Omega)\bigr]=0.
\end{align}
Equivalently, we have
\begin{align}
\Pr\!\bigl[W|\underline{\bm\alpha}\rangle \text{ is non-Gaussian}\bigr]=1.
\end{align}
This completes the proof.
\end{proof}

Note that there may exist alternative proofs of the activation theorem, for example via cumulant-based witnesses or operator-algebraic arguments. However, such approaches typically involve subtle analyticity issues, and making all steps fully rigorous can be delicate. For this reason, we do not pursue those routes here. Instead, we present below a direct argument that avoids these technical caveats.

\begin{remark}[Why analyticity issue appears in alternative proof with cumulant or operator-algebra]
A key step in the proof of Theorem~\ref{prop:measure-zero-bargmann-multimode} is a
\emph{uniqueness} principle: if a family of witnesses vanishes on a set of coherent amplitudes
of positive area, then it must vanish on the whole connected region.
To make such a step mathematically legitimate, one needs some regularity that prevents a
function from having a ``large'' zero set without being identically zero.

In the cumulant approach, the natural witnesses are the higher-order cumulants
$\kappa_{r,s}$ of the output state.
For a fixed unitary $W$, each map
\begin{align}
\alpha \longmapsto \kappa_{r,s}\!\bigl(W|\underline{\alpha}\rangle\!\langle\underline{\alpha}|W^\dagger\bigr)
\end{align}
is built from matrix elements of $W$ and from derivatives of a logarithm of the characteristic
function.
In infinite dimension, none of these operations is automatically smooth in the parameter
$\alpha$ without additional control.
In particular, taking a logarithm introduces a branch choice, and taking high derivatives
requires uniform convergence bounds.
Thus, to apply a uniqueness theorem, one typically assumes (or proves) that these witnesses are
real analytic in $(\Re\alpha,\Im\alpha)$.
Real analyticity is exactly the condition that upgrades ``zero on a set of positive area'' into
``zero everywhere''.

By contrast, the Bargmann--Fock route avoids imposing analyticity as an \emph{assumption} because
the coherent kernel vectors $k_\alpha(z)=e^{\alpha z}$ depend \emph{entirely} on $\alpha$, and a
bounded linear map $T$ preserves this holomorphic dependence.
One then constructs a differential \emph{Gaussianity witness} $\mathsf J[f]$ that vanishes
identically if and only if $f$ is a quadratic exponential, and applies the identity theorem for
holomorphic functions.
In other words, the Bargmann method obtains the needed uniqueness mechanism ``for free'' from
the built-in holomorphic structure of coherent states, whereas cumulant or operator-algebra
methods must explicitly assume (or work hard to prove) enough analyticity to justify the same
uniqueness step.
\end{remark}

\subsection{Learning the remaining part of the overall unitary}\label{supp:Learning the remaining part of the overall unitary}

\subsubsection{A ``probe-independent'' output structure}

We now consider the learning of the remaining layers of \(U_{\rm ge}\) after the counter-rotation procedures described in \ref{supp:Learning of the general Gaussian layer and physical counter-rotation} and \ref{supp:Learning of the second passive-Gaussian layer}. 
In particular, starting from the probe state \( |\underline{\bs\alpha^{(\ell)}}\rangle \), we first reconstruct the effective Gaussian counter rotation \(U_{\widetilde S_{\bs\alpha^{(\ell)}}}^\dagger\). 
We then switch to a second probe \( |\underline{\bs\alpha^{(h)}}\rangle \) and reconstruct the effective passive-Gaussian counter rotation \(U^\dagger_{\widetilde O_{2,\bs\alpha^{(\ell)},\bs\alpha^{(h)}}}\). 
With the subscripts
\begin{align}
(\ell,\ h),
\end{align}
and these counter rotations fixed, Lemma~\ref{lem:s9} implies that, for an arbitrary probe \( |\underline{\bs\alpha^{(k)}}\rangle \), the resulting output state takes the following form after counter rotation.
\begin{align}
|\widetilde \Psi^{\rm prod}_{\bs\alpha^{(\ell)},\bs\alpha^{(h)},\bs\alpha^{(k)}}\>&=U^\dag_{\widetilde O_{2,\bs\alpha^{(l)},\bs\alpha^{(h)}}} U_{\widetilde S_{\bs\alpha^{(\ell)}}}^\dag  U_{\rm ge} |\underline{\bs\alpha^{(k)}}\>\\
&= U^\dag_{\widetilde O_{2,\bs\alpha^{(l)},\bs\alpha^{(h)}}} U_{\widetilde S_{\bs\alpha^{(\ell)}}}^\dag  U_{S} \left(\bigotimes_{j=1}^m W_j\right) U_{O} |\underline{\bs\alpha^{(k)}}\>\\
&\simeq  U^\dag_{\widetilde O_{2,\bs\alpha^{(l)},\bs\alpha^{(h)}}} U_{ O_{2,{\bs\alpha}^{(\ell)}}} \left(\bigotimes_{j=1}^m U_{ S_{j,{\bs \alpha}^{(\ell)}}}^\dag \right)\left(\bigotimes_{i=1}^m W_i\right) U_{O} |\underline{\bs\alpha^{(k)}}\>\\
&\simeq 
U_{{\pi^{'-1}}}
\left[
\left(\bigotimes_{j\in \mathcal N^*} U_{R(-\vartheta_j)}^{\bs\alpha^{(\ell)},\bs\alpha^{(h)}}\right)
\otimes
U_{\map G^*}^{\bs\alpha^{(\ell)},\bs\alpha^{(h)}}
\right]
U_{\pi^{-1}} \left(\bigotimes_{j=1}^m U_{ S_{j,{\bs \alpha}^{(\ell)}}}^\dag \right)\left(\bigotimes_{i=1}^m W_i\right) U_{O} |\underline{\bs\alpha^{(k)}}\>\\
&=| \Psi^{\rm prod}_{\bs\alpha^{(\ell)},\bs\alpha^{(h)},\bs\alpha^{(k)}}\>\label{eqs147}
\end{align}
where the first approximate equality follows from Lemma \ref{lem:Gaussian-disentangling for different input}, \(U_{\pi^{'-1}}\) and \(U_{\pi^{-1}}\) denote the permutation unitaries associated with the permutations \(\pi^{'-1}\) and \(\pi^{-1}\), respectively, \(\map N^* \subset [m]\) denotes the set of modes associated with non-Gaussian single-mode unitaries, namely, \(\pi^{-1} j\in \map N^*\) if and only if \(W_j\) is non-Gaussian, \(\map G^*=[m]\setminus \map N^*\) denotes the complementary set of modes, $\{U^{\bs\alpha^{(\ell)},\bs\alpha^{(h)}}_{R(-\vartheta_j)}\}$ denotes single-mode phase rotation unitaries, $U_{\map G^*}^{\bs\alpha^{(\ell)},\bs\alpha^{(h)}}$ denotes a Gaussian unitary acting on $\map G^*$, the second approximate equality uses Theorem \ref{thm:gauge-general-gaussian-probes} and omit the permutation's dependence on $\bs\alpha^{(\ell)}$ and $\bs\alpha^{(h)}$ because all the non-Gaussian unitaries $U_{S_{j,\bs\alpha^{(\ell)}}}^\dag W_j$ produces non-Gaussian states with unit probability (see Theorem \ref{prop:measure-zero-bargmann-multimode}), and this feature is independent of $\{U_{S_{j,\bs\alpha^{(\ell)}}}^\dag\}$ and $|\underline{\bs\alpha^{(h)}}\>$.

Then, our task is to learn 
\begin{align}\label{def:U_rest}
U_{\rm rest}=U_{{\pi^{'-1}}}
\left[
\left(\bigotimes_{j\in \mathcal N^*} U^{\bs\alpha^{(\ell)},\bs\alpha^{(h)}}_{R(-\vartheta_j)}\right)
\otimes
U_{\map G^*}^{\bs\alpha^{(\ell)},\bs\alpha^{(h)}}
\right]
U_{\pi^{-1}} \left(\bigotimes_{j=1}^m U_{ S_{j,{\bs \alpha}^{(\ell)}}}^\dag \right)\left(\bigotimes_{i=1}^m W_i\right) U_{O},
\end{align}
and reconstruct the original unitary as 
\begin{align}
\widetilde U_{\rm ge}=U_{\widetilde S_{\bs\alpha^{(\ell)}}} U_{\widetilde O_{2,\bs\alpha^{(l)},\bs\alpha^{(h)}}} \widetilde U_{\rm rest},  
\end{align}
where $\widetilde U_{\rm rest}$ denotes the reconstructed unitary after the learning process for  $U_{\rm rest}$.

\subsubsection{Error-independent identification of the non-Gaussian output modes}

Because of Theorem \ref{prop:measure-zero-bargmann-multimode}, the state \( |\Psi^{\rm prod}_{\bs\alpha^{(\ell)},\bs\alpha^{(h)},\bs\alpha^{(k)}}\rangle =U_{\rm rest}|\underline{\bs\alpha^{(k)}}\>\) is always a product of a \( |\map G^*| \)-mode Gaussian state and \( |\map N^*| \) single-mode non-Gaussian states, conditioned on two unknown permutation processes. In particular, for any output mode that is non-Gaussian and decoupled from the other modes under a given probe coherent state \(\ket{\underline{\bs\alpha^{(k)}}}\), the same mode remains non-Gaussian and decoupled with probability one when the probe is changed randomly to another coherent state \(\ket{\underline{\bs\alpha^{(k')}}}\).  Therefore, we can implement the following steps: 
\begin{enumerate}
\item[(RE1)] \emph{Input state preparation and probing.} Prepare \(M_{k'}\) copies of each distinct coherent probe state
\(\ket{\underline{\bs\alpha^{(k')}}}\) for \(k'=1,\dots,K_{\rm wid}\) with \(K_{\rm wid}
=
\left\lceil
\log(1/\delta_{\rm wid})/\log 4
\right\rceil\). Send them through
\(U_{\rm rest}\). For each \(k'\in\{1,\dots,K_{\rm wid}\}\), the coherent amplitudes
\(\{\alpha_j^{(k')}\}_{j=1}^m\) of \(\ket{\underline{\bs\alpha^{(k')}}}\) are sampled
independently and uniformly from the disk
\begin{align}
\{\alpha\in\mathbb C \mid |\alpha|^2\le E_{\rm prep}\},
\end{align}
where \(\alpha_j^{(k')}\) denotes the amplitude of the \(j\)-th mode, \(E_{\rm prep}\) is a photon-number bound for the unscaled
coherent base probes.
\item[(RE2)] \emph{Local tomography for reduced states.} Implement local tomography \cite{lvovsky2009continuous,becker2024classical,mele2025learning} to reconstruct reduced states \(\{\widetilde \rho_j^{(k')}\}\) of \( |\Psi^{\rm prod}_{\bs\alpha^{(\ell)},\bs\alpha^{(h)},\bs\alpha^{(k')}}\rangle \) up to an error bound: 
\begin{align}
\frac 1 2 \left\|\widetilde \rho_j^{(k')} - \Tr_{j'\neq j}\left[|\Psi^{\rm prod}_{\bs\alpha^{(\ell)},\bs\alpha^{(h)},\bs\alpha^{(k')}}\rangle\<\Psi^{\rm prod}_{\bs\alpha^{(\ell)},\bs\alpha^{(h)},\bs\alpha^{(k')}}|\right]\right\|_1 \le \epsilon_j',\ \ \ \ \forall j=1,\cdots,m.
\end{align}

\item[(RE3)]
\emph{An error-independent identification of the non-Gaussian output modes.}

We identify the modes whose reduced states remain close to pure states for all probe choices \(k'=1,\dots,K\). In particular, we define the minimal set of approximate pure reduced states 
\begin{align}
\map X_{\rm pure}
:=&
\bigcap_{k'=1}^{K}
\left\{
j\in [m]
\,\middle|\,
\left\|\widetilde \rho_j^{(k')}\right\|_\infty 
\ge
1-\epsilon_j'
\right\}\\
\equiv& \bigcap_{k'=1}^{K}
\left\{
j\in [m]
\,\middle|\,
\min_{\ket{\phi}\in\mathcal H,\ \|\phi\|=1}
\frac12
\left\|
\widetilde \rho_j^{(k')}-|\phi\>\<\phi|
\right\|_1
\le
\epsilon_j'
\right\}.
\label{eq:Xpure-definition}
\end{align}
In this way, by classically computing the operator norm of each reconstructed reduced state, one can directly obtain its distance to the set of pure states and determine whether it should be included in \(\map X_{\rm pure}\).

As illustrated in Fig.~\ref{fig:diagram_Psi_prod}, whenever an output mode is found to have a reduced state that remains pure, it can originate from one of three sources: 
\emph{(i) a correctly identified single-mode non-Gaussian state, 
(ii) a correctly identified single-mode Gaussian state, or 
(iii) a misidentified mixed reduced state from the multimode Gaussian state}. Therefore, every output mode whose output state factorizes as a single-mode non-Gaussian pure state tensored with the remaining system must be included in \(\map X_{\rm pure}\), even with tomography errors $\{\epsilon_j'\}$.

Naively, we can also detect the non-Gaussian output modes by evaluating the Gaussinity of the reduced states. Here, a detailed comparison between this classification method and the approach based on Gaussianity test is given in Remark \ref{rem: Purity test vs Gaussianity test} and Fig.~\ref{fig:diagram_Psi_prod}. 
\end{enumerate}

\begin{figure}[t]
\centering\includegraphics[width=0.58\textwidth,trim=100 0 0 0,clip,angle=0]{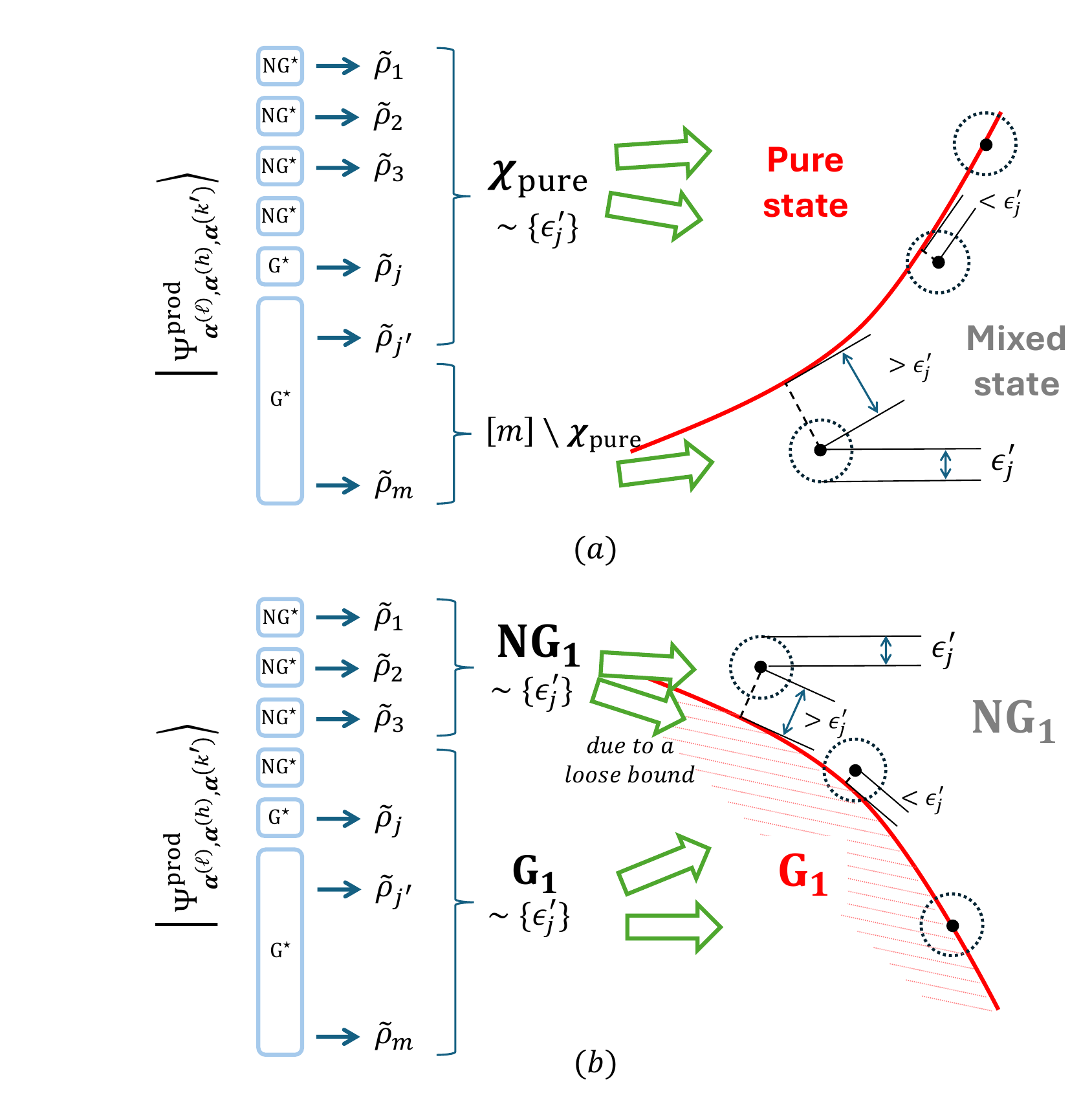}
\caption{\textbf{Identification of non-Gaussian output modes from local tomography.}
Under the perturbation assumption, the state
\(\ket{\Psi^{\rm prod}_{\bs\alpha^{(\ell)},\bs\alpha^{(h)},\bs\alpha^{(k')}}}\)
is exactly a product of \(|\map N_{\rm re}^\star|\) single-mode non-Gaussian states and one
\(|\map G_{\rm re}^\star|\)-mode Gaussian state. Blocks labeled by \(\mathrm{NG}^*\) denote the non-Gaussian states within
\(\ket{\Psi^{\rm prod}_{\bs\alpha^{(\ell)},\bs\alpha^{(h)},\bs\alpha^{(k')}}}\),
whereas blocks labeled by \(\mathrm{G}^*\) denote Gaussian states. Local tomography on the reduced states
\(\{\rho_j\}_{j=1}^m\) yields reconstructed single-mode states
\(\{\widetilde\rho_j\}_{j=1}^m\), where each \(\widetilde\rho_j\) is accurate up to trace-distance error \(\epsilon_j'\). In the trace-distance geometry of state space, each \(\widetilde\rho_j\) can be viewed as the center of an \(\epsilon_j'\)-ball.
(a) \emph{Purity test.} The red curve denotes the set of all pure states.
Using Eq.~(\ref{eq:Xpure-definition}), we identify the mode set \(\map X_{\rm pure}\). Every output mode carrying a non-Gaussian pure state, marked by \(\mathrm{NG}^*\), must belong to \(\map X_{\rm pure}\). 
(b) \emph{Gaussianity test.}
The red shaded region denotes the set of single-mode Gaussian states \(\mathsf G_1\). For each \(j=1,\dots,m\), we compute the relative entropy from
\(\widetilde\rho_j\) to the Gaussian state with the same displacement vector
and covariance matrix, which yields a loose bound on its trace distance to
the set of single-mode Gaussian states. We then assign \(j\) to
\(\map N_{\rm re}\) or \(\map G_{\rm re}\) accordingly. In this step, some single-mode non-Gaussian states may be misidentified as Gaussian, whereas every mode belonging to the multimode Gaussian part is always classified correctly.}\label{fig:diagram_Psi_prod}
\end{figure}

\begin{remark}[Gaussianity test does not work]\label{rem: Purity test vs Gaussianity test}
In addition to the non-Gaussian state classification method shown in step (RE3), we can also directly detect the Gaussianity of the reduced states of \(\ket{\Psi^{\rm prod}_{\bs\alpha^{(\ell)},\bs\alpha^{(h)},\bs\alpha^{(k')}}}\). For each reconstructed reduced state \(\widetilde\rho_j\), let
\(\tau_j\in\mathsf G_1\) denote the associate Gaussian state having the same
displacement vector and covariance matrix as \(\widetilde\rho_j\).
By Ref.~\cite{marian2013relative}, this state is the unique optimizer of the
relative-entropy projection onto the single-mode Gaussian manifold, namely
\begin{align}
\Delta_{{\rm re},j}
:=
\inf_{\sigma\in\mathsf G_1}
D\!\left(\widetilde\rho_j\middle\|\sigma\right)
=
D\!\left(\widetilde\rho_j\middle\|\tau_j\right).
\end{align}
Furthermore, by the quantum Pinsker inequality~\cite{hiai1981sufficiency}, we have 
\begin{align}
\inf_{\sigma\in\mathsf G_1}
\frac12\|\widetilde\rho_j-\sigma\|_1
\le
\frac12\|\widetilde\rho_j-\tau_j\|_1
\le
\sqrt{\frac{\Delta_{{\rm re},j}}{2}},
\end{align}
where \(D(\cdot\|\cdot)\) is defined with the natural logarithm.
Therefore, we can set the threshold as follows
\begin{align}
\Delta_{{\rm re},j}\le 2(\epsilon_j')^2,
\end{align}
which indicates the condition
\begin{align}
\inf_{\sigma\in\mathsf G_1}
\frac12\|\widetilde\rho_j-\sigma\|_1
\le
\epsilon_j'.
\end{align}
If the above condition is satisfied, we assign \(j\in \map G_{\rm re}\). Otherwise, we place \(j\) into the remaining set $\map N_{\rm re}:=[m]\setminus \map G_{\rm re}$. The procedure for identifying the modes of \(\map N_{\rm re}\) and \(\map G_{\rm re}\) is illustrated in Fig.~\ref{fig:diagram_Psi_prod}. (b). Upon identifying the Gaussianity of all reduced states \(\{\rho_j\}\) of
\(
\ket{\Psi^{\rm prod}_{\bs\alpha^{(\ell)},\bs\alpha^{(h)},\bs\alpha^{(k)}}},
\)
only three cases can arise: (i) a non-Gaussian reduced state is correctly identified as non-Gaussian, (ii) a non-Gaussian reduced state is misidentified as Gaussian, or (iii) a reduced state from the multimode Gaussian part is identified correctly.  Nevertheless, because the bound converting relative entropy into trace distance is loose, some non-Gaussian states that remain well separated from the Gaussian set in trace distance may be classified as Gaussian.

Next, we relate the observed sets of reduced states \(\map N_{\rm re}\) and \(\map G_{\rm re}\) to the true partition induced by the exact factorization of
\(
\ket{\Psi^{\rm prod}_{\bs\alpha^{(\ell)},\bs\alpha^{(h)},\bs\alpha^{(k)}}},
\) denoted by the partition
\((\map N_{\rm re}^\star,\map G_{\rm re}^\star)\). First note that, reduced states alone do not suffice to determine whether a multimode state is Gaussian. For example, the non-Gaussian state
\(
\ket{\psi_\phi}
=
\sqrt{1-\lambda^2}\sum_{n\ge 0}\lambda^n e^{i\phi n^2}\ket{n,n}
\)
with \(\phi\neq 0\) and the Gaussian two-mode squeezed vacuum
\(
\ket{\psi_{\rm TMSV}}
=
\sqrt{1-\lambda^2}\sum_{n\ge 0}\lambda^n \ket{n,n}
\)
have exactly the same single-mode reduced states. While, within the present structural class, the only possible ambiguity in \(\map G_{\rm re}\) is the inclusion of some misidentified single-mode non-Gaussian states, and no genuinely multimode non-Gaussian component can be hidden inside the subsystem indexed by \(\map G_{\rm re}\). Nevertheless, the corresponding single unitaries might still be away from an approximate Gaussian unitaries, although each reduced state $\{\tilde \rho_j'\}$ of
\(\ket{\widetilde \Psi^{\rm Gau}_{\bs\alpha^{(\ell)},\bs\alpha^{(h)},\bs\alpha^{(k)}}}\) are close to Gaussian states. For instance, we could have a non-Gaussian unitaries $e^{i\theta (a^\dag a)^2}D(-\gamma)$ that transforms a fixed coherent state to a Gaussian state. 

In the next subsection, we introduce a protocol that first learns and
counter-rotates the single-mode non-Gaussian unitaries. By contrast, a
protocol based on the Gaussianity test would generally require an additional
round of learning and counter-rotation, since some non-Gaussian output modes
may be misclassified as belonging to the Gaussian sector. The purity test
avoids this issue. Despite tomography errors, it always includes all
non-Gaussian output modes in \(\map X_{\rm pure}\), making
\(\map X_{\rm pure}\) a reliable starting point for the subsequent steps.
\end{remark}

In the next subsection, we will introduce a protocol to partially learn the non-Gaussian local unitaries associated with $\map X_{\rm pure}$.

\subsubsection{Single-mode unitary reconstruction from state tomography}\label{supp:Single-mode unitary reconstruction from state tomography}

Naively, one might hope to learn the local non-Gaussian unitaries and the multimode Gaussian unitary of $U_{\rm rest}U_O^\dag$ in Eq. (\ref{def:U_rest}) efficiently by preparing different coherent probe states to have different $\{\tilde \rho_j\}$. However, the first passive-Gaussian layer $U_O$ obstructs this strategy, because it mixes the input modes and thereby makes the effective coherent inputs seen by each local non Gaussian unitary and by the Gaussian multimode unitary of $U_{\rm rest}U_O^\dag$ completely unknown. Thus, a natural question arises as follows
\begin{quote}
\emph{whether it is possible to reconstruct a non-Gaussian unitary from unknown coherent probe states.}
\end{quote}

In this subsection, we will show that, while the unknown coherent input states prevent identification of the unitary itself, it does not prevent reconstruction of the unitary modulo the natural ambiguity given by a phase rotation. More precisely, we will show that, for an arbitrary unitary channel $\map W$, there exists a learning protocol which, using experiments with distinct coherent probe states, outputs a unitary channel $\map V$ such that the distance below between $\map W$ and $\map V$ vanishes as the round of experiments increases, 
\begin{align}
d_{\diamond,E}^{\mathrm{coh}}(\mathcal W,\mathcal V)
:=
\inf_{\theta\in\mathbb R}
\bigl\|
\mathcal W-\mathcal V\circ \mathcal C_{\theta}
\bigr\|_{\diamond}^{E},\label{eqs158}
\end{align}
where $\mathcal C_{\theta}(\rho)=e^{-i\theta a^\dagger a}\rho\, e^{i\theta a^\dagger a}$ refers to a phase-rotation  channel. Here, the learner cannot measure the input coherent state.  Instead, he can prepare coherent states  $|\underline{s e^{i\vartheta}\alpha}\rangle$ and $\alpha=\sqrt{\mu}\,e^{i\theta_0}$ for chosen \(s\ge0\) and
\(\vartheta\in\mathbb R\), where both \(\mu>0\) and \(\theta_0\) are unknown.

Specifically, given an unknown single-mode unitary $W$ and a set of coherent probe states $\{|\underline{\alpha_k}\>\}_{k=1}^K$, we consider the following learning steps:

\begin{enumerate}
\item[(STU1)] \emph{Calibration of input intensity.}
The learner first estimates the unknown intensity
\(\mu=|\alpha|^2\) without measuring the input probes.  For a finite
calibration set \(\mathcal S_{\rm cal}\subset(0,\infty)\), the learner prepares input coherent states $|0\rangle$ 
and $|\underline{s\alpha}\rangle$ for $s\in\mathcal S_{\rm cal}$, sends them through \(W_\star\), and tomographically estimates the output
overlaps $F_s
:=
\left|
\langle 0|W_\star^\dag W_\star|\underline{s\alpha}\rangle
\right|^2$. By unitarity, we have $F_s
=
|\langle 0|\underline{s\alpha}\rangle|^2
=
e^{-s^2\mu}$. Thus, once the calibration set contains a scale with \(s^2\mu=\Theta(1)\), the
quantity \(-\log F_s/s^2\) gives a stable estimate of \(\mu\).  The unknown
phase \(\theta_0\) is not estimated.

\item[(STU2)] \emph{Fourier-ring probe generation.}
Using the calibrated intensity \(\widehat\mu\), the learner chooses $s_q:=\sqrt{\frac{q}{\widehat\mu}}$ and $\theta_\ell:=\frac{2\pi\ell}{L}$, and prepares the coherent probes
\begin{align}
|\underline{s_q e^{i\theta_\ell}\alpha}\rangle
=
\left|\underline{
\sqrt{\lambda q}\,
e^{i(\theta_0+\theta_\ell)}}
\right\rangle,
\qquad
q=1,\ldots,J,\quad \ell=0,\ldots,L-1,
\end{align}
where $\lambda:=\frac{\mu}{\widehat\mu}$ is close to \(1\) on the calibration-successful event.  The vacuum probe
\(|0\rangle\) is also used.  Hence the actual input amplitudes remain unknown,
but their relative Fourier-ring geometry is known up to a controlled radial
calibration error.

\item[(STU3)] \emph{Tomography of all output states.}
Each probe is sent through the unknown unitary: $|\psi_0\rangle:=W_\star|0\rangle$ and $|\psi_{q,\ell}\rangle
:=
W_\star
\left|\underline{
\sqrt{\lambda q}\,
e^{i(\theta_0+\theta_\ell)}}
\right\rangle$. Using the output-state tomography primitive under the stated output energy
bound, the learner obtains pure-state estimates
\begin{align}
|u_0\rangle\langle u_0|
\approx
|\psi_0\rangle\langle\psi_0|,
\qquad
|u_{q,\ell}\rangle\langle u_{q,\ell}|
\approx
|\psi_{q,\ell}\rangle\langle\psi_{q,\ell}|.
\end{align}

\item[(STU4)] \emph{Local phase synchronization.}
Tomography determines each output state only up to an arbitrary phase.  These
phases are synchronized using a local graph on the Fourier-ring probes: edges
connect neighboring angles, neighboring radii, and the vacuum to the first
ring.  Along every such edge, the corresponding coherent-state overlap has a
constant lower bound.  Therefore, the relative phases can be propagated across
the graph with only polynomial accumulation of error. This produces aligned representatives satisfying, for some common irrelevant
phase \(\omega\), $|\widehat\phi_0\rangle
\approx
e^{i\omega}W_\star|0\rangle$ and $|\widehat\phi_{q,\ell}\rangle
\approx
e^{i\omega}W_\star
\left|\underline{
\sqrt{\lambda q}\,
e^{i(\theta_0+\theta_\ell)}}
\right\rangle$.

\item[(STU5)] \emph{Classically approximate output states with Fock probe states.}
Define the gauge-fixed unitary $W_g:=W_\star e^{i\theta_0\hat n}$. The unknown phase \(\theta_0\) of the coherent source is therefore absorbed
into an input-side phase rotation. The vacuum gives the first column: $|\widehat v_0\rangle
:=
\Pi_{\le R}|\widehat\phi_0\rangle$. For \(q=1,\ldots,J\), the learner applies the Fourier filter
\begin{align}
|\widehat v_q\rangle
:=
\frac{1}{L\,c_q(q)}
\sum_{\ell=0}^{L-1}
e^{-iq\theta_\ell}
\Pi_{\le R}|\widehat\phi_{q,\ell}\rangle,
\qquad
c_q(q)=e^{-q/2}\frac{q^{q/2}}{\sqrt{q!}},\ \ \Pi_{\le R}:=\sum_{j=0}^{R}|j\rangle\!\langle j|.
\end{align}
The Fourier-ring identity implies $|\widehat v_q\rangle
\approx
e^{i\omega}\Pi_{\le R}W_g|q\rangle$ with $\{|q\>,q=0,\ldots,J\}$ being Fock states. Here \(J\), \(R\), and $L$ are chosen in the theorem as
\begin{align}
J=\Theta\!\left(\frac{N}{\epsilon^2}\right),
\qquad
R=\max\left\{J,\Theta\!\left(\frac{N_{\rm dyn}}{\epsilon^2}\right)\right\},\qquad
L:=\Theta\!\left(J+\log\left(\frac 1\epsilon\right)\right)
\end{align}
where $N_{\rm dyn}$ is a dynamical energy bound defined as $\sup_{\substack{
|\varphi\rangle\in\mathcal H_{\le J}\\
\|\varphi\|=1}}
\langle\varphi|W_\star^\dagger\hat n W_\star|\varphi\rangle
\le N_{\rm dyn}$, $N=E-1/2$ and $\epsilon$ were defined in Eq. (\ref{eqs158}). The normalization is stable because $c_q(q)^{-1}=O(q^{1/4})$.

\item[(STU6)] \emph{Polar completion.}
Collect the recovered columns into $\widehat A_R
:=
\bigl[
|\widehat v_0\rangle,\ldots,|\widehat v_J\rangle
\bigr]$. Take the polar part $\widehat Q_R
:=
\widehat A_R
(\widehat A_R^\dagger\widehat A_R)^{-1/2}$, extend \(\widehat Q_R\) to a unitary \(\widehat V_R\) on
\(\mathcal H_{\le R}\). In particular, since \(R\ge J\), choose an orthonormal completion
\(\{|\chi_{J+1}\rangle,\ldots,|\chi_R\rangle\}\) of
\(\{\widehat Q_R|0\rangle,\ldots,\widehat Q_R|J\rangle\}\) in
\(\mathcal H_{\le R}\), and define
\begin{align}
\widehat V_R
:=
\sum_{j=0}^{J}\widehat Q_R|j\rangle\langle j|
+
\sum_{j=J+1}^{R}|\chi_j\rangle\langle j|.
\end{align}
Finally, we can define $\widehat U:=\widehat V_R\oplus I_{>R}$ and obtain the reconstructed unitary channel   $\widehat{\mathcal W}(\rho):=\widehat U\rho\widehat U^\dagger$.

\end{enumerate}

Before showing the main theorem of efficient unitary learning, let us look at six useful lemmas, which are useful in the derivation of the corresponding query complexity. 

\begin{lemma}[Energy-constrained pure-state tomography (special cases of Theorem~S36 of \cite{mele2025learning})]
\label{lem:moment-tomography-primitive}
Let \(|\psi\rangle\) be a one-mode pure state satisfying the energy constriant $\langle \psi|\hat n|\psi\rangle\le N_{\rm out}$. For any tomography accuracy \(\varepsilon_{\rm tom}\in(0,1)\) and failure
probability \(p\in(0,1)\), there is a tomography procedure using
\begin{align}
M_{\rm tom}(N_{\rm out},\varepsilon_{\rm tom},p)
:=
\left\lceil
2^{21}
\left(
\frac{eN_{\rm out}}{\varepsilon_{\rm tom}^2}+2e
\right)
\varepsilon_{\rm tom}^{-2}
\log\frac{4}{p}
\right\rceil
\end{align}
copies of \(|\psi\rangle\), which returns a normalized vector
\(|\widetilde\psi\rangle\) satisfying $\frac12
\left\|
|\widetilde\psi\rangle\!\langle\widetilde\psi|
-
|\psi\rangle\!\langle\psi|
\right\|_1
\le \varepsilon_{\rm tom}$ with probability at least \(1-p\).
\end{lemma}

\begin{lemma}[Learning of calibrated coherent input states]
\label{lem:output-only-calibration}
Consider the single-mode states generated by applying the same unitary on vacuum or
coherent states $|\psi_0\rangle := W_\star |0\rangle,$ $|\psi_s\rangle := W_\star |\underline{s\alpha}\rangle,$ $|\alpha|=\sqrt{\mu}$, for calibration scales \(s\in\mathcal S_{\rm cal}\subset(0,\infty)\), where
\(\mathcal S_{\rm cal}\) is finite and contains at least one \(s\) satisfying $\frac14\le s^2\mu\le 1$. Assume the energy constraints $\langle \psi_0|\hat n|\psi_0\rangle\le N_{\rm out}$ and $\langle \psi_s|\hat n|\psi_s\rangle\le N_{\rm out}$ for $s\in\mathcal S_{\rm cal}$.

Then, there is a learning
protocol based on output tomography such that, with probability at least
\(1-p_{\rm fail}\), it returns \(\widehat\mu\) satisfying
\begin{align}
|\widehat\mu-\mu|\le \zeta\mu ,\ \ \ \zeta\in(0,1/2).
\end{align}
Its number of channel uses is at most
\begin{align}
(1+|\mathcal S_{\rm cal}|)
M_{\rm tom}\!\left(
N_{\rm out},
\frac{\zeta}{128e^2},
\frac{p_{\rm fail}}{1+|\mathcal S_{\rm cal}|}
\right).
\end{align}
\end{lemma}

\begin{proof} From state tomography, we can reconstruct \(|\psi_0\rangle\) and each
\(|\psi_s\rangle\), \(s\in\mathcal S_{\rm cal}\), to trace-distance accuracy
\begin{align}
\varepsilon_{\rm cal}:=\frac{\zeta}{128e^2}.
\end{align}
More explicitly, we have the following bounds on the tomography-successful event.
\begin{align}
\begin{cases}
\frac12
\left\|
|\widetilde\psi_0\rangle\!\langle\widetilde\psi_0|
-
|\psi_0\rangle\!\langle\psi_0|
\right\|_1
\le \varepsilon_{\rm cal},
\\[0.6em]
\frac12
\left\|
|\widetilde\psi_s\rangle\!\langle\widetilde\psi_s|
-
|\psi_s\rangle\!\langle\psi_s|
\right\|_1
\le \varepsilon_{\rm cal},
\qquad
s\in\mathcal S_{\rm cal}.
\end{cases}
\end{align}
By the union bound and Lemma~\ref{lem:moment-tomography-primitive}, all these
tomography calls succeed with probability at least \(1-p_{\rm fail}\).

For each \(s\in\mathcal S_{\rm cal}\), let us denote the estimated fidelity
 $\widehat F_s:=|\langle \widetilde\psi_0|\widetilde\psi_s\rangle|^2$ and its true value $F_s
:=
|\langle \psi_0|\psi_s\rangle|^2
=
|\langle 0|\underline{s\alpha}\rangle|^2
=
e^{-s^2\mu}$, where we used the unitarity of \(W_\star\). On the tomography-successful event, we have
\begin{align}
|\widehat F_s-F_s|
&=
\left|
|\langle \widetilde\psi_0|\widetilde\psi_s\rangle|^2
-
|\langle \psi_0|\psi_s\rangle|^2
\right|
\\
&=
\Bigl|
\Tr\!\left[
\left(
|\widetilde\psi_0\rangle\!\langle\widetilde\psi_0|
-
|\psi_0\rangle\!\langle\psi_0|
\right)
|\widetilde\psi_s\rangle\!\langle\widetilde\psi_s|
\right]
\\
&\qquad+
\Tr\!\left[
|\psi_0\rangle\!\langle\psi_0|
\left(
|\widetilde\psi_s\rangle\!\langle\widetilde\psi_s|
-
|\psi_s\rangle\!\langle\psi_s|
\right)
\right]
\Bigr|
\\
&\le
\left\|
|\widetilde\psi_0\rangle\!\langle\widetilde\psi_0|
-
|\psi_0\rangle\!\langle\psi_0|
\right\|_1
+
\left\|
|\widetilde\psi_s\rangle\!\langle\widetilde\psi_s|
-
|\psi_s\rangle\!\langle\psi_s|
\right\|_1
\\
&\le
4\varepsilon_{\rm cal},
\label{eqs198}
\end{align}
where we used H\"older's inequality. For each \(s\) with \(\widehat F_s>0\), we can define an estimate of the exponent
\begin{align}
\widehat x_s:=-\log \widehat F_s .
\end{align}
We select any scale \(\widehat s\in\mathcal S_{\rm cal}\) satisfying
\begin{align}
\widehat x_{\widehat s}\in\left[\frac18,2\right].
\label{eqs200}
\end{align}
Such a scale exists on the tomography-successful event. Indeed, by the
assumption on \(\mathcal S_{\rm cal}\), there exists
\(s^\circ\in\mathcal S_{\rm cal}\) such that $\frac14\le s^{\circ\,2}\mu\le 1$. If we set $x_{s^\circ}:=s^{\circ\,2}\mu$, we have $x_{s^\circ}\in\left[\frac14,1\right]$ and $F_{s^\circ}=e^{-x_{s^\circ}}\in[e^{-1},e^{-1/4}]$. Using Eq.~\eqref{eqs198} and the assumption $4\varepsilon_{\rm cal}
=
\frac{\zeta}{32e^2}
<
\frac{1}{64e^2}$, we obtain
\begin{align}
\widehat F_{s^\circ}
&\ge
e^{-1}-\frac{1}{64e^2}
> e^{-2},
\\
\widehat F_{s^\circ}
&\le
e^{-1/4}+\frac{1}{64e^2}
\le e^{-1/8}.
\end{align}
Therefore, we have $\widehat x_{s^\circ}
=
-\log \widehat F_{s^\circ}
\in
\left[\frac18,2\right]$. Hence \(s^\circ\) itself satisfies the selection rule, so the desired scale
\(\widehat s\) exists.

Now define the estimates $\widehat\mu:=\frac{\widehat x_{\widehat s}}{\widehat s^2}$ and $x_{\widehat s}:=\widehat s^{\,2}\mu$. For the selected scale \(\widehat s\), Eq.~\eqref{eqs200} implies
\begin{align}
\widehat F_{\widehat s}
=
e^{-\widehat x_{\widehat s}}
\ge e^{-2}.
\end{align}
Together with Eq.~\eqref{eqs198} and $4\varepsilon_{\rm cal}
=
\frac{\zeta}{32e^2}
<
\frac{1}{64e^2}
=
\frac{e^{-2}}{64}$, we get
\begin{align}
F_{\widehat s}
\ge
\widehat F_{\widehat s}-4\varepsilon_{\rm cal}
\ge
e^{-2}-\frac{e^{-2}}{64}
>
\frac12 e^{-2}.
\end{align}
Given also the condition \(\widehat F_{\widehat s}\ge e^{-2}\), we have
\begin{align}
\min\{F_{\widehat s},\widehat F_{\widehat s}\}
\ge
\frac12 e^{-2}.
\end{align}
By the mean-value theorem applied to the map \(x\mapsto \log x\), we have 
\begin{align}
|\widehat x_{\widehat s}-x_{\widehat s}|
&=
|\log F_{\widehat s}-\log\widehat F_{\widehat s}|
\\
&\le
\frac{
|\widehat F_{\widehat s}-F_{\widehat s}|
}{
\min\{F_{\widehat s},\widehat F_{\widehat s}\}
}
\\
&\le
2e^2\cdot 4\varepsilon_{\rm cal}
=
8e^2\varepsilon_{\rm cal}
=
\frac{\zeta}{16}.
\end{align}
Moreover, since \(\widehat x_{\widehat s}\ge 1/8\), we have
\begin{align}
x_{\widehat s}
\ge
\widehat x_{\widehat s}
-
|\widehat x_{\widehat s}-x_{\widehat s}|
\ge
\frac18-\frac{\zeta}{16}
\ge
\frac{3}{32},
\end{align}
where the last inequality uses \(\zeta<1/2\). Consequently, we have 
\begin{align}
\frac{|\widehat\mu-\mu|}{\mu}
&=
\frac{
\left|
\frac{\widehat x_{\widehat s}}{\widehat s^2}
-
\frac{x_{\widehat s}}{\widehat s^2}
\right|
}{
\frac{x_{\widehat s}}{\widehat s^2}
}
\\
&=
\frac{|\widehat x_{\widehat s}-x_{\widehat s}|}{x_{\widehat s}}
\\
&\le
\frac{\zeta/16}{3/32}
=
\frac{2\zeta}{3}
\le
\zeta .
\end{align}
Thus, we have $|\widehat\mu-\mu|
\le
\zeta\mu$.
\end{proof}

\begin{lemma}[Fourier-ring synthesis of a Fock vector]
\label{lem:output-only-fourier-synthesis}
The Fourier-ring superposition of coherent states
\begin{align}
|\Phi_{q,L}(\lambda,\theta_0)\rangle
:=
\frac{1}{L\,c_q(q)}
\sum_{\ell=0}^{L-1}
e^{-iq\theta_\ell}
\left|\underline{
\sqrt{\lambda q}\,
e^{i(\theta_0+\theta_\ell)}}
\right\rangle ,
\qquad
c_q(q):=e^{-q/2}\frac{q^{q/2}}{\sqrt{q!}},\ \ \theta_\ell:=\frac{2\pi\ell}{L},\ \ \ell=0,\ldots,L-1,
\end{align}
synthesizes the Fock state \(|q\rangle\), up to an
unknown coherent-gauge phase \(e^{iq\theta_0}\), with error
\begin{align}
\left\|
|\Phi_{q,L}(\lambda,\theta_0)\rangle
-
e^{iq\theta_0}|q\rangle
\right\|_2
\le
q|\lambda-1|^2+\varepsilon_{\rm al}.
\end{align}

Here, we assume \(q\ge1\), \(\lambda\in[1/2,2]\), 
\(\varepsilon_{\rm al}\in(0,1)\), and $L
\ge
(2e-3)q
+
2\log\frac1{\varepsilon_{\rm al}}
+
\frac12\log(q+1)
+
1$. For all \(q\le J\), it is enough to choose $L\ge C\left(J+\log\frac1{\varepsilon_{\rm al}}\right)$ with a universal constant \(C>0\).
\end{lemma}

\begin{proof}
The idea is that the angular Fourier average selects the \(q\)-th Fock
coefficient of a coherent state.  The only remaining errors are the radius
mismatch \(\lambda\neq 1\) and the finite Fourier aliasing terms. Recall the coherent state has an expansion
\begin{align}
\left|\underline{
\sqrt{\lambda q}\,
e^{i(\theta_0+\theta_\ell)}}
\right\rangle
=
\sum_{n=0}^{\infty}
c_n(\lambda q)\,
e^{in(\theta_0+\theta_\ell)}
|n\rangle ,
\end{align}
where $\{c_n(x):=e^{-x/2}\frac{x^{n/2}}{\sqrt{n!}}\}$ denote the amplitudes in Fock basis. Substituting this expansion into the definition of
\(|\Phi_{q,L}(\lambda,\theta_0)\rangle\), we obtain
\begin{align}
|\Phi_{q,L}(\lambda,\theta_0)\rangle&=\frac{1}{L\,c_q(q)}
\sum_{\ell=0}^{L-1}
e^{-iq\theta_\ell}
\left|\underline{
\sqrt{\lambda q}\,
e^{i(\theta_0+\theta_\ell)}}
\right\rangle\\
&=
\sum_{n=0}^{\infty}
\frac{c_n(\lambda q)}{c_q(q)}
e^{in\theta_0}
\left(
\frac1L
\sum_{\ell=0}^{L-1}
e^{i(n-q)\theta_\ell}
\right)
|n\rangle .
\end{align}
Given the relation \(\theta_\ell=2\pi\ell/L\), the discrete Fourier average satisfies
\begin{align}
\frac1L
\sum_{\ell=0}^{L-1}
e^{i(n-q)\theta_\ell}
=
\begin{cases}
1, & n\equiv q \pmod L,\\
0, & n\not\equiv q \pmod L .
\end{cases}
\end{align}
Thus the only Fock levels that survive are $n=q,\ q+L,\ q+2L,\ldots$. Therefore, we have 
\begin{align}
|\Phi_{q,L}(\lambda,\theta_0)\rangle
=
\frac{c_q(\lambda q)}{c_q(q)}
e^{iq\theta_0}|q\rangle
+
\sum_{t\ge1}
\frac{c_{q+tL}(\lambda q)}{c_q(q)}
e^{i(q+tL)\theta_0}
|q+tL\rangle .
\end{align}

We now compare this vector with \(e^{iq\theta_0}|q\rangle\).  Since the Fock
vectors are orthonormal, the error is controlled by two terms:
\begin{align}
\left\|
|\Phi_{q,L}(\lambda,\theta_0)\rangle
-
e^{iq\theta_0}|q\rangle
\right\|_2
\le
\left|
\frac{c_q(\lambda q)}{c_q(q)}-1
\right|
+
\left(
\sum_{t\ge1}
\left|
\frac{c_{q+tL}(\lambda q)}{c_q(q)}
\right|^2
\right)^{1/2}.
\end{align}

We first bound the main coefficient.  A direct calculation gives
\begin{align}
A=\log\frac{c_q(\lambda q)}{c_q(q)}
=
\frac q2\bigl(\log\lambda-(\lambda-1)\bigr).\label{eqs221}
\end{align}
Within the range \(\lambda\in[1/2,2]\), we can control this term explicitly.  Let us denote $f(\lambda):=\log\lambda-(\lambda-1)$. Then, we have 
\begin{align}
f(1)=0,\qquad f'(1)=0,\qquad f''(\lambda)=-\frac1{\lambda^2}.
\end{align}
By the Lagrange remainder of  Taylor's theorem \cite{zorich2016mathematical}, for some \(\xi\) between \(1\) and \(\lambda\), we have 
\begin{align}
f(\lambda)
=
-\frac{(\lambda-1)^2}{2\xi^2}.
\end{align}
Since \(\xi\in[1/2,2]\), we have \(1/\xi^2\le 4\). Therefore, we have 
\begin{align}
|\log\lambda-(\lambda-1)|
=
|f(\lambda)|
\le
2|\lambda-1|^2 .
\end{align}
Consequently, we have 
\begin{align}
\left|
\log\frac{c_q(\lambda q)}{c_q(q)}
\right|
\le
q|\lambda-1|^2 .
\end{align}
Since \(\log\lambda\le \lambda-1\) for every \(\lambda>0\), we have $A\le 0$. Hence, we have 
\begin{align}
0<
\frac{c_q(\lambda q)}{c_q(q)}
=
e^A
\le 1,
\end{align}
and therefore
\begin{align}
\left|
\frac{c_q(\lambda q)}{c_q(q)}-1
\right|
=
1-e^A
\le
-A
=
|A|.
\end{align}
Using the previous bound $|A|
\le
q|\lambda-1|^2$, we obtain
\begin{align}
\left|
\frac{c_q(\lambda q)}{c_q(q)}-1
\right|
\le
q|\lambda-1|^2 .
\end{align}

It remains to bound the aliasing tail.  Recall that we have $|c_n(x)|^2
=
e^{-x}\frac{x^n}{n!}$. This is exactly the probability mass function of a Poisson random variable with
mean \(x\).  Hence, if we have \(Z_x\sim{\rm Pois}(x)\), then 
$|c_n(x)|^2
=
\Pr[Z_x=n]$. Therefore, we have 
\begin{align}
\sum_{t\ge1}
|c_{q+tL}(\lambda q)|^2
&\le
\sum_{n\ge q+L}
|c_n(\lambda q)|^2
\\
&=
\Pr[Z_{\lambda q}\ge q+L].
\end{align}
Since \(\lambda\le2\), we have \(\lambda q\le2q\).  The Poisson tail is
monotone increasing in its mean: if \(Z_{\lambda q}\sim{\rm Pois}(\lambda q)\)
and \(Z'\sim{\rm Pois}(2q-\lambda q)\) are independent, then
\(Z_{\lambda q}+Z'\sim{\rm Pois}(2q)\), and hence we have 
\begin{align}
\Pr[Z_{\lambda q}\ge q+L]
\le
\Pr[Z_{2q}\ge q+L].
\end{align}

We now bound the latter tail directly.  Let \(Z_{2q}\sim{\rm Pois}(2q)\).  By
Markov's inequality applied to \(e^{Z_{2q}}\), we have 
\begin{align}
\Pr[Z_{2q}\ge q+L]
&=
\Pr[e^{Z_{2q}}\ge e^{q+L}]
\\
&\le
e^{-(q+L)}\mathbb E[e^{Z_{2q}}].
\end{align}
Since \(Z_{2q}\sim{\rm Pois}(2q)\), we have $\Pr[Z_{2q}=k]
=
e^{-2q}\frac{(2q)^k}{k!}$ for $k=0,1,2,\ldots$. Therefore, we have 
\begin{align}
\mathbb E[e^{Z_{2q}}]
&=
\sum_{k=0}^{\infty}
e^k\,\Pr[Z_{2q}=k]
\\
&=
\sum_{k=0}^{\infty}
e^k e^{-2q}\frac{(2q)^k}{k!}
\\
&=
e^{-2q}
\sum_{k=0}^{\infty}
\frac{(2qe)^k}{k!}
\\
&=
e^{-2q}e^{2qe}
=
\exp\!\bigl(2q(e-1)\bigr).
\end{align}
Thus, we have 
\begin{align}
\Pr[Z_{2q}\ge q+L]
\le
\exp\!\left((2e-3)q-L\right).
\end{align}

It remains to control the normalization factor.  Recall the definition 
$|c_q(q)|^{-1}
=
e^{q/2}\frac{\sqrt{q!}}{q^{q/2}}$. We can apply the standard Stirling upper bound \cite{robbins1955remark}
\begin{align}
q!\le
\sqrt{2\pi q}\left(\frac{q}{e}\right)^q e^{1/(12q)}
\le
e\sqrt q\left(\frac{q}{e}\right)^q,
\qquad q\ge1 .
\end{align}
Taking square roots gives $\sqrt{q!}
\le
\sqrt e\, q^{1/4}
\left(\frac{q}{e}\right)^{q/2}$. Therefore, we have 
\begin{align}
|c_q(q)|^{-1}
&=
e^{q/2}\frac{\sqrt{q!}}{q^{q/2}}
\\
&\le
e^{q/2}
\frac{
\sqrt e\, q^{1/4}
\left(\frac{q}{e}\right)^{q/2}
}{
q^{q/2}
}
\\
&=
\sqrt e\, q^{1/4}
\\
&\le
\sqrt e\, (q+1)^{1/4}.\label{eqs243}
\end{align}
We now choose the ring size \(L\) so that the aliasing contribution is at most
a prescribed accuracy \(\varepsilon_{\rm al}\in(0,1)\).  Specifically, take
\(L\) large enough to satisfy
\begin{align}
L
\ge
(2e-3)q
+
2\log\frac1{\varepsilon_{\rm al}}
+
\frac12\log(q+1)
+
1 .
\label{eq:fourier-ring-L-choice}
\end{align}
Then the previously obtained Poisson tail estimate gives
\begin{align}
\Pr[Z_{2q}\ge q+L]
&\le
\exp\!\left((2e-3)q-L\right)
\\
&\le
\exp\!\left(
-2\log\frac1{\varepsilon_{\rm al}}
-\frac12\log(q+1)
-1
\right)
\\
&=
e^{-1}(q+1)^{-1/2}\varepsilon_{\rm al}^{\,2}.
\label{eq:fourier-ring-poisson-tail-final}
\end{align}

Next, we combine this tail bound with the normalization factor
\(|c_q(q)|^{-1}\).  From Eq. (\ref{eqs243}), we have 
\begin{align}
\left(
\sum_{t\ge1}
\left|
\frac{c_{q+tL}(\lambda q)}{c_q(q)}
\right|^2
\right)^{1/2}
&\le
|c_q(q)|^{-1}
\left(
\sum_{t\ge1}
|c_{q+tL}(\lambda q)|^2
\right)^{1/2}
\\
&\le
\sqrt e\,(q+1)^{1/4}
\left(
e^{-1}(q+1)^{-1/2}
\varepsilon_{\rm al}^{\,2}
\right)^{1/2}
\\
&=
\varepsilon_{\rm al}.
\label{eq:fourier-ring-alias-bound}
\end{align}
Thus the entire aliasing tail contributes at most
\(\varepsilon_{\rm al}\).

Combining this aliasing estimate with the main-coefficient bound $\left|
\frac{c_q(\lambda q)}{c_q(q)}-1
\right|
\le
q|\lambda-1|^2$, we obtain
\begin{align}
\left\|
|\Phi_{q,L}(\lambda,\theta_0)\rangle
-
e^{iq\theta_0}|q\rangle
\right\|_2
\le
q|\lambda-1|^2
+
\varepsilon_{\rm al}.
\end{align}
This proves that the Fourier-ring superposition approximates the Fock vector
\(|q\rangle\), up to the unknown phase \(e^{iq\theta_0}\), with an error
controlled by the radius mismatch \(q|\lambda-1|^2\) and the chosen aliasing
accuracy \(\varepsilon_{\rm al}\).

Finally, since
\begin{align}
\log(q+1)\le q,
\qquad q\ge1,
\end{align}
the explicit condition \eqref{eq:fourier-ring-L-choice} is implied by the
simpler scaling condition
\begin{align}
L
\ge
C\left(
q+\log\frac1{\varepsilon_{\rm al}}
\right)
\end{align}
for a sufficiently large universal constant \(C>0\).  Hence the lemma follows.
\end{proof}

\begin{lemma}[Local phase synchronization on the Fourier-ring graph]
\label{lem:output-only-phase-sync} Consider states generated by applying the same unitary on coherent states  $|\psi_0\rangle:=W_\star|0\rangle$ and $|\psi_{q,\ell}\rangle
:=
W_\star
\left|\underline{
\sqrt{\lambda q}\,
e^{i(\theta_0+\theta_\ell)}}
\right\rangle$ for  $q=1,\ldots,J,$ $
\ell=0,\ldots,L-1,$ $
\theta_\ell=\frac{2\pi\ell}{L}$, $\lambda\in[1/2,2]$, and $L\ge 8J$.

Suppose output tomography returns normalized states $|u_0\rangle$ and $|u_{q,\ell}\rangle$ such that
\begin{align}
\begin{cases}
\frac12
\left\|
|u_0\rangle\!\langle u_0|
-
|\psi_0\rangle\!\langle\psi_0|
\right\|_1
\le
\varepsilon_{\rm tom},\\
\frac12
\left\|
|u_{q,\ell}\rangle\!\langle u_{q,\ell}|
-
|\psi_{q,\ell}\rangle\!\langle\psi_{q,\ell}|
\right\|_1
\le
\varepsilon_{\rm tom}
\end{cases}
\end{align}
for all \(q,\ell\).  Assume also that the radius calibration error satisfies $|\lambda-1|\le \zeta$. Then, provided \(\varepsilon_{\rm tom}+\zeta\) is smaller than a universal
constant, there is an explicit phase-synchronization procedure which outputs
aligned states $|\widehat\phi_0\rangle$ and $|\widehat\phi_{q,\ell}\rangle$
such that, for some common phase \(\omega\in\mathbb R\), the following conditions are satisfied
\begin{align}
\begin{cases}
\left\|
|\widehat\phi_0\rangle
-
e^{i\omega}|\psi_0\rangle
\right\|_2
\le
C(J+L)(\varepsilon_{\rm tom}+\zeta)\\
\max_{q,\ell}
\left\|
|\widehat\phi_{q,\ell}\rangle
-
e^{i\omega}|\psi_{q,\ell}\rangle
\right\|_2
\le
C(J+L)(\varepsilon_{\rm tom}+\zeta)
\end{cases},
\end{align}
where \(C>0\) is a univeral constant.
\end{lemma}

\begin{proof}
Let us define the true input amplitudes $\alpha_0:=0$ and $\alpha_{q,\ell}:=
\sqrt{\lambda q}\,e^{i(\theta_0+\theta_\ell)} $. We also define the nominal amplitudes by setting \(\lambda=1\) and
dropping the unknown common phase \(\theta_0\), $\bar\alpha_0:=0$ and $\bar\alpha_{q,\ell}:=
\sqrt q\,e^{i\theta_\ell}$. The nominal overlaps $\bar g_{uv}:=\langle \underline{\bar\alpha_u}|\underline{\bar\alpha_v}\rangle$ are known to the learner, while the true overlaps are $g_{uv}:=\langle \underline{\alpha_u}|\underline{\alpha_v}\rangle
=
\langle \psi_u|\psi_v\rangle$, where the last equality follows from unitarity of \(W_\star\).

Now we synchronize phases on the graph, where vertices are $0
\ \text{and}\ 
(q,\ell)$ for $q=1,\ldots,J$ and $\ell=0,\ldots,L-1$, root edge is \(0\sim(1,0)\), the angular edges are  $(q,\ell)\sim(q,\ell+1 \!\!\!\pmod L)$, the radial edges are  $(q,\ell)\sim(q+1,\ell)$ for $q=1,\ldots,J-1$. We first show that every edge has a constant overlap lower bound.  For an
angular edge, we have 
\begin{align}
|\alpha_{q,\ell+1}-\alpha_{q,\ell}|^2
&=
2\lambda q\bigl(1-\cos(2\pi/L)\bigr)
\\
&\le
\lambda q\left(\frac{2\pi}{L}\right)^2
\le
\frac{8\pi^2 J}{L^2}=\frac{\pi^2}{8J}
\le
\frac{\pi^2}{8},
\end{align}
where we used \(\lambda\le2\) and \(L\ge8J\).  For a radial edge, we have 
\begin{align}
|\alpha_{q+1,\ell}-\alpha_{q,\ell}|^2
=
\lambda(\sqrt{q+1}-\sqrt q)^2
\le 2.
\end{align}
For the root edge \(0\sim(1,0)\), we have
\begin{align}
|\alpha_{1,0}-\alpha_0|^2=\lambda\le 2.
\end{align}
Since coherent-state overlaps satisfy the condition $|\langle \underline{\alpha}|\underline{\beta}\rangle|
=
e^{-|\alpha-\beta|^2/2}$, there is a universal constant \(c_0>0\) such that
\begin{align}
|g_{uv}|\ge c_0
\end{align}
for every edge \(u\sim v\).

The learner uses the nominal overlaps obtained by setting \(\lambda=1\).  We
now show that these nominal overlaps are close to the true input overlaps when
\(|\lambda-1|\le \zeta\). For every vertex \(v\), let \(\bar\alpha_v\) denote the nominal amplitude,
namely $\bar\alpha_0:=0$ and $\bar\alpha_{q,\ell}:=\sqrt q\,e^{i\theta_\ell}$. The true amplitude is $\alpha_v(\lambda)
=
e^{i\theta_0}\sqrt{\lambda}\,\bar\alpha_v$. The common phase \(e^{i\theta_0}\) cancels in coherent-state overlaps.  Hence,
for every edge \(u\sim v\),
\begin{align}
g_{uv}(\lambda)
:=
\langle \underline{\alpha_u(\lambda)}|\underline{\alpha_v(\lambda)}\rangle
=
\exp\!\left[
\lambda
\left(
-\frac{|\bar\alpha_u|^2+|\bar\alpha_v|^2}{2}
+
\bar\alpha_u^*\bar\alpha_v
\right)
\right].
\end{align}
The nominal overlap is $\bar g_{uv}=g_{uv}(1)$. Differentiating in \(\lambda\), we get
\begin{align}
\frac{d}{d\lambda}g_{uv}(\lambda)
=
h_{uv}e^{\lambda h_{uv}},
\qquad
h_{uv}:=
-\frac{|\bar\alpha_u|^2+|\bar\alpha_v|^2}{2}
+
\bar\alpha_u^*\bar\alpha_v .
\end{align}
Given the relation  $\operatorname{Re} h_{uv}
=
-\frac12|\bar\alpha_u-\bar\alpha_v|^2
\le 0$, we have
\begin{align}
|e^{\lambda h_{uv}}|\le 1,
\qquad
\lambda\in[1/2,2].
\end{align}
Therefore, we have 
\begin{align}
\left|
\frac{d}{d\lambda}g_{uv}(\lambda)
\right|
\le
|h_{uv}|.
\end{align}

We now bound \(h_{uv}\) on each type of edge.  For an angular edge
\(u=(q,\ell)\), \(v=(q,\ell+1)\), writing \(\Delta:=2\pi/L\), we have
\begin{align}
h_{uv}
=
q(e^{i\Delta}-1),
\end{align}
and hence
\begin{align}
|h_{uv}|
=
q|e^{i\Delta}-1|
\le
q\Delta
=
\frac{2\pi q}{L}
\le
\frac{2\pi J}{L}
\le
\frac{\pi}{4},
\end{align}
where we used the assumption \(L\ge 8J\).

For a radial edge \(u=(q,\ell)\), \(v=(q+1,\ell)\), the two amplitudes have the
same phase, and thus
\begin{align}
h_{uv}
=
-\frac12(\sqrt{q+1}-\sqrt q)^2.
\end{align}
Therefore, we have 
\begin{align}
|h_{uv}|
=
\frac12(\sqrt{q+1}-\sqrt q)^2
\le
\frac12.
\end{align}

For the root edge \(0\sim(1,0)\), we have
\begin{align}
h_{uv}=-\frac12,
\qquad
|h_{uv}|=\frac12.
\end{align}

Thus, on every edge, we have 
\begin{align}
\sup_{\lambda\in[1/2,2]}
\left|
\frac{d}{d\lambda}g_{uv}(\lambda)
\right|
\le C
\end{align}
for a universal constant \(C>0\).  By the mean value theorem \cite{zorich2016mathematical}, we have 
\begin{align}
|g_{uv}(\lambda)-\bar g_{uv}|
=
|g_{uv}(\lambda)-g_{uv}(1)|
\le
C|\lambda-1|
\le
C\zeta .
\end{align} 

Now we convert tomography error into vector error.  For each vertex \(v\), the
pure-state trace-distance identity gives
\begin{align}
\frac12
\left\|
|u_v\rangle\!\langle u_v|
-
|\psi_v\rangle\!\langle\psi_v|
\right\|_1
=
\sqrt{1-|\langle u_v|\psi_v\rangle|^2}
\le
\varepsilon_{\rm tom}.
\end{align}
Choosing the phase \(\xi_v\in\mathbb R\) so that
$e^{i\xi_v}\langle u_v|\psi_v\rangle
=
\left|\langle u_v|\psi_v\rangle\right|$, we have 
\begin{align}
\left\|
|u_v\rangle
-
e^{i\xi_v}|\psi_v\rangle
\right\|_2^2
&=
2-
2\operatorname{Re}\!\left(
e^{i\xi_v}\langle u_v|\psi_v\rangle
\right)
\\
&=
2\left(
1-\left|\langle u_v|\psi_v\rangle\right|
\right)
\\
&\le
2\left(
1-\left|\langle u_v|\psi_v\rangle\right|^2
\right)
\\
&\le
2\varepsilon_{\rm tom}^2.
\end{align}
For an edge \(a\sim b\), let us define $h_{ab}':=\langle u_a|u_b\rangle$ and $g_{ab}:=\langle \psi_a|\psi_b\rangle$. From the vector estimate, for each vertex \(v\) there exists a vector $|e_v\rangle:=|u_v\rangle-e^{i\xi_v}|\psi_v\rangle$ such that $\|e_v\|_2\le \sqrt2\,\varepsilon_{\rm tom}$. 
Therefore, for the edge \(a\sim b\), we have 
\begin{align}
h_{ab}'
&=
\langle u_a|u_b\rangle
\\
&=
\left(
e^{-i\xi_a}\langle\psi_a|+\langle e_a|
\right)
\left(
e^{i\xi_b}|\psi_b\rangle+|e_b\rangle
\right)
\\
&=
e^{i(\xi_b-\xi_a)}
\langle\psi_a|\psi_b\rangle
+
e^{-i\xi_a}\langle\psi_a|e_b\rangle
+
e^{i\xi_b}\langle e_a|\psi_b\rangle
+
\langle e_a|e_b\rangle
\\
&=
e^{i(\xi_b-\xi_a)}g_{ab}
+
e^{-i\xi_a}\langle\psi_a|e_b\rangle
+
e^{i\xi_b}\langle e_a|\psi_b\rangle
+
\langle e_a|e_b\rangle .
\end{align}
By Cauchy--Schwarz and \(\|\psi_a\|=\|\psi_b\|=1\), we have 
\begin{align}
\left|
h_{ab}'
-
e^{i(\xi_b-\xi_a)}g_{ab}
\right|
&\le
|\langle\psi_a|e_b\rangle|
+
|\langle e_a|\psi_b\rangle|
+
|\langle e_a|e_b\rangle|
\\
&\le
\|e_b\|_2+\|e_a\|_2+\|e_a\|_2\|e_b\|_2
\\
&\le
2\sqrt2\,\varepsilon_{\rm tom}
+
2\varepsilon_{\rm tom}^2 .
\end{align}
In particular, since \(\varepsilon_{\rm tom}\le 1\), this implies
\begin{align}
\left|
h_{ab}'
-
e^{i(\xi_b-\xi_a)}g_{ab}
\right|
\le
C'\varepsilon_{\rm tom}
\end{align}
for a universal constant \(C'>0\). 

We now compare the phase of the measured overlap with the phase of the nominal
input overlap.  Fix an edge \(u\sim v\), and write $\Delta_{uv}:=\xi_v-\xi_u$, $
b_{uv}:=h_{uv}'=\langle u_u|u_v\rangle$, and $w_{uv}:=e^{i\Delta_{uv}}\bar g_{uv}$. Here \(b_{uv}\) is the measured overlap between the tomographic representatives,
while \(w_{uv}\) is the nominal overlap with the true tomography phases
\(\xi_u,\xi_v\) inserted. We claim that \(b_{uv}\) and \(w_{uv}\) are close.  Indeed, we have 
\begin{align}
|b_{uv}-w_{uv}|
&=
\left|
h_{uv}'
-
e^{i(\xi_v-\xi_u)}\bar g_{uv}
\right|
\\
&\le
\left|
h_{uv}'
-
e^{i(\xi_v-\xi_u)}g_{uv}
\right|
+
\left|
e^{i(\xi_v-\xi_u)}
(g_{uv}-\bar g_{uv})
\right|
\\
&\le
C'\varepsilon_{\rm tom}
+
C\zeta
\\
&\le
C_1(\varepsilon_{\rm tom}+\zeta),
\label{eq:edge-overlap-close}
\end{align}
where we used the tomography-overlap estimate and the calibration estimate
\(|g_{uv}-\bar g_{uv}|\le C\zeta\).

Moreover, every true edge overlap has a constant lower bound
\(|g_{uv}|\ge c_0\), and \(|g_{uv}-\bar g_{uv}|\le C\zeta\).  Thus, for
\(\zeta\) sufficiently small, we have 
\begin{align}
|\bar g_{uv}|\ge \frac{c_0}{2}.
\end{align}
Therefore, we have 
\begin{align}
|w_{uv}|=|\bar g_{uv}|\ge \frac{c_0}{2}.
\end{align}
If \(\varepsilon_{\rm tom}+\zeta\) is sufficiently small, then
Eq.~\eqref{eq:edge-overlap-close} also implies \(b_{uv}\neq0\). We now pass from overlap closeness to phase closeness.  For any nonzero
complex numbers \(b,w\), one has 
$\left|
\frac{w}{|w|}-\frac{b}{|b|}
\right|
\le
\frac{2|w-b|}{|w|}$. Applying this with \(b=b_{uv}\) and \(w=w_{uv}\), we obtain
\begin{align}
\left|
\frac{w_{uv}}{|w_{uv}|}
-
\frac{b_{uv}}{|b_{uv}|}
\right|
&\le
\frac{2|w_{uv}-b_{uv}|}{|w_{uv}|}
\\
&\le
C_2(\varepsilon_{\rm tom}+\zeta).
\end{align}
Finally, given the relations $\frac{w_{uv}}{|w_{uv}|}
=
e^{i(\xi_v-\xi_u)}
\frac{\bar g_{uv}}{|\bar g_{uv}|}$ and $\frac{b_{uv}}{|b_{uv}|}
=
\frac{h_{uv}'}{|h_{uv}'|}$, we get
\begin{align}
\left|
\frac{\bar g_{uv}}{|\bar g_{uv}|}
\frac{|h_{uv}'|}{h_{uv}'}
e^{i(\xi_v-\xi_u)}
-
1
\right|
&=
\left|
\frac{w_{uv}}{|w_{uv}|}
\frac{|b_{uv}|}{b_{uv}}
-
1
\right|
\\
&=
\left|
\frac{w_{uv}}{|w_{uv}|}
-
\frac{b_{uv}}{|b_{uv}|}
\right|
\\
&\le
C_2(\varepsilon_{\rm tom}+\zeta).
\label{eq:local-edge-phase-error}
\end{align}

We now define the actual synchronization procedure.  Choose a spanning tree of
the graph rooted at \(0\).  Set
$|\widehat\phi_0\rangle:=|u_0\rangle$. Suppose \(u\) is the parent of \(v\) in the tree, and the phase of \(u\) has
already been fixed.  Write 
$|\widehat\phi_u\rangle=e^{i\tau_u}|u_u\rangle$. We choose \(\tau_v\) by the rule
\begin{align}
e^{i(\tau_v-\tau_u)}
=
\frac{\bar g_{uv}}{|\bar g_{uv}|}
\frac{|h_{uv}'|}{h_{uv}'},
\end{align}
and set $|\widehat\phi_v\rangle:=e^{i\tau_v}|u_v\rangle$. This rule says that the measured edge overlap is rotated so that its phase
matches the known nominal input-overlap phase.

Let \(\omega:=\xi_0\).  Eq. 
\eqref{eq:local-edge-phase-error} shows that each edge contributes at most
\(C_2(\varepsilon_{\rm tom}+\zeta)\) to the phase error relative to the true
global phase.  Along any path in the spanning tree, these errors add.  Since
the graph diameter is at most \(J+L\), for every vertex \(v\), we have 
\begin{align}
\left|
e^{i(\tau_v+\xi_v-\omega)}-1
\right|
\le
C_2(J+L)(\varepsilon_{\rm tom}+\zeta).
\end{align}
Finally, we have 
\begin{align}
\left\|
|\widehat\phi_v\rangle
-
e^{i\omega}|\psi_v\rangle
\right\|_2
&=
\left\|
e^{i\tau_v}|u_v\rangle
-
e^{i\omega}|\psi_v\rangle
\right\|_2
\\
&\le
\left|
e^{i(\tau_v+\xi_v-\omega)}-1
\right|
+
\left\|
|u_v\rangle-e^{i\xi_v}|\psi_v\rangle
\right\|_2
\\
&\le
C_2(J+L)(\varepsilon_{\rm tom}+\zeta).
\end{align}
This bound includes the root vertex \(0\).  Therefore it holds for
\(|\widehat\phi_0\rangle\) and uniformly for all
\(|\widehat\phi_{q,\ell}\rangle\), completing the proof.
\end{proof}

\begin{lemma}[Recovery of Fock columns by Fourier filtering]
\label{lem:output-only-column-recovery}
Let us denote $W_g:=W_\star e^{i\theta_0\hat n}$. Assume that the local phase synchronization step has produced aligned vectors $|\widehat\phi_0\rangle,$ and $
|\widehat\phi_{q,\ell}\rangle,$ for $q=1,\ldots,J$, $\ell=0,\ldots,L-1$, such that, for some common phase \(\omega\in\mathbb R\),
\begin{align}
\begin{cases}
\left\|
|\widehat\phi_0\rangle
-
e^{i\omega}W_\star|0\rangle
\right\|_2
\le
\Delta_{\rm ph},\\
\max_{q,\ell}
\left\|
|\widehat\phi_{q,\ell}\rangle
-
e^{i\omega}
W_\star
\left|\underline{
\sqrt{\lambda q}\,
e^{i(\theta_0+\theta_\ell)}}
\right\rangle
\right\|_2
\le
\Delta_{\rm ph},
\end{cases}
\end{align}
with $\Delta_{\rm ph}
:=
C_{\rm ph}(J+L)(\varepsilon_{\rm tom}+\zeta)$. Assume also that the Fourier-ring synthesis lemma \ref{lem:output-only-fourier-synthesis} gives, for every
\(q=1,\ldots,J\), $\left\|
|\Phi_{q,L}(\lambda,\theta_0)\rangle
-
e^{iq\theta_0}|q\rangle
\right\|_2
\le q\zeta^2+\varepsilon_{\rm al}$, where $|\Phi_{q,L}(\lambda,\theta_0)\rangle
:=
\frac{1}{L\,c_q(q)}
\sum_{\ell=0}^{L-1}
e^{-iq\theta_\ell}
\left|\underline{
\sqrt{\lambda q}\,
e^{i(\theta_0+\theta_\ell)}}
\right\rangle$ is the approximated Fock state. Define $|\widehat v_0\rangle
:=
\Pi_{\le R}|\widehat\phi_0\rangle$, and, for \(q=1,\ldots,J\), $|\widehat v_q\rangle
:=
\frac{1}{L\,c_q(q)}
\sum_{\ell=0}^{L-1}
e^{-iq\theta_\ell}
\Pi_{\le R}|\widehat\phi_{q,\ell}\rangle .$ Let us denote $\widehat A_R
:=
\bigl[
|\widehat v_0\rangle,\ldots,|\widehat v_J\rangle
\bigr]$ and $A_R
:=
\Pi_{\le R}W_g\Pi_{\le J}$.

Then there is a universal constant \(C>0\) such that
\begin{align}
\left\|
\widehat A_R-e^{i\omega}A_R
\right\|_\infty
\le
C(J+1)^{3/4}(J+L)(\varepsilon_{\rm tom}+\zeta)
+
(J+1)^{3/2}\zeta^2
+
\sqrt{J+1}\,\varepsilon_{\rm al}.
\end{align}
\end{lemma}

\begin{proof}
The matrix \(A_R\) has columns $A_R|q\rangle
=
\Pi_{\le R}W_g|q\rangle$ for $q=0,\ldots,J$. Thus it suffices to control the error of each recovered column
\(|\widehat v_q\rangle\). First, let us consider the case with \(q=0\).  Since we have $W_g|0\rangle
=
W_\star e^{i\theta_0\hat n}|0\rangle
=
W_\star|0\rangle$, and since \(\Pi_{\le R}\) is a contraction, the phase-synchronization bound gives the following bound
\begin{align}
\left\|
|\widehat v_0\rangle
-
e^{i\omega}\Pi_{\le R}W_g|0\rangle
\right\|_2
&=
\left\|
\Pi_{\le R}|\widehat\phi_0\rangle
-
e^{i\omega}\Pi_{\le R}W_\star|0\rangle
\right\|_2
\\
&\le
\left\|
|\widehat\phi_0\rangle
-
e^{i\omega}W_\star|0\rangle
\right\|_2
\\
&\le
\Delta_{\rm ph}.
\label{eq:column-q0-error}
\end{align}

Now fix \(q\in\{1,\ldots,J\}\).  We compare the recovered column
\(|\widehat v_q\rangle\) with $e^{i\omega}\Pi_{\le R}W_\star|\Phi_{q,L}(\lambda,\theta_0)\rangle$. Using the definition of \(|\widehat v_q\rangle\) and
\(|\Phi_{q,L}(\lambda,\theta_0)\rangle\), we have
\begin{align}
&|\widehat v_q\rangle
-
e^{i\omega}
\Pi_{\le R}W_\star|\Phi_{q,L}(\lambda,\theta_0)\rangle
\\
&=
\frac{1}{L\,c_q(q)}
\sum_{\ell=0}^{L-1}
e^{-iq\theta_\ell}
\Pi_{\le R}
\left(
|\widehat\phi_{q,\ell}\rangle
-
e^{i\omega}
W_\star
\left|\underline{
\sqrt{\lambda q}\,
e^{i(\theta_0+\theta_\ell)}}
\right\rangle
\right).
\end{align}
Taking norms, using the triangle inequality, and using that
\(\Pi_{\le R}\) is a contraction, we obtain
\begin{align}
&\left\|
|\widehat v_q\rangle
-
e^{i\omega}
\Pi_{\le R}W_\star|\Phi_{q,L}(\lambda,\theta_0)\rangle
\right\|_2
\\
&\le
\frac{1}{L\,|c_q(q)|}
\sum_{\ell=0}^{L-1}
\left\|
|\widehat\phi_{q,\ell}\rangle
-
e^{i\omega}
W_\star
\left|\underline{
\sqrt{\lambda q}\,
e^{i(\theta_0+\theta_\ell)}}
\right\rangle
\right\|_2
\\
&\le
|c_q(q)|^{-1}\Delta_{\rm ph}.
\end{align}
By the normalization estimate from Eq. (\ref{eqs243}), we have the relation $|c_q(q)|^{-1}
\le
C(q+1)^{1/4}
\le
C(J+1)^{1/4}$. Hence, we have 
\begin{align}
\left\|
|\widehat v_q\rangle
-
e^{i\omega}
\Pi_{\le R}W_\star|\Phi_{q,L}(\lambda,\theta_0)\rangle
\right\|_2
\le
C(J+1)^{1/4}\Delta_{\rm ph}.
\label{eq:tomography-column-error}
\end{align}

Next, let us replace the synthesized Fourier-ring state by the target Fock vector.
Since \(W_\star\) and \(\Pi_{\le R}\) are contractions, we have the bounds
\begin{align}
&\left\|
e^{i\omega}
\Pi_{\le R}W_\star|\Phi_{q,L}(\lambda,\theta_0)\rangle
-
e^{i\omega}
\Pi_{\le R}W_\star e^{iq\theta_0}|q\rangle
\right\|_2
\\
&\le
\left\|
|\Phi_{q,L}(\lambda,\theta_0)\rangle
-
e^{iq\theta_0}|q\rangle
\right\|_2
\\
&\le
q\zeta^2+\varepsilon_{\rm al}
\\
&\le
J\zeta^2+\varepsilon_{\rm al}.
\label{eq:synthesis-column-error}
\end{align}
where we applied Lemma \ref{lem:output-only-fourier-synthesis}.

Finally, we have the relation $W_\star e^{iq\theta_0}|q\rangle
=
W_\star e^{i\theta_0\hat n}|q\rangle
=
W_g|q\rangle$. Combining \eqref{eq:tomography-column-error} and
\eqref{eq:synthesis-column-error}, we get, for every \(q=1,\ldots,J\),
\begin{align}
\left\|
|\widehat v_q\rangle
-
e^{i\omega}\Pi_{\le R}W_g|q\rangle
\right\|_2
\le
C(J+1)^{1/4}\Delta_{\rm ph}
+
J\zeta^2
+
\varepsilon_{\rm al}.
\label{eq:columnwise-error}
\end{align}
The same bound also holds for \(q=0\), after increasing the universal constant,
because \(\Delta_{\rm ph}\le (J+1)^{1/4}\Delta_{\rm ph}\).

Let us denote 
\begin{align}
\Delta_{\rm col}
:=
C(J+1)^{1/4}\Delta_{\rm ph}
+
J\zeta^2
+
\varepsilon_{\rm al}.
\end{align}
Then every column of
\(\widehat A_R-e^{i\omega}A_R\) has Hilbert-space norm at most
\(\Delta_{\rm col}\).  Therefore, we have 
\begin{align}
\left\|
\widehat A_R-e^{i\omega}A_R
\right\|_\infty
&\le
\left\|
\widehat A_R-e^{i\omega}A_R
\right\|_F
\\
&\le
\sqrt{J+1}\,\Delta_{\rm col}
\\
&\le
C(J+1)^{3/4}\Delta_{\rm ph}
+
(J+1)^{3/2}\zeta^2
+
\sqrt{J+1}\,\varepsilon_{\rm al}.
\end{align}
Substituting $\Delta_{\rm ph}
=
C_{\rm ph}(J+L)(\varepsilon_{\rm tom}+\zeta)$ gives
\begin{align}
\left\|
\widehat A_R-e^{i\omega}A_R
\right\|_\infty
\le
C(J+1)^{3/4}(J+L)(\varepsilon_{\rm tom}+\zeta)
+
(J+1)^{3/2}\zeta^2
+
\sqrt{J+1}\,\varepsilon_{\rm al}.
\end{align}
This proves the final displayed
bound.
\end{proof}

\begin{lemma}[Polar completion and energy-constrained error]
\label{lem:output-only-polar-completion}
Consider a one-mode unitary $U$ satisfying $\mathsf N_U(J)
:=
\sup_{\substack{|\varphi\rangle\in\mathcal H_{\le J}\\ \|\varphi\|=1}}
\langle\varphi|U^\dagger\hat n U|\varphi\rangle
\le N_{\rm dyn}$. Let $A_R:=\Pi_{\le R}U\Pi_{\le J}$, and assume $\frac{N_{\rm dyn}}{R+1}\le \frac14$. Suppose that an estimated column matrix \(\widehat A_R\) satisfies
\begin{align}
\inf_{\omega\in\mathbb R}
\left\|
\widehat A_R-e^{i\omega}A_R
\right\|_\infty
\le \tau,
\end{align}
where \(\tau>0\) is smaller than a universal constant.  Define the polar
isometry $\widehat Q_R
:=
\widehat A_R
(\widehat A_R^\dagger\widehat A_R)^{-1/2}$, extend \(\widehat Q_R\) to a unitary \(\widehat V_R\) on
\(\mathcal H_{\le R}\), and set 
$\widehat U:=\widehat V_R\oplus I_{>R}$.

Then, we have 
\begin{align}
\left\|
\widehat U\,\boldsymbol{\cdot}\,\widehat U^\dagger
-
U\,\boldsymbol{\cdot}\,U^\dagger
\right\|_{\diamond}^E
\le
4\sqrt{\frac{N}{J+1}}
+
4\sqrt{\frac{N_{\rm dyn}}{R+1}}
+
C\tau,
\end{align}
where \(C>0\) is a universal constant, $N=E-1/2$ is the average photon number of the input state in the definition of energy constrained diamond norm. 
\end{lemma}

\begin{proof}
Here, we note that the role of \(A_R\) is to store the first \(J+1\) columns of \(U\), truncated
to the output subspace \(\mathcal H_{\le R}\).  The dynamical energy assumption
implies that this truncation loses only a small norm.  Explicitly, we have 
\begin{align}
A_R^\dagger A_R
&=
\Pi_{\le J}U^\dagger\Pi_{\le R}U\Pi_{\le J}
\\
&=
I_{\le J}
-
\Pi_{\le J}U^\dagger\Pi_{>R}U\Pi_{\le J}.
\end{align}
For every normalized \(|\varphi\rangle\in\mathcal H_{\le J}\), Markov's
inequality for the number operator gives
\begin{align}
\|\Pi_{>R}U|\varphi\rangle\|_2^2
\le
\frac{1}{R+1}
\langle\varphi|U^\dagger \hat n U|\varphi\rangle
\le
\frac{N_{\rm dyn}}{R+1},\label{eqs334}
\end{align}
where $\hat n=a^\dag a$ denotes the photon number operator.  Consequently, we have 
\begin{align}
\|A_R^\dagger A_R-I_{\le J}\|_\infty
\le
\frac{N_{\rm dyn}}{R+1}
\le
\frac14.\label{eqs334s}
\end{align}
Thus we have \(A_R^\dagger A_R\succeq \frac34 I_{\le J}\). So the polar isometry
\begin{align}
Q_R:=A_R(A_R^\dagger A_R)^{-1/2}
\end{align}
is well-defined. Let us set $\alpha_R:=\frac{N_{\rm dyn}}{R+1}$. We first decompose the exact low-input action of \(U\) into its part inside
\(\mathcal H_{\le R}\) and its output tail:
\begin{align}
U\Pi_{\le J}
=
\Pi_{\le R}U\Pi_{\le J}
+
\Pi_{>R}U\Pi_{\le J}
=
A_R+\Pi_{>R}U\Pi_{\le J}.
\end{align}
Hence we have 
\begin{align}
\|U\Pi_{\le J}-Q_R\|_\infty
\le
\|\Pi_{>R}U\Pi_{\le J}\|_\infty
+
\|A_R-Q_R\|_\infty .
\end{align}
From Eq. (\ref{eqs334}), we have $\|\Pi_{>R}U\Pi_{\le J}\|_\infty
\le
\sqrt{\alpha_R}$. For the second term, let us write $M_R:=A_R^\dagger A_R$. From Eq. (\ref{eqs334s}), the eigenvalues of \(M_R\) lie in $[1-\alpha_R,1]$. Since $Q_R=A_RM_R^{-1/2}$, we have 
\begin{align}
\|A_R-Q_R\|_\infty
=
\|Q_R(M_R^{1/2}-I)\|_\infty
=
\|M_R^{1/2}-I\|_\infty .
\end{align}
Because the eigenvalues of \(M_R^{1/2}\) lie in $[\sqrt{1-\alpha_R},1]$, we get
\begin{align}
\|A_R-Q_R\|_\infty
\le
1-\sqrt{1-\alpha_R}
\le
\alpha_R
\le
\sqrt{\alpha_R},
\end{align}
where the last inequality uses \(\alpha_R\le 1\). Combining the two bounds gives
\begin{align}
\|U\Pi_{\le J}-Q_R\|_\infty
\le
2\sqrt{\alpha_R}
=
2\sqrt{\frac{N_{\rm dyn}}{R+1}}.
\label{eq:polar-exact-isometry-error}
\end{align}

We next compare the polar isometry of \(\widehat A_R\) with that of \(A_R\).
Choose a phase \(\omega_0\in\mathbb R\) such that 
\begin{align}
\|\widehat A_R-e^{i\omega_0}A_R\|_\infty\le \tau.
\end{align}
It is convenient to remove this common phase before applying the polar map.  Define $\widetilde A_R:=e^{-i\omega_0}\widehat A_R $. Then we have $\|\widetilde A_R-A_R\|_\infty\le \tau$. Let us denote $G_R:=A_R^\dagger A_R$ and $\widetilde G_R:=\widetilde A_R^\dagger \widetilde A_R$. Since \(A_R^\dagger A_R\succeq \frac34 I\) and \(\tau\) is sufficiently small,
we also have $\widetilde G_R\succeq \frac12 I$. Indeed, for every unit vector \(x\), we have 
\begin{align}
\begin{cases}
\|\widetilde A_R x\|_2
\ge
\|A_Rx\|_2-\|(\widetilde A_R-A_R)x\|_2
\ge
\sqrt{\frac34}-\tau,\\
\|\widetilde A_R x\|_2
\le
\|A_Rx\|_2+\|(\widetilde A_R-A_R)x\|_2
\le
1+\tau,
\end{cases}\label{eqs342}
\end{align}
which is at least \(1/\sqrt2\) for all sufficiently small universal \(\tau\). Now define the polar isometries $Q_R:=A_R G_R^{-1/2}$ and $\widetilde Q_R:=\widetilde A_R\widetilde G_R^{-1/2}$. Since \(\widetilde A_R=e^{-i\omega_0}\widehat A_R\), its polar isometry is $\widetilde Q_R
=
e^{-i\omega_0}
\widehat A_R(\widehat A_R^\dagger\widehat A_R)^{-1/2}
=
e^{-i\omega_0}\widehat Q_R$. Thus it is enough to prove $\|\widetilde Q_R-Q_R\|_\infty\le C\tau$. We first compare the Gram matrices.  Since \(\|A_R\|_\infty\le1\) and
\(\|\widetilde A_R-A_R\|_\infty\le\tau\), we have 
\begin{align}
\|\widetilde G_R-G_R\|_\infty
&=
\|\widetilde A_R^\dagger\widetilde A_R-A_R^\dagger A_R\|_\infty
\\
&\le
\|(\widetilde A_R-A_R)^\dagger\widetilde A_R\|_\infty
+
\|A_R^\dagger(\widetilde A_R-A_R)\|_\infty
\\
&\le
\tau(1+\tau)+\tau
\\
&\le
3\tau ,
\end{align}
for \(\tau\le1\), where we applied the triangle inequality and Eq. (\ref{eqs342}).  Since both \(G_R\) and \(\widetilde G_R\) are bounded below
by a positive constant, the inverse-square-root map is Lipschitz on this
region:
\begin{align}
\|\widetilde G_R^{-1/2}-G_R^{-1/2}\|_\infty
\le
C\|\widetilde G_R-G_R\|_\infty
\le
C\tau .\label{eqs347}
\end{align}
This standard estimate follows from the functional calculus for positive
matrices, for example from the integral representation of \(X^{-1/2}\) and the
resolvent identity \cite{higham2008functions}. Therefore, we have 
\begin{align}
\|\widetilde Q_R-Q_R\|_\infty
&=
\|\widetilde A_R\widetilde G_R^{-1/2}
-
A_RG_R^{-1/2}\|_\infty
\\
&\le
\|(\widetilde A_R-A_R)\widetilde G_R^{-1/2}\|_\infty
+
\|A_R(\widetilde G_R^{-1/2}-G_R^{-1/2})\|_\infty
\\
&\le
\|\widetilde A_R-A_R\|_\infty
\|\widetilde G_R^{-1/2}\|_\infty
+
\|A_R\|_\infty
\|\widetilde G_R^{-1/2}-G_R^{-1/2}\|_\infty
\\
&\le
C\tau .
\label{eq:polar-perturbation-bound}
\end{align}
where we used the triangle inequality, submultiplicativity of the operator
norm, $\|\widetilde A_R-A_R\|_\infty\le \tau$, $
\|A_R\|_\infty\le 1$, $
\|\widetilde G_R^{-1/2}\|_\infty\le \sqrt2$, 
and Eq. (\ref{eqs347}). Finally, since \(\widetilde Q_R=e^{-i\omega_0}\widehat Q_R\), we obtain $\|\widehat Q_R-e^{i\omega_0}Q_R\|_\infty
=
\|\widetilde Q_R-Q_R\|_\infty
\le
C\tau$.

We now combine the two estimates.  Since \(\widehat V_R\) extends the isometry
\(\widehat Q_R\), the full-space unitary \(\widehat U=\widehat V_R\oplus I_{>R}\)
satisfies $\widehat U\Pi_{\le J}=\widehat Q_R $. Therefore, we have 
\begin{align}
\|U\Pi_{\le J}-e^{-i\omega_0}\widehat U\Pi_{\le J}\|_\infty
&=
\|U\Pi_{\le J}-e^{-i\omega_0}\widehat Q_R\|_\infty
\\
&\le
\|U\Pi_{\le J}-Q_R\|_\infty
+
\|Q_R-e^{-i\omega_0}\widehat Q_R\|_\infty
\\
&\le
2\sqrt{\frac{N_{\rm dyn}}{R+1}}
+
C\tau .
\label{eq:low-energy-unitary-comparison}
\end{align}

We now lift this low-input-subspace estimate to the energy-constrained diamond
norm.  Let \(|\Psi\rangle\) be an arbitrary purification of an input state
whose signal oscillator energy is at most \(E=N+\frac12\).  Equivalently, the
signal photon-number expectation is at most \(N\).  Given the relation $I-\Pi_{\le J}
=
\Pi_{>J}
\le
\frac{\hat n}{J+1}$, we have
$\|(\Pi_{>J}\otimes I)|\Psi\rangle\|_2^2
\le
\frac{1}{J+1}
\langle \Psi|(\hat n\otimes I)|\Psi\rangle
\le
\frac{N}{J+1}$. Hence, we have 
\begin{align}
\|(\Pi_{>J}\otimes I)|\Psi\rangle\|_2
\le
\sqrt{\frac{N}{J+1}}.
\label{eq:input-tail-bound}
\end{align}
Let us write $|\Psi_{\le J}\rangle:=(\Pi_{\le J}\otimes I)|\Psi\rangle$ and $|\Psi_{>J}\rangle:=(\Pi_{>J}\otimes I)|\Psi\rangle $. Then, we have 
\begin{align}
&\left\|
(U\otimes I)|\Psi\rangle
-
(e^{-i\omega_0}\widehat U\otimes I)|\Psi\rangle
\right\|_2
\\
&\le
\left\|
(U\otimes I)|\Psi_{\le J}\rangle
-
(e^{-i\omega_0}\widehat U\otimes I)|\Psi_{\le J}\rangle
\right\|_2
\\
&\quad+
\left\|
(U\otimes I)|\Psi_{>J}\rangle
-
(e^{-i\omega_0}\widehat U\otimes I)|\Psi_{>J}\rangle
\right\|_2 .
\end{align}
For the low-photon component, Eq.~\eqref{eq:low-energy-unitary-comparison}
gives
\begin{align}
\left\|
(U\otimes I)|\Psi_{\le J}\rangle
-
(e^{-i\omega_0}\widehat U\otimes I)|\Psi_{\le J}\rangle
\right\|_2
\le
\left(
2\sqrt{\frac{N_{\rm dyn}}{R+1}}
+
C\tau
\right)
\|\Psi_{\le J}\|_2 .
\end{align}
Since we have \(\|\Psi_{\le J}\|_2\le1\), this is at most $2\sqrt{\frac{N_{\rm dyn}}{R+1}}
+
C\tau $. For the high-photon component, both \(U\) and \(\widehat U\) are unitary, so we have 
\begin{align}
\left\|
(U\otimes I)|\Psi_{>J}\rangle
-
(e^{-i\omega_0}\widehat U\otimes I)|\Psi_{>J}\rangle
\right\|_2
\le
2\|\Psi_{>J}\|_2
\le
2\sqrt{\frac{N}{J+1}},
\end{align}
where we used Eq.~\eqref{eq:input-tail-bound}.  Combining the two pieces gives
\begin{align}
\left\|
(U\otimes I)|\Psi\rangle
-
(e^{-i\omega_0}\widehat U\otimes I)|\Psi\rangle
\right\|_2
\le
2\sqrt{\frac{N}{J+1}}
+
2\sqrt{\frac{N_{\rm dyn}}{R+1}}
+
C\tau .
\end{align}

For normalized vectors \(|\Phi\rangle\) and \(|\Xi\rangle\), we use the elementary pure-state bound $\left\|
|\Phi\rangle\!\langle\Phi|
-
|\Xi\rangle\!\langle\Xi|
\right\|_1
\le
2\,
\bigl\|
|\Phi\rangle-|\Xi\rangle
\bigr\|_2$. Indeed, we have  $\left\|
|\Phi\rangle\!\langle\Phi|
-
|\Xi\rangle\!\langle\Xi|
\right\|_1
=
2\sqrt{1-|\langle\Phi|\Xi\rangle|^2},$ while $1-|\langle\Phi|\Xi\rangle|^2
\le
2\bigl(1-\operatorname{Re}\langle\Phi|\Xi\rangle\bigr)
=
\bigl\|
|\Phi\rangle-|\Xi\rangle
\bigr\|_2^2$. Therefore we have 
\begin{align}
&\left\|
(U\otimes I)|\Psi\rangle\!\langle\Psi|(U^\dagger\otimes I)
-
(e^{-i\omega_0}\widehat U\otimes I)
|\Psi\rangle\!\langle\Psi|
(e^{i\omega_0}\widehat U^\dagger\otimes I)
\right\|_1
\\
&\le
4\sqrt{\frac{N}{J+1}}
+
4\sqrt{\frac{N_{\rm dyn}}{R+1}}
+
C\tau .
\end{align}
Taking the supremum over all purifications with signal energy at most \(E\)
gives
\begin{align}
\left\|
e^{-i\omega_0}\widehat U\,\boldsymbol{\cdot}\,
e^{i\omega_0}\widehat U^\dagger
-
U\,\boldsymbol{\cdot}\,U^\dagger
\right\|_{\diamond}^E
\le
4\sqrt{\frac{N}{J+1}}
+
4\sqrt{\frac{N_{\rm dyn}}{R+1}}
+
C\tau .
\end{align}
Finally, the global phase \(e^{-i\omega_0}\) has no effect on the induced
unitary channel.  Hence
\begin{align}
\left\|
\widehat U\,\boldsymbol{\cdot}\,\widehat U^\dagger
-
U\,\boldsymbol{\cdot}\,U^\dagger
\right\|_{\diamond}^E
\le
4\sqrt{\frac{N}{J+1}}
+
4\sqrt{\frac{N_{\rm dyn}}{R+1}}
+
C\tau .
\end{align}
\end{proof}

Now, a theorem for efficient learning is given as follows.

\begin{theorem}[Efficient unitary learning from unknown coherent probe states]
\label{thm:output-only-efficient-unknown-coherent}
Consider the learning process of an unknown single-mode unitary channel $\mathcal W_\star(\rho):=W_\star\rho W_\star^\dagger$. The learner can prepare coherent states $|\underline{s e^{i\vartheta}\alpha}\rangle$ with $\alpha=\sqrt{\mu}\,e^{i\theta_0}$ for any chosen scale \(s\ge0\) and phase
\(\vartheta\in\mathbb R\), where \(\mu\in [\mu_{\min},\mu_{\max}]\) and \(\theta_0\) are unknown. Meanwhile, the learner cannot measure the probe states. Let \(E=N+\frac12\) be the input-energy constraint in the
energy-constrained diamond norm. Assume that every output state queried by the protocol has mean photon number
at most \(N_{\rm out}\). We additionally assume the dynamical photon-number bound
$
\sup_{\substack{
|\varphi\rangle\in
\mathcal H_{\le \left\lceil C_J N/\epsilon^2\right\rceil}\\
\|\varphi\|=1}}
\langle\varphi|W_\star^\dagger \hat n W_\star|\varphi\rangle<\infty$, 
for a sufficiently large universal constant \(C_J>0\) and a constant $\epsilon$.

Then there exists an output-only learning protocol returning an estimator
\(\widehat{\mathcal W}\) such that
\begin{align}
\Pr\!\left[\inf_{\theta\in\mathbb R}
\bigl\|
\widehat{\mathcal W}-\mathcal W_\star\circ \mathcal C_{\theta}
\bigr\|_{\diamond}^E
\le
\epsilon
\right]
\ge
1-\delta ,
\end{align}
where $\map C_\theta(\cdot)=e^{-i\theta a^\dag a }\cdot e^{i\theta a^\dag a }$ denotes the phase-rotation channel. The total number of channel uses satisfies
\begin{align}
M_{\rm tot}
\le
C(N_{\rm out}+1)
\frac{\Lambda_\epsilon^{7}}{\epsilon^4}
\left[
(1+S_{\rm cal})
\log\frac{12(1+S_{\rm cal})}{\delta}
+
(1+\Lambda_\epsilon^2)
\log\frac{12(1+\Lambda_\epsilon^2)}{\delta}
\right],
\end{align}
where \(C>0\) is a universal constant, 
$\Lambda_\epsilon
:=
1+\frac{N}{\epsilon^2}+\log\frac1\epsilon$, and $S_{\rm cal}
=
O\!\left(
\log\frac{\mu_{\max}}{\mu_{\min}}
\right)$ are control parameters.  
\end{theorem}

\begin{proof}
Internally, the protocol sets the parameters as 
$J:=\max\left\{1,\left\lceil C_J N/\epsilon^2\right\rceil\right\},$ $L:=\left\lceil C_L(J+\log(1/\epsilon))\right\rceil$, $R:=\max\left\{
J,
\left\lceil C_R N_{\rm dyn}/\epsilon^2\right\rceil
\right\}$, $\varepsilon_{\rm tr}:=
\min\left\{
\frac1{100},
\frac{\epsilon}{C_\star(J+1)^{3/4}(J+L)}
\right\}$, $\varepsilon_{\rm cal}:=\frac{\varepsilon_{\rm tr}}{128e^2}$, and $\varepsilon_{\rm al}:=\frac{\epsilon}{C_{\rm al}\sqrt{J+1}}$. We also  define $W_g:=W_\star e^{i\theta_0\hat n}$ with a chosen phase $\theta_0$, and assume the following condition
\begin{align}
\mathsf N_{W_g}(J)
:=
\sup_{\substack{
|\varphi\rangle\in\mathcal H_{\le J}\\
\|\varphi\|=1}}
\langle\varphi|W_g^\dagger\hat n W_g|\varphi\rangle
\le N_{\rm dyn}.
\end{align}
The protocol begins by estimating the unknown coherent intensity
\(\mu=|\alpha|^2\) from output states only.  We apply
Lemma~\ref{lem:output-only-calibration} with $\zeta=\varepsilon_{\rm tr}$ and $p_{\rm fail}=\frac{\delta}{3}$. Since the calibration tomography accuracy is $\varepsilon_{\rm cal}
=
\frac{\varepsilon_{\rm tr}}{128e^2}$, Lemma~\ref{lem:moment-tomography-primitive} gives the following calibration
cost:
\begin{align}
M_{\rm cal}
\le
(1+|\mathcal S_{\rm cal}|)
\left\lceil
2^{21}
\left(
\frac{eN_{\rm out}}{\varepsilon_{\rm cal}^{2}}+2e
\right)
\varepsilon_{\rm cal}^{-2}
\log\frac{12(1+|\mathcal S_{\rm cal}|)}{\delta}
\right\rceil .
\end{align}
With probability at least \(1-\delta/3\), the calibration returns
\(\widehat\mu\) such that
\begin{align}
|\widehat\mu-\mu|
\le
\varepsilon_{\rm tr}\mu .
\label{eq:thm-calibration-good}
\end{align}
Let us define $\lambda:=\frac{\mu}{\widehat\mu}$. Because of the assumption \(\varepsilon_{\rm tr}\le 1/100\), Eq.~\eqref{eq:thm-calibration-good}
implies
\begin{align}
|\lambda-1|
&=
\left|
\frac{\mu}{\widehat\mu}-1
\right|
=
\frac{|\widehat\mu-\mu|}{\widehat\mu}
\\
&\le
\frac{\varepsilon_{\rm tr}\mu}{(1-\varepsilon_{\rm tr})\mu}
=
\frac{\varepsilon_{\rm tr}}{1-\varepsilon_{\rm tr}}
\le
2\varepsilon_{\rm tr}.
\label{eq:thm-lambda-close}
\end{align}
In particular, we can set \(\lambda\in[1/2,2]\).

Using the calibrated value \(\widehat\mu\), the learner prepares the
Fourier-ring probes by choosing
\begin{align}
s_q:=\sqrt{\frac{q}{\widehat\mu}},
\qquad
\theta_\ell:=\frac{2\pi\ell}{L},
\qquad
q=1,\ldots,J,\quad
\ell=0,\ldots,L-1 .
\end{align}
Since the actual unknown amplitude is
\(\alpha=\sqrt{\mu}e^{i\theta_0}\), the true input probe is
\begin{align}
|\underline{s_qe^{i\theta_\ell}\alpha}\rangle
&=
\left|
\underline{\sqrt{\frac{q}{\widehat\mu}}\,
e^{i\theta_\ell}
\sqrt{\mu}e^{i\theta_0}}
\right\rangle
\\
&=
\left|\underline{
\sqrt{\lambda q}\,
e^{i(\theta_0+\theta_\ell)}}
\right\rangle .
\label{eq:thm-actual-ring-probe}
\end{align}
The protocol also uses the vacuum probe \(|0\rangle\).  We denote the
corresponding output states by $|\psi_0\rangle:=W_\star|0\rangle$ and $|\psi_{q,\ell}\rangle
:=
W_\star
\left|\underline{
\sqrt{\lambda q}\,
e^{i(\theta_0+\theta_\ell)}}
\right\rangle$. By assumption, all these queried output states have mean photon number at most
\(N_{\rm out}\).

The learner then tomographs the \(1+JL\) training output states $|\psi_0\rangle$ and $|\psi_{q,\ell}\rangle$ for $
q=1,\ldots,J,\,\ell=0,\ldots,L-1$ to trace-distance accuracy \(\varepsilon_{\rm tr}\).  Each tomography call is
assigned failure probability $\frac{\delta}{3(1+JL)}$. By Lemma~\ref{lem:moment-tomography-primitive}, this training tomography stage
uses at most
\begin{align}
M_{\rm tr}
\le
(1+JL)
\left\lceil
2^{21}
\left(
\frac{eN_{\rm out}}{\varepsilon_{\rm tr}^{2}}+2e
\right)
\varepsilon_{\rm tr}^{-2}
\log\frac{12(1+JL)}{\delta}
\right\rceil
\end{align}
channel uses.  By the union bound, with probability at least \(1-\delta/3\),
all training tomography calls succeed simultaneously.  On this event, there
are normalized representatives $|u_0\rangle$ and $|u_{q,\ell}\rangle$ such that
\begin{align}
\begin{cases}
\frac12
\left\|
|u_0\rangle\!\langle u_0|
-
|\psi_0\rangle\!\langle\psi_0|
\right\|_1
\le
\varepsilon_{\rm tr},\\
\frac12
\left\|
|u_{q,\ell}\rangle\!\langle u_{q,\ell}|
-
|\psi_{q,\ell}\rangle\!\langle\psi_{q,\ell}|
\right\|_1
\le
\varepsilon_{\rm tr}
\end{cases}
\end{align}
for every \(q,\ell\).

We now condition on the event that both calibration and all training tomography
calls succeed.  By Eq.~\eqref{eq:thm-lambda-close}, the radius calibration
error obeys $|\lambda-1|
\le
2\varepsilon_{\rm tr}$. Therefore Lemma~\ref{lem:output-only-phase-sync}, applied with $\varepsilon_{\rm tom}=\varepsilon_{\rm tr}$ and $\zeta=2\varepsilon_{\rm tr}$ produces aligned vectors $|\widehat\phi_0\rangle$, $
|\widehat\phi_{q,\ell}\rangle$, and a common phase \(\omega\in\mathbb R\) satisfying
\begin{align}
\begin{cases}
\left\|
|\widehat\phi_0\rangle
-
e^{i\omega}|\psi_0\rangle
\right\|_2
\le
C(J+L)\varepsilon_{\rm tr}\\
\max_{q,\ell}
\left\|
|\widehat\phi_{q,\ell}\rangle
-
e^{i\omega}|\psi_{q,\ell}\rangle
\right\|_2
\le
C(J+L)\varepsilon_{\rm tr}
\end{cases},
\label{eq:thm-phase-sync-ring}
\end{align}
where $C$ denotes a universal constant. 

The Fourier-ring synthesis lemma is now applied uniformly for
\(q=1,\ldots,J\).  Indeed, the definition of \(L=L_\epsilon\) guarantees
both \(L\ge 8J\) and, for every \(q\le J\), we have 
\begin{align}
L
\ge
(2e-3)q
+
2\log\frac1{\varepsilon_{\rm al}}
+
\frac12\log(q+1)
+
1 .
\end{align}
Thus Lemma~\ref{lem:output-only-fourier-synthesis} gives, for every
\(q=1,\ldots,J\),
\begin{align}
\left\|
|\Phi_{q,L}(\lambda,\theta_0)\rangle
-
e^{iq\theta_0}|q\rangle
\right\|_2
\le
q|\lambda-1|^2+\varepsilon_{\rm al}
\le
4J\varepsilon_{\rm tr}^2+\varepsilon_{\rm al}.
\label{eq:thm-fourier-ring-error}
\end{align}

We construct the recovered truncated Fock columns as $|\widehat v_0\rangle
:=
\Pi_{\le R}|\widehat\phi_0\rangle$, and, for \(q=1,\ldots,J\), $|\widehat v_q\rangle
:=
\frac{1}{L\,c_q(q)}
\sum_{\ell=0}^{L-1}
e^{-iq\theta_\ell}
\Pi_{\le R}|\widehat\phi_{q,\ell}\rangle $. Let us denote $\widehat A_R
:=
\bigl[
|\widehat v_0\rangle,\ldots,|\widehat v_J\rangle
\bigr]$ and $A_R
:=
\Pi_{\le R}W_g\Pi_{\le J}$. Applying Lemma~\ref{lem:output-only-column-recovery}, with $\varepsilon_{\rm tom}=\varepsilon_{\rm tr},$ and $\zeta=2\varepsilon_{\rm tr}$, and with the aliasing tolerance \(\varepsilon_{\rm al}\), yields
\begin{align}
\left\|
\widehat A_R-e^{i\omega}A_R
\right\|_\infty
\le\;&
C(J+1)^{3/4}(J+L)\varepsilon_{\rm tr}
\\
&+
C(J+1)^{3/2}\varepsilon_{\rm tr}^2
+
\sqrt{J+1}\,\varepsilon_{\rm al}.
\label{eq:thm-column-error}
\end{align}
We now check that the right-hand side is \(\mathcal O(\epsilon)\).  By the definition of
\(\varepsilon_{\rm tr}\), we have 
\begin{align}
C(J+1)^{3/4}(J+L)\varepsilon_{\rm tr}
\le
\frac{C}{C_\star}\epsilon .
\end{align}
Moreover, since \(0<\epsilon<1\) and \(J+L\ge 1\), we have 
\begin{align}
(J+1)^{3/2}\varepsilon_{\rm tr}^2
&\le
(J+1)^{3/2}
\left(
\frac{\epsilon}
{C_\star (J+1)^{3/4}(J+L)}
\right)^2
\\
&=
\frac{\epsilon^2}{C_\star^2(J+L)^2}
\le
\frac{\epsilon}{C_\star^2}.
\end{align}
Finally, by the choice $\varepsilon_{\rm al}
=
\frac{\epsilon}{C_{\rm al}\sqrt{J+1}}$, we have
\begin{align}
\sqrt{J+1}\,\varepsilon_{\rm al}
=
\frac{\epsilon}{C_{\rm al}}.
\end{align}
Hence, by choosing \(C_\star\) and \(C_{\rm al}\) sufficiently large, we obtain
\begin{align}
\left\|
\widehat A_R-e^{i\omega}A_R
\right\|_\infty
\le
\tau_\epsilon,
\qquad
\tau_\epsilon:=c_{\rm pol}\epsilon,
\label{eq:thm-tau-epsilon}
\end{align}
where \(c_{\rm pol}>0\) is a sufficiently small universal constant.

On this good event, let us define the polar isometry $\widehat Q_R
:=
\widehat A_R
(\widehat A_R^\dagger\widehat A_R)^{-1/2}$. Eq.~\eqref{eq:thm-tau-epsilon} and the smallness of \(c_{\rm pol}\)
ensure that the inverse square root is well-defined.  Extend \(\widehat Q_R\)
to a unitary \(\widehat V_R\) on \(\mathcal H_{\le R}\). In particular, since \(R\ge J\), choose an orthonormal completion
\(\{|\chi_{J+1}\rangle,\ldots,|\chi_R\rangle\}\) of
\(\{\widehat Q_R|0\rangle,\ldots,\widehat Q_R|J\rangle\}\) in
\(\mathcal H_{\le R}\), and define
\begin{align}
\widehat V_R
:=
\sum_{j=0}^{J}\widehat Q_R|j\rangle\langle j|
+
\sum_{j=J+1}^{R}|\chi_j\rangle\langle j|.
\end{align} 
Then, we define the
full-space unitary $\widehat U:=\widehat V_R\oplus I_{>R}$ and $\widehat{\mathcal W}(\rho):=\widehat U\rho\widehat U^\dagger$. Since \(W_g=W_\star e^{i\theta_0\hat n}\) satisfies the condition $\mathsf N_{W_g}(J)\le N_{\rm dyn}$, and since \(R\ge C_R N_{\rm dyn}/\epsilon^2\), Lemma
\ref{lem:output-only-polar-completion}, applied with \(U=W_g\), gives
\begin{align}
\left\|
\widehat{\mathcal W}
-
\mathcal W_g
\right\|_{\diamond}^E
\le
4\sqrt{\frac{N}{J+1}}
+
4\sqrt{\frac{N_{\rm dyn}}{R+1}}
+
C\tau_\epsilon,
\label{eq:thm-diamond-gauge-fixed}
\end{align}
where $\mathcal W_g(\rho):=W_g\rho W_g^\dagger$ denotes the unitary channel of $U_g$. Then, the choices $J
\ge
C_J\frac{N}{\epsilon^2}$, $R
\ge
C_R\frac{N_{\rm dyn}}{\epsilon^2}$, $\tau_\epsilon=c_{\rm pol}\epsilon$ imply, for sufficiently large \(C_J,C_R\) and sufficiently small \(c_{\rm pol}\),
that
\begin{align}
4\sqrt{\frac{N}{J+1}}
+
4\sqrt{\frac{N_{\rm dyn}}{R+1}}
+
C\tau_\epsilon
\le
\epsilon .
\end{align}
Thus we have 
\begin{align}
\left\|
\widehat{\mathcal W}
-
\mathcal W_g
\right\|_{\diamond}^E
\le
\epsilon .
\label{eq:thm-gauge-fixed-final}
\end{align}

It remains to return from the gauge-fixed unitary \(W_g\) to the original
unitary \(W_\star\).  Since we define $W_g=W_\star e^{i\theta_0\hat n}$, the difference between \(W_g\) and \(W_\star\) is an input-side phase-space
rotation, which is precisely part of the coherent gauge quotiented out by
\(d_{\diamond,E}^{\rm coh}\).  Therefore, we have 
\begin{align}
d_{\diamond,E}^{\rm coh}
(\widehat{\mathcal W},\mathcal W_\star)
\le
\left\|
\widehat{\mathcal W}
-
\mathcal W_g
\right\|_{\diamond}^E
\le
\epsilon .
\end{align}

The calibration stage fails with probability at most \(\delta/3\), and the
training tomography stage fails with probability at most \(\delta/3\).  All
remaining operations are deterministic on the good event.  Hence the success
probability is at least $1-\frac{\delta}{3}-\frac{\delta}{3}
\ge
1-\delta$.

Finally, the total number of channel uses is the sum of the calibration
tomography cost and the Fourier-ring training tomography cost.  The calibration
cost is \(M_{\rm cal}\), and the training cost is \(M_{\rm tr}\).  Therefore
\begin{align}
M_{\rm tot}
\le\;&
(1+|\mathcal S_{\rm cal}|)
\left\lceil
2^{21}
\left(
\frac{eN_{\rm out}}{\varepsilon_{\rm cal}^{2}}+2e
\right)
\varepsilon_{\rm cal}^{-2}
\log\frac{12(1+|\mathcal S_{\rm cal}|)}{\delta}
\right\rceil
\\
&+
(1+JL)
\left\lceil
2^{21}
\left(
\frac{eN_{\rm out}}{\varepsilon_{\rm tr}^{2}}+2e
\right)
\varepsilon_{\rm tr}^{-2}
\log\frac{12(1+JL)}{\delta}
\right\rceil .\label{eqs385}
\end{align}
This is the claimed sample-complexity bound.

If a coarse intensity window
\(\mu\in[\mu_{\min},\mu_{\max}]\) is available, one may choose the calibration
set dyadically.  For example, choose squared scales \(s_k^2\) forming a
factor-two grid from order \(1/\mu_{\max}\) to order \(1/\mu_{\min}\).  Then,
for every \(\mu\in[\mu_{\min},\mu_{\max}]\), one grid point satisfies
\begin{align}
\frac14\le s_k^2\mu\le 1,
\end{align}
and the number of grid points is
\begin{align}
\mathcal O\!\left(\log\frac{\mu_{\max}}{\mu_{\min}}\right).
\end{align}

We now rewrite this bound directly in terms of the accuracy parameters.  Set
\begin{align}
S_{\rm cal}:=|\mathcal S_{\rm cal}|,
\qquad
\Lambda_\epsilon
:=
1+\frac{N}{\epsilon^2}+\log\frac1\epsilon .
\end{align}
By the choices of \(J_\epsilon\) and \(L_\epsilon\), there is a universal
constant \(C>0\) such that the conditions $J_\epsilon+1\le C\Lambda_\epsilon$, $
J_\epsilon+L_\epsilon\le C\Lambda_\epsilon$, and $
1+J_\epsilon L_\epsilon\le C\Lambda_\epsilon^2$ are satisfied. Moreover, from the definition of \(\varepsilon_{\rm tr}\) and
\(\varepsilon_{\rm cal}\), we have $\varepsilon_{\rm tr}^{-1}
\le
C\frac{\Lambda_\epsilon^{7/4}}{\epsilon}$ and $\varepsilon_{\rm cal}^{-1}
\le
C\frac{\Lambda_\epsilon^{7/4}}{\epsilon}$. Since, for \(\varepsilon\in(0,1)\), we have 
\begin{align}
2^{21}
\left(
\frac{eN_{\rm out}}{\varepsilon^2}+2e
\right)
\varepsilon^{-2}
\log\frac4p
\le
C(N_{\rm out}+1)\varepsilon^{-4}\log\frac4p.
\end{align}
Hence, Eq.~\eqref{eqs385} implies
\begin{align}
M_{\rm tot}
\le
C(N_{\rm out}+1)
\frac{\Lambda_\epsilon^{7}}{\epsilon^4}
\left[
(1+S_{\rm cal})
\log\frac{12(1+S_{\rm cal})}{\delta}
+
(1+\Lambda_\epsilon^2)
\log\frac{12(1+\Lambda_\epsilon^2)}{\delta}
\right].
\label{eq:explicit-channel-use-bound-Lambda}
\end{align}

The dynamical parameter \(N_{\rm dyn}\) enters through the output cutoff used in
the estimator as follows
\begin{align}
R_\epsilon
=
\max\left\{
J_\epsilon,
\left\lceil C_R\frac{N_{\rm dyn}}{\epsilon^2}\right\rceil
\right\}
\le
C\left(
1+\frac{N+N_{\rm dyn}}{\epsilon^2}
\right).
\label{eq:explicit-output-cutoff-bound}
\end{align}
Thus the number of channel uses is given by
Eq.~\eqref{eq:explicit-channel-use-bound-Lambda}, while the finite-dimensional
post-processing and the final estimator act on a cutoff dimension
\(R_\epsilon+1\), which is polynomial in \(N,N_{\rm dyn}\), and \(1/\epsilon\).

Eq.~\eqref{eq:explicit-output-cutoff-bound} shows that the
protocol uses polynomially many channel calls and polynomial-dimensional
post-processing in
\begin{align}
N,\quad N_{\rm dyn},\quad N_{\rm out},\quad \frac1\epsilon,\quad
\log\frac1\delta,\quad S_{\rm cal}.
\end{align}

Finally, if a coarse intensity window
\(\mu\in[\mu_{\min},\mu_{\max}]\) is known, one may choose the calibration set
dyadically, with
\begin{align}
S_{\rm cal}
=
O\!\left(\log\frac{\mu_{\max}}{\mu_{\min}}\right).
\end{align}
Substituting this value of \(S_{\rm cal}\) into the previous display gives the
corresponding explicit bound in terms of the coarse intensity window.
\end{proof}

In the current single-mode learning steps (STU1)-(STU6) with unknown input coherent probes, the channel-use complexity depends on how the unknown coherent amplitude enters, through the size of the
calibration scale set $|\mathcal S_{\rm cal}|
=
O\!\left(
\log\frac{\mu_{\max}}{\mu_{\min}}
\right)$, where \(\mu_{\min}\le |\alpha|^2\le \mu_{\max}\) is a coarse intensity window
for the unknown coherent input state of the corresponding single-mode unitary. Due to the first passive-Gaussian layer of $U_{\rm ge}$ in Eq. (\ref{def:U_three layer with S}), we need to ensure that the ratio $\mu_{\max}/\mu_{\min}$ of the coherent
amplitudes entering the selected single-mode non-Gaussian gates is not
double-exponentially large.  Here, the following lemma
shows that this holds with high probability under the random preparation in
step~(RE1).

\begin{lemma}[Polynomial-size calibration window after passive mixing]
\label{lem:selected-mode-calibration-window}
Let \(U\in \mathrm{U}(m)\) be the interferometric matrix of an \(m\)-mode passive
Gaussian unitary.  Let the input coherent amplitudes
\(\alpha_1,\ldots,\alpha_m\) be independent and identically distributed
according to the uniform probability distribution on $D_{\sqrt{E_{\rm prep}}}
:=
\{\alpha\in\mathbb C:\ |\alpha|^2\le E_{\rm prep}\}$. For each output mode \(k\in[m]\), let us denote the output amplitude $\beta_k
:=
\sum_{j=1}^m U_{kj}\alpha_j$. Let \(\mathcal I\subseteq[m]\) be any nonempty set of selected output modes.

Then, with probability at least \(1-\delta\),
all selected output intensities satisfy $\mu_-(\delta)
\le
|\beta_k|^2
\le
\mu_+(\delta)$ for $k\in\mathcal I$ with $\mu_-(\delta)
:=
\frac{E_{\rm prep}\,\delta}{2m|\mathcal I|}$ and $\mu_+(\delta)
:=
4E_{\rm prep}\log\frac{8|\mathcal I|}{\delta}$. Consequently, we have 
\begin{align}
\log\frac{\mu_+(\delta)}{\mu_-(\delta)}
=
O\!\left(
1+
\log\frac{m|\mathcal I|}{\delta}
+
\log\log\frac{8|\mathcal I|}{\delta}
\right).
\end{align}
\end{lemma}

\begin{proof}
We prove an upper bound on the largest selected intensity and a lower bound on
the smallest selected intensity, and then combine the two estimates.

Let us first fix an output mode \(k\) and denote the output coherent amplitude $\beta_k
=
\sum_{j=1}^m U_{kj}\alpha_j$. Since the distribution of each \(\alpha_j\) is rotationally symmetric around
the origin, \(\mathbb E[\alpha_j]=0\).  Hence
\(\Re(U_{kj}\alpha_j)\) and \(\Im(U_{kj}\alpha_j)\) are centered real random
variables.  Also, for every realization of \(\alpha_j\), we have 
\begin{align}
|\Re(U_{kj}\alpha_j)|
\le
|U_{kj}|\sqrt{E_{\rm prep}},
\qquad
|\Im(U_{kj}\alpha_j)|
\le
|U_{kj}|\sqrt{E_{\rm prep}}.
\end{align}
The variables \(\{\Re(U_{kj}\alpha_j)\}_{j=1}^m\) are independent.  Applying
Hoeffding's inequality to
\(\Re\beta_k=\sum_{j=1}^m\Re(U_{kj}\alpha_j)\), we obtain, for every \(s>0\),
\begin{align}
\Pr\!\left[
|\Re\beta_k|\ge s
\right]
&\le
2\exp\!\left(
-\frac{2s^2}
{\sum_{j=1}^m(2|U_{kj}|\sqrt{E_{\rm prep}})^2}
\right)
\\
&=
2\exp\!\left(
-\frac{s^2}{2E_{\rm prep}\sum_{j=1}^m|U_{kj}|^2}
\right)
\\
&=
2\exp\!\left(
-\frac{s^2}{2E_{\rm prep}}
\right),
\end{align}
because the \(k\)-th row of \(U\) has unit Euclidean norm.  The same argument
gives
\begin{align}
\Pr\!\left[
|\Im\beta_k|\ge s
\right]
\le
2\exp\!\left(
-\frac{s^2}{2E_{\rm prep}}
\right).
\end{align}
If \(|\beta_k|^2\ge t\), then at least one of
\(|\Re\beta_k|\ge\sqrt{t/2}\) or
\(|\Im\beta_k|\ge\sqrt{t/2}\) must hold.  Therefore, by the union bound in Lemma \ref{lem:union bound}, we have 
\begin{align}
\Pr\!\left[
|\beta_k|^2\ge t
\right]
&\le
\Pr\!\left[
|\Re\beta_k|\ge \sqrt{t/2}
\right]
+
\Pr\!\left[
|\Im\beta_k|\ge \sqrt{t/2}
\right]
\\
&\le
4\exp\!\left(
-\frac{t}{4E_{\rm prep}}
\right).
\end{align}
Taking another union bound over \(k\in\mathcal I\), we get
\begin{align}
\Pr\!\left[
\max_{k\in\mathcal I}|\beta_k|^2\ge t
\right]
\le
4|\mathcal I|
\exp\!\left(
-\frac{t}{4E_{\rm prep}}
\right).
\end{align}
With $t=\mu_+(\delta)
=
4E_{\rm prep}\log\frac{8|\mathcal I|}{\delta}$, the right-hand side equals $4|\mathcal I|
\exp\!\left(
-\log\frac{8|\mathcal I|}{\delta}
\right)
=
\frac{\delta}{2}$. Thus, we have 
\begin{align}
\Pr\!\left[
\max_{k\in\mathcal I}|\beta_k|^2>\mu_+(\delta)
\right]
\le
\frac{\delta}{2}.
\label{eq:selected-upper-tail-new}
\end{align}

We now prove the lower-tail estimate. Fix \(k\in\mathcal I\) and let \(u>0\).
Since the \(k\)-th row of \(U\) has unit Euclidean norm, i.e., $\sum_{j=1}^m |U_{kj}|^2=1$,  at least one entry in this row must satisfy $|U_{kj_\star}|^2\ge \frac1m$. Let us fix such an index \(j_\star=j_\star(k)\). We now condition on all input amplitudes except \(\alpha_{j_\star}\).  Under this
conditioning, $B:=\sum_{j\neq j_\star}U_{kj}\alpha_j$ is a fixed complex number, while \(\alpha_{j_\star}\) remains uniformly
distributed on \(D_{\sqrt{E_{\rm prep}}}\), by independence. Therefore, we have 
\begin{align}
\beta_k
=
U_{kj_\star}\alpha_{j_\star}+B .
\end{align}
Let us set $Z:=U_{kj_\star}\alpha_{j_\star}$. Conditionally on the other amplitudes, \(Z\) is uniformly distributed on the
disk centered at the origin with radius $|U_{kj_\star}|\sqrt{E_{\rm prep}}$, whose area is $\pi E_{\rm prep}|U_{kj_\star}|^2$.

Now consider the event \(|\beta_k|^2\le u\). Since \(\beta_k=Z+B\), this event
is equivalent to $|Z+B|\le \sqrt u$. Thus, conditionally on the other amplitudes, the event \(|\beta_k|^2\le u\)
occurs only when the uniform random point \(Z\) falls inside the disk of radius
\(\sqrt u\) centered at \(-B\). Intersecting this disk with the support of \(Z\)
can only reduce its area. Hence
\begin{align}
\Pr\!\left[
|\beta_k|^2\le u
\,\middle|\,
\{\alpha_j:j\neq j_\star\}
\right]
\le
\frac{\pi u}{\pi E_{\rm prep}|U_{kj_\star}|^2}.
\end{align}
Using the condition \(|U_{kj_\star}|^2\ge 1/m\), we obtain
\begin{align}
\Pr\!\left[
|\beta_k|^2\le u
\,\middle|\,
\{\alpha_j:j\neq j_\star\}
\right]
\le
\frac{mu}{E_{\rm prep}}.
\end{align}

The right-hand side is deterministic and does not depend on the values of the
conditioned amplitudes $\{\alpha_j,j\neq j_\star\}$. Therefore, by the tower property of conditional
expectation, we have 
\begin{align}
\Pr\!\left[
|\beta_k|^2\le u
\right]
=
\mathbb E\!\left[
\Pr\!\left[
|\beta_k|^2\le u
\,\middle|\,
\{\alpha_j:j\neq j_\star\}
\right]
\right]
\le
\mathbb E\!\left[
\frac{mu}{E_{\rm prep}}
\right]
=
\frac{mu}{E_{\rm prep}},
\end{align}
where the average is taken over $\{\alpha_j,j\neq j_\star\}$ that appear in $B$. 

Finally, applying the union bound over all selected modes \(k\in\mathcal I\),
we get
\begin{align}
\Pr\!\left[
\min_{k\in\mathcal I}|\beta_k|^2\le u
\right]
\le
\sum_{k\in\mathcal I}
\Pr\!\left[
|\beta_k|^2\le u
\right]
\le
\frac{m|\mathcal I|u}{E_{\rm prep}}.
\end{align}
Therefore, choosing $u=\mu_-(\delta)
=
\frac{E_{\rm prep}\delta}{2m|\mathcal I|}$ gives
\begin{align}
\Pr\!\left[
\min_{k\in\mathcal I}|\beta_k|^2\le \mu_-(\delta)
\right]
\le
\frac{\delta}{2}.\label{eq:selected-lower-tail-new}
\end{align}

Combining Eqs.~\eqref{eq:selected-upper-tail-new} and
\eqref{eq:selected-lower-tail-new}, we obtain, with probability at least
\(1-\delta\),
\begin{align}
\mu_-(\delta)
\le
|\beta_k|^2
\le
\mu_+(\delta),
\qquad
k\in\mathcal I.
\end{align}

It remains to compute the size of the logarithmic window.  By the definitions
of \(\mu_-(\delta)\) and \(\mu_+(\delta)\),
\begin{align}
\frac{\mu_+(\delta)}{\mu_-(\delta)}
&=
\frac{
4E_{\rm prep}\log(8|\mathcal I|/\delta)
}{
E_{\rm prep}\delta/(2m|\mathcal I|)
}
\\
&=
\frac{8m|\mathcal I|}{\delta}
\log\frac{8|\mathcal I|}{\delta}.
\end{align}
Therefore, we have 
\begin{align}
\log\frac{\mu_+(\delta)}{\mu_-(\delta)}
&=
\log\frac{8m|\mathcal I|}{\delta}
+
\log\log\frac{8|\mathcal I|}{\delta}.
\end{align}
This proves the claimed scaling and completes the proof.
\end{proof}

\subsubsection{Learning of the left multimode-Gaussian unitary}\label{supp:learning step 4-last Gaussian layer}

As established in step (RE3), we identify a set of output modes
\(\map X_{\rm pure}\) that stably yield pure reduced states, with
\(\map N_{\rm re}^\star \subseteq \map X_{\rm pure}\). Here
\(\map N_{\rm re}^\star\) denotes the modes corresponding to the
single-mode non-Gaussian states in the true partition of the underlying
\(m\)-mode state in the ideal case without tomography errors, namely when
\(\epsilon_j'=0\) for all \(j\). Meanwhile, all misidentified modes corresponding to the multi-mode Gaussian state belonging to $[m]\setminus\map N_{\rm re}^\star$ are included in $\chi_{\rm pure}$. On these accounts, the next steps of the protocol for learning $U_{\rm rest}$ in Eq. (\ref{def:U_rest}) will be as follows   
\begin{enumerate}

\item [(RE4)] \emph{Single-mode unitary learning.} 


As required in Step~(STU1), we prepare the vacuum together with
\(K_{\rm wid}\) independent coherent base probes, whose amplitudes are sampled
as in Lemma~\ref{lem:selected-mode-calibration-window} with  $K_{\rm wid}
:=
\left\lceil
\frac{\log(1/\delta_{\rm wid})}{\log 4}
\right\rceil$. Apply Lemma~\ref{lem:selected-mode-calibration-window} with failure
probability \(1/4\) and with the selected set
\(\mathcal I=\map X_{\rm pure}\). For each base probe, with probability at
least \(3/4\), all selected input intensities satisfy $\mu_-^{(0)}
\le
\left|(U_O\bs\alpha)_j\right|^2
\le
\mu_+^{(0)}, j\in\map X_{\rm pure}$, with $\mu_-^{(0)}:=
\frac{E_{\rm prep}}{8m|\map X_{\rm pure}|}$ and $\mu_+^{(0)}:=
4E_{\rm prep}
\log\!\left(32|\map X_{\rm pure}|\right)$. Hence, with probability at least \(1-\delta_{\rm wid}\), at least one of the
\(K_{\rm wid}\) base probes satisfies this window simultaneously for all
selected modes. We use
\([\mu_-^{(0)},\mu_+^{(0)}]\) as the common coarse calibration window in
Theorem~\ref{thm:output-only-efficient-unknown-coherent-main}. Consequently, we have $S_{\rm cal}
=
\mathcal O\!\left(
\log\frac{\mu_+^{(0)}}{\mu_-^{(0)}}
\right)
=
O\!\left(
\log\!\left(m|\map X_{\rm pure}|\right)
\right)$.

Then, we apply identical attenuation channels and phase rotations to the
initial coherent probe states. Since these operations commute \cite{brod2020classical,serafini2023quantum} with the first
passive-Gaussian layer \(U_O\) in Eq.~\eqref{def:U_three layer with S}, they may
be equivalently regarded as operations applied at the output ports of \(U_O\).
By suitably choosing the attenuation ratios and phase-rotation angles, the
coherent amplitudes after \(U_O\) on the modes \(j\in\chi_{\rm pure}\) can be
made to satisfy the requirement of step (STU2), thereby forming the desired
Fourier-ring geometry.

Apply steps (STU3)-(STU6) to all the modes $j\in \map X_{\rm pure}$ to obtain approximate single-mode unitaries:  
\begin{align}
d_{\diamond,E^{**}}^{\mathrm{coh}}\left(\widetilde W_j',W_j'\right)\le \epsilon_j,\ \ \ j\in \map X_{\rm pure},
\end{align}
where  $W_j'=U^{\bs\alpha^{(\ell)},\bs\alpha^{(h)}}_{R(-\vartheta_{\pi'j})}
U^\dag_{S_{\pi\pi'j,\bs\alpha^{(\ell)}}} W_{\pi\pi'j}$ for $j\in \map N_{\rm re}^\star$ and $\widetilde W_j'$ is an approximate single-mode Gaussian unitary for $j\in \map X_{\rm pure}\setminus \map N_{\rm re}^\star$. 
\item[(RE5)] \emph{Gaussian unitary learning via local tomography.} Use the \(2m+1\) coherent probes \(\{|\underline{\bs \alpha^{(g)}}\rangle\}_{g=0}^{2m}\) from the vacuum-shared protocol of Ref.~\cite{fanizza2025efficient}, as specified in Proposition~\ref{thm:structured-product-output-learning}. Send each probe through \(\widetilde U_{\rm rest}\).  Conduct local tomography to reconstruct the reduced states $\{\tilde \rho_j^{(g)},j\in \map X_{\rm pure}\}$ for $|\Psi^{\rm prod}_{\bs\alpha^{(\ell)},\bs\alpha^{(h)},\bs\alpha^{(g)}}\>$. Implement the efficient Gaussian-state learning protocol \cite{mele2025learning} for $\widetilde \Psi_{[m]\setminus \map X_{\rm pure}}=\Tr_{\map X_{\rm pure}}[|\Psi^{\rm prod}_{\bs\alpha^{(\ell)},\bs\alpha^{(h)},\bs\alpha^{(g)}}\>\<\Psi^{\rm prod}_{\bs\alpha^{(\ell)},\bs\alpha^{(h)},\bs\alpha^{(g)}}|]$. The tomography errors are as follows:
\begin{align}
\begin{cases}
\frac 1 2 \left\|\widetilde \rho_j^{(g)} - \Tr_{j'\neq j}\left[|\Psi^{\rm prod}_{\bs\alpha^{(\ell)},\bs\alpha^{(h)},\bs\alpha^{(g)}}\rangle\<\Psi^{\rm prod}_{\bs\alpha^{(\ell)},\bs\alpha^{(h)},\bs\alpha^{(g)}}|\right]\right\|_1 \le \epsilon_j',\ \ \ \ \forall j\in\map X_{\rm pure}\\
\frac 1 2 \left\|\widetilde \Psi_{[m]\setminus \map X_{\rm pure}}- \Tr_{\map X_{\rm pure}}\left[|\Psi^{\rm prod}_{\bs\alpha^{(\ell)},\bs\alpha^{(h)},\bs\alpha^{(g)}}\rangle\<\Psi^{\rm prod}_{\bs\alpha^{(\ell)},\bs\alpha^{(h)},\bs\alpha^{(g)}}|\right]\right\|_1 \le \left(m-\left|\map X_{\rm pure}\right|\right)\epsilon_G'
\end{cases}.
\end{align}

Reconstruct an approximate Gaussian state as 
\begin{align}
\widetilde \Psi_{G}^{(g)}&= \bigotimes_{j\in \map X_{\rm pure}} \widetilde W_j'^{\dag}\,\tilde \rho_j^{(g)}\,\widetilde W_j'\otimes \widetilde \Psi_{[m]\setminus \map X_{\rm pure}}.
\end{align}

From the input-output relation, we can efficiently reconstruct the approximate Gaussian unitary  (see Proposition \ref{thm:structured-product-output-learning})
\begin{align}
\widetilde U_{\rm final}&= \left(\bigotimes_{j\in \map X_{\rm pure}}\widetilde W_j^{'\dag} \otimes I_{\map G_{\rm re}}\right)\widetilde U_{\rm rest}
\end{align}

Note that the counter-rotation $\{\widetilde W'^\dag_j\}$ might includes some single-mode Gaussian unitary acting after the Gaussian mixing unitary $U_{\pi'\map G_{\rm re}^\star}^{\bs\alpha^{(\ell)},\bs\alpha^{(h)}}$, which is from the degeneracy of local covariance matrices (see \ref{supp:Learning of the second passive-Gaussian layer}). This phenomenon does not make the protocol fail as the corresponding unitaries $\{\widetilde W_j'^{\dag}\}$  transform a joint Gaussian unitary to another Gaussian unitary, which can be further identified via Gaussian unitary learning.

\item[(RE6)] \emph{Final assemble.} Reconstruct the estimate of the full unitary $U_{\rm ge}$ as follows: 
\begin{align}
\widetilde U_{\rm ge}=U_{\widetilde S_{\bs\alpha^{(\ell)}}} U_{\widetilde O_{2,\bs\alpha^{(l)},\bs\alpha^{(h)}}} \left[\bigotimes_{j\in \chi_{\rm pure}}\widetilde W_j^{'} \otimes I_{\map G_{\rm re}\setminus \chi_{\rm pure}}\right]\widetilde U_{\rm final}.
\end{align}
\end{enumerate}

\begin{proposition}[Structured Gaussian unitary learning from local tomography]
\label{thm:structured-product-output-learning}
Consider the learning task for an $m$-mode Gaussian unitary as follows:
\begin{align}\label{eqs251}
U_{\rm final}
=
\left(
\bigotimes_{j\in J} U_j
\;\otimes\;
U_R
\right)U_O ,
\end{align}
where $J\subseteq [m]$, $|J|=\mathcal O(m)$, each $U_j$ is a single-mode
Gaussian unitary consisting of a displacement and a phase rotation, $U_R$ is an
$(m-|J|)$-mode Gaussian unitary, and $U_O$ is passive. 
For every probe $|\underline{x^{(a)}}\>$ with $\<\underline{x^{(a)}}|\widehat H_m|\underline{x^{(a)}}\>\le mE, (a=0,\cdots,2m)$, the output $\rho^{(a)}
:=
\mathcal U_{\rm final}\bigl(|\underline{x^{(a)}}\rangle\langle \underline{x^{(a)}}|\bigr)=\left(
\bigotimes_{j\in J}\rho^{(a)}_j
\right)
\otimes
\rho^{(a)}_R$ satisfies the energy constraints 
\begin{align}
\Tr[\rho^{(a)}_j\widehat E_1]\le E',
\qquad
\Tr[\rho^{(a)}_R\widehat E_r]\le rE',\ \ (R:=[m]\setminus J, r:=|R|, E'\ge 1/2).
\label{eq:local-output-photon-bound}
\end{align}

Then, there exists a local-tomography-based learning protocol that returns an estimator $\widehat{\mathcal U}_{\rm final}$ such that
\begin{align}
\Pr\!\left[
\left\|
\widehat{\mathcal U}_{\rm final}-\mathcal U_{\rm final}
\right\|_{\diamond}^{mE}
\le
\epsilon
\right]
\ge
1-\delta .
\end{align}
The corresponding number of channel queries is
\begin{align}
\mathsf M_{\rm tot}
=
\mathcal O\!\left(
\frac{
m^{16}(E'+1/2)^{10}(mE-m/2+1)^4(8mE'+1)^2
}{
(E-1/2)^2\epsilon^8
}
\log\!\frac{m^4}{\delta}
\right).
\label{eq:corrected-query-bound-factorized}
\end{align}
\end{proposition}

\begin{proof}
Set parameters $N\coloneqq E-\frac12$ and $N'\coloneqq E'-\frac12$. Thus \(mN\) is the total input photon-number budget and \(N'\) is the per-mode output photon-number budget. Let us first define the parametrization of the overall Gaussian unitary and the
corresponding probe family. Without loss of generality, we can write the Gaussian unitary
$U_{\rm final}$ in the form $U_{\rm final}:=
 D_{d}  U_S$ for some displacement vector ${d}\in\mathbb R^{2m}$ and some symplectic
matrix $S\in\mathrm{Sp}_{2m}(\mathbb R)$ satisfying $\|S\|_\infty\le z$. Here, we adopt the fixed coherent probe family appearing
in the vacuum-shared protocol \cite{fanizza2025efficient}, namely
\begin{align}
x^{(0)}:=0,
\qquad
x^{(k)}:=\eta e_k,
\qquad
k=1,\dots,2m,
\end{align}
where $e_k$ is the $k$th standard basis vector of $\mathbb R^{2m}$ and
$\eta=\Theta(\sqrt{mN})$ is chosen so that the input photon number is at most
$mN$. For definiteness, and since all constant factors are absorbed in the
asymptotic notation, we may take $\eta=\sqrt{mN}$. Let us denote the output state as  $\rho^{(a)}
:=
\mathcal U_{\rm final}\bigl(|\underline{x^{(a)}}\rangle\langle \underline{x^{(a)}}|\bigr)$ with $a=0,\dots,2m$. By the structural assumption in Eq.~\eqref{eqs251}, each $\rho^{(a)}$ factorizes
as
\begin{align}
\rho^{(a)}
=
\left(
\bigotimes_{j\in J}\rho^{(a)}_j
\right)
\otimes
\rho^{(a)}_R .
\end{align}

Now, we implement local tomography on the output product states. Let us set  $\delta_0
:=
\frac{\delta}{(2m+1)(|J|+1)}$ and denote the energy operator $\widehat E_n
:=\widehat N_n+\frac n2 I$ with $\widehat N_n=\sum_{j=1}^n a_j^\dag a_j$. For an $n$-mode pure Gaussian state satisfying $\Tr[\psi\widehat E_n]\le n\mathcal E$, tomography accuracy $\varepsilon$, and failure probability $\delta_0$,
Theorem~S59 of \cite{mele2025learning} gives the sufficient sample complexity
\begin{align}
\mathsf M_{\rm pG}(n,\mathcal E,\varepsilon,\delta_0)
:=
(n+3)\left\lceil
217600\, n^2 \mathcal E^2(24n^2 \mathcal E^2+3n)\,
\varepsilon^{-4}
\log\!\left(
\frac{2(2n^2+3n)}{\delta_0}
\right)
\right\rceil .
\label{eq:Mpg-structured}
\end{align}
Applying this theorem separately to every factor of every probe
output, we use $\mathsf M_{\rm pG}\!\left(1,N'+\frac12,\epsilon'_j,\delta_0\right)$ copies for the single-mode state $\rho^{(a)}_j$ and $\mathsf M_{\rm pG}\!\left(r,N'+\frac12,r\epsilon',\delta_0\right)$ copies for the $r$-block state $\rho^{(a)}_R$. Given that we have $2m+1$ different probes, we can take a union bound over all
$(2m+1)(|J|+1)$ tomography processes. Then, we obtain that with probability at least
$1-\delta$, simultaneously for all probes $a=0,\dots,2m$,
\begin{align}
\frac12\left\|
\widehat\rho^{(a)}_j-\rho^{(a)}_j
\right\|_1
\le
\epsilon'_j,
\qquad
j\in J,
\label{eq:single-mode-trace-err-corrected}
\end{align}
and
\begin{align}
\frac12\left\|
\widehat\rho^{(a)}_R-\rho^{(a)}_R
\right\|_1
\le
r\epsilon' .
\label{eq:block-trace-err-corrected}
\end{align}

Therefore, the total number of copies, equivalently channel queries, used by
the protocol is
\begin{align}
\mathsf M_{\rm tot}
=
(2m+1)\left(
\sum_{j\in J}
\mathsf M_{\rm pG}\!\left(1,N'+\frac12,\epsilon'_j,\delta_0\right)
+
\mathsf M_{\rm pG}\!\left(r,N'+\frac12,r\epsilon',\delta_0\right)
\right).
\label{eq:structured-total-copies-exact-corrected}
\end{align}

Then, let us connect the trace-distance error from local tomography to the
distance in first moments. The estimator-energy bound used in the proof of Theorem~S59 of \cite{mele2025learning} implies (see Eq.~(S324) of \cite{mele2025learning}): $\Tr[\widehat\rho\,\widehat E_n]
\le
2\Tr[\rho\,\widehat E_n]$. Hence, using the photon-number assumption in
\eqref{eq:local-output-photon-bound}, we obtain
\begin{align}
\Tr[\widehat\rho^{(a)}_j\widehat E_1]
\le
2\Tr[\rho^{(a)}_j\widehat E_1]
\le
2\left(N'+\frac12\right)
=
2N'+1,
\qquad
j\in J,
\label{eq:single-mode-est-energy-corrected}
\end{align}
and
\begin{align}
\Tr[\widehat\rho^{(a)}_R\widehat E_r]
\le
2\Tr[\rho^{(a)}_R\widehat E_r]
\le
2r\left(N'+\frac12\right)
=
2rN'+r .
\label{eq:block-est-energy-corrected}
\end{align}

Let us denote the first moments $\mu^{(a)}_j:=\overline{\bs r}(\rho^{(a)}_j)$, $\widehat\mu^{(a)}_j:=\overline{\bs r}(\widehat\rho^{(a)}_j)$ for $j\in J$, $\mu^{(a)}_R:=\overline{\bs r}(\rho^{(a)}_R),$ and $\widehat\mu^{(a)}_R:=\overline{\bs r}(\widehat\rho^{(a)}_R)$.

Assume the condition
\begin{align}
\max\Bigl\{\max_{j\in J}\epsilon'_j,\; r\epsilon'\Bigr\}<\frac1{200}.
\end{align}
Since both $\rho^{(a)}_j$ and $\widehat\rho^{(a)}_j$ have
$\widehat E_1$-energy at most $2N'+1$ by
\eqref{eq:single-mode-est-energy-corrected}, Theorem~S52 of
\cite{mele2025learning} gives
\begin{align}
\left\|
\widehat\mu^{(a)}_j-\mu^{(a)}_j
\right\|_2
\le
200\,\epsilon'_j\sqrt{4(2N'+1)+1}
=
200\,\epsilon'_j\sqrt{8N'+5},
\qquad
j\in J.
\label{eq:single-mode-moment-bound-corrected}
\end{align}
Likewise, since both $\rho^{(a)}_R$ and $\widehat\rho^{(a)}_R$ have
$\widehat E_r$-energy at most $2rN'+r$ by
\eqref{eq:block-est-energy-corrected}, the same theorem yields
\begin{align}
\left\|
\widehat\mu^{(a)}_R-\mu^{(a)}_R
\right\|_2
\le
200\,r\epsilon'\sqrt{4(2rN'+r)+1}
=
200\,r\epsilon'\sqrt{8rN'+4r+1}.
\label{eq:block-moment-bound-corrected}
\end{align}

Now concatenate the first moments according to the mode ordering $J\,|\,R$ and
define $\mu^{(a)}
:=
\left(
(\mu^{(a)}_j)_{j\in J},
\mu^{(a)}_R
\right)$ and $\widehat\mu^{(a)}
:=
\left(
(\widehat\mu^{(a)}_j)_{j\in J},
\widehat\mu^{(a)}_R
\right)$. Combining Eqs. \eqref{eq:single-mode-moment-bound-corrected} and
\eqref{eq:block-moment-bound-corrected} gives, for every probe $a$,
\begin{align}
\left\|
\widehat\mu^{(a)}-\mu^{(a)}
\right\|_2
\le
200\,\Gamma,
\label{eq:global-moment-bound-corrected}
\end{align}
with
\begin{align}
\Gamma
:=
\sqrt{
(8N'+5)\sum_{j\in J}(\epsilon'_j)^2
+
r^2(8rN'+4r+1)(\epsilon')^2
}.
\label{eq:Gamma-def-corrected}
\end{align}

Then, we recover the global displacement vector and the global symplectic
matrix. Note that the action of a Gaussian unitary
$G_{d,S}= D_{d} U_S$ on first moments
is affine: $\overline{\bs r}\!\left(\mathcal G_{d,S}(\rho)\right)
=
S\overline{\bs r}(\rho)+\bs\xi$. Hence, for the chosen coherent probes, we have
\begin{align}
\mu^{(0)}=d,
\qquad
\mu^{(k)}=d+\eta Se_k,
\qquad
k=1,\dots,2m .
\label{eq:probe-moment-identities-corrected}
\end{align}
Let us define the raw estimators $\widehat d:=\widehat\mu^{(0)}$ and $\widehat A e_k
:=
\frac{\widehat\mu^{(k)}-\widehat\mu^{(0)}}{\eta}$ for $k=1,\dots,2m$. Then Eqs.  \eqref{eq:global-moment-bound-corrected} and
\eqref{eq:probe-moment-identities-corrected} imply that
\begin{align}
\|\widehat d-d\|_2
\le
200\,\Gamma ,
\label{eq:d-error-structured-corrected}
\end{align}
and, for every $k=1,\dots,2m$,
\begin{align}
\|(\widehat A-S)e_k\|_2
&=
\frac1\eta
\left\|
(\widehat\mu^{(k)}-\mu^{(k)})
-
(\widehat\mu^{(0)}-\mu^{(0)})
\right\|_2
\nonumber\\
&\le
\frac{400\,\Gamma}{\eta}.
\end{align}
Therefore, we have
\begin{align}
\|\widehat A-S\|_\infty
\le
\|\widehat A-S\|_{\rm F}
\le
\frac{400\sqrt{2m}\,\Gamma}{\eta}
:=
\alpha .
\label{eq:A-error-structured-corrected}
\end{align}

Now, we implement a symplectic regularization protocol. To enforce exact
symplecticity while staying close to the raw estimate, we define
\begin{align}
\widehat S
:=
\bigl(-\Omega\widehat A^\top\Omega\widehat A\bigr)^{-1/2}\widehat A .
\label{eq:sympl-regularization-structured-corrected}
\end{align}
For $\|S\|_\infty \le z$, if we have
\begin{align}
(2z+1)\alpha<\frac12,
\label{eq:smallness-regularization-corrected}
\end{align}
then Lemma~4.4 of \cite{fanizza2025efficient} yields
$\widehat S\in\mathrm{Sp}_{2m}(\mathbb R)$ together with
\begin{align}
\|\widehat S-S\|_\infty
\le
9z^2\alpha .
\label{eq:S-error-structured-corrected}
\end{align}
Using Eq. \eqref{eq:A-error-structured-corrected}, condition
\eqref{eq:smallness-regularization-corrected} is ensured by
\begin{align}
\Gamma
<
\frac{\eta}{800(2z+1)\sqrt{2m}}.
\label{eq:Gamma-smallness-corrected}
\end{align}

Next, let us connect the parameter error to energy-constrained diamond-norm
error. Lemma~6.3 of \cite{fanizza2025efficient}, applied with total input
photon-number parameter $\bar n=mN$ and target parameter $\epsilon/2$, implies
that if
\begin{align}
\|\widehat S-S\|_\infty
\le
\frac{\epsilon^2}{10368\,m z(mN+1)},
\qquad
\|\widehat d-d\|_2
\le
\frac{\epsilon}{4\sqrt2\sqrt{z^2mN+1}},
\label{eq:our-thresholds-corrected}
\end{align}
then
\begin{align}
\left\|
\widehat{\mathcal U}_{\rm final}-\mathcal U_{\rm final}
\right\|_{\diamond}^{mE}
\le
\epsilon.
\end{align}
Using Eqs. \eqref{eq:d-error-structured-corrected} and
\eqref{eq:S-error-structured-corrected}, a sufficient condition for
Eq. \eqref{eq:our-thresholds-corrected} is
\begin{align}
200\,\Gamma
&\le
\frac{\epsilon}{4\sqrt2\sqrt{z^2mN+1}},
\label{eq:disp-threshold-expanded-corrected}
\\
9z^2\alpha
&\le
\frac{\epsilon^2}{10368\,m z(mN+1)}.
\label{eq:sympl-threshold-expanded-corrected}
\end{align}
Substituting Eqs. \eqref{eq:A-error-structured-corrected} into
\eqref{eq:sympl-threshold-expanded-corrected} gives
\begin{align}
\Gamma
\le
\frac{\eta\epsilon^2}
{37324800\sqrt2\,m^{3/2}z^3(mN+1)}.
\label{eq:Gamma-symplectic-corrected}
\end{align}
Likewise, Eq. \eqref{eq:disp-threshold-expanded-corrected} is equivalent to
\begin{align}
\Gamma
\le
\frac{\epsilon}
{800\sqrt2\,\sqrt{z^2mN+1}}.
\label{eq:Gamma-displacement-corrected}
\end{align}

Consequently, a sufficient compatibility condition on the local tomography
errors is
\begin{align}
\Gamma
\le
\Gamma_\star
:=
\min\left\{
\frac{\epsilon}{800\sqrt2\,\sqrt{z^2mN+1}},
\;
\frac{\eta\epsilon^2}
{37324800\sqrt2\,m^{3/2}z^3(mN+1)},
\;
\frac{\eta}{800(2z+1)\sqrt{2m}}
\right\},
\label{eq:Gamma-compatibility-corrected}
\end{align}
where $\Gamma$ is defined in Eq. \eqref{eq:Gamma-def-corrected}, together with $\max\Bigl\{\max_{j\in J}\epsilon'_j,\;r\epsilon'\Bigr\}<\frac1{200}$. Under these conditions, Lemma~6.3 of \cite{fanizza2025efficient} yields $\left\|
\widehat{\mathcal U}_{\rm final}-\mathcal U_{\rm final}
\right\|_{\diamond}^{mE}
\le
\epsilon$.

Finally, let us summarize the scalings in query complexity by simplifying
\eqref{eq:structured-total-copies-exact-corrected}. By
\eqref{eq:Mpg-structured}, for fixed universal constants, we have
\begin{align}
\mathsf M_{\rm pG}\!\left(1,N'+\frac12,\epsilon'_j,\delta_0\right)
=
\mathcal O\!\left(
\frac{(N'+1)^4}{(\epsilon'_j)^4}
\log\!\frac1{\delta_0}
\right),
\end{align}
and
\begin{align}
\mathsf M_{\rm pG}\!\left(r,N'+\frac12,r\epsilon',\delta_0\right)
=
\mathcal O\!\left(
\frac{r(N'+1)^4}{(\epsilon')^4}
\log\!\frac{r^2}{\delta_0}
\right).
\end{align}
Thus, using the condition  $\delta_0=\frac{\delta}{(2m+1)(|J|+1)}$ with $r\le m$, Eq.~\eqref{eq:structured-total-copies-exact-corrected} gives
\begin{align}
\mathsf M_{\rm tot}
=
\mathcal O\!\left(
m\log\!\frac{m^3(|J|+1)}{\delta}
\left[
\sum_{j\in J}\frac{(N'+1)^4}{(\epsilon'_j)^4}
+
\frac{r(N'+1)^4}{(\epsilon')^4}
\right]
\right).
\label{eq:Mtot-before-explicit-choice-corrected}
\end{align}

To make the dependence on the target diamond error $\epsilon$ explicit, it is
sufficient to choose the tomography accuracies so that
\begin{align}
(8N'+5)\sum_{j\in J}(\epsilon'_j)^2
\le
\frac{\Gamma_\star^2}{4},
\qquad
r^2(8rN'+4r+1)(\epsilon')^2
\le
\frac{\Gamma_\star^2}{4}.
\label{eq:explicit-epsilon-choice-1-corrected}
\end{align}
For instance, one may take
\begin{align}
\epsilon'_j
=
\frac{\Gamma_\star}{2\sqrt{|J|(8N'+5)}},
\qquad
j\in J,
\qquad
\epsilon'
=
\frac{\Gamma_\star}{2r\sqrt{8rN'+4r+1}}.
\label{eq:explicit-epsilon-choice-2-corrected}
\end{align}
Then, we have 
\begin{align}
\Gamma^2
=
(8N'+5)\sum_{j\in J}(\epsilon'_j)^2
+
r^2(8rN'+4r+1)(\epsilon')^2
\le
\frac{\Gamma_\star^2}{2},
\end{align}
and hence $\Gamma\le \Gamma_\star$. Therefore all conditions in Eq. 
\eqref{eq:Gamma-compatibility-corrected} are satisfied, provided the local
trace-distance errors are also below $1/200$ as above.

Substituting Eq. \eqref{eq:explicit-epsilon-choice-2-corrected} into Eq. 
\eqref{eq:structured-total-copies-exact-corrected} gives
\begin{align}
\mathsf M_{\rm tot}
&\le
C_0\,m
\log\!\left(
\frac{C_1\,m^3(|J|+1)}{\delta}
\right)
\left[
\sum_{j\in J}
\frac{(N'+1)^4\,|J|^2(8N'+5)^2}{\Gamma_\star^4}
+
\frac{r^5(N'+1)^4(8rN'+4r+1)^2}{\Gamma_\star^4}
\right]
\nonumber\\
&\le
C_0\,m
\log\!\left(
\frac{C_1\,m^3(|J|+1)}{\delta}
\right)
\frac{(N'+1)^4}{\Gamma_\star^4}
\left[
|J|^3(8N'+5)^2
+
r^5(8rN'+4r+1)^2
\right],
\label{eq:Mtot-explicit-Gammastar-corrected}
\end{align}
for universal constants $C_0,C_1>0$.

In the small-error regime, the second term in the definition of
$\Gamma_\star$ is the dominant one. Therefore, we can choose
\begin{align}
\Gamma_\star^{-4}
=
\mathcal O\!\left(
\frac{m^6z^{12}(mN+1)^4}{\eta^4\epsilon^8}
\right).
\end{align}
Substituting this into Eq. 
\eqref{eq:Mtot-explicit-Gammastar-corrected} yields
\begin{align}
\mathsf M_{\rm tot}
&=
\mathcal O\!\left(
\frac{
m^7 z^{12}(mN+1)^4 (N'+1)^4
}{
\eta^4\epsilon^8
}
\log\!\frac{m^3(|J|+1)}{\delta}
\left[
|J|^3(8N'+5)^2
+
r^5(8rN'+4r+1)^2
\right]
\right).
\label{eq:Mtot-small-error-z-corrected}
\end{align}

Since the coherent probes in the learning protocol are chosen with
$\eta=\Theta(\sqrt{mN})$, we have $\eta^4=\Theta(m^2N^2)$. Moreover, applying
the photon-number identity (Lemma S55 of \cite{mele2025learning}) $\Tr[\rho\widehat N_m]
=
\frac{\Tr[V(\rho)-I]}4
+
\frac{\|m(\rho)\|_2^2}{2}$ to the vacuum output $\rho^{(0)}$ gives
\begin{align}
\Tr[\rho^{(0)}\widehat N_m]
=
\frac{\Tr[SS^\top-I]}4+\frac{\|d\|_2^2}{2}.
\end{align}
Using the output photon-number bound
$\Tr[\rho^{(0)}\widehat N_m]\le mN'$ and dropping the nonnegative displacement
term, we obtain
\begin{align}
\|S\|_\infty
\le
\|S\|_{\rm F}
=
\sqrt{\Tr[SS^\top]}
\le
\sqrt{4mN'+2m}
=
\mathcal O\!\bigl(\sqrt{m(N'+1)}\bigr).
\label{eq:S-bound-from-output-photon-corrected}
\end{align}
Thus $z^{12}=\mathcal O(m^6(N'+1)^6)$. If $|J|=\mathcal O(m)$ and
$r=\mathcal O(m)$, then
\begin{align}
|J|^3(8N'+5)^2
+
r^5(8rN'+4r+1)^2
=
\mathcal O\!\left(
m^5(8mN'+4m+1)^2
\right).
\end{align}
Combining these estimates in
Eq. \eqref{eq:Mtot-small-error-z-corrected} yields
\begin{align}
\mathsf M_{\rm tot}
=
\mathcal O\!\left(
\frac{
m^{16}(N'+1)^{10}(mN+1)^4(8mN'+4m+1)^2
}{
N^2\epsilon^8
}
\log\!\frac{m^4}{\delta}
\right).
\end{align}

In the same small-error regime, the explicit choice in Eq. 
\eqref{eq:explicit-epsilon-choice-2-corrected} may be summarized as
\begin{align}
\epsilon'_j
&=
\mathcal O\!\left(
\frac{
\sqrt{N}\,\epsilon^2
}{
m^{3}(N'+1)^2(mN+1)
}
\right),
\qquad
j\in J,
\\
\epsilon'
&=
\mathcal O\!\left(
\frac{
\sqrt{N}\,\epsilon^2
}{
m^{4}(N'+1)^2(mN+1)
}
\right).
\end{align}
This completes the proof.
\end{proof}

\subsection{Summary of the learning protocol}\label{supp:final protocol of GE layer}

Let us summarize the learning protocol for the Gaussian-entanglable unitary $U_{\rm ge}=U_{S} \left(\bigotimes_{j=1}^m W_j\right) U_{O}$, which proceeds in the following steps.
\begin{enumerate}
\item[Step 1.] \emph{Estimate $U_S$ and physical counter-rotation.} Probe with multiple copies of \textbf{a coherent state} $|\underline{\bs\alpha^{(\ell)}}\>$.  Perform \textbf{heterodyne measurement} on the Gaussian-entanglable state
\begin{align}
|\Psi_{\bs\alpha^{(\ell)}}\>:=U_{\rm ge} |\underline{\bs\alpha^{(\ell)}}\>.
\end{align}
Estimate the covariance matrix $V_{\bs\alpha^{(\ell)}}$ and identify the Gaussian unitary $U_{\widetilde S_{\bs \alpha^{(\ell)}}}^\dag$ that can partially undo the effect of $U_{S}$ (see step (GD2) in \ref{supp:Learning of the general Gaussian layer and physical counter-rotation}). If the symplectic eigenvalues of $V_{\bs\alpha^{(\ell)}}$ are nondegenerate, skip Step 2. 

Physically apply $U_{\widetilde S_{\bs \alpha^{(\ell)}}}^\dag$ in the following steps. 

\item[Step 2.] \emph{Learning the second passive-layer and classical counter-rotation.} Perform \textbf{heterodyne measurement and shadow tomography} protocol of Ref. \cite{zhao2025complexity} on the passive-separable state 
\begin{align}
|\Psi_{\bs\alpha,\bs\alpha'}^{\rm ps}\>&= U_{\widetilde S_{\bs \alpha^{(\ell)}}}^\dag U_{\rm ge} |\underline{\bs\alpha^{(h)}}\>.
\end{align}
Reconstruct the passive-Gaussian unitary $U^\dag_{\widetilde O_{2,\bs\alpha^{(l)},\bs\alpha^{(h)}}}$ that can transform the overall state into a product of a multi-mode Gaussian state and single-mode non-Gaussian states. 

In the follow-up steps, always apply a classical-rotation $\bs \gamma\to W_{\widetilde O_{2,\bs\alpha^{(l)},\bs\alpha^{(h)}}}\bs\gamma$ on the heterodyne measurement results, with $W_O$ representing the mixing unitary  of a symplectic matrix $O$, to effectively implement a counter rotation $U^\dag_{\widetilde O_{2,\bs\alpha^{(l)},\bs\alpha^{(h)}}}$ to the output states. 

\emph{(Note i).} By Lemma~\ref{prop:single-mode-symplectic-zero-measure-energy}, with probability one over the choice of the coherent probe $\bs\alpha$, the passive-Gaussian unitary $U_{O_{2,\bs\alpha}}$ in Eq.~(\ref{eqs10}) acts on a probe-independent partition of modes. Therefore, with probability one, one can determine whether $U_{O_{2,\bs\alpha}}$ contains a nontrivial passive mixing structure, or reduces merely to permutations and single-mode phase rotations in Step 1. In this case, skip this step. 

\emph{(Note ii).} The shadow tomography protocol \cite{zhao2025complexity}  constructs an $\epsilon$-covering net over the family of passive-separable states with squared energy bounded by $E'$, and tests each candidate state using a global fidelity witness built from the fidelities between the locally reconstructed reduced states and the corresponding reduced target state. Since the $\epsilon$-covering net has size $L=\exp(\mathcal O(m))$ and each candidate fidelity witness is a local observable, shadow tomography estimates all $L$ corresponding observables in $\mathcal O(\log L)=\mathcal O(m)$ measurement rounds. Finally, the reconstruction error in trace distance $\epsilon_{\rm ps}$ vanishes as the error in the fidelity witness goes to zero. The overall sample complexity is $\textbf{poly}\left(m,E',\frac 1 {\epsilon_{\rm ps}},\log \frac 1 {\delta_{\rm ps}}\right)$, where $\delta_{\rm ps}$ is the failure probability. 

\emph{(Note iii).} Based on the properties of the reconstructed local states before the passive layer, we can reconstruct $U_{\widetilde O_{2,\bs\alpha^{(\ell)},\bs\alpha^{(h)}}}$ using Theorem~\ref{thm:gauge-general-gaussian-probes}. Precisely, if there are no Gaussian local states before the passive layer with an identical covariance matrix up to a phase rotation, then we can reconstruct the passive layer up to a permutation and single-mode phase rotations. Otherwise, if there are multiple Gaussian local states with an identical covariance matrix up to a phase rotation, each block of such states introduces an additional passive mixing between these modes as an additional degree of freedom. 

\item[Step 3.] \emph{Learning the remaining non-Gaussian layer.} Probe with multiple copies of {$K_{\rm wid}
=
\mathcal O(\log(1/\delta_{\rm wid}))$ distinct coherent states} $\{|\underline{\bs\alpha^{(k')}}\rangle\}_{k'=1}^K$ (RE1). Coarse identifying 
the modes associated with single-mode non-Gaussian state through \textbf{local tomography} (RE2) and \textbf{purity witness} (RE3). In particular, we identify the set of pure modes $\map X_{\rm pure}$. We then apply the \textbf{tomography-based} step (RE4) to learn all the single-mode non-Gaussian unitaries $\{W_j',j\in\chi_{\rm pure}\}$ included in remaining unitary $U_{\rm rest}$.

On the calibration-successful event, the Fourier-ring construction uses
\(q\le J\) and \(\lambda\le2\). Since the unscaled coherent base probe has
total photon number at most \(mE_{\rm prep}\), its scaled versions satisfy $\Tr\!\left(
\widehat H_m\rho_{\rm in}
\right)
\le
mE_{\rm probe}$ with $E_{\rm probe}
:=
\frac12
+
\frac{2J E_{\rm prep}}{\mu_-^{(0)}}
=
\frac12
+
16Jm|\map X_{\rm pure}|$. Thus \(E_{\rm probe}\) is polynomial in
\(m,E^{**},1/\epsilon\).

In the next step, we always implement the counter-rotation step in (RE5) on the \textbf{heterodyne tomography} results.

\item[Step 4.] \emph{Learning the remaining Gaussian part layer.} After counter-rotation of step 3, we can reconstruct approximately Gaussian output states
\begin{align}
|\Psi^{\rm prod}_{\bs\alpha^{(\ell)},\bs\alpha^{(h)},\bs\alpha^{(k')}}\>&=\left[\left(\bigotimes_{j\in \map X_{\rm pure}}W_j^{'\dag}\right)\otimes I_{[m]\setminus\map X_{\rm pure}}\right]U^\dag_{\widetilde O_{2,\bs\alpha^{(l)},\bs\alpha^{(h)}}} U_{\widetilde S_{\bs\alpha^{(\ell)}}}^\dag  U_{\rm ge} |\underline{\bs\alpha^{(k')}}\rangle.
\end{align}

Based on \textbf{local tomography} with non-Gaussian unitary counter-rotation, we can reconstruct $\widetilde U_{\rm final}\simeq\left[\left(\bigotimes_{j\in \map X_{\rm pure}}W_j^{'\dag}\right)\otimes I_{[m]\setminus\map X_{\rm pure}}\right]U^\dag_{\widetilde O_{2,\bs\alpha^{(l)},\bs\alpha^{(h)}}} $ $U_{\widetilde S_{\bs\alpha^{(\ell)}}}^\dag  U_{\rm ge}$.

\item[Step 5.] \emph{Assemble.} Return the reconstructed overall unitary 
\begin{align}
\widetilde U_{\rm ge}= U_{\widetilde S_{\bs\alpha^{(\ell)}}}   U_{\widetilde O_{2,\bs\alpha^{(l)},\bs\alpha^{(h)}}} \left[\left(\bigotimes_{j\in \map X_{\rm pure}}W_j^{'}\right)\otimes I_{[m]\setminus\map X_{\rm pure}}\right]\widetilde U_{\rm final}.
\end{align}
\end{enumerate}

In addition, the following remark concerns the experimental resources required by the learning protocol.

\begin{remark}[Resources and physical relevance]
\label{rem:resources-feasibility}
The learning protocol above relies only on off-the-shelf continuous-variable laboratory operations and classical postprocessing.   Concretely, the required resources are:

\begin{itemize}
\item \textbf{State preparation.}
Preparation of \(\mathcal O\left(\textbf{\em poly}\left(m\right)\right)\) distinct coherent states
\(
\{\ket{\underline{\bm\alpha^{(k)}}}=\bigotimes_{j=1}^m \ket{\underline{\alpha_j^{(k)}}}\}_{k=1}^{\mathcal O\left(\textbf{\em poly}\left(m\right)\right)},
\)
each with \(M\) copies, and each satisfying the finite-energy constraint
\(
\|\bm\alpha^{(k)}\|_2^2 \le m\left(E_{\rm probe}-\frac12\right).
\)

\item \textbf{Measurements.}
The learning protocol relies only on local heterodyne detection over independent copies. 

\item \textbf{Physical counter-rotation.} When \(U_{\rm ge}\) is passive-Gaussian entanglable, namely, when \(U_{\rm ge}\) takes the form of \(U_{\rm pge}\) in Eq.~(\ref{def:U_three layer_PGE}) or, equivalently, when \(U_S\) in Eq.~(\ref{def:U_three layer with S}) is passive, no physical unitary operation is required.

\item \textbf{Shadow tomography.} Based on Lemma~\ref{prop:single-mode-symplectic-zero-measure-energy}, we can identify with probability one the case in which the covariance matrix of the  output state of $U_{\rm ge}$ has nondegenerate symplectic eigenvalues regardless of the probe \(|\underline{\bs\alpha}\>\). In this case, \(U_{O_{2,\bs\alpha^{(l)},\bs\alpha^{(h)}}}\) reduces to an unknown permutation unitary. Therefore, Step~2 is unnecessary, and hence the protocol does not require shadow tomography of multimode non-Gaussian states at any stage.
\item \textbf{One-direction learning.} Our protocol does not reply on the use of $U_{\rm ge}^\dag$ or $U_{\rm ge}^{\mathsf T}$, which could be non-physical in optical setting. 
\end{itemize}

Overall, the protocol is experimentally feasible because it uses only coherent probe states and heterodyne measurements,
together with classical data processing and, Gaussian unitaries.
All nontrivial learning is delegated to classical computation from measurement data, while the quantum hardware remains
simple and fully compatible with current multimode photonic and microwave CV platforms \cite{serafini2023quantum},.
\end{remark}

\subsection{A brief proof that the query complexity grows polynomially}\label{supp:query complexity of GE layer}

We now analyze the query complexity of the above learning protocol, namely, how many queries to the unitary \(U_{\rm ge}\) are required. Specifically, the following theorem is given:

\begin{theorem}[The query complexity of learning Gaussian-entangleable unitaries is polynomial]
\label{thm:query-complexity-GE-final}
Consider a Gaussian-entanglable unitary $U_{\rm ge}
=
U_S\Bigl(\bigotimes_{j=1}^m W_j\Bigr)U_O$, where \(U_S\) is an arbitrary \(m\)-mode Gaussian unitary, \(U_O\) is an arbitrary \(m\)-mode passive-Gaussian unitary, and each \(W_j\) is an arbitrary single-mode unitary. We further assume that the available probe state \(\ket{\Psi_{\rm in}}\) and the corresponding output state \(\ket{\Psi_{\rm out}}=U_{\rm ge}\ket{\Psi_{\rm in}}\) satisfy the following energy constraints $\Tr\left\{\left[ \left(\sum_{j=1}^m a_j^\dag a_j+\frac m 2\right)\right] |\Phi_{\rm in}\>\<\Phi_{\rm in}|\right\}\le  mE_{\rm probe}$ and $\sqrt{\Tr\left[ \left(\sum_{j=1}^m a_j^\dag a_j+\frac m 2 \right)^2|\Psi_{\rm out}\>\<\Psi_{\rm out}|\right]}\le  mE_{\rm II}$. Assume additionally a finite uniform intermediate evaluation-energy bound
\(E^{**}\) for the inputs to the effective single-mode channels appearing in
the exact factorization used below. Let us denote $E^*=\max\{E_{\rm probe},E_{\rm II},E^{**}\}$. 

Then, there exists a learning protocol that, with success probability at least \(1-\delta\), outputs a unitary \(\widetilde U_{\rm ge}\) such that $\|\mathcal U_{\rm ge}-\widetilde{\mathcal U}_{\rm ge}\|_{\diamond}^{mE}\le \epsilon .$ The total query complexity is 
\begin{align}
M=\textbf{\em poly}\!\Bigl(
m,\,
E^*,\,
1/\epsilon,\,
\log(1/\delta)
\Bigr).
\end{align}
\end{theorem}

\begin{proof}
By the perturbative error-propagation assumption stated in the learning task (see \ref{supp:learning task of ge unitary}), we analyze each step assuming that all preceding steps are exact, and the total reconstruction error is the sum of the errors contributed by the individual steps. Therefore, we can fix $\epsilon,\delta\in(0,1)$, and choose error and failure probability bounds
\begin{align}
\epsilon
=
\epsilon_{\rm gd}
+
\epsilon_{\rm ps}
+
\epsilon_{\rm ng}
+
\epsilon_{\rm G},
\qquad
\delta
=
\delta_{\rm gd}
+
\delta_{\rm ps}
+
\delta_{\rm ng}
+
\delta_{\rm G}+
\delta_{\rm wid},
\label{eq:budget-final-proof}
\end{align}
where \(\epsilon_{\rm gd}\), \(\epsilon_{\rm ps}\), \(\epsilon_{\rm ng}\), and \(\epsilon_{\rm G}\) denote the reconstruction errors arising in the corresponding steps, while \(\delta_{\rm gd}\), \(\delta_{\rm ps}\), \(\delta_{\rm ng}\), and \(\delta_{\rm G}\) denote the associated failure probabilities, to be specified explicitly later.

Let us first describe the exact factorization of $U_{\rm ge}$ selected by the noiseless protocol for the probes $\ket{\underline{\bs\alpha^{(\ell)}}}$ in Step 1, $\ket{\underline{\bs\alpha^{(h)}}}$ in Step 2, $\ket{\underline{\bs\alpha^{(k')}}}$ in Step 3 and the \(2m+1\) coherent probes in the final Gaussian unitary learning (see \ref{supp:final protocol of GE layer}). By Lemma~\ref{lem:Gaussian-disentangling for different input}, in the ideal limit of Step~1 there exists a Gaussian unitary, denoted here by
\(
U_{S,\bs\alpha^{(\ell)}},
\)
such that
\begin{align}
U_{S,\bs\alpha^{(\ell)}}^\dag U_{\rm ge}
\end{align}
maps every later coherent probe used in the protocol to a passive-separable output family.
Next, in the ideal limit of Step~2, there exists a passive-Gaussian unitary
\(
U_{O_2,\bs\alpha^{(\ell)},\bs\alpha^{(h)}}
\)
such that
\begin{align}
U_{\rm rest}=U_{O_2,\bs\alpha^{(\ell)},\bs\alpha^{(h)}}^\dag
U_{S,\bs\alpha^{(\ell)}}^\dag
U_{\rm ge}
\end{align}
maps the chosen probe family to a product of a multimode Gaussian state and single-mode non-Gaussian states.
Then, by the local reconstruction step (RE1)-(RE4), on the event that every genuinely non-Gaussian mode is activated (probability one shown in Theorem \ref{prop:measure-zero-bargmann-multimode}), there exist single-mode unitaries
\(
\{W'_j\}_{j\in\chi_{\rm pure}}
\)
such that
\begin{align}
U_{\rm final}=
\left[
\left(\bigotimes_{j\in\chi_{\rm pure}}W'_j\right)\otimes I_{\map G}
\right]
U_{\rm rest},
\label{eq:exact-rest-decomposition}
\end{align}
where \(U_{\rm final}\) is an \(m\)-mode Gaussian unitary.
This is exactly the residual unitary learned in Step~(RE5). Define the uniform intermediate evaluation-energy bound $E^{**}
:=
\max_{j\in\chi_{\rm pure}}
\sup_{\Tr(\widehat H_m\rho)\le mE}
\Tr\!\left[
\widehat H_1\,
\Tr_{\neq j}
\!\left(
\mathcal U_{\rm final}(\rho)
\right)
\right]
<\infty$. Thus \(E^{**}\) bounds the energy entering every effective local channel
\(\mathcal W'_j\) under the global evaluation-energy constraint \(mE\). Finally, Step~(RE6) reconstructs \(U_{\rm ge}\) from
the final estimate as follows
\begin{align}
\widetilde U_{\rm ge}
=
U_{\widetilde S_{\bs\alpha^{(\ell)}}}\,
U_{\widetilde O_{2,\bs\alpha^{(\ell)},\bs\alpha^{(h)}}}\,\left[
\left(\bigotimes_{j\in\chi_{\rm pure}}\widetilde W'_j\right)\otimes I_{\map G}
\right]
\widetilde U_{\rm final}.
\label{eq:def-final-estimate-proof}
\end{align}

Let us denote $\mathcal W'_{\chi_{\rm pure}}:=\left[\left(\bigotimes_{j\in\chi_{\rm pure}}\mathcal W'_j\right)\otimes \mathcal I_{\map G}\right].$ Then Eq.~\eqref{eq:exact-rest-decomposition} implies $\mathcal U_{\rm ge}
=
\mathcal U_{S_{\bs\alpha^{(\ell)}}}
\circ
\mathcal U_{O_{2,\bs\alpha^{(\ell)},\bs\alpha^{(h)}}}
\circ
\mathcal W'_{\chi_{\rm pure}}
\circ
\mathcal U_{\rm final}$. For simplicity of notation, let us further introduce the intermediate channels
\begin{align}
\mathcal A_0
&:=
\mathcal U_{S_{\bs\alpha^{(\ell)}}}
\circ
\mathcal U_{O_{2,\bs\alpha^{(\ell)},\bs\alpha^{(h)}}}
\circ
\mathcal W'_{\chi_{\rm pure}}
\circ
\mathcal U_{\rm final}
=
\mathcal U_{\rm ge},
\\
\mathcal A_1
&:=
\mathcal U_{\widetilde S_{\bs\alpha^{(\ell)}}}
\circ
\mathcal U_{O_{2,\bs\alpha^{(\ell)},\bs\alpha^{(h)}}}
\circ
\mathcal W'_{\chi_{\rm pure}}
\circ
\mathcal U_{\rm final},
\\
\mathcal A_2
&:=
\mathcal U_{\widetilde S_{\bs\alpha^{(\ell)}}}
\circ
\mathcal U_{\widetilde O_{2,\bs\alpha^{(\ell)},\bs\alpha^{(h)}}}
\circ
\mathcal W'_{\chi_{\rm pure}}
\circ
\mathcal U_{\rm final},
\\
\mathcal A_3
&:=
\mathcal U_{\widetilde S_{\bs\alpha^{(\ell)}}}
\circ
\mathcal U_{\widetilde O_{2,\bs\alpha^{(\ell)},\bs\alpha^{(h)}}}
\circ
\widetilde{\mathcal W}'_{\chi_{\rm pure}}
\circ
\mathcal U_{\rm final},
\\
\mathcal A_4
&:=
\mathcal U_{\widetilde S_{\bs\alpha^{(\ell)}}}
\circ
\mathcal U_{\widetilde O_{2,\bs\alpha^{(\ell)},\bs\alpha^{(h)}}}
\circ
\widetilde{\mathcal W}'_{\chi_{\rm pure}}
\circ
\widetilde{\mathcal U}_{\rm final}
=
\widetilde{\mathcal U}_{\rm ge}.
\end{align}
Hence, by the triangle inequality for the energy-constrained diamond norm,
\begin{align}
\|\mathcal U_{\rm ge}-\widetilde{\mathcal U}_{\rm ge}\|_{\diamond}^{mE}
&=
\|\mathcal A_0-\mathcal A_4\|_{\diamond}^{mE}
\nonumber\\
&\le
\|\mathcal A_0-\mathcal A_1\|_{\diamond}^{mE}
+
\|\mathcal A_1-\mathcal A_2\|_{\diamond}^{mE}
+
\|\mathcal A_2-\mathcal A_3\|_{\diamond}^{mE}
+
\|\mathcal A_3-\mathcal A_4\|_{\diamond}^{mE}.
\label{eq:main-telescope-proof}
\end{align}

We now bound the four terms separately. For the first term, by definition \(\epsilon_{\rm gd}\) is the contribution of Step~1 to the final reconstruction error under the perturbative assumption, namely when all later steps are exact. Therefore, we have 
\begin{align}
\|\mathcal A_0-\mathcal A_1\|_{\diamond}^{mE}&\le \left\|\mathcal U_{ S_{\bs\alpha^{(\ell)}}}-\mathcal U_{\widetilde S_{\bs\alpha^{(\ell)}}}\right\|_\diamond^{mE'}\\
&\le \epsilon_{\rm gd},
\label{eq:bound-step-gd}
\end{align}
where $E'$ denotes the average energy per mode before the application of $\mathcal U_{ S_{\bs\alpha^{(\ell)}}}$. This finite error $\epsilon_{\rm gd}$ is achieved by combining the moment-estimation protocol (Theorem S53 of \cite{mele2025learning}), the symplectic reconstruction protocol (see Eqs. (S84) and (S160)~ of \cite{zhao2025complexity}), and the Gaussian Solovay--Kitaev theorem (Theorem 6 of~\cite{becker2021energy}).

For the second term, by definition \(\epsilon_{\rm ps}\) is the contribution of Step~2 to the final reconstruction error under the same first-order approximation. Hence, we have 
\begin{align}
\|\mathcal A_1-\mathcal A_2\|_{\diamond}^{mE}&\le \left\|\mathcal U_{ \widetilde S_{\bs\alpha^{(\ell)}}}\circ \mathcal U_{O_{2,\bs\alpha^{(\ell)},\bs\alpha^{(h)}}}-\mathcal U_{ \widetilde S_{\bs\alpha^{(\ell)}}}\circ \mathcal U_{\widetilde O_{2,\bs\alpha^{(\ell)},\bs\alpha^{(h)}}}\right\|_\diamond^{mE'}\\
&= \left\|\mathcal U_{O_{2,\bs\alpha^{(\ell)},\bs\alpha^{(h)}}}-\mathcal U_{\widetilde O_{2,\bs\alpha^{(\ell)},\bs\alpha^{(h)}}}\right\|_\diamond^{mE'}\\
&\le \epsilon_{\rm ps},
\label{eq:bound-step-ps}
\end{align}
where the energy $E'$ in the  first inequality is identical to that in Eq. (\ref{eq:bound-step-gd}) as the operation $U_{O_{2,\bs\alpha^{(\ell)},\bs\alpha^{(h)}}}$ is passive, where the equality uses post-composition by a unitary channel and the fact that the passive unitary in this step preserves the relevant energy bound.

For the third term, write
\(
\chi_{\rm pure}=\{j_1,\dots,j_{|\chi_{\rm pure}|}\}
\)
and define
\begin{align}
\mathcal V^{(r)}
:=
\left[
\left(\bigotimes_{s<r}\widetilde{\mathcal W}'_{j_s}\right)
\otimes
\left(\bigotimes_{s\ge r}\mathcal W'_{j_s}\right)
\otimes
\mathcal I_{[m]\setminus\chi_{\rm pure}}
\right],
\qquad
r=1,\dots,|\chi_{\rm pure}|+1.
\end{align}
Then, we have 
\(
\mathcal V^{(1)}=\mathcal W'_{\chi_{\rm pure}}
\)
and
\(
\mathcal V^{(|\chi_{\rm pure}|+1)}=\widetilde{\mathcal W}'_{\chi_{\rm pure}}
\). Applying the triangle inequality to the telescoping decomposition (see Eq. (S99) of \cite{zhao2025complexity}) yields
\begin{align}
\|\mathcal A_2-\mathcal A_3\|_{\diamond}^{mE}
&\le
\sum_{r=1}^{|\chi_{\rm pure}|}
\Bigl\|
\mathcal U_{\widetilde S_{\bs\alpha^{(\ell)}}}
\circ
\mathcal U_{\widetilde O_{2,\bs\alpha^{(\ell)},\bs\alpha^{(h)}}}
\circ
\mathcal V^{(r)}
\circ
\mathcal U_{\rm final}
\nonumber\\
&\hspace{8em}
-
\mathcal U_{\widetilde S_{\bs\alpha^{(\ell)}}}
\circ
\mathcal U_{\widetilde O_{2,\bs\alpha^{(\ell)},\bs\alpha^{(h)}}}
\circ
\mathcal V^{(r+1)}
\circ
\mathcal U_{\rm final}
\Bigr\|_{\diamond}^{mE}.
\end{align}
Since post-composition by unitary channels does not increase the energy-constrained diamond norm, each summand is bounded by
\(
\|\mathcal W'_{j_r}-\widetilde{\mathcal W}'_{j_r}\|_{\diamond}^{E^{**}}
\le \epsilon_{j_r}
\) where $E^{**}$ is the uniform intermediate evaluation-energy bound
(see the parameter ``$\mu_{\max}$'' in Lemma \ref{lem:selected-mode-calibration-window}). Therefore, we have 
\begin{align}
\|\mathcal A_2-\mathcal A_3\|_{\diamond}^{mE}
\le
\sum_{j\in\chi_{\rm pure}}\epsilon_j
=
\epsilon_{\rm ng}.
\label{eq:bound-step-ng}
\end{align}

For the fourth term, again using invariance of the diamond norm under unitary post-composition,
\begin{align}
\|\mathcal A_3-\mathcal A_4\|_{\diamond}^{mE}
=
\|\mathcal U_{\rm final}-\widetilde{\mathcal U}_{\rm final}\|_{\diamond}^{mE}
\le
\epsilon_{\rm G},
\label{eq:bound-step-gaussian}
\end{align}
where \(\epsilon_{\rm G}\) is the reconstruction error of the Gaussian unitary \(U_{\rm final}\) learned in Step~(RE5).

Substituting Eqs.~\eqref{eq:bound-step-gd}, \eqref{eq:bound-step-ps}, \eqref{eq:bound-step-ng}, and \eqref{eq:bound-step-gaussian} into Eq.~\eqref{eq:main-telescope-proof}, we obtain
\begin{align}
\|\mathcal U_{\rm ge}-\widetilde{\mathcal U}_{\rm ge}\|_{\diamond}^{mE}
\le
\epsilon_{\rm gd}
+
\epsilon_{\rm ps}
+
\epsilon_{\rm ng}
+
\epsilon_{\rm G}
=
\epsilon.
\end{align}
In the proposed learning protocol, the errors $\epsilon_{\rm gd}$, $\epsilon_{\rm ps}$, $\epsilon_{\rm ng}$, and $\epsilon_{\rm G}$ can all be reduced to arbitrarily small values with a polynomial measurement rounds (see Secs.~\ref{supp:Learning of the general Gaussian layer and physical counter-rotation}, \ref{supp:Learning of the second passive-Gaussian layer}, and \ref{supp:Learning the remaining part of the overall unitary}).

Then, it remains to control the failure probability.
By the activation theorem (Theorem \ref{prop:measure-zero-bargmann-multimode}) together with the repeated random coherent probing step, the event that every genuinely non-Gaussian local unitary is activated with probability at least \(1-\delta_{\rm act}\simeq 1\), independent of learning steps 3 and 4.
Conditioned on this event, the failures of Step~1, Step~2, the single-mode reconstructions \(\{\widetilde W'_j\}_{j\in\map N}\), and the Gaussian reconstruction of \(U_{\rm final}\) occur with probabilities at most
\(
\delta_{\rm gd},
\delta_{\rm ps},
\{\delta_j\}_{j\in\map N},
\delta_{\rm G}
\),
respectively.
Hence the union bound in Lemma \ref{lem:union bound} gives
\begin{align}
\Pr\!\left[
\|\mathcal U_{\rm ge}-\widetilde{\mathcal U}_{\rm ge}\|_{\diamond}^{mE}\le \epsilon
\right]
\ge
1-
\delta_{\rm gd}
-
\delta_{\rm ps}
-
\sum_{j\in\chi_{\rm pure}}\delta_j
-
\delta_{\rm G}-\delta_{\rm wid}
\ge
1-\delta,
\end{align}
where $\delta_{\rm wid}$ is the failure probability introduced in Lemma \ref{lem:selected-mode-calibration-window}.

Finally, the total query complexity is obtained by summing the costs of the four stages. Step~1 has complexity
\(
M_{\rm gd}
=
\textbf{poly}\!\left(m,E_{\rm II},1/\epsilon_{\rm gd},\log(1/\delta_{\rm gd})\right)
\)
which follows from covariance-matrix learning in Theorem S53 of Ref.~\cite{mele2025learning}, together with the subsequent analysis of the error induced by state counter-rotation in Ref.~\cite{zhao2025complexity}. Given Lemma S22 of \cite{zhao2025complexity} and the fact that second-moment energy bounds can be used as an energy upper bound, the energy in the second step is smaller than $E_{\rm II}$. Therefore, Step~2 has complexity
\(
M_{\rm ps}
=
\textbf{poly}\!\left(m,E_{\rm II},1/\epsilon_{\rm ps},\log(1/\delta_{\rm ps})\right)
\)
from passive-separable state shadow tomography \cite{zhao2025complexity}. For each \(j\in\chi_{\rm pure}\), the local reconstruction of \(W'_j\) has cost
\(
M_j=\textbf{poly}\!\left(E^{**},1/\epsilon_j,\log(1/\delta_j)\right).
\)
Finally, Step~(RE5) learns the Gaussian unitary \(U_{\rm final}\) using \(2m+1\) distinct coherent probe settings, each repeated as required by Proposition~\ref{thm:structured-product-output-learning}. Combining the vacuum-shared probe construction of Ref.~\cite{fanizza2025efficient} with the local-tomography analysis above, its query cost satisfies
\(
M_{\rm G}
=
\textbf{poly}\!\left(m,E,E_{\rm II},1/\epsilon_{\rm G},\log(1/\delta_{\rm G})\right).
\) Finally, let $E^{*}=\max\{E_{\rm probe},E_{\rm II},E^{**}\}$,  we have 
\begin{align}
M_{\rm tot}
=
M_{\rm gd}
+
M_{\rm ps}
+
\sum_{j\in\map N}M_j
+
M_{\rm G}
=
\textbf{poly}\!\left(m,E^{*},1/\epsilon,\log(1/\delta)\right).\label{eqs471}
\end{align}
\end{proof}

\section{Learning $t$-doped  Gaussian unitaries}\label{supp:learning t doped unitary}

\subsection{Learning task}

In this supplemental note, we  introduce a learning protocol for arbitrary $t$-doped Gaussian unitaries. Without loss of the generality, we can represent them as follows
\begin{align}\label{supp_def:U_doped}
U_{\rm doped}&= D_{t,(m)} U_{S_{t,(m)}}\left(\Pi_{i=0}^{t-1} W_{i,(\kappa)} D_{i,(m)} U_{S_{i,(m)}}\right),
\end{align}
where $\{D_{i,(m)}\}_{i=0}^{t}$ are $m$-mode displacement operators, $\{U_{S_{i,(m)}}\}_{i=0}^{t}$ are $m$-mode passive-Gaussian unitaries associated with the symplectic  matrices $\{S_{i,(m)}\}_{i=0}^{t}$, respectively, $\{W_{i,(\kappa)}\}_{i=0}^{t-1}$ are $\kappa$-mode local non-Gaussian unitaries. Equivalently, by using the fact that each Gaussian layer maps an \(m\)-mode displacement operator to another \(m\)-mode displacement operator, together with the compression theory of \(t\)-doped Gaussian unitaries \cite{mele2025learning}, we can rewrite \(U_{\rm doped}\) as follows: 
\begin{align}\label{supp_def:U_doped_three_leyer}
U_{\rm doped}&=U_{S} \left(U_{(\kappa t)}\otimes D_{(m-\kappa t)}\right) U_{O}
\end{align}
where $U_{(\kappa t)}$ is a $\kappa t$-mode possibly non-Gaussian unitary, $D_{(m-\kappa t)}$ denotes an $(m-\kappa t)$-mode displacement operator.

Without loss of generality, we consider the same constraints on energy and its second moment
\begin{align}\label{supp:three energy constraints t doped}
\begin{cases}
\Tr\left\{\left[ \left(\sum_{j=1}^m a_j^\dag a_j+\frac m 2\right)\right] |\Phi_{\rm in}\>\<\Phi_{\rm in}|\right\}\le & mE_{\rm probe}
\\[0.7em]
\sqrt{\Tr\left\{\left[ \left(\sum_{j=1}^m a_j^\dag a_j+\frac m 2 \right)^2\right]|\Psi_{\rm out}\>\<\Psi_{\rm out}|\right\}}\le & mE_{\rm II}
\end{cases},
\end{align}
where $|\Psi_{\rm out}\>=U_{\rm doped}|\Phi_{\rm in}\>$ refers to the output state, $|\Phi_{\rm in}\>$ denotes the input state. Then, the goal is to reconstruct \(\widetilde U_{\rm doped}\) such that $\|\mathcal U_{\rm doped}-\widetilde{\mathcal U}_{\rm doped}\|_\diamond^{mE}\le \epsilon$ with success probability at least \(1-\delta\).

\subsection{Learning of the Gaussian layer}\label{supp:Learning of the Gaussian layer-t doped}

Similar to Steps (GD1)-(GD3) in \ref{supp:Learning of the general Gaussian layer and physical counter-rotation}, we introduce the following procedure to reconstruct \(U_{\widetilde S_{\bs \alpha^{(\ell)}}}\), which partially counter-rotates the action of \(U_S\) in Eq.~(\ref{supp_def:U_doped_three_leyer}).
\begin{enumerate}
\item [(GD1')] Probe $U_{\rm doped}$ by a multimode-coherent state $|\underline{\bs\alpha^{(\ell)}}\>$. 
\item [(GD2')] Perform heterodyne measurement to estimate the covariance matrix $V_{\bs\alpha^{(\ell)}}$ of the output state $|\Psi_{\bs\alpha^{(\ell)}}\>:=U_{\rm doped}|\underline{\bs\alpha^{(\ell)}}\>$. Using the same method as Step (GD2) in \ref{supp:Learning of Gaussian entanglable unitaries}, construct the symplectic matrix $\widetilde S_{\bs\alpha^{(\ell)}}$ whose inverse can symplectically diagonalize $V_{\bs\alpha^{(\ell)}}$. On the other hand, $V_{\bs\alpha^{(\ell)}}$ can be written as 
$V_{\bs\alpha^{(\ell)}}= S_{\bs\alpha^{(\ell)}}
\left[\left(\Lambda_{\bs\alpha^{(\ell)}}\oplus I_{(m-\kappa t)}\right)\otimes  I^{(2)}\right] 
S^T_{\bs\alpha^{(\ell)}}$ where $ \Lambda_{\bs\alpha^{(\ell)}} =\text{diag}(\lambda_1,\cdots,\lambda_m)$ is a $m\times m$ diagonal matrix. Therefore, the matrix $S_{\bs\alpha^{(\ell)}}$ is associated with the true symplectic matrix $S$ as follows
\begin{align}
S_{\bs\alpha^{(\ell)}}=S \left(S_{(\kappa t),{\bs\alpha^{(\ell)}}}\oplus I_{(m-\kappa t)}\right),
\end{align}
where $S_{(\kappa t),{\bs\alpha^{(\ell)}}}$ denote a $\kappa t$-mode symplectic matrix. Therefore, one can use the estimated symplectic matrix $\widetilde S_{\bs\alpha^{(\ell)}}$ as an approximation of $S_{\bs\alpha^{(\ell)}}$, where the gauge structure consists of a passive-Gaussian unitary arising from the degeneracy of the symplectic eigenvalues, together with a permutation that sorts the symplectic eigenvalues.
\item [(GD3')] Physically implement $U_{\widetilde S_{\bs\alpha^{(\ell)}}}$ on the following experiments, when the probes are probably changed. 
\end{enumerate}

It is straightforward to verify that the learned Gaussian unitary \(U^\dag_{\widetilde S_{\bs\alpha^{(\ell)}}}\) exactly cancels the squeezing part \(U_S\), and that the remaining transformation is a passive-Gaussian unitary \(U_{O_{2,\bs\alpha^{(\ell)}}}\), which is determined solely by \(\bs\alpha^{(\ell)}\), similar to Eq. (\ref{eqs10}).

Experimentally, Lemma~\ref{prop:single-mode-symplectic-zero-measure-energy}, together with the fact that pure Gaussian states achieve the minimal symplectic eigenvalues, enables one to efficiently identify the special situation in which exactly \(\kappa t\) symplectic eigenvalues are strictly larger than \(1\). In this case, the gauge passive-Gaussian unitary is block diagonal with respect to the partition between the first \(\kappa t\) modes and the remaining \(m-\kappa t\) modes (see a detailed proof in \cite{zhao2025complexity}). 

Finally, after enough rounds of heterodyne measurements, one can establish an estimate of the unitary $U_{\widetilde S_{\bs\alpha^{(\ell)}}}$ which converges to a Gaussian unitary $U_{ S_{\bs\alpha^{(\ell)}}}$ as (see also Eq. (\ref{eqs29}))
\begin{align}
\left\|\map U_{ S_{\bs\alpha^{(\ell)}}}-\map U_{\widetilde S_{\bs\alpha^{(\ell)}}}\right\|_\diamond^{mE_{\rm II}}&\le \epsilon_{\rm gd}.
\end{align} 
The query-complexity analysis follows the same argument as in \ref{supp:Learning of Gaussian entanglable unitaries}.

\subsection{Learning of the second passive layer}

After physically applying the Gaussian unitary \(U_{\widetilde S^\dag_{\bs\alpha^{(\ell)}}}\)  (see \ref{supp:Learning of the Gaussian layer-t doped}), the output state becomes a generalized passive-separable state across the first \(\kappa t\) modes and the remaining \(m-\kappa t\) modes: 
\begin{align}\label{eqs186}
U^\dag_{\widetilde S_{\bs\alpha^{(\ell)}}}U_{\rm doped}|\underline{\bs \alpha^{(h)}}\>\simeq U_{O_{2,\bs\alpha^{(\ell)}}}\left(U^\dag_{S_{(\kappa t),\bs\alpha^{(\ell)}}}U_{(\kappa t)}\otimes D_{(m-\kappa t)}\right)U_O|\underline{\bs \alpha^{(h)}}\>
\end{align}
for an alternative probe $|\underline{\bs\alpha^{(h)}}\>$. In this subsection, we will introduce the learning protocol for the layer $U_{O_{2,\bs\alpha^{(\ell)}}}$.

Before showing the explicit protocol, we need to generalize the quantum D-S theorem in \ref{supp:Learning of the second passive-Gaussian layer} and the activation theorem in \ref{supp:activation theorem} to the case with $t$-doped unitary learning. 

\subsubsection{Generalization of the quantum DS theorem}

Let us look at a useful proposition for the subsequent discussion about gauge structure. For simplicity, we temporarily consider a bipartition into a $t$-mode subsystem and an $(m-t)$-mode subsystem.

\begin{proposition}[Cross-block coherent state preservation]
\label{prop:cross-block-coherent-preservation}
Define the partition of modes $A=\{1,\ldots,t\}, B=\{t+1,\ldots,m\}$. Let $U_O$ be an $m$-mode passive-Gaussian unitary, and let
$U\in\mathrm U(m)$ be its mode-mixing matrix, $U_O^\dagger \bm a\, U_O = U\,\bm a,$ with $\bm a=(a_1,\ldots,a_m)^{\mathsf T}$. Write $\bm a_A=(a_1,\ldots,a_t)^{\mathsf T},$ and $\bm a_B=(a_{t+1},\ldots,a_m)^{\mathsf T}$, and let $U_{B,A}\in\mathbb C^{(m-t)\times t}$ be the submatrix of $U$ with rows indexed by $B$ and columns indexed by $A$. Let $U_{B,A}=X\Sigma Y^\dagger$ be its singular-value decomposition, and let \(V_A\) be the \(t\)-mode passive-Gaussian unitary on the modes in \(A\) determined by \(Y^\dagger\). 

Assume that there exist pure states $|\psi\rangle_A$, $|\phi\rangle_A$, and coherent amplitudes
$\bm\alpha,\bm\beta\in\mathbb C^{m-t}$ such that
\begin{align}
U_O\Bigl(|\psi\rangle_A\otimes |\underline{\bm\alpha}\rangle_B\Bigr)
=
|\phi\rangle_A\otimes |\underline{\bm\beta}\rangle_B,
\end{align}
where $|\underline{\bs\alpha}\>$ denotes a coherent state in the $(m-t)$-mode. Then, there exist a coherent amplitude \(\bm\mu\in\mathbb C^r\) with $r=\textbf{\em rank}(\Sigma$) and a pure state \(|\chi\rangle\) on the remaining \(t-r\) modes such that
\begin{align}\label{eqs106}
|\psi\rangle_A
=
V_A^\dag \left(|\underline{\bm\mu}\rangle_{1,\ldots,r}\otimes |\chi\rangle_{r+1,\ldots,t}\right).
\end{align}
\end{proposition}

\begin{proof}
Let us define $|0\rangle_B:=|0\rangle^{\otimes(m-t)}$, $|\underline{\bm\alpha}\rangle_B=D_B(\bm\alpha)|0\rangle_B$, $|\underline{\bm\beta}\rangle_B=D_B(\bm\beta)|0\rangle_B$, where $D_B(\cdot)$ denotes the multimode displacement operator on the modes in $B$. In addition, we can introduced an operator 
\begin{align}
\widetilde U
:=
\bigl(I_A\otimes D_B(-\bm\beta)\bigr)\,
U_O\,
\bigl(I_A\otimes D_B(\bm\alpha)\bigr).
\end{align}

Then, we have 
\begin{align}
\widetilde U\Bigl(|\psi\rangle_A\otimes |0\rangle_B\Bigr)
=
|\phi\rangle_A\otimes |0\rangle_B.
\end{align}

Fix any $k\in B$. Since the $k$th output mode is vacuum, we have 
\begin{align}
a_k\,\widetilde U\Bigl(|\psi\rangle_A\otimes |0\rangle_B\Bigr)=0.
\end{align}
Equivalently, we have 
\begin{align}
\widetilde U^\dagger a_k \widetilde U
\Bigl(|\psi\rangle_A\otimes |0\rangle_B\Bigr)=0.
\end{align}
Applying the definitions 
\begin{align}
U_O^\dagger a_k U_O=\sum_{j=1}^m U_{kj}a_j,
\qquad
D(\xi)^\dagger a\,D(\xi)=a+\xi,
\end{align}
we obtain the following relation 
\begin{align}
\widetilde U^\dagger a_k \widetilde U
=
\sum_{j\in A}U_{kj}a_j
+
\sum_{\ell\in B}U_{k\ell}(a_\ell+\alpha_\ell)
-\beta_k.
\end{align}
Acting on $|\psi\rangle_A\otimes |0\rangle_B$ and using the fact $a_\ell|0\rangle_B=0$ for all $\ell\in B$, we get the following equation: 
\begin{align}
\left(\sum_{j\in A}U_{kj}a_j\right)|\psi\rangle_A
=
\left(\beta_k-\sum_{\ell\in B}U_{k\ell}\alpha_\ell\right)|\psi\rangle_A
\end{align}
Therefore, we have 
\begin{align}
U_{B,A}\,\bm a_A\,|\psi\rangle_A=\bm\lambda\,|\psi\rangle_A
\end{align}
for some vector $\bm\lambda\in\mathbb C^{m-t}$. Here, we can take a singular-value decomposition $U_{B,A}=X\Sigma Y^\dagger$, where $X\in\mathrm U(m-t)$, $Y\in\mathrm U(t)$, and $\Sigma=
\begin{pmatrix}
\operatorname{diag}(s_1,\ldots,s_r) & 0\\
0 & 0
\end{pmatrix},$ for $s_1,\ldots,s_r>0$.

By defining the new annihilation operators $\bm b:=Y^\dagger \bm a_A$, we have 
\begin{align}
\Sigma\,\bm b\,|\psi\rangle_A = X^\dagger\bm\lambda\,|\psi\rangle_A.
\end{align}
Hence, for each $j=1,\ldots,r$, we have 
\begin{align}
b_j|\psi\rangle_A=\mu_j|\psi\rangle_A,
\qquad
\mu_j:=\frac{(X^\dagger\bm\lambda)_j}{s_j}.
\end{align}

Let $V_A$ be the $t$-mode passive-Gaussian unitary implementing $Y^\dagger$, namely
\begin{align}
V_A^\dagger \bm a_A V_A = Y^\dagger \bm a_A=\bm b.
\end{align}
We have 
\begin{align}
a_j\,V_A|\psi\rangle_A
=
V_A b_j|\psi\rangle_A
=
\mu_j\,V_A|\psi\rangle_A,
\qquad
j=1,\ldots,r.
\end{align}
Thus the first $r$ modes of $V_A|\psi\rangle_A$ are coherent states. Since $V_A|\psi\rangle_A$ is pure, there exists a pure state $|\chi\rangle$ on the remaining $t-r$ modes such that
\begin{align}
V_A|\psi\rangle_A
=
|\underline{\bm\mu}\rangle_{1,\ldots,r}\otimes |\chi\rangle_{r+1,\ldots,t},
\end{align}
which gives Eq. (\ref{eqs106}). 
\end{proof}

Here, we have a remark about the new definition of mixing: 

\begin{remark}[Difference between the mode-mixing and block-mixing] The original mixing condition $\forall j\in[m],\ \exists\, k\neq \ell \text{ such that } U_{kj}\neq 0 \text{ and } U_{\ell j}\neq 0$ is a \emph{modewise spreading condition}. It says that each input mode is sent to at least two output modes. This is the right notion for Proposition~\ref{prop:DS-passive-general}, where both the input and the output are products across \emph{single modes}. By contrast, the condition $\operatorname{rank}(U_{B,A})=r$ is a \emph{cross-block rank condition}. It measures how many linearly independent directions of the input block $A$ are seen from the output block $B$. Equivalently, it counts how many independent annihilation constraints on $|\psi\rangle_A$ are enforced by requiring the output on $B$ to remain coherent states.

Thus, the first condition is \emph{combinatorial}, while the second is \emph{algebraic}. The first asks whether each individual mode spreads out. The second asks whether the whole block $A$ leaks into $B$ in enough independent directions. In general, neither implies the other. The original condition may hold even if $U_{B,A}=0$, for instance when the modes in $A$ spread only inside the output block $A$. Then $B$ sees nothing about $A$. Conversely, it is possible that $\operatorname{rank}(U_{B,A})=t$, even though the original condition fails, since full column rank only says that the map from $A$ to $B$ is injective. The key message is that, once the first $t$ modes are allowed to form a genuinely correlated multimode state, the relevant notion of mixing is no longer single-mode spreading, but cross-block visibility measured by $\operatorname{rank}(U_{B,A})$.
\end{remark}

Proposition~\ref{prop:cross-block-coherent-preservation} shows that if a nontrivial passive-Gaussian unitary preserves the product structure across two state blocks, then the input state must satisfy strong structural constraints. The following lemma establishes the converse direction relevant to our setting, namely, that specific structural features of the product input state are reflected in the corresponding learned passive-Gaussian unitary.

\begin{lemma}[Gauge structure of the passive-Gaussian unitary in $t$-doped states]
\label{thm:gauge-learned-CDB}
Define the partition of modes $A=\{1,\ldots,t\},$ and $B=\{t+1,\ldots,m\}$ in an $m$-mode quantum system. Fix an $m$-mode passive-Gaussian unitary $U_O$ with mode-mixing matrix
$U\in\mathrm U(m)$, defined by $U_O^\dagger \bm a\,U_O = U\,\bm a,$ for $\bm a=(a_1,\ldots,a_m)^{\mathsf T}$. Assume that a learning procedure returns an $m$-mode passive-Gaussian unitary
$U_{\widetilde O}$ and a decomposition of the learned input block $A=C\sqcup D$ such that the following condition is satisfied 
\begin{equation}
\label{eq:learned-CDB-main}
U_O\Bigl(|\psi\rangle_A\otimes |\underline{\bm\alpha}\rangle_B\Bigr)
\simeq 
U_{\widetilde O}\Bigl(|\underline{\bm\mu}\rangle_C\otimes |\chi\rangle_D\otimes |\underline{\bm\beta}\rangle_B\Bigr),
\end{equation}
where $|\underline{\bm\alpha}\rangle_B$, $|\underline{\bm\beta}\rangle_B$, and $|\underline{\bm\mu}\rangle_C$ are coherent states, and
$|\chi\rangle_D$ is a pure state with no coherent direction, namely
\begin{equation}
\label{eq:no-coherent-direction-main}
\left(\sum_{j\in D} c_j a_j\right)|\chi\rangle_D=\lambda |\chi\rangle_D
\quad\Longrightarrow\quad
c_j=0 \text{ for all } j\in D.
\end{equation}

Define the passive disentangler $V\coloneqq U_{\widetilde O}^\dagger U_O$, and let $W\in\mathrm U(m)$ be its mode-mixing matrix, $V^\dagger \bm a\,V = W\,\bm a$. Then, we have $U=\widetilde U\,W,$ where $\widetilde U$ is the mode-mixing matrix of $U_{\widetilde O}$, and the residual gauge satisfies
\begin{equation}
\label{eq:WDB-zero-main}
W_{D,B}=0.
\end{equation}
Equivalently, in the row order $C,D,B$ and column order $A,B$, we have 
\begin{equation}
\label{eq:W-block-form-main}
W=
\begin{pmatrix}
W_{C,A} & W_{C,B}\\
W_{D,A} & 0\\
W_{B,A} & W_{B,B}
\end{pmatrix}.
\end{equation}
Thus the only universal constraint is that the true coherent block $B$ cannot mix into the learned non-coherent sector $D$. In the phase space, we have 
\begin{equation}
\widetilde O = O\,O(W)^{\mathsf T},
\end{equation}
where 
$O(X)=
\begin{pmatrix}
\Re X & -\Im X\\
\Im X & \Re X
\end{pmatrix}$ is the corresponding symplectic-orthogonal matrix of $X$. Here, $O(W)$ has the same zero-block structure induced by \eqref{eq:WDB-zero-main}.
\end{lemma}

\begin{proof}
By definition of $V=U_{\widetilde O}^\dagger U_O$, the identity
\eqref{eq:learned-CDB-main} is equivalent to
\begin{equation}
\label{eq:reverse-disentangled-CDB-main}
V^\dagger\Bigl(|\underline{\bm\mu}\rangle_C\otimes |\chi\rangle_D\otimes |\underline{\bm\beta}\rangle_B\Bigr)
=
|\psi\rangle_A\otimes |\underline{\bm\alpha}\rangle_B.
\end{equation}

Fix any output index $k\in B$. Since the $k$th mode on the right-hand side of
\eqref{eq:reverse-disentangled-CDB-main} is coherent with amplitude $\alpha_k$, we have
\begin{equation}
a_k\,|\psi\rangle_A\otimes |\underline{\bm\alpha}\rangle_B
=
\alpha_k\,|\psi\rangle_A\otimes |\underline{\bm\alpha}\rangle_B.
\end{equation}
Applying $a_k$ to both sides of \eqref{eq:reverse-disentangled-CDB-main} and multiplying by $V$ on the left give the following equation 
\begin{equation}
\left(\sum_{j=1}^m (W^\dagger)_{kj} a_j\right)
\left(|\underline{\bm\mu}\rangle_C\otimes |\chi\rangle_D\otimes |\underline{\bm\beta}\rangle_B\right)
=
\alpha_k
\Bigl(|\underline{\bm\mu}\rangle_C\otimes |\chi\rangle_D\otimes |\underline{\bm\beta}\rangle_B\Bigr).
\end{equation}
Since $|\underline{\bm\mu}\rangle_C$ and $|\underline{\bm\beta}\rangle_B$ are coherent states, the contributions from the modes in $C$ and $B$ reduce to scalar multiples. Hence there exists a scalar $\lambda_k\in\mathbb C$ such that
\begin{equation}
\label{eq:D-eigen-main}
\left(\sum_{j\in D} (W^\dagger)_{kj} a_j\right)|\chi\rangle_D
=
\lambda_k |\chi\rangle_D.
\end{equation}
By \eqref{eq:no-coherent-direction-main}, the only annihilation-operator linear combination on $D$ admitting $|\chi\rangle_D$ as an eigenstate is the zero operator. Therefore, we have 
\begin{equation}
(W^\dagger)_{kj}=0
\qquad
\text{for every } k\in B,\ j\in D.
\end{equation}
Equivalently, we have 
\begin{equation}
W_{D,B}=0.
\end{equation}
This proves \eqref{eq:WDB-zero-main}, and \eqref{eq:W-block-form-main} is just the corresponding block form.

Finally, from the definition  $V=U_{\widetilde O}^\dagger U_O$ 
and the Heisenberg relations $U_O^\dagger \bm a\,U_O = U\,\bm a,$ $U_{\widetilde O}^\dagger \bm a\,U_{\widetilde O} = \widetilde U\,\bm a,$ and $V^\dagger \bm a\,V = W\,\bm a$, we obtain $W=\widetilde U^\dagger U,$ and $U=\widetilde U\,W$.

Passing to quadratures gives
\begin{equation}
O=\widetilde O\,O(W),
\qquad
\widetilde O = O\,O(W)^{\mathsf T},
\end{equation}
since $O(W)$ is orthogonal. This completes the proof.
\end{proof}

Note that, in practice, once a state has been reconstructed in the form $U_{\widetilde O}\Bigl(|\underline{\bm\mu}\rangle_C\otimes |\chi\rangle_D\otimes |\underline{\bm\beta}\rangle_B\Bigr)$, while the true target is known to be of the form $U_O\Bigl(|\psi\rangle_A\otimes |\underline{\bm\alpha}\rangle_B\Bigr)$, one can automatically update the reconstruction to the former form, since it has a more specialized input. In other words, the state $|\psi\>_A$ is alternatively represented as a passive-separable state $|\psi\>_A=U_{\widetilde O_{A}}\left(|\underline{\bs\mu}\>_C\otimes |\chi\>_D\right)$. On this account, the overall reconstructed passive-Gaussian unitary $U_{\widetilde O}$ can be used to approximate $U_O U_{O_A}$, where $U_{O_A}$ is the unitary that $U_{\widetilde O_A}$ converges into. Therefore, the following corollary suffices to use in learning protocols.

\begin{corollary}[Block-diagonal gauge structure of the learned Gaussian-passive unitary in $t$-doped states]
\label{cor:block-diagonal-gauge-no-C} Under Lemma~\ref{thm:gauge-learned-CDB}, we further assume that $C=\varnothing$ and $A=D$. Equivalently, the learned input takes the form
\begin{equation}
\label{eq:learned-DB-main}
U_O\Bigl(|\psi\rangle_A\otimes |\underline{\bm\alpha}\rangle_B\Bigr)
\simeq
U_{\widetilde O}\Bigl(|\chi\rangle_A\otimes |\underline{\bm\beta}\rangle_B\Bigr),
\end{equation}
where $|\underline{\bm\alpha}\rangle_B$ and $|\underline{\bm\beta}\rangle_B$ are coherent states, and
$|\chi\rangle_A$ is a pure state with no coherent direction, namely
\begin{equation}
\left(\sum_{j\in A} c_j a_j\right)|\chi\rangle_A=\lambda |\chi\rangle_A
\quad\Longrightarrow\quad
c_j=0 \text{ for all } j\in A.
\end{equation}
Then the residual gauge is block diagonal with respect to the bipartition $A\sqcup B$, namely $W_{A,B}=0$, $W_{B,A}=0$. Hence, we have 
\begin{equation}
W = W_A \oplus W_B
\end{equation}
for some $W_A\in \mathrm U(t)$ and $W_B\in \mathrm U(m-t)$. Equivalently, we have 
\begin{equation}
U=\widetilde U\,(W_A\oplus W_B).
\end{equation}
In phase space, the learned symplectic matrix differs from the true one by a block-diagonal right gauge,
\begin{equation}
\widetilde O
=
O
\Bigl(O(W_A)\oplus O(W_B)\Bigr)^{\mathsf T}.
\end{equation}
\end{corollary}

\begin{proof}
Since $C=\varnothing$, Proposition~\ref{thm:gauge-learned-CDB} gives
\begin{equation}
W_{A,B}=0.
\end{equation}
Now write $W$ in $2\times 2$ block form with respect to the bipartition $A\sqcup B$,
\begin{equation}
W=
\begin{pmatrix}
W_{A,A} & W_{A,B}\\
W_{B,A} & W_{B,B}
\end{pmatrix}.
\end{equation}
Because $W$ is unitary and $W_{A,B}=0$, the relation $WW^\dagger = I$ implies
\begin{equation}
W_{A,A}W_{B,A}^\dagger = 0.
\end{equation}
On the other hand, from the condition $W^\dagger W = I$ and $W_{A,B}=0$, we obtain
\begin{equation}
W_{A,A}^\dagger W_{A,A} = I_A,
\end{equation}
so $W_{A,A}$ is unitary. Hence it is invertible, and therefore we have 
\begin{equation}
W_{B,A}=0.
\end{equation}
Thus, we have 
\begin{equation}
W=
\begin{pmatrix}
W_{A,A} & 0\\
0 & W_{B,B}
\end{pmatrix}
=
W_A\oplus W_B,
\end{equation}
where $W_A:=W_{A,A}\in\mathrm U(t)$ and $W_B:=W_{B,B}\in\mathrm U(m-t)$.

The identity $U=\widetilde U\,W$ therefore becomes
\begin{equation}
U=\widetilde U\,(W_A\oplus W_B).
\end{equation}
Passing to quadratures gives
\begin{equation}
\widetilde O
=
O
\Bigl(O(W_A)\oplus O(W_B)\Bigr)^{\mathsf T}.
\end{equation}
This completes the proof.
\end{proof}

\subsubsection{Generalization of the activation theorem}

In addition to the gauge structure, we need a complementary result describing how coherent probe states determine the partition between coherent and noncoherent states before the passive-Gaussian unitary. Since the gauge analysis is built on this distinction, it is essential to understand how this partition varies with the initial coherent probe. Inspired by Theorem~\ref{prop:measure-zero-bargmann-multimode}, we state the following corollary.

\begin{corollary}[Random multimode coherent probes almost never remain coherent unless the unitary preserves coherent states]
\label{cor:measure-zero-coherent-multimode}
Let $m\in\mathbb N$, let $\Omega\subset\mathbb C^m$ be a nonempty connected open set, and let $W$ be an $m$-mode unitary.
For $\bm\alpha=(\alpha_1,\dots,\alpha_m)\in\mathbb C^m$, write $\ket{\underline{\bm\alpha}}\coloneqq \bigotimes_{j=1}^m \ket{\underline{\alpha_j}}$. Define
\begin{align}
\mathcal C_W(\Omega)
\coloneqq
\bigl\{
\bm\alpha\in\Omega \,\big|\, W\ket{\underline{\bm\alpha}}\ \text{is an $m$-mode coherent state}
\bigr\}.
\end{align}
Then the following are equivalent.
\begin{enumerate}
\item $\mathcal C_W(\Omega)$ has positive $2m$-dimensional Lebesgue measure in $\Omega$.
\item There exist $\phi\in\mathbb R$, $\bm\beta\in\mathbb C^m$, and a passive-Gaussian unitary $U_O$ with
\begin{align}
U_O^\dagger \bm a\,U_O = O\,\bm a,
\qquad
O\in\mathrm U(m),
\qquad
\bm a=(a_1,\dots,a_m)^{\mathsf T},
\end{align}
such that
\begin{align}
W=e^{i\phi}D(\bm\beta)\,U_O.
\end{align}
\end{enumerate}
In particular, if $W$ is not of the form $e^{i\phi}D(\bm\beta)U_O$, then $\mathcal C_W(\Omega)$ has Lebesgue measure zero.

Consequently, if $\bm\alpha$ is drawn from any probability distribution on $\Omega$ that is absolutely continuous with respect to Lebesgue measure on $\mathbb C^m\simeq\mathbb R^{2m}$, then
\begin{align}
\Pr\!\bigl[W\ket{\underline{\bm\alpha}}\ \text{is a coherent state}\bigr]=0,
\qquad
\Pr\!\bigl[W\ket{\underline{\bm\alpha}}\ \text{is not a coherent state}\bigr]=1.
\end{align}
\end{corollary}

\begin{proof}
Assume first that \(\mathcal C_W(\Omega)\) has positive \(2m\)-dimensional Lebesgue measure. Since every \(m\)-mode coherent state is a pure Gaussian state, we have $\mathcal C_W(\Omega)\subset \mathcal E_W(\Omega)$, where \(\mathcal E_W(\Omega)\) is the set in Theorem~\ref{prop:measure-zero-bargmann-multimode}, i.e., the set of complex amplitudes that correspond to Gaussian output states from $W$. Hence \(\mathcal E_W(\Omega)\) also has positive Lebesgue measure. By Theorem~\ref{prop:measure-zero-bargmann-multimode}, \(W\) is an \(m\)-mode Gaussian unitary. Therefore there exist matrices \(X,Y\in\mathbb C^{m\times m}\) and a vector \(\bm\beta\in\mathbb C^m\) such that $W^\dagger \bm a\,W = X\bm a + Y\bm a^\dagger + \bm\beta$, with \(\bm a=(a_1,\dots,a_m)^{\mathsf T}\) and \(\bm a^\dagger=(a_1^\dagger,\dots,a_m^\dagger)^{\mathsf T}\). Fix any \(\bm\alpha\in\mathcal C_W(\Omega)\). By definition, \(W\ket{\underline{\bm\alpha}}\) is an \(m\)-mode coherent state, so there exists \(\bm\eta(\bm\alpha)\in\mathbb C^m\) such that
\begin{align}
\bm a\,W\ket{\underline{\bm\alpha}}=\bm\eta(\bm\alpha)\,W\ket{\underline{\bm\alpha}}.
\end{align}
Further applying \(W^\dagger\) gives $\bigl(X\bm a + Y\bm a^\dagger + \bm\beta\bigr)\ket{\underline{\bm\alpha}}
=
\bm\eta(\bm\alpha)\ket{\underline{\bm\alpha}}$. Using \(a_k\ket{\underline{\bm\alpha}}=\alpha_k\ket{\underline{\bm\alpha}}\) for all \(k\), we obtain
\begin{align}
Y\bm a^\dagger\ket{\underline{\bm\alpha}}
=
\bigl(\bm\eta(\bm\alpha)-X\bm\alpha-\bm\beta\bigr)\ket{\underline{\bm\alpha}}.
\end{align}
Looking at the \(j\)-th component, there exists \(c_j\in\mathbb C\) such that
\begin{align}
\sum_{k=1}^m Y_{jk}a_k^\dagger\ket{\underline{\bm\alpha}}
=
c_j\ket{\underline{\bm\alpha}}.
\end{align}
Now fix any \(\ell\in\{1,\dots,m\}\) and apply \(a_\ell-\alpha_\ell\) to both sides. Since $(a_\ell-\alpha_\ell)\ket{\underline{\bm\alpha}}=0$ and $(a_\ell-\alpha_\ell)a_k^\dagger\ket{\underline{\bm\alpha}}
=
[a_\ell,a_k^\dagger]\ket{\underline{\bm\alpha}}
+
a_k^\dagger(a_\ell-\alpha_\ell)\ket{\underline{\bm\alpha}}
=
\delta_{\ell k}\ket{\underline{\bm\alpha}}$, we get
\begin{align}
Y_{j\ell}\ket{\underline{\bm\alpha}}=0.
\end{align}
Since \(\ket{\underline{\bm\alpha}}\neq 0\), it follows that \(Y_{j\ell}=0\). As \(j\) and \(\ell\) are arbitrary, we conclude that $Y=0$.

Hence, we have 
\begin{align}
W^\dagger \bm a\,W = X\bm a+\bm\beta.
\end{align}
Because \(W\) is unitary, the transformed annihilation operators satisfy the canonical commutation relations, so we have $[X\bm a,(X\bm a)^\dagger]=I_m$. Equivalently, we have  $XX^\dagger=I_m$. Since \(X\) is square, it follows that \(X\in\mathrm U(m)\). Write \(O\coloneqq X\), and let \(U_O\) be the passive-Gaussian unitary determined by $U_O^\dagger \bm a\,U_O = O\bm a$. Define $U\coloneqq D(\bm\beta)\,U_O$. Then, we have 
\begin{align}
U^\dagger \bm a\,U
=
U_O^\dagger D(\bm\beta)^\dagger \bm a\,D(\bm\beta)U_O
=
U_O^\dagger(\bm a+\bm\beta)U_O
=
O\bm a+\bm\beta
=
W^\dagger \bm a\,W.
\end{align}
Therefore, with \(V\coloneqq U^\dagger W\), we have $V^\dagger \bm a\,V=\bm a$. Equivalently, we have $V a_j = a_j V$, $V a_j^\dagger = a_j^\dagger V$ for all $j=1,\dots,m$. Let \(\ket{\bm 0}\) be the multimode vacuum. Since \(a_j\ket{\bm 0}=0\) for all \(j\), we have
\begin{align}
a_j V\ket{\bm 0}=V a_j\ket{\bm 0}=0
\qquad
\text{for all }j.
\end{align}
By uniqueness of the vacuum, there exists \(\phi\in\mathbb R\) such that $V\ket{\bm 0}=e^{i\phi}\ket{\bm 0}$. Now let \(\ket{\bm n}\) be any multimode Fock basis vector. Since \(\ket{\bm n}\) is obtained by applying creation operators to \(\ket{\bm 0}\), the commutation relations above imply
\begin{align}
V\ket{\bm n}=e^{i\phi}\ket{\bm n}
\qquad
\text{for all }\bm n\in\mathbb N^m.
\end{align}
Hence \(V=e^{i\phi}I\), and therefore we have 
\begin{align}
W=e^{i\phi}D(\bm\beta)\,U_O.
\end{align}
This proves \((1)\Rightarrow(2)\).

Conversely, assume $W=e^{i\phi}D(\bm\beta)\,U_O$ with \(U_O^\dagger \bm a\,U_O=O\bm a\) and \(O\in\mathrm U(m)\). Let \(\bm\alpha\in\Omega\). Since
\begin{align}
a_j\,U_O\ket{\underline{\bm\alpha}}
=
U_O\bigl(U_O^\dagger a_j U_O\bigr)\ket{\underline{\bm\alpha}}
=
U_O\sum_{k=1}^m O_{jk}a_k\ket{\underline{\bm\alpha}}
=
(O\bm\alpha)_j\,U_O\ket{\underline{\bm\alpha}},
\end{align}
the state \(U_O\ket{\underline{\bm\alpha}}\) is an \(m\)-mode coherent state. Applying the displacement \(D(\bm\beta)\) preserves coherent states. So \(W\ket{\underline{\bm\alpha}}\) is also an \(m\)-mode coherent state. Thus every \(\bm\alpha\in\Omega\) belongs to \(\mathcal C_W(\Omega)\), and hence $\mathcal C_W(\Omega)=\Omega$. Since \(\Omega\) is a nonempty open subset of \(\mathbb C^m\simeq\mathbb R^{2m}\), it has positive Lebesgue measure. This proves \((2)\Rightarrow(1)\).

The final probabilistic statement follows immediately, because any set of Lebesgue measure zero has probability zero under any probability distribution that is absolutely continuous with respect to Lebesgue measure.
\end{proof}

It is worth noting that Corollary \ref{cor:measure-zero-coherent-multimode} is not sufficient to identify the activation relation in the case of $t$-doped, as the output of the $\kappa t$-mode unitary in Eq. (\ref{supp_def:U_doped_three_leyer}) could be a product of coherent states and incoherent states. After reconstructing the generalized passive-separable state in Eq. (\ref{eqs186}) using the shadow tomography protocol in \cite{zhao2025complexity}, whether the coherent input state of $U_{\widetilde O_{2,\bs\alpha^{(h)}}}$ is the output of $U_{(\kappa t)}$ is  unknown. Therefore, Corollary \ref{cor:measure-zero-coherent-multimode} alone does not indicate the structure of the $\kappa t$-mode unitary from the output. In the following, we provide two lemmas, a proposition, and two corollaries to address the activation theorem in this special case.

\begin{lemma}[Vector-valued coherent-factor witness]
\label{lem:vector-valued-coherent-factor-witness}
Let \(\mathcal K\) be a Hilbert space, and let $F:\mathbb C^r\to \mathcal K$ be a nonzero entire \(\mathcal K\)-valued function. For \(1\le j,k\le r\), let us define a witness function as follows
\begin{align}
\mathsf C_{jk}[F](\bm z)
\coloneqq
F(\bm z)\otimes \partial_j\partial_kF(\bm z)
-
\partial_jF(\bm z)\otimes \partial_kF(\bm z)
\in
\mathcal K\widehat\otimes\mathcal K,
\end{align}
where \(\partial_j=\partial/\partial z_j\). Then the following are equivalent.
\begin{enumerate}
\item $\mathsf C_{jk}[F]\equiv 0
\qquad
\text{for all }1\le j,k\le r$.
\item There exist \(\bm\eta\in\mathbb C^r\) and a nonzero vector \(\xi\in\mathcal K\) such that $F(\bm z)
=
e^{\bm\eta^{\mathsf T}\bm z}\xi$ $\text{for all }\bm z\in\mathbb C^r$.
\end{enumerate}
\end{lemma}

\begin{proof}
The implication \((2)\Rightarrow(1)\) is immediate, since we have the relations 
\begin{align}
\partial_jF(\bm z)=\eta_jF(\bm z),
\qquad
\partial_j\partial_kF(\bm z)=\eta_j\eta_kF(\bm z).
\end{align}

Now, let us prove \((1)\Rightarrow(2)\). Let us denote $U_F\coloneqq
\{\bm z\in\mathbb C^r\mid F(\bm z)\neq0\}$. Since we have \(F\not\equiv0\), \(U_F\) is nonempty and open. Fix \(\bm z\in U_F\). From the relation
\begin{align}
F(\bm z)\otimes \partial_j^2F(\bm z)
=
\partial_jF(\bm z)\otimes \partial_jF(\bm z),
\end{align}
and the fact that \(x\otimes y=u\otimes u\) with \(x\neq0\) implies
\(u\in\operatorname{span}\{x\}\), we obtain
\begin{align}
\partial_jF(\bm z)\in\operatorname{span}\{F(\bm z)\}.
\end{align}
Indeed, if \(u\notin\operatorname{span}\{x\}\), choose a linear functional
\(\ell\) with \(\ell(x)=0\) and \(\ell(u)=1\). Applying \(\ell\otimes I\) to
\(x\otimes y=u\otimes u\) gives \(0=u\), a contradiction. 
Thus, on \(U_F\), there are holomorphic functions \(g_j\) such that
\begin{align}
\partial_jF(\bm z)=g_j(\bm z)F(\bm z),
\qquad
j=1,\dots,r.
\end{align}
Substituting this into \(\mathsf C_{jk}[F]=0\), we get
\begin{align}
0
&=
F\otimes \partial_j\partial_kF-\partial_jF\otimes\partial_kF
\nonumber\\
&=
F\otimes\partial_j(g_kF)-g_jF\otimes g_kF
\nonumber\\
&=
F\otimes\bigl((\partial_jg_k)F+g_jg_kF\bigr)-g_jg_kF\otimes F
\nonumber\\
&=
(\partial_jg_k)\,F\otimes F,
\end{align}
where the third equality follows from the Leibniz rule
\(\partial_j(g_kF)=(\partial_jg_k)F+g_k\partial_jF
=(\partial_jg_k)F+g_jg_kF\), together with
\(g_jF\otimes g_kF=g_jg_kF\otimes F\); the fourth equality follows from
the bilinearity of the tensor product, since
\(F\otimes(g_jg_kF)=g_jg_kF\otimes F\), so the two \(g_jg_kF\otimes F\)
terms cancel. Since \(F(\bm z)\neq0\) on \(U_F\), it follows that
\begin{align}
\partial_jg_k=0
\qquad
\text{on }U_F
\end{align}
for all \(j,k\). Hence each \(g_k\) is constant on every connected component of \(U_F\).

Let \(U_0\) be one connected component of \(U_F\). On \(U_0\), there exists
\(\bm\eta\in\mathbb C^r\) such that
\begin{align}
\partial_jF=\eta_jF
\qquad
\text{for all }j=1,\dots,r.
\end{align}
Therefore, we have 
\begin{align}
F(\bm z)=e^{\bm\eta^{\mathsf T}\bm z}\xi
\qquad
\text{on }U_0
\end{align}
for some nonzero \(\xi\in\mathcal K\). The right-hand side extends continuously and never vanishes. Hence no boundary point of \(U_0\) can be a zero of \(F\). Thus \(U_0\) is both open and closed in \(\mathbb C^r\). Since \(\mathbb C^r\) is connected, we have $U_0=\mathbb C^r$. This proves
\begin{align}
F(\bm z)=e^{\bm\eta^{\mathsf T}\bm z}\xi
\qquad
\text{for all }\bm z\in\mathbb C^r.
\end{align}
\end{proof}

\begin{lemma}[Positive-measure coherent output block propagates to all coherent inputs]
\label{lem:positive-measure-coherent-block-global}
Let us set integers \(y\ge1\),  \(0\le x<y\), and $r\coloneqq y-x$. Write the output Hilbert space as $\mathcal H_y
=
\mathcal H_A\otimes\mathcal H_B$, where \(A\) contains \(x\) modes and \(B\) contains \(r\) modes. Let
\(\Omega\subset\mathbb C^y\) be a nonempty connected open set, and let \(W\) be a \(y\)-mode unitary. Define
\begin{align}
\mathcal C_{W}^{A|B}(\Omega)
\coloneqq
\left\{
\bm\alpha\in\Omega
\,\middle|\,
\exists\,\ket{\psi_{\bm\alpha}}\in\mathcal H_A,\ 
\exists\,\bm\eta_{\bm\alpha}\in\mathbb C^r
\text{ such that }
W\ket{\underline{\bm\alpha}}
=
\ket{\psi_{\bm\alpha}}_A\otimes\ket{\underline{\bm\eta_{\bm\alpha}}}_B
\right\}.
\end{align}
If \(\mathcal C_{W}^{A|B}(\Omega)\) has positive \(2y\)-dimensional Lebesgue measure in \(\Omega\), then
\begin{align}
W\ket{\underline{\bm\alpha}}
=
\ket{\psi_{\bm\alpha}}_A\otimes\ket{\underline{\bm\eta_{\bm\alpha}}}_B
\end{align}
for every \(\bm\alpha\in\mathbb C^y\), for suitable
\(\ket{\psi_{\bm\alpha}}\in\mathcal H_A\) and \(\bm\eta_{\bm\alpha}\in\mathbb C^r\).
\end{lemma}

\begin{proof}
Similar to the discussion in \ref{supp:Learning of the second passive-Gaussian layer}, we use unnormalized coherent states
\begin{align}
\|{\bm\alpha}\rangle
\coloneqq
e^{\sum_{\ell=1}^y \alpha_\ell a_\ell^\dagger}\ket{\bm0},
\qquad
\langle \bm z\|
\coloneqq
\bra{\bm0}e^{\sum_{j=1}^r z_j b_j},
\end{align}
where \(\bm b=(b_1,\dots,b_r)^{\mathsf T}\) denotes the annihilation operators of the \(B\)-block. For
\(\bm\alpha\in\mathbb C^y\) and \(\bm z\in\mathbb C^r\), define the \(\mathcal H_A\)-valued Bargmann function
\begin{align}
F_{\bm\alpha}(\bm z)
\coloneqq
(I_A\otimes\langle \bm z\|)\,W\|{\bm\alpha}\rangle .
\end{align}
Expanding in the Fock bases gives
\begin{align}
F_{\bm\alpha}(\bm z)
=
\sum_{\bm n\in\mathbb N^r}
\sum_{\bm p\in\mathbb N^y}
\frac{\bm z^{\bm n}\bm\alpha^{\bm p}}{\sqrt{\bm n!\,\bm p!}}\,
\bm v_{\bm n,\bm p},
\end{align}
where $\bm v_{\bm n,\bm p}
\coloneqq
(I_A\otimes\langle \bm n|)W\ket{\bm p}
\in\mathcal H_A$ denotes the matrix elements of $W$ satisfying \(\|\bm v_{\bm n,\bm p}\|_2^2
=\|(I_A\otimes\langle \bm n|)W|\bm p\rangle\|_2^2
=\langle \bm p|W^\dagger(I_A\otimes|\bm n\rangle\langle\bm n|)W|\bm p\rangle
\le \langle \bm p|W^\dagger W|\bm p\rangle=1\), where the norm is the Hilbert-space norm on \(\mathcal H_A\).

Hence the series converges absolutely and locally uniformly on compact subsets of
\(\mathbb C^y\times\mathbb C^r\), because for every closed polydisc $|\alpha_\ell|\le R_\ell$ and $|z_j|\le S_j$, we have
\begin{align}
\sum_{\bm n,\bm p}
\left\|
\frac{\bm z^{\bm n}\bm\alpha^{\bm p}}{\sqrt{\bm n!\,\bm p!}}\,
\bm v_{\bm n,\bm p}
\right\|_2
&\le
\prod_{j=1}^r
\left(
\sum_{n_j\ge0}\frac{S_j^{n_j}}{\sqrt{n_j!}}
\right)
\prod_{\ell=1}^y
\left(
\sum_{p_\ell\ge0}\frac{R_\ell^{p_\ell}}{\sqrt{p_\ell!}}
\right)
<\infty,
\end{align}
where we used the relations $|\bm z^{\bm n}|=\prod_{j=1}^r |z_j|^{n_j}\le \prod_{j=1}^r S_j^{n_j},$ $|\bm\alpha^{\bm p}|=\prod_{\ell=1}^y |\alpha_\ell|^{p_\ell}
\le \prod_{\ell=1}^y R_\ell^{p_\ell}$, $\bm n! = \prod_{j=1}^r n_j!,$ and $\bm p! = \prod_{\ell=1}^y p_\ell!$, so that the nonnegative multi-index sums factor into products of one-dimensional sums.
Each one-dimensional series converges by the ratio test. Therefore, the map \((\bm\alpha,\bm z)\mapsto F_{\bm\alpha}(\bm z)\) is jointly entire.

Now let us denote \(\bm\alpha\in\mathcal C_{W}^{A|B}(\Omega)\). Then, we have $W\ket{\underline{\bm\alpha}}
=
\ket{\psi_{\bm\alpha}}_A\otimes\ket{\underline{\bm\eta_{\bm\alpha}}}_B$. Equivalently, for unnormalized coherent states, we have 
\begin{align}
W\|{\bm\alpha}\rangle
=
\|\chi_{\bm\alpha}\>\otimes\|{\bm\eta_{\bm\alpha}}\rangle_B
\end{align}
for some unnormalized state $\|\chi_{\bm\alpha}\>\in\mathcal H_A$. Therefore, we have 
\begin{align}
F_{\bm\alpha}(\bm z)
=
e^{\bm\eta_{\bm\alpha}^{\mathsf T}\bm z}\|\chi_{\bm\alpha}\>.
\end{align}
Using Lemma~\ref{lem:vector-valued-coherent-factor-witness}, we have 
\begin{align}
\mathsf C_{jk}[F_{\bm\alpha}](\bm z)=0
\qquad
\text{for all }\bm z\in\mathbb C^r,\ 1\le j,k\le r.
\end{align}

Fix \(u,v\in\mathcal H_A\), \(\bm z\in\mathbb C^r\), and \(1\le j,k\le r\). Define
\begin{align}
G_{u,v;\bm z,j,k}(\bm\alpha)
\coloneqq
\left\langle
u\otimes v,\,
\mathsf C_{jk}[F_{\bm\alpha}](\bm z)
\right\rangle .
\end{align}
Since \(F_{\bm\alpha}(\bm z)\) is jointly entire, \(G_{u,v;\bm z,j,k}\) is holomorphic in
\(\bm\alpha\). It vanishes on \(\mathcal C_{W}^{A|B}(\Omega)\), which has positive
\(2y\)-dimensional Lebesgue measure in \(\Omega\). By the identity theorem \cite{Ahlfors1979ComplexAnalysis} for holomorphic functions of several complex variables, we have 
\begin{align}
G_{u,v;\bm z,j,k}(\bm\alpha)=0
\qquad
\text{for all }\bm\alpha\in\mathbb C^y.
\end{align}
Since \(u,v,\bm z,j,k\) were arbitrary, we get
\begin{align}
\mathsf C_{jk}[F_{\bm\alpha}](\bm z)=0
\qquad
\text{for all }\bm\alpha\in\mathbb C^y,\ \bm z\in\mathbb C^r,\ 1\le j,k\le r.
\end{align}
Applying Lemma~\ref{lem:vector-valued-coherent-factor-witness} to
\(\bm z\mapsto F_{\bm\alpha}(\bm z)\) for each fixed \(\bm\alpha\), we obtain
\begin{align}
F_{\bm\alpha}(\bm z)
=
e^{\bm\eta(\bm\alpha)^{\mathsf T}\bm z}\|\chi_{\bm\alpha}\>
\end{align}
for some \(\bm\eta(\bm\alpha)\in\mathbb C^r\) and some unnormalized state \(\|\chi_{\bm\alpha}\>\in\mathcal H_A\). Since the \(r\)-mode Bargmann transform is injective, this is equivalent to $W\|{\bm\alpha}\rangle
=
\|\chi_{\bm\alpha}\>\otimes\|\bm\eta(\bm\alpha)\rangle_B$. After normalization, this gives
\begin{align}
W\ket{\underline{\bm\alpha}}
=
\ket{\psi_{\bm\alpha}}_A\otimes\ket{\underline{\bm\eta(\bm\alpha)}}_B
\end{align}
for every \(\bm\alpha\in\mathbb C^y\).
\end{proof}

\begin{proposition}[Structure of unitaries with a coherent output block]
\label{prop:coherent-output-block-structure}
Let us denote integers \(y\ge1\), \(0\le x<y\), and \(r=y-x\). Let us denote  the Hilbert space $\mathcal H_y=\mathcal H_A\otimes\mathcal H_B$, where \(A\) contains \(x\) modes and \(B\) contains \(r\) modes. Let \(W\) be a \(y\)-mode unitary. Assume that for every
\(\bm\alpha\in\mathbb C^y\), there exist
\(\ket{\psi_{\bm\alpha}}\in\mathcal H_A\) and
\(\bm\eta(\bm\alpha)\in\mathbb C^r\) such that
\begin{align}
W\ket{\underline{\bm\alpha}}
=
\ket{\psi_{\bm\alpha}}_A\otimes\ket{\underline{\bm\eta(\bm\alpha)}}_B.
\end{align}
Then there exist an \(x\)-mode unitary \(U_A\), a vector \(\bm\beta\in\mathbb C^r\), and a
\(y\)-mode passive-Gaussian unitary \(U_{O_u}\), with $U_{O_u}^\dagger \bm a\,U_{O_u}=O_u\bm a$, $
O_u\in {\rm U}(y)$, such that
\begin{align}
W
=
(U_A\otimes D_B(\bm\beta))\,U_{O_u}.
\end{align}
\end{proposition}

\begin{proof}
As in the proof of
Lemma~\ref{lem:positive-measure-coherent-block-global}, let us define $F_{\bm\alpha}(\bm z)
\coloneqq
(I_A\otimes\langle \bm z\|)\,W\|{\bm\alpha}\rangle$. By assumption, we have  $F_{\bm\alpha}(\bm z)
=
e^{\bm\eta(\bm\alpha)^{\mathsf T}\bm z}\|\chi_{\bm\alpha}\>$ for some unnornalized state \(\|\chi_{\bm\alpha}\>\in\mathcal H_A\). We first show that
\(\bm\eta\) is entire. Let us fix \(\bm\alpha_0\in\mathbb C^y\). Choose \(u\in\mathcal H_A\) such that $\langle u,\chi_{\bm\alpha_0}\rangle\neq0$. Then \(\langle u,F_{\bm\alpha}(\bm0)\rangle\) is nonzero in a neighborhood of
\(\bm\alpha_0\), and on this neighborhood we have 
\begin{align}
\eta_j(\bm\alpha)
=
\frac{
\left\langle u,\partial_{z_j}F_{\bm\alpha}(\bm0)\right\rangle
}{
\left\langle u,F_{\bm\alpha}(\bm0)\right\rangle
}.
\end{align}
Since \(F_{\bm\alpha}(\bm z)\) is jointly entire, each \(\eta_j\) is holomorphic near
\(\bm\alpha_0\). As \(\bm\alpha_0\) was arbitrary, the map $\bm\eta:\mathbb C^y\to\mathbb C^r$ is entire. Next, by unitarity of \(W\), we have 
\begin{align}
\left|\langle \underline{\bm\gamma}|\underline{\bm\alpha}\rangle\right|
&=
\left|
\langle \underline{\bm\gamma}|W^\dag W|\underline{\bm\alpha}\rangle
\right|
\nonumber\\
&=
\left|
\langle \psi_{\bm\gamma}|\psi_{\bm\alpha}\rangle
\right|
\left|
\langle \underline{\bm\eta(\bm\gamma)}|\underline{\bm\eta(\bm\alpha)}\rangle
\right|
\nonumber\\
&\le
\left|
\langle \underline{\bm\eta(\bm\gamma)}|\underline{\bm\eta(\bm\alpha)}\rangle
\right|.
\end{align}
Using the relation  $\left|\langle \underline{\bm u}|\underline{\bm v}\rangle\right|
=
\exp\!\left(-\frac12\|\bm u-\bm v\|^2_2\right)$, we obtain $\|\bm\eta(\bm\alpha)-\bm\eta(\bm\gamma)\|_2
\le
\|\bm\alpha-\bm\gamma\|_2\,
\text{for all }\bm\alpha,\bm\gamma\in\mathbb C^y$. Thus \(\bm\eta\) is entire and globally Lipschitz. Hence its complex Jacobian is bounded: $\|D\bm\eta(\bm\alpha)\mathbf v\|_2
\le
\|\mathbf v\|_2\,\text{for all }\bm\alpha\in\mathbb C^y,\ \mathbf v\in\mathbb C^y$.

Every entry of \(D\bm\eta\) is therefore a bounded entire function, and Liouville's theorem gives that
\(D\bm\eta\) is constant. Hence there exist a complex matrix 
\(K\in\mathbb C^{r\times y}\) and a complex vector \(\bm\beta\in\mathbb C^r\) such that
\begin{align}
\bm\eta(\bm\alpha)=K\bm\alpha+\bm\beta.
\end{align}

We now show that \(K\) is a coisometry: $KK^\dagger=I_r$. For \(\bm\lambda\in\mathbb C^r\), let \(D_B(\bm\lambda)\) be the \(r\)-mode displacement operator. For all
\(\bm\alpha,\bm\gamma\in\mathbb C^y\), we have 
\begin{align}
&\left\langle \underline{\bm\gamma}\right|
W^\dagger(I_A\otimes D_B(\bm\lambda))W
\left|\underline{\bm\alpha}\right\rangle
\nonumber\\
&\quad =
\langle \psi_{\bm\gamma}|\psi_{\bm\alpha}\rangle
\,
\<\underline{K\bm\gamma+\bm\beta}|
D_B(\bm\lambda)
|\underline{K\bm\alpha+\bm\beta}\rangle .
\end{align}
Given the relation $\langle \psi_{\bm\gamma}|\psi_{\bm\alpha}\rangle
=
\frac{\langle \underline{\bm\gamma}|\underline{\bm\alpha}\rangle}
{\langle\underline{ K\bm\gamma+\bm\beta}|\underline{K\bm\alpha+\bm\beta}\rangle}$, we get
\begin{align}
&\left\langle \underline{\bm\gamma}\right|
W^\dagger(I_A\otimes D_B(\bm\lambda))W
\left|\underline{\bm\alpha}\right\rangle
\nonumber\\
&\quad =
\exp\!\left(
-\frac12\|\bm\lambda\|_2^2
+
\bm\lambda^{\mathsf T}\overline{\bm\beta}
-
\bm\lambda^\dagger\bm\beta
+
(K^\dagger\bm\lambda)^{\mathsf T}\overline{\bm\gamma}
-
(K^\dagger\bm\lambda)^\dagger\bm\alpha
\right)
\langle \underline{\bm\gamma}|\underline{\bm\alpha}\rangle .
\end{align}
On the other hand, we have 
\begin{align}
\langle \underline{\bm\gamma}|D(\,K^\dagger\bm\lambda\,)|\underline{\bm\alpha}\rangle
=
\exp\!\left(
-\frac12\|K^\dagger\bm\lambda\|_2^2
+
(K^\dagger\bm\lambda)^{\mathsf T}\overline{\bm\gamma}
-
(K^\dagger\bm\lambda)^\dagger\bm\alpha
\right)
\langle \underline{\bm\gamma}|\underline{\bm\alpha}\rangle .
\end{align}
Because coherent states form a total set, the preceding identities imply
\begin{align}
W^\dagger(I_A\otimes D_B(\bm\lambda))W
=
c(\bm\lambda)\,D(K^\dagger\bm\lambda),
\end{align}
with $c(\bm\lambda)
=
\exp\!\left(
-\frac12\bigl(\|\bm\lambda\|_2^2-\|K^\dagger\bm\lambda\|_2^2\bigr)
+
\bm\lambda^{\mathsf T}\overline{\bm\beta}
-
\bm\lambda^\dagger\bm\beta
\right)$. The left-hand side is unitary, and \(D(K^\dagger\bm\lambda)\) is unitary. Hence we have  $|c(\bm\lambda)|=1
\,
\text{for all }\bm\lambda\in\mathbb C^r$. The last two terms in the exponent of \(c(\bm\lambda)\) are purely imaginary, so we have 
\begin{align}
\|\bm\lambda\|_2^2=\|K^\dagger\bm\lambda\|_2^2
\qquad
\text{for all }\bm\lambda\in\mathbb C^r.
\end{align}
Equivalently, we have  $KK^\dagger=I_r$. Since \(K\in\mathbb C^{r\times y}\) has orthonormal rows, we may complete its rows to a unitary matrix
\(O_u\in {\rm U}(y)\). Let us write $O_u
=
\begin{pmatrix}
P\\
K
\end{pmatrix}$ and $P\in\mathbb C^{x\times y}$. Let \(U_{O_u}\) be the passive-Gaussian unitary satisfying $U_{O_u}^\dagger\bm a\,U_{O_u}=O_u\bm a$. Define $G
\coloneqq
(I_A\otimes D_B(\bm\beta))U_{O_u}$. From the displacement relation $D_B(\bm\beta)^\dagger\bm b\,D_B(\bm\beta)=\bm b+\bm\beta$ and from the fact that the last \(r\) rows of \(O_u\) are \(K\), we obtain $G^\dagger\bm b\,G
=
K\bm a+\bm\beta$. Equivalently, for all \(\bm\lambda\in\mathbb C^r\), we have $G^\dagger(I_A\otimes D_B(\bm\lambda))G
=
W^\dagger(I_A\otimes D_B(\bm\lambda))W$. Let us set $R\coloneqq W G^\dagger$. Then we have 
\begin{align}
R^\dagger(I_A\otimes D_B(\bm\lambda))R
=
I_A\otimes D_B(\bm\lambda)
\qquad
\text{for all }\bm\lambda\in\mathbb C^r.
\end{align}
Thus \(R\) commutes with all \(r\)-mode Weyl displacement operators. By the irreducibility of the Weyl representation on the \(r\)-mode Fock space, its commutant is $\mathcal B(\mathcal H_A)\otimes I_B$. Therefore there exists a unitary \(U_A\) on \(\mathcal H_A\) such that $R=U_A\otimes I_B$. Consequently, we have 
\begin{align}
W
=
R G
=
(U_A\otimes I_B)(I_A\otimes D_B(\bm\beta))U_{O_u}
=
(U_A\otimes D_B(\bm\beta))U_{O_u}.
\end{align}
This proves the proposition.
\end{proof}

\begin{corollary}[Random coherent probes almost never produce a coherent output block unless the unitary has block-passive structure]
\label{cor:random-coherent-block-factorization}
Let us set integers \(y\ge1\), \(0\le x<y\), and $r\coloneqq y-x$. Let $\mathcal H_y=\mathcal H_A\otimes\mathcal H_B$ be a Hilbert space, where \(A\) contains \(x\) output modes and \(B\) contains \(r\) output modes. Let
\(\Omega\subset\mathbb C^y\) be a nonempty connected open set, and let \(W\) be a \(y\)-mode unitary. Define
\begin{align}
\mathcal C_{W}^{A|B}(\Omega)
\coloneqq
\left\{
\bm\alpha\in\Omega
\,\middle|\,
\exists\,\ket{\psi_{\bm\alpha}}\in\mathcal H_A,\ 
\exists\,\bm\eta_{\bm\alpha}\in\mathbb C^r
\text{ such that }
W\ket{\underline{\bm\alpha}}
=
\ket{\psi_{\bm\alpha}}_A\otimes\ket{\underline{\bm\eta_{\bm\alpha}}}_B
\right\}.
\end{align}
Then the following are equivalent.
\begin{enumerate}
\item \(\mathcal C_{W}^{A|B}(\Omega)\) has positive \(2y\)-dimensional Lebesgue measure in \(\Omega\).
\item There exist an \(x\)-mode unitary \(U_A\), a vector \(\bm\beta\in\mathbb C^r\), and a
\(y\)-mode passive-Gaussian unitary \(U_{O_u}\), with $U_{O_u}^\dagger\bm a\,U_{O_u}=O_u\bm a,
\,
O_u\in {\rm U}(y)$, such that
\begin{align}
W
=
(U_A\otimes D_B(\bm\beta))U_{O_u}.
\end{align}
\end{enumerate}
Consequently, if \(\bm\alpha\) is drawn from any probability distribution on \(\Omega\) that is absolutely continuous with respect to Lebesgue measure on
\(\mathbb C^y\simeq\mathbb R^{2y}\), then
\begin{align}
\Pr\!\left[
W\ket{\underline{\bm\alpha}}
\text{ factorizes as an \(x\)-mode state tensored with an \(r\)-mode coherent state}
\right]
=
0
\end{align}
unless \(W\) has the above block-passive structure.

On the other hand, if \(W\) has the above block-passive structure, then $\mathcal C_W^{A|B}(\Omega)=\Omega$. Consequently, for any probability distribution of \(\bm\alpha\) supported on \(\Omega\), not necessarily absolutely continuous, we have
\begin{align}
\Pr\!\left[
W\ket{\underline{\bm\alpha}}
\text{ factorizes as an \(x\)-mode state tensored with an \(r\)-mode coherent state}
\right]
=
1.
\end{align}
\end{corollary}

\begin{proof}
We first prove \((2)\Rightarrow(1)\). Suppose we have  $W=(U_A\otimes D_B(\bm\beta))U_{O_u}$. Since \(U_{O_u}\) is passive-Gaussian, it maps every \(y\)-mode coherent state to a \(y\)-mode coherent state: $U_{O_u}\ket{\underline{\bm\alpha}}
=
\ket{\underline{(O_u\bm\alpha)_A}}_A\otimes\ket{\underline{(O_u\bm\alpha)_B}}_B$. Hence we have 
\begin{align}
W\ket{\underline{\bm\alpha}}
=
U_A\ket{\underline{(O_u\bm\alpha)_A}}_A
\otimes
D_B(\bm\beta)\ket{\underline{(O_u\bm\alpha)_B}}_B.
\end{align}
The second tensor factor is coherent for every \(\bm\alpha\in\Omega\), so we have $\mathcal C_{W}^{A|B}(\Omega)=\Omega$. Since \(\Omega\) is nonempty and open, it has positive \(2y\)-dimensional Lebesgue measure.

We now prove \((1)\Rightarrow(2)\). Assume that
\(\mathcal C_{W}^{A|B}(\Omega)\) has positive Lebesgue measure. By
Lemma~\ref{lem:positive-measure-coherent-block-global}, for every
\(\bm\alpha\in\mathbb C^y\), there exist
\(\ket{\psi_{\bm\alpha}}\in\mathcal H_A\) and
\(\bm\eta_{\bm\alpha}\in\mathbb C^r\) such that
\begin{align}
W\ket{\underline{\bm\alpha}}
=
\ket{\psi_{\bm\alpha}}_A\otimes\ket{\underline{\bm\eta_{\bm\alpha}}}_B.
\end{align}
Applying Proposition~\ref{prop:coherent-output-block-structure}, we get
\begin{align}
W
=
(U_A\otimes D_B(\bm\beta))U_{O_u}
\end{align}
for some \(x\)-mode unitary \(U_A\), some \(\bm\beta\in\mathbb C^r\), and some \(y\)-mode passive-Gaussian unitary \(U_{O_u}\). This proves the equivalence.

It remains to justify the measure-zero statement. The set
\(\mathcal C_{W}^{A|B}(\Omega)\) is measurable. Indeed, by Lemma~\ref{lem:vector-valued-coherent-factor-witness}, membership in
\(\mathcal C_{W}^{A|B}(\Omega)\) is equivalent to the vanishing of the vector-valued witnesses $\mathsf C_{jk}[F_{\bm\alpha}](\bm z)$ for all \(1\le j,k\le r\) and all \(\bm z\in\mathbb C^r\). Since the witness functions are continuous in
\(\bm\alpha\), it is enough to impose their vanishing on the countable dense set $\bm z\in(\mathbb Q+i\mathbb Q)^r$ and against a countable dense set of test vectors in
\(\mathcal H_A\widehat\otimes\mathcal H_A\). Thus
\(\mathcal C_{W}^{A|B}(\Omega)\) is Borel measurable.

If \(W\) is not of the form $(U_A\otimes D_B(\bm\beta))U_{O_u}$, then the equivalence just proved implies that
\(\mathcal C_{W}^{A|B}(\Omega)\) cannot have positive Lebesgue measure. Since it is measurable, we get
\begin{align}
\operatorname{Leb}_{2y}\!\left(\mathcal C_{W}^{A|B}(\Omega)\right)=0.
\end{align}
The final probabilistic statement follows immediately from absolute continuity of the sampling distribution.
\end{proof}

\subsubsection{Gauge structure after counter-rotation}\label{supp:learning the second layer of t doped}

Now, we are ready to look at the learning protocol for the second passive layer $U_{O_{2,\bs\alpha^{(\ell)}}}$ in the passive-separable state showed in Eq. (\ref{eqs186}). First, we can apply the shadow tomography protocol for generalized passive-separable states \cite{zhao2025complexity}. Then, for a fixed trace-distance error $\epsilon_{\rm ps}'$ and a failure probability $\delta_{\rm ps}$, with 
\begin{align}
\textbf{poly}(m,E_{\rm II},\epsilon_{\rm ps}',\log(1/\delta_{\rm ps}))+ \textbf{exp}(\kappa t)
\end{align}
rounds of heterodyne measurement, we are able to reconstruct a state: 
\begin{align}\label{eqs581}
|\widetilde \Psi_{\rm II}\>=U_{\widetilde O_{2,\bs\alpha^{(\ell)},\bs\alpha^{(h)}}}\left(|\widetilde \chi_{(x)}\>\otimes |\underline{\widetilde {\bs\alpha}_{(m-x)}}\>\right)\simeq U_{O_{2,\bs\alpha^{(\ell)}}}\left(U^\dag_{S_{(\kappa t),\bs\alpha^{(\ell)}}}U_{(\kappa t)}\otimes D_{(m-\kappa t)}\right)U_O|\underline{\bs \alpha^{(h)}}\>
\end{align}
where $|\widetilde \chi_{(x)}\>$ denotes a $x$-mode non-coherent state with $x\le \kappa t$, $|\underline{\widetilde {\bs\alpha}_{(m-x)}}\>$ is a $(m-x)$-mode coherent state. Similar to Eq. (\ref{eqs67}), we can reconstruct the second passive-Gaussian layer with an error 
\begin{align}
\left\|
\mathcal U_{O_{2,\bs\alpha^{(\ell)}}}
-
\mathcal U_{\widetilde O_{2,\bs\alpha^{(\ell)}}}
\right\|_{\diamond}^{mE_{\rm II}}\le \epsilon_{\rm ps}.
\end{align}
Here, the relation between the error $\epsilon_{\rm ps}$ and $\epsilon_{\rm ps}'$ is shown in \cite{zhao2025complexity}. 

Without loss of generality, we denote the true passive-Gaussian unitary that the unitary $U_{\widetilde O_{2,\bs\alpha^{(\ell)},\bs\alpha^{(h)}}}$ in Eq. (\ref{prop:coherent-output-block-structure}) converges into as $U_{O_{2,\bs\alpha^{(\ell)},\bs\alpha^{(h)}}}:=U_{O_{2,\bs\alpha^{(\ell)}}}\left(U^\dag_{O_{(\kappa t)}}\otimes I_{(m-\kappa t)}\right)$, where $U_{O_{(\kappa t)}}$ denotes the passive-Gaussian unitary that transforms the output state of $U^\dag_{S_{(\kappa t),\bs\alpha^{(\ell)}}}U_{(\kappa t)}$ into an $x$-mode incoherent state and $(\kappa t-x)$-mode coherent states. Given Corollary \ref{cor:block-diagonal-gauge-no-C}, we have the following relation:
\begin{align}
U_{\widetilde O_{2,\bs\alpha^{(\ell)},\bs\alpha^{(h)}}}&\simeq U_{O_{2,\bs\alpha^{(\ell)},\bs\alpha^{(h)}}}\left[U_{O_{x}}\otimes U_{O_{(m-x)}}\right].
\end{align}
Furthermore, by Corollary \ref{cor:random-coherent-block-factorization}, we have the following relation with probability one, when replacing one input coherent state into other coherent states: 
\begin{align}
U_{O_{(\kappa t)}}U^\dag_{S_{(\kappa t),\bs\alpha^{(\ell)}}}U_{(\kappa t)}&= (U_{(x)}\otimes D_{(\kappa t -x)}(\bm\beta))U_{O'_{(\kappa t)}},
\end{align}
where $U_{(x)}$ is an $x$-mode unitary, $D_{(\kappa t -x)}(\bm\beta)$ denotes a $(\kappa t-x)$-mode displacement operator, $U_{O'_{(\kappa t)}}$ refers to a passive-Gaussian unitary. 

In the next steps, we always assume that the experimenter implements a counter-rotation over the heterodyne measurement results. The resulting state will be: 
\begin{align}
|\widetilde \Psi_{\rm III}\>&=U_{\widetilde O_{2,\bs\alpha^{(\ell)},\bs\alpha^{(h)}}}^\dag U_{O_{2,\bs\alpha^{(\ell)}}}\left(U^\dag_{S_{(\kappa t),\bs\alpha^{(\ell)}}}U_{(\kappa t)}\otimes D_{(m-\kappa t)}\right)U_O|\underline{\bs \alpha^{(k)}}\>\\
&=U_{\widetilde O_{2,\bs\alpha^{(\ell)},\bs\alpha^{(h)}}}^\dag U_{O_{2,\bs\alpha^{(\ell)}}}\left(U^\dag_{O_{(\kappa t)}}\otimes I_{(m-\kappa t)}\right)\left(U_{O_{(\kappa t)}}U^\dag_{S_{(\kappa t),\bs\alpha^{(\ell)}}}U_{(\kappa t)}\otimes D_{(m-\kappa t)}\right)U_O|\underline{\bs \alpha^{(k)}}\>\\
&\simeq \left[U_{O_{x}}^\dag \otimes U_{O_{(m-x)}}^\dag \right]\left(U_{O_{(\kappa t)}}U^\dag_{S_{(\kappa t),\bs\alpha^{(\ell)}}}U_{(\kappa t)}\otimes D_{(m-\kappa t)}\right)U_O|\underline{\bs \alpha^{(k)}}\>\\
&=  \left[ U_{(x)}^\star \otimes D_{(m -x)}^\star\right]U_{O^\star}|\underline{\bs \alpha^{(k)}}\>,\label{eqs589}
\end{align}
where $U_{(x)}^\star$ denotes an $x$-mode unitary, $D_{(m -x)}^\star$ denotes an $(m -x)$-mode displacement operator, $U_{O^\star}$ refers to an $m$ mode passive-Gaussian unitary.

\subsection{Learning of the remaining layers}

\subsubsection{Identification of output mode types}\label{supp:Identification of output mode types_doped}

In \ref{supp:learning the second layer of t doped}, we show that a shadow tomography protocol could reconstruct a passive-Gaussian unitary. By classical counter-rotation with this unitary, the learning task reduces to learning
\begin{align}\label{U_rest_t_doped}
U_{\rm t,rest}=&\left[ U_{(x)}^\star \otimes D_{(m -x)}^\star\right]U_{O^\star},
\end{align}
where $U_{(x)}^\star$ is an $x$-mode unitary, $D_{(m -x)}^\star$ denotes an $(m-x)$-mode displacement operator, $U_{O^\star}$ refers to an $m$-mode passive-Gaussian unitary. 

In the meantime, the output state of $U_{\rm t,rest}$ (see Eq.~(\ref{eqs589})) stably factorizes into an \(x\)-mode incoherent state and an \((m-x)\)-mode coherent state, due to Corollary \ref{cor:random-coherent-block-factorization}. Now, the next task becomes to identify which output mode corresponds to \(U_{(x)}^\star\) and which does not. Here, we propose to  identify the output modes with the following steps:

\begin{enumerate}
\item[(RE1')] \emph{Moment estimation.} Probe $U_{\rm t,rest}$ with a coherent state $|\underline{\bs\alpha^{(k)}}\>$. Perform heterodyne measurements on the output state to estimate its displacement vector \(\bs\xi_j\) and covariance matrix \( V_j\) of $j$-th mode with $j=1,\cdots,m$, yielding the corresponding estimates \(\widetilde{\bs\xi}_j\) and \(\widetilde{ V}_j\), respectively.
\item[(RE2')] \emph{Identification of coherent modes.} \textbf{If}, for the reduced state $\rho_j$ at the $j$-th mode, the estimated covariance-matrix entry satisfies
\begin{align}
\|V_j-V_{\rm vac}\|_{\max}
&:=
\max_{i,j\in\{q,p\}}
\left|
(V_j)_{ij}
-
\delta_{ij}
\right|
\leq
\epsilon_{\rm cov},
\end{align}
where $V_{\rm vac}$ denotes the covariance matrix of the vacuum state. Then, we have 
\begin{align}
\frac12
\min_{\alpha\in\mathbb C}
\left\|\rho_j-|\underline{\alpha}\rangle\langle\underline{\alpha}|
\right\|_1
&\leq
\frac12
\left\|
\rho_j-D(\bs\xi_j)|0\rangle\langle0|D^\dag (\bs\xi_j)
\right\|_1\\
&\le \sqrt{1-\<0|D^\dag (\bs\xi_j)\rho_j D (\bs\xi_j)|0\>}\\
&\le \sqrt{\Tr\left[a^\dag a\cdot D^\dag (\bs\xi_j)\rho_j D (\bs\xi_j)\right]}\\
&=
\sqrt{\frac{(V_j)_{qq}+(V_j)_{pp}-2}{4}}
\\
&\leq
\sqrt{\frac{
\left(1+\epsilon_{\rm cov}\right)
+
\left(1+\epsilon_{\rm cov}\right)
-
2
}{4}}
\\
&\leq   \sqrt{\frac{\epsilon_{\rm cov}} 2},\label{eqs593}
\end{align}
where $D(\bs\xi)$ represents the displacement operator associated with a displacement vector $\bs\xi$, the second inequality follows from the Fuchs–van de Graaf inequality, the third inequality is obtained using the relation $I-|0\>\<0|\le a^\dag a$. 

Therefore, in this case, we classify the $j$-th output mode as the mode that stably output coherent states, i.e., $j\in \chi_{\rm coh}$. \textbf{Otherwise}, we classify the $j$-th mode as belonging to $[m]\setminus\chi_{\rm coh}$.

To compute the sample complexity for this step, we can directly apply Theorem S53 of \cite{mele2025learning}. In particular, we can estimate $\widetilde V$ then satisfies
$\|\widetilde V_j-V_j\|_{\max}\le \|\widetilde V_j-V_j\|_{\infty}\leq \epsilon_{\rm cov}/2$; hence, whenever
$\|\widetilde V_j-I_2\|_{\max}\leq \epsilon_{\rm cov}/2$, one certifies
\begin{align}
\|V_j-V_{\rm vac}\|_{\max}
=
\|V_j-I_2\|_{\max}
\leq
\|\widetilde V_j-I_2\|_{\max}
+
\|\widetilde V_j-V_j\|_{\max}
\leq
\epsilon_{\rm cov}
\end{align}
with probability at least $1-\delta_{\rm cov}$, where $V_{\rm vac}$ denotes the covariance matrix of the vacuum state. This step requires 
\begin{align}\label{eqs595}
M_{\rm cov}
=
4\left\lceil
68\log\!\left(\frac{10}{\delta_{\rm cov}}\right)
\frac{800(8E_{\rm II}^2+3)}{\epsilon_{\rm cov}^2}
\right\rceil ,
\end{align}
number of state copies, where $E_{\rm II}$ is defined by the moment constraint $\sqrt{\operatorname{Tr}[\rho (a^\dag a )^2]}\leq E_{\rm II}$. 
\end{enumerate}

Note that Steps (RE1') and (RE2') can only identify the output modes of $U_{(x)}^\star$ subject to a finite error. Different probe states will lead to different errors. Nevertheless, we propose to repeat Steps (RE1') and (RE2') throughout the remaining protocol. Given that each iteration requires estimating only \(\mathcal O(1)\) parameters, it does not influence the efficiency of the overall protocol. 

Another issue is that the set of identified coherent-state output modes may be larger than the true set of output modes that stably produce coherent states. In Sec.~\ref{supp:Learning of the remaining Gaussian unitary_doped}, we will show that this does not affect the correctness of the remaining protocol.

\subsubsection{Learning of the local unitary that might not preserve coherent states}\label{supp:Learning of the local unitary that might not preserve coherent states_doped}

Let us look at the learning protocol for the unknown $x$-mode unitary $U^\star_{(x)}$ in Eq. (\ref{U_rest_t_doped}). Due to the existence of the first passive-Gaussian layer $U_{O^\star}$, the input coherent state will be unknown. Here, we introduce a protocol that can partially reconstruct a multimode unitary with unknown coherent probe states.  In simple terms, the protocol probes the unknown unitary with different unknown input states, reconstructs the corresponding output states via quantum tomography, computes their pairwise overlaps to infer the pairwise overlaps of the input coherent amplitudes, reconstructs the input probe set from the overlaps, and finally learns the unitary from the input--output relation.

Before showing the main theorem, let us look at some useful lemmas. 

\begin{lemma}[Multimode coherent overlaps and Bargmann phases]
\label{lem:xmode-overlap-bargmann}
For \(\bs \alpha,\bs \beta\in\mathbb C^x\), the \(x\)-mode coherent states satisfy $\langle \underline{\bs \alpha}|\underline{\bs \beta}\rangle
  =
  \exp\left(
-\frac12\|\bs \alpha\|_2^2
-\frac12\|\bs \beta\|_2^2
+
\bs \alpha^\dagger\bs \beta
  \right)$. Consequently, we have 
\begin{align}
\frac 1 2 \left\|
  |\underline{\bs \alpha}\rangle\langle\underline{\bs \alpha}|-
  |\underline{\bs \beta}\rangle\langle\underline{\bs \beta}|
\right\|_1
  =
  \sqrt{1-e^{-\|\bs \alpha-\bs \beta\|_2^2}}
  \le
  \|\bs \alpha-\bs \beta\|_2 .
\end{align}
Moreover, for the third-order Bargmann invariant $B(\bs \alpha,\bs \beta,\bs \gamma)
  =
  \langle \underline{\bs \alpha}|\underline{\bs \beta}\rangle
  \langle \underline{\bs \beta}|\underline{\bs \gamma}\rangle
  \langle \underline{\bs \gamma}|\underline{\bs \alpha}\rangle $, we have 
\begin{align}
\arg B(\bs \alpha,\bs \beta,\bs \gamma)
  =
  \operatorname{Im}
  \left(
(\bs \alpha-\bs \gamma)^\dagger(\bs \beta-\bs \gamma)
  \right)
  \quad
  \operatorname{mod} 2\pi .
\end{align}
\end{lemma}

\begin{proof}
The \(x\)-mode coherent state has the Fock expansion $|\underline{\bs \alpha}\rangle
  =
  e^{-\|\alpha\|_2^2/2}
  \sum_{\boldsymbol n\in\mathbb N^x}
  \frac{\alpha^{\boldsymbol n}}{\sqrt{\boldsymbol n!}}
  |\boldsymbol n\rangle$, where
$\boldsymbol n!
  =
  n_1!\cdots n_x!$ and $\alpha^{\boldsymbol n}
  =
  \alpha_1^{n_1}\cdots\alpha_x^{n_x}$. Therefore, we have 
\begin{align}
\langle\underline{\bs \alpha}|\underline{\bs \beta}\rangle
  &=
  e^{-(\|\alpha\|_2^2+\|\beta\|_2^2)/2}
  \prod_{\ell=1}^x
  \left(
\sum_{n_\ell=0}^{\infty}
\frac{(\overline{\alpha_\ell}\beta_\ell)^{n_\ell}}{n_\ell!}
  \right) \\
  &=
  \exp\left(
-\frac12\|\bs \alpha\|_2^2
-\frac12\|\bs \beta\|_2^2
+
\bs \alpha^\dagger\bs \beta
  \right).
\end{align}
Taking the squared modulus gives $|\langle\underline{\bs \alpha}|\underline{\bs \beta}\rangle|^2=
  \exp\left(-\|\bs \alpha-\bs \beta\|_2^2\right)$. The trace-distance formula follows from the pure-state identity $\frac 1 2 
\left\|
  |\psi\rangle\langle\psi|-
  |\phi\rangle\langle\phi|
\right\|_1
  =
  \sqrt{1-|\langle\psi|\phi\rangle|^2}$.

For the Bargmann phase, the real exponential prefactors in the three coherent
overlaps do not contribute to the argument. Hence we have 
\begin{align}
\arg B(\bs \alpha,\bs \beta,\bs \gamma)
  &=
  \operatorname{Im}
  \left(
\bs \alpha^\dagger\bs \beta
+
\bs \beta^\dagger\bs \gamma
+
\bs \gamma^\dagger\bs \alpha
  \right)   \\
  &=
  \operatorname{Im}
  \left(
(\bs \alpha-\bs \gamma)^\dagger(\bs \beta-\bs \gamma)
  \right)
  \quad
  \operatorname{mod} 2\pi .
\end{align}
This proves the lemma.
\end{proof}

The previous lemma shows that distances alone are not enough in more than one
mode.  The next lemma is the exact gauge-fixing statement.  

\begin{lemma}[Exact \(x\)-mode coherent gauge reconstruction]
\label{lem:xmode-exact-gauge}
Let us denote an $x$-mode pure state as $|\psi_k\rangle
  =
  W_\star|\underline{\bs \alpha_k}\rangle$, $\rho_k
  =
  |\psi_k\rangle\langle\psi_k|$ for $1\le k\le d$ and
\(\alpha_k\in\mathbb C^x\), where \(W_\star\) is an \(x\)-mode unitary.  Let us set the differences as $\bs v_k
  =
  \bs \alpha_k-\bs \alpha_1$ for $1\le k\le d$. Assume that \(v_2,\ldots,v_d\) span \(\mathbb C^x\).  Also assume
the non-aliasing condition $\operatorname{Im}(\bs v_i^\dagger \bs v_j)
  \in
  (-\pi+\zeta,\pi-\zeta)
  \,
  \text{for all } i,j\ge 2$ for some \(\zeta>0\). The reconstruction in this lemma uses not only the pairwise quantities
\(\operatorname{Tr}(\rho_i\rho_j)\), but also the Bargmann triple products $B_{ij1}:=\operatorname{Tr}(\rho_i\rho_j\rho_1)$. Without these triple-product phases, the amplitudes are determined only up to
\(\mathbb R^{2x}\rtimes O(2x)\), and the resulting gauge need not be physically
implementable.

Then the density matrices \(\rho_1,\ldots,\rho_d\) determine a set of amplitudes $\bs \beta_1,\ldots,\bs \beta_d\in\mathbb C^x$ such that $\bs \beta_1=0$ and that there exists a unitary matrix \(R\in {\rm U}(x)\) satisfying $\bs \beta_k
  =
  R(\bs \alpha_k-\bs \alpha_1)$ for $1\le k\le d$. Equivalently, there exists a physical coherent gauge $G
  =
  D(\bs \gamma)\Gamma(R)$ with $D(\bs \gamma)$ being a displacement operator, $\Gamma(R)$ being a passive-Gaussian unitary and $\bs \gamma
  =
  -R\bs \alpha_1$, such that $G|\underline{\bs \alpha_k}\rangle
  =
  e^{i\chi_k}|\underline{\bs \beta_k}\rangle$ for suitable phases \(\chi_k\).  If we define $W_g
  =
  W_\star G^\dagger$, then, after choosing the output representatives $|\phi_k\rangle
  =
  e^{-i\chi_k}|\psi_k\rangle$, one has the exact gauge-fixed relation
\begin{align}
|\phi_k\rangle
  =
  W_g|\underline{\bs \beta_k}\rangle,
  \qquad
1\le k\le d.
\end{align}
\end{lemma}

\begin{proof}
First, pairwise transition probabilities are computable from the density
matrices:
\begin{align}
\operatorname{Tr}(\rho_i\rho_j)
  =
  |\langle\psi_i|\psi_j\rangle|^2
  =
  |\langle\underline{\bs \alpha_i}|\underline{\bs \alpha_j}\rangle|^2 .
\end{align}
By Lemma~\ref{lem:xmode-overlap-bargmann}, we have 
\begin{align}
\|\bs \alpha_i-\bs \alpha_j\|_2^2
  =
  -\log\operatorname{Tr}(\rho_i\rho_j).
\end{align}
Thus the real Euclidean distance matrix of the point set in
\(\mathbb C^x\simeq\mathbb R^{2x}\) is known.  From distances alone, classical
Euclidean distance geometry reconstructs the set only up to $\mathbb R^{2x}\rtimes {\rm O}(2x)$. This is the full metric gauge. To reduce the ambiguity, we use the third-order Bargmann invariants $B_{ij1}
  =
  \operatorname{Tr}(\rho_i\rho_j\rho_1)
  =
  \langle\psi_i|\psi_j\rangle
  \langle\psi_j|\psi_1\rangle
  \langle\psi_1|\psi_i\rangle$. Because \(W_\star\) is unitary, we have 
\begin{align}
B_{ij1}
  =
  \langle\underline{\bs \alpha_i}|\underline{\bs \alpha_j}\rangle
  \langle\underline{\bs \alpha_j}|\underline{\bs \alpha_1}\rangle
  \langle\underline{\bs \alpha_1}|\underline{\bs \alpha_i}\rangle .
\end{align}
Lemma~\ref{lem:xmode-overlap-bargmann} gives $\arg B_{ij1}
  =
  \operatorname{Im}
  \left(
(\bs \alpha_i-\bs \alpha_1)^\dagger(\bs \alpha_j-\bs \alpha_1)
  \right)
  =
  \operatorname{Im}(\bs v_i^\dagger \bs v_j)
  \,
  \operatorname{mod} 2\pi$. The non-aliasing assumption chooses a unique branch, so the actual imaginary
part is recovered. On the other hand, the real part of \(\bs v_i^\dagger \bs v_j\) is recovered from distances:
\begin{align}
\operatorname{Re}(\bs v_i^\dagger \bs v_j)
  =
  \frac12
  \left(
\|\bs v_i\|_2^2+\|\bs v_j\|_2^2-\|\bs v_i-\bs v_j\|_2^2
  \right).
\end{align}
Therefore the Hermitian Gram matrix $H
  =
  (H_{ij})_{2\le i,j\le d}$ with $H_{ij}
  =
  \bs v_i^\dagger \bs v_j$ is exactly determined. This is the point at which the non-physical reflection branch is removed.
If one used distances only, replacing \(H\) by \(\overline H\) would give the
conjugate realization \(\bs v_i\mapsto R\overline{\bs v_i}\), which is the
multi-mode analogue of the one-mode reflection.  That branch reverses all
Bargmann phases.  Since the phases of \(B_{ij1}\) determine
\(\operatorname{Im}(\bs v_i^\dagger\bs v_j)\), the conjugate branch is excluded.
Hence the remaining linear freedom is complex-linear unitary, \(R\in U(x)\),
rather than a general element of \(O(2x)\). Since \(\bs v_2,\ldots,\bs v_d\) span \(\mathbb C^x\), the matrix \(H\) has rank \(x\). Let us choose any factorization $H
  =
  Z^\dagger Z$ for $Z\in\mathbb C^{x\times(d-1)}$. 
  
Let \(\bs \beta_i\) be the column of \(Z\) corresponding to \(i\ge 2\), and set
\(\bs \beta_1=0\).  Then we have 
\begin{align}
\bs \beta_i^\dagger\bs \beta_j
  =
  \bs v_i^\dagger \bs v_j
  \qquad
  \text{for all } i,j\ge 2.
\end{align}
Since both \(\{\bs \beta_i\}_{i\ge2}\) and \(\{\bs v_i\}_{i\ge2}\) span
\(\mathbb C^x\) and have the same Hermitian Gram matrix, there exists
\(R\in {\rm U}(x)\) such that
\begin{align}
\bs \beta_i
  =
  R\bs v_i
  =
  R(\bs \alpha_i-\bs \alpha_1).
\end{align}

Finally, with \(\bs \gamma=-R\bs \alpha_1\), the Gaussian unitary
\begin{align}
G
  =
  D(\bs \gamma)\Gamma(R)
\end{align}
acts on coherent states as
\begin{align}
G|\underline{\bs \alpha_k}\rangle
  =
  e^{i\chi_k}|\underline{R\bs \alpha_k+\bs \gamma}\rangle
  =
  e^{i\chi_k}|\underline{R(\bs \alpha_k-\bs \alpha_1)}\rangle
  =
  e^{i\chi_k}|\underline{\bs \beta_k}\rangle .
\end{align}
Thus we have 
\begin{align}
W_g|\underline{\bs \beta_k}\rangle
  =
  W_\star G^\dagger|\underline{\bs \beta_k}\rangle
  =
  e^{-i\chi_k}W_\star|\underline{\bs \alpha_k}\rangle
  =
  e^{-i\chi_k}|\psi_k\rangle
  =
  |\phi_k\rangle .
\end{align}
This proves the claim.
\end{proof}

The next
lemma states the corresponding stability statement.  

\begin{lemma}[Stable recovery of the gauge-fixed amplitudes]
\label{lem:xmode-stable-gauge}
Keep the notation and assumptions of Lemma~\ref{lem:xmode-exact-gauge}.  Let us denote  $\widetilde\rho_k
  =
|\widetilde\psi_k\rangle\langle\widetilde\psi_k|$ be tomographic estimates satisfying $\frac12\|\widetilde\rho_k-\rho_k\|_1
  \le
  \eta$ for $1\le k\le d$. We define the distances $d_{ij}
  =
  \|\bs \alpha_i-\bs \alpha_j\|_2$, $D_\star
  =
  \max_{i<j}d_{ij}$, $d_{\min}
  =
  \min_{i<j}d_{ij}$. Assume \(d_{\min}>0\) and $4\eta
  <
  \min\left\{
e^{-D_\star^2},
1-e^{-d_{\min}^2}
  \right\}$. Let us denote $\widehat p_{ij}
  =
  \operatorname{Tr}(\widetilde\rho_i\widetilde\rho_j)$ and $\widehat d_{ij}
  =
  \sqrt{-\log \widehat p_{ij}}$. Then we have 
\begin{align}
|\widehat d_{ij}-d_{ij}|
  \le
  \Delta_{\rm dist}(\eta)
\end{align}
for all \(i<j\), where
\begin{align}
\Delta_{\rm dist}(\eta)
  :=
  \max_{i<j}
  \max
  \left\{
  d_{ij}
  -
  \sqrt{-\log(e^{-d_{ij}^2}+4\eta)},
  \sqrt{-\log(e^{-d_{ij}^2}-4\eta)}
  -
  d_{ij}
  \right\}.
\end{align}

Furthermore, suppose the condition $12\eta
  <
  e^{-3D_\star^2/2}$ and the branch margin is not crossed: $12e^{3D_\star^2/2}\eta
  <
  \frac{\zeta}{2}$. For \(2\le i<j\le d\), define $\operatorname{Re}\widehat H_{ij}
=
\frac12
\left(
\widehat d_{i1}^2+\widehat d_{j1}^2-\widehat d_{ij}^2
\right)$ and $\operatorname{Im}\widehat H_{ij}
=
\operatorname{unwrap}_{(-\pi+\zeta,\pi-\zeta)}
\left(\arg \widetilde B_{ij1}\right)$. For the diagonal, set $\widehat H_{ii}:=\widehat d_{i1}^2$, and for \(j<i\), set
$\widehat H_{ji}:=\overline{\widehat H_{ij}}$. Then the Hermitian Gram matrix \(H=(\bs v_i^\dagger \bs v_j)_{2\le i,j\le d}\) can be
estimated by a matrix \(\widehat H\) satisfying
\begin{align}
\max_{i,j\ge2}|\widehat H_{ij}-H_{ij}|
  \le
  \Delta_H(\eta),
\end{align}
where
\begin{align}
\Delta_H(\eta)
  :=
  \frac32
  \left(2D_\star+\Delta_{\rm dist}(\eta)\right)
  \Delta_{\rm dist}(\eta)
  +
  12e^{3D_\star^2/2}\eta .
\end{align}

Finally, fix an index set \(J\subset\{2,\ldots,d\}\) of size \(x\) such that
\(H_{J,J}\) is positive definite, and use the corresponding Cholesky chart to
factor nearby Gram matrices.  Then there exist constants $L_{\rm geom}^{(x)}<\infty$ and $\eta_0>0$, depending only on the exact probe geometry and the chosen chart, such that for
all \(0<\eta\le \eta_0\),
\begin{align}
\max_{1\le k\le d}\|\bs{\widehat\beta}_k-\bs \beta_k\|_2
  \le
  L_{\rm geom}^{(x)}\Delta_H(\eta).
\end{align}
\end{lemma}

\begin{proof}
For the overlap estimates, we have 
\begin{align}
\left|
\operatorname{Tr}(\widetilde\rho_i\widetilde\rho_j)
-
\operatorname{Tr}(\rho_i\rho_j)
\right|
  &\le
  \left|
  \operatorname{Tr}\bigl((\widetilde\rho_i-\rho_i)\widetilde\rho_j\bigr)
  \right|
  +
  \left|
  \operatorname{Tr}\bigl(\rho_i(\widetilde\rho_j-\rho_j)\bigr)
  \right|\\
  &\le
  \|\widetilde\rho_i-\rho_i\|_1
  +
  \|\widetilde\rho_j-\rho_j\|_1  \\
  &\le
  4\eta ,\label{eqs646}
\end{align}
where we applied the triangle inequality and the H\"older's inequality. Since we have $\operatorname{Tr}(\rho_i\rho_j)
  =
  e^{-d_{ij}^2}$, let us write $p_{ij}
  :=
  e^{-d_{ij}^2}$ and $f(p)
  :=
  \sqrt{-\log p}$. Then we have \(d_{ij}=f(p_{ij})\).  From Eq.~\eqref{eqs646}, we have
\begin{align}
p_{ij}-4\eta
  \le
  \widehat p_{ij}
  \le
  p_{ij}+4\eta .
\end{align}
The assumption $4\eta
  <
  \min\left\{
e^{-D_\star^2},
1-e^{-d_{\min}^2}
  \right\}$ implies, for every \(i<j\), $0<p_{ij}-4\eta
  \,\text{and}\,
p_{ij}+4\eta<1$. Hence all logarithms below are well defined.  Since \(f(p)=\sqrt{-\log p}\) is
strictly decreasing on \((0,1)\), we obtain
\begin{align}
f(p_{ij}+4\eta)
  \le
  f(\widehat p_{ij})
  \le
  f(p_{ij}-4\eta).
\end{align}
Equivalently, we have 
\begin{align}
\sqrt{-\log(e^{-d_{ij}^2}+4\eta)}
  \le
  \widehat d_{ij}
  \le
  \sqrt{-\log(e^{-d_{ij}^2}-4\eta)} .
\end{align}
Since \(d_{ij}=f(p_{ij})\), this gives
\begin{align}
|\widehat d_{ij}-d_{ij}|
  \le
  \max
  \left\{
  d_{ij}
  -
  \sqrt{-\log(e^{-d_{ij}^2}+4\eta)},
  \sqrt{-\log(e^{-d_{ij}^2}-4\eta)}
  -
  d_{ij}
  \right\}.
\end{align}
Taking the maximum over all \(i<j\) proves
\begin{align}
|\widehat d_{ij}-d_{ij}|
  \le
  \Delta_{\rm dist}(\eta).
\end{align}

For the triple products, let us define $\widetilde B_{ij1}
  =
  \operatorname{Tr}(\widetilde\rho_i\widetilde\rho_j\widetilde\rho_1)$ and $B_{ij1}
  =
  \operatorname{Tr}(\rho_i\rho_j\rho_1)$. Similar to Eq. (\ref{eqs646}), by inserting and subtracting one factor at a time, we have 
\begin{align}
|\widetilde B_{ij1}-B_{ij1}|
  \le
  6\eta .
\end{align}
Moreover, we have 
\begin{align}
|B_{ij1}|
  =
  |\langle\underline{\bs \alpha_i}|\underline{\bs \alpha_j}\rangle|
  |\langle\underline{\bs \alpha_j}|\underline{\bs \alpha_1}\rangle|
  |\langle\underline{\bs \alpha_1}|\underline{\bs \alpha_i}\rangle|
  \ge
  e^{-3D_\star^2/2}.
\end{align}
We use the elementary Lipschitz estimate for the argument map away from the
origin:
\begin{align}
|\arg z-\arg w|
  \le
  \frac{|z-w|}{\min_{t\in[0,1]} |w+t(z-w)|}.
\end{align}
In particular, if \(|z-w|\le |w|/2\), then
\begin{align}
|\arg z-\arg w|
  \le
  \frac{2|z-w|}{|w|}.
\end{align}
This gives
\begin{align}
|\arg \widetilde B_{ij1}-\arg B_{ij1}|
  \le
  12e^{3D_\star^2/2}\eta .
\end{align}
Here, the condition \(12\eta<e^{-3D_\star^2/2}\) allows us to apply the phase
perturbation bound, while the branch-margin condition $12e^{3D_\star^2/2}\eta
  <
  \frac{\zeta}{2}$ ensures that the same unwrapping branch is used as in the exact reconstruction.

The real part of \(H_{ij}\) is reconstructed from three squared distances:
\begin{align}
\operatorname{Re}H_{ij}
  =
  \frac12
  \left(
d_{i1}^2+d_{j1}^2-d_{ij}^2
  \right).
\end{align}
From the already proved distance estimate, for every relevant pair \(a,b\) we have $|\widehat d_{ab}-d_{ab}|
  \le
  \Delta_{\rm dist}(\eta)$. Since we have \(d_{ab}\le D_\star\), this implies the following relation
\begin{align}
\widehat d_{ab}
  \le
  d_{ab}+\Delta_{\rm dist}(\eta)
  \le
  D_\star+\Delta_{\rm dist}(\eta) .
\end{align}
Therefore, we have 
\begin{align}
|\widehat d_{ab}^2-d_{ab}^2|
  &=
  |\widehat d_{ab}-d_{ab}|\,
  |\widehat d_{ab}+d_{ab}|  \\
  &\le
  \Delta_{\rm dist}(\eta)\,
  \left(
D_\star+\Delta_{\rm dist}(\eta)+D_\star
  \right)\\
  &=
  \left(
2D_\star+\Delta_{\rm dist}(\eta)
  \right)\Delta_{\rm dist}(\eta) .
\end{align}
Let us denote $A_{\rm dist}(\eta)
  :=
  \left(
2D_\star+\Delta_{\rm dist}(\eta)
  \right)
  \Delta_{\rm dist}(\eta)$. For \(i,j\ge2\), the reconstructed real part is $\operatorname{Re}\widehat H_{ij}
  =
  \frac12
  \left(
\widehat d_{i1}^2
+
\widehat d_{j1}^2
-
\widehat d_{ij}^2
  \right)$, with the convention \(\widehat d_{ii}=d_{ii}=0\). Hence, for \(i\neq j\), we have 
\begin{align}
\left|
  \operatorname{Re}\widehat H_{ij}
  -
  \operatorname{Re}H_{ij}
\right|
&\le
\frac12
\left(
  |\widehat d_{i1}^2-d_{i1}^2|
  +
  |\widehat d_{j1}^2-d_{j1}^2|
  +
  |\widehat d_{ij}^2-d_{ij}^2|
\right)\\
&\le
\frac32
A_{\rm dist}(\eta) \\
&=
\frac32
\left(
  2D_\star+\Delta_{\rm dist}(\eta)
\right)
\Delta_{\rm dist}(\eta).
\end{align}
When \(i=j\), we have $\operatorname{Re}H_{ii}
  =
  d_{i1}^2,$ and $\operatorname{Re}\widehat H_{ii}
  =
  \widehat d_{i1}^2$, so
\begin{align}
\left|
  \operatorname{Re}\widehat H_{ii}
  -
  \operatorname{Re}H_{ii}
\right|
  \le
  A_{\rm dist}(\eta)
  \le
  \frac32 A_{\rm dist}(\eta).
\end{align}
Thus the same bound holds uniformly for all \(i,j\ge2\):
\begin{align}
\left|
  \operatorname{Re}\widehat H_{ij}
  -
  \operatorname{Re}H_{ij}
\right|
  \le
  \frac32
  \left(
2D_\star+\Delta_{\rm dist}(\eta)
  \right)
  \Delta_{\rm dist}(\eta).
\end{align}
The imaginary part is reconstructed from the unwrapped Bargmann phase, whose
error is at most \(12e^{3D_\star^2/2}\eta\).  This proves the entrywise bound on
\(\widehat H-H\).

It remains to pass from the Gram matrix to coordinates.  Since
\(H_{J,J}\) is positive definite, the Cholesky factor of \(H_{J,J}\) is locally
a \(C^1\) function of \(H_{J,J}\).  Once the \(x\) basis columns have been fixed,
the remaining columns are recovered by solving a nonsingular linear system.
Thus the entire coordinate map $H
  \longmapsto
  (\bs \beta_1,\ldots,\bs \beta_d)$ is \(C^1\) in a neighborhood of the exact \(H\).  The mean value theorem gives a
finite local Lipschitz constant \(L_{\rm geom}^{(x)}\), proving the final claim.
\end{proof}

The next lemma is a direct \(x\)-mode photon-number cutoff statement.  

\begin{lemma}[Direct \(x\)-mode photon-number cutoff]
\label{lem:xmode-cutoff}
For an \(x\)-mode unitary \(U\), let us define
\begin{align}
\mathsf N_U^{(x)}(q)
  :=
  \sup_{\substack{
|\varphi\rangle\in\mathcal H_{\le q}^{(x)}\\
\|\varphi\|=1
  }}
  \langle\varphi|U^\dagger\hat n_x U|\varphi\rangle,
\end{align}
with $\mathcal H_{\le q}^{(x)}
  =
  \operatorname{span}
  \left\{
|\boldsymbol n\rangle:
|\boldsymbol n|\le q
  \right\}$. Let \(\Pi_{\le q}^{(x)}\) be the projector onto this subspace.  For a photon-number
budget \(N\), define the constrained diamond norm $\|\Phi\|_{\diamond}^{(x,N)}
  =
  \sup_{\rho_{AR}:\operatorname{Tr}(\hat n_x \rho_A)\le N}
  \left\|
(\Phi\otimes\operatorname{id}_R)(\rho_{AR})
  \right\|_1$. Fix \(0<\varepsilon<1\), and set $q
  =
  \left\lceil
\frac{256N}{\varepsilon^2}
  \right\rceil$. Assume the conditions $\mathsf N_U^{(x)}(q)
  \le
  N_{\rm dyn}$ and $r
  \ge
  \max
  \left\{
q,
\left\lceil
  \frac{256N_{\rm dyn}}{\varepsilon^2}
\right\rceil
  \right\}$.

Then there exists a unitary \(V_r\) on \(\mathcal H_{\le r}^{(x)}\) such that,
for every unitary \(V_{>r}\) on the orthogonal complement,
\begin{align}
\left\|
  U\,\boldsymbol{\cdot}\,U^\dagger
  -
  (V_r\oplus V_{>r})\,\boldsymbol{\cdot}\,(V_r\oplus V_{>r})^\dagger
\right\|_{\diamond}^{(x,N)}
  \le
  \frac{\varepsilon}{2}.
\end{align}
\end{lemma}

\begin{proof}
Let us denote the projectors $P_q
  =
  \Pi_{\le q}^{(x)}$ and $P_r
  =
  \Pi_{\le r}^{(x)}$. Let \(\rho_{AR}\) be any state satisfying $\operatorname{Tr}(\hat n_x \rho_A)
  \le
  N$ with $\hat n_x $ being the photon number operator of $x$ modes. 

First, we analyze the input tail.  Since every vector in
\(I-P_q\) has total photon number at least \(q+1\), Markov's inequality gives the relations
\begin{align}
\operatorname{Tr}\!\left[(I-P_q)\rho_A\right]
  \le
  \frac{\operatorname{Tr}(\hat n_x \rho_A)}{q+1}
  \le
  \frac{N}{q+1}
  \le
  \frac{\varepsilon^2}{256}.
\end{align}
Equivalently, we have  $\operatorname{Tr}\!\left[
  \bigl((I-P_q)\otimes I_R\bigr)\rho_{AR}
\right]
  \le
  \frac{\varepsilon^2}{256}$. We define the projected, generally subnormalized, state as follows
\begin{align}
\sigma_{AR}
  :=
  (P_q\otimes I_R)\rho_{AR}(P_q\otimes I_R).
\end{align}
By the gentle measurement lemma \cite{wilde2017qit}, we have 
\begin{align}
\|\rho_{AR}-\sigma_{AR}\|_1
  \le
  2\sqrt{\frac{\varepsilon^2}{256}}
  =
  \frac{\varepsilon}{8}.
\label{eq:xmode-input-gentle}
\end{align}

Next, we construct a finite-dimensional unitary approximation to \(U\) on the
input subspace \(\mathcal H_{\le q}^{(x)}\).  For every normalized vector
\(|\varphi\rangle\in\mathcal H_{\le q}^{(x)}\), the assumption
\(\mathsf N_U^{(x)}(q)\le N_{\rm dyn}\) gives
\begin{align}
\langle\varphi|U^\dagger \hat n_x  U|\varphi\rangle
  \le
  N_{\rm dyn}.
\end{align}
Since every vector in \(I-P_r\) has total photon number at least \(r+1\), another
application of Markov's inequality yields
\begin{align}
\|(I-P_r)U|\varphi\rangle\|_2^2
  \le
  \frac{
\langle\varphi|U^\dagger \hat n_x  U|\varphi\rangle
  }{r+1}
  \le
  \frac{N_{\rm dyn}}{r+1}
  \le
  \frac{\varepsilon^2}{256}.
\end{align}
Taking the supremum over all unit vectors in \(\mathcal H_{\le q}^{(x)}\), we
obtain
\begin{align}
\|(I-P_r)UP_q\|_{\infty}
  \le
  \frac{\varepsilon}{16}.
\label{eq:xmode-output-tail-op}
\end{align}

Now define $A
  :=
  P_rUP_q:
  \mathcal H_{\le q}^{(x)}
  \longrightarrow
  \mathcal H_{\le r}^{(x)}$. Then we have 
\begin{align}
A^\dagger A
  =
  P_qU^\dagger P_rUP_q
  =
  P_q
  -
  P_qU^\dagger(I-P_r)UP_q .
\end{align}
Hence, by Eq.~\eqref{eq:xmode-output-tail-op}, we have 
\begin{align}
\|A^\dagger A-P_q\|_{\infty}
  =
  \|P_qU^\dagger(I-P_r)UP_q\|_{\infty}
  =
  \|(I-P_r)UP_q\|_{\infty}^2
  \le
  \frac{\varepsilon^2}{256}.
\label{eq:xmode-AA-close}
\end{align}
In particular, \(A^\dagger A\) is strictly positive on
\(\mathcal H_{\le q}^{(x)}\), so \(A\) has a polar decomposition
\begin{align}
A
  =
  S(A^\dagger A)^{1/2},
\end{align}
where \(S:\mathcal H_{\le q}^{(x)}\to\mathcal H_{\le r}^{(x)}\) is an isometry. We now bound \(A-S\).  On \(\mathcal H_{\le q}^{(x)}\), the eigenvalues of
\(A^\dagger A\) lie in the region $\left[
  1-\frac{\varepsilon^2}{256},
  1
\right]$. Therefore the eigenvalues of \((A^\dagger A)^{1/2}\) lie in $\left[
  \sqrt{1-\frac{\varepsilon^2}{256}},
  1
\right]$, and so we have 
\begin{align}
\|(A^\dagger A)^{1/2}-P_q\|_{\infty}
  \le
  \frac{\varepsilon^2}{256}
  \le
  \frac{\varepsilon}{16},
\end{align}
where we used \(0<\varepsilon<1\).  Since we have $A-S
  =
  S\left((A^\dagger A)^{1/2}-P_q\right)$, we get
\begin{align}
\|A-S\|_{\infty}
  \le
  \frac{\varepsilon}{16}.
\label{eq:xmode-polar-error}
\end{align}

Because \(r\ge q\), the domain subspace
\(\mathcal H_{\le q}^{(x)}\) is contained in \(\mathcal H_{\le r}^{(x)}\).
The isometry \(S\) can therefore be extended to a unitary \(V_r\) on
\(\mathcal H_{\le r}^{(x)}\): choose an orthonormal basis of
\(\mathcal H_{\le r}^{(x)}\ominus\mathcal H_{\le q}^{(x)}\) and map it to an
orthonormal basis of
\(\mathcal H_{\le r}^{(x)}\ominus S\mathcal H_{\le q}^{(x)}\).  This defines a
unitary \(V_r\) satisfying $V_rP_q
  =
  SP_q$. Let \(V_{>r}\) be any unitary on the orthogonal complement of
\(\mathcal H_{\le r}^{(x)}\), and set $V
  =
  V_r\oplus V_{>r}$. Then we have 
\begin{align}
\|(U-V)P_q\|_{\infty}
  &=
  \|UP_q-SP_q\|_{\infty}  \\
  &\le
  \|(I-P_r)UP_q\|_{\infty}
  +
  \|P_rUP_q-SP_q\|_{\infty}  \\
  &=
  \|(I-P_r)UP_q\|_{\infty}
  +
  \|A-S\|_{\infty}   \\
  &\le
  \frac{\varepsilon}{16}
  +
  \frac{\varepsilon}{16}
  =
  \frac{\varepsilon}{8}.
\label{eq:xmode-UminusV-on-q}
\end{align}

We now compare the two unitary channels on the truncated state
\(\sigma_{AR}\).  Since \(\sigma_{AR}\) is supported on
\(\mathcal H_{\le q}^{(x)}\) on the system register, we may insert \(P_q\) next
to every occurrence of \(U-V\).  Let us denote $\Delta
  :=
  (U-V)P_q$. Then we have 
\begin{align}
&(U\otimes I_R)\sigma_{AR}(U^\dagger\otimes I_R)
-
(V\otimes I_R)\sigma_{AR}(V^\dagger\otimes I_R) \\
&\quad =
(\Delta\otimes I_R)\sigma_{AR}(U^\dagger\otimes I_R)
+
(V\otimes I_R)\sigma_{AR}(\Delta^\dagger\otimes I_R).
\end{align}
Taking the trace norm and using Hölder's inequality, we have 
\begin{align}
&\left\|
(U\otimes I_R)\sigma_{AR}(U^\dagger\otimes I_R)
-
(V\otimes I_R)\sigma_{AR}(V^\dagger\otimes I_R)
\right\|_1   \\
&\quad\le
\|\Delta\|_{\infty}\|\sigma_{AR}\|_1\|U^\dagger\|_{\infty}
+
\|V\|_{\infty}\|\sigma_{AR}\|_1\|\Delta^\dagger\|_{\infty}   \\
&\quad\le
2\|\Delta\|_{\infty} \\
&\quad\le
\frac{\varepsilon}{4},
\label{eq:xmode-truncated-channel-error}
\end{align}
because \(\|\sigma_{AR}\|_1=\operatorname{Tr}(\sigma_{AR})\le1\), and
\(\|U\|_{\infty}=\|V\|_{\infty}=1\).

Finally, we return from \(\sigma_{AR}\) to the original state \(\rho_{AR}\).
Using Eq.~\eqref{eq:xmode-input-gentle} and unitary invariance of the trace
norm,
\begin{align}
&\left\|
(U\otimes I_R)\rho_{AR}(U^\dagger\otimes I_R)
-
(V\otimes I_R)\rho_{AR}(V^\dagger\otimes I_R)
\right\|_1   \\
&\quad\le
\left\|
(U\otimes I_R)(\rho_{AR}-\sigma_{AR})(U^\dagger\otimes I_R)
\right\|_1   \nonumber \\
&\qquad+
\left\|
(U\otimes I_R)\sigma_{AR}(U^\dagger\otimes I_R)
-
(V\otimes I_R)\sigma_{AR}(V^\dagger\otimes I_R)
\right\|_1   \nonumber \\
&\qquad+
\left\|
(V\otimes I_R)(\sigma_{AR}-\rho_{AR})(V^\dagger\otimes I_R)
\right\|_1   \\
&\quad\le
\frac{\varepsilon}{8}
+
\frac{\varepsilon}{4}
+
\frac{\varepsilon}{8}
=
\frac{\varepsilon}{2}.
\end{align}
This bound holds for every reference system \(R\), every admissible input state
\(\rho_{AR}\), and every choice of the complement unitary \(V_{>r}\).  Taking
the supremum over all such \(\rho_{AR}\) proves
\begin{align}
\left\|
  U\,\boldsymbol{\cdot}\,U^\dagger
  -
  (V_r\oplus V_{>r})\,\boldsymbol{\cdot}\,
  (V_r\oplus V_{>r})^\dagger
\right\|_{\diamond}^{(x,N)}
  \le
  \frac{\varepsilon}{2}.
\end{align}
\end{proof}
We now introduce the finite coherent coefficient matrix only at the point where
it is needed.  

\begin{lemma}[Multimode coherent coefficient matrix and perturbation]
\label{lem:xmode-coherent-vandermonde}
For \(r\in\mathbb N\), let us denote $\mathcal I_r^{(x)}
  =
  \left\{
\boldsymbol n\in\mathbb N^x:
|\boldsymbol n|\le r
  \right\}$ and $d_x(r)
  =
  |\mathcal I_r^{(x)}|
  =
  \binom{x+r}{x}$. For \(\bs z\in\mathbb C^x\), define a function $c_r^{(x)}(\bs z)
  =
  \left(
e^{-\|\bs z\|_2^2/2}
\frac{\bs z^{\boldsymbol n}}{\sqrt{\boldsymbol n!}}
  \right)_{\boldsymbol n\in\mathcal I_r^{(x)}}
  \in
  \mathbb C^{d_x(r)}$. Let $d
  =
  d_x(r)$, and suppose that $\bs \beta_1,\ldots,\bs \beta_d\in\mathbb C^x$ are unisolvent for polynomials in \(x\) complex variables of total degree at
most \(r\).  

Then the matrix $C
  =
  \left[
c_r^{(x)}(\bs \beta_1)\ \cdots\ c_r^{(x)}(\bs \beta_d)
  \right]$ is invertible. Moreover, if we assume \(\|\bs z\|_2,\|\bs w\|_2\le R+1\), then we have  $\|c_r^{(x)}(\bs z)-c_r^{(x)}(\bs w)\|_2
  \le
  L_C^{(x)}\|\bs z-\bs w\|_2$, where one may take $L_C^{(x)}
  =
  \sqrt{2x}\,(\sqrt r+R+1)$. Consequently, if the conditions $\max_k\|\bs {\widehat\beta}_k-\bs \beta_k\|_2
  \le
  \Delta_{\bs \beta}$ and \(\|\bs \beta_k\|_2\le R\) are fulfilled, then, provided \(\Delta_{\bs \beta}\le1\), we have 
\begin{align}
\|\widehat C-C\|_{\infty}
  \le
  \sqrt d\,L_C^{(x)}\Delta_{\bs \beta} .
\end{align}
If we further assume $\sqrt d\,L_C^{(x)}\Delta_{\bs \beta}
  \le
  \frac{1}{2\kappa_C}$ and $\kappa_C
  =
  \|C^{-1}\|_{\infty}$, then \(\widehat C\) is invertible and $\|\widehat C^{-1}\|_{\infty}
  \le
  2\kappa_C$.
\end{lemma}

\begin{proof}
Let us first prove the invertibility of \(C\).  Define the polynomial evaluation matrix 
$V_{\boldsymbol n,k}
  =
  \frac{\bs \beta_k^{\boldsymbol n}}{\sqrt{\boldsymbol n!}},$ with $\boldsymbol n\in\mathcal I_r^{(x)}$ and $1\le k\le d$. Then we have 
\begin{align}
C
  =
  V
  \operatorname{diag}
  \left(
e^{-\|\bs \beta_1\|_2^2/2},
\ldots,
e^{-\|\bs \beta_d\|_2^2/2}
  \right).
\end{align}
The diagonal factor is invertible.  Hence it is enough to show that \(V\) is
invertible.

Suppose \(a=(a_{\boldsymbol n})_{\boldsymbol n\in\mathcal I_r^{(x)}}\) satisfies $a^\dagger V
  =
  0 $. Equivalently, for every \(1\le k\le d\), we have $0
  =
  \sum_{\boldsymbol n\in\mathcal I_r^{(x)}}
  \overline{a_{\boldsymbol n}}
  \frac{\bs \beta_k^{\boldsymbol n}}{\sqrt{\boldsymbol n!}} $. Define the holomorphic polynomial $p(\bs z)
  =
  \sum_{\boldsymbol n\in\mathcal I_r^{(x)}}
  \overline{a_{\boldsymbol n}}
  \frac{\bs z^{\boldsymbol n}}{\sqrt{\boldsymbol n!}}$. This polynomial has total degree at most \(r\), and the preceding equations say $p(\bs \beta_k)
  =
  0$ for $1\le k\le d$. By unisolvency, the only polynomial of total degree at most \(r\) vanishing at
all \(\bs \beta_k\)'s is the zero polynomial.  Therefore \(p=0\), hence
\(a_{\boldsymbol n}=0\) for every \(\boldsymbol n\).  Thus \(V\), and therefore
\(C\), is invertible.

We next prove the Lipschitz estimate.  For
\(\boldsymbol n\in\mathcal I_r^{(x)}\), write $f_{\boldsymbol n}(\bs z)
  =
  e^{-\|\bs z\|_2^2/2}
  \frac{\bs z^{\boldsymbol n}}{\sqrt{\boldsymbol n!}}$, so that $c_r^{(x)}(\bs z)
  =
  \left(f_{\boldsymbol n}(\bs z)\right)_{\boldsymbol n\in\mathcal I_r^{(x)}}$. Write $z_\ell
  =
  q_\ell+ip_\ell$ for $1\le \ell\le x$. We bound the real partial derivatives of the vector \(c_r^{(x)}(\bs z)\).

Let us fix \(\ell\).  Differentiating with respect to \(q_\ell\), we get
\begin{align}
\partial_{q_\ell} f_{\boldsymbol n}(\bs z)
  =
  e^{-\|\bs z\|_2^2/2}
  \left(
\frac{n_\ell \bs z^{\boldsymbol n-e_\ell}}{\sqrt{\boldsymbol n!}}
-
q_\ell
\frac{\bs z^{\boldsymbol n}}{\sqrt{\boldsymbol n!}}
  \right),
\end{align}
where the first term is interpreted as zero if \(n_\ell=0\).  Hence we have 
\begin{align}
\left\|
  \partial_{q_\ell} c_r^{(x)}(\bs z)
\right\|_2
  \le
  A_\ell(\bs z)+|q_\ell|\,
  \left\|c_r^{(x)}(\bs z)\right\|_2,
\end{align}
with $A_\ell(\bs z)^2
  =
  e^{-\|\bs z\|_2^2}
  \sum_{\substack{\boldsymbol n\in\mathcal I_r^{(x)}\\ n_\ell\ge1}}
  \frac{
n_\ell^2
|\bs z^{\boldsymbol n-e_\ell}|^2
  }{
\boldsymbol n!
  }$. Then, we can set \(\boldsymbol m=\boldsymbol n-e_\ell\).  We will have  $\frac{n_\ell^2}{\boldsymbol n!}
  =
  \frac{m_\ell+1}{\boldsymbol m!}$. Therefore, we have 
\begin{align}
A_\ell(\bs z)^2
  &=
  e^{-\|\bs z\|_2^2}
  \sum_{|\boldsymbol m|\le r-1}
  (m_\ell+1)
  \frac{|\bs z^{\boldsymbol m}|^2}{\boldsymbol m!}\\
  &\le
  r\,
  e^{-\|\bs z\|_2^2}
  \sum_{|\boldsymbol m|\le r-1}
  \frac{|\bs z^{\boldsymbol m}|^2}{\boldsymbol m!}\\
  &\le
  r .
\end{align}
Also, from the relation $\left\|c_r^{(x)}(\bs z)\right\|_2^2
  =
  e^{-\|\bs z\|_2^2}
  \sum_{|\boldsymbol n|\le r}
  \frac{|\bs z^{\boldsymbol n}|^2}{\boldsymbol n!}
  \le
  e^{-\|\bs z\|_2^2}
  \sum_{\boldsymbol n\in\mathbb N^x}
  \frac{|\bs z^{\boldsymbol n}|^2}{\boldsymbol n!}
  =
  1$, we have 
\begin{align}
\left\|
  \partial_{q_\ell} c_r^{(x)}(\bs z)
\right\|_2
  \le
  \sqrt r+|q_\ell|
  \le
  \sqrt r+\|\bs z\|_2 .
\end{align}

The derivative with respect to \(p_\ell\) is analogous:
\begin{align}
\partial_{p_\ell} f_{\boldsymbol n}(\bs z)
  =
  e^{-\|\bs z\|_2^2/2}
  \left(
i\frac{n_\ell \bs z^{\boldsymbol n-e_\ell}}{\sqrt{\boldsymbol n!}}
-
p_\ell
\frac{\bs z^{\boldsymbol n}}{\sqrt{\boldsymbol n!}}
  \right).
\end{align}
The same argument gives
\begin{align}
\left\|
  \partial_{p_\ell} c_r^{(x)}(\bs z)
\right\|_2
  \le
  \sqrt r+\|\bs z\|_2 .
\end{align}

Thus the real Jacobian of \(c_r^{(x)}\), viewed as a map
\(\mathbb R^{2x}\to\mathbb C^{d_x(r)}\), satisfies
\begin{align}
\left\|
  D c_r^{(x)}(\bs z)
\right\|_\infty
  \le
  \left(
\sum_{\ell=1}^x
\left\|
  \partial_{q_\ell} c_r^{(x)}(\bs z)
\right\|_2^2
+
\sum_{\ell=1}^x
\left\|
  \partial_{p_\ell} c_r^{(x)}(\bs z)
\right\|_2^2
  \right)^{1/2}
  \le
  \sqrt{2x}\,(\sqrt r+\|\bs z\|_2).
\end{align}
Now suppose \(\|\bs z\|_2,\|\bs w\|_2\le R+1\).  For \(0\le t\le1\), define $\bs z_t
  =
  \bs w+t(\bs z-\bs w)$. By convexity of the Euclidean norm, we have $\|\bs z_t\|_2
  \le
  R+1 $. Therefore, we have 
\begin{align}
c_r^{(x)}(\bs z)-c_r^{(x)}(\bs w)
  =
  \int_0^1
  D c_r^{(x)}(\bs z_t)[\bs z-\bs w]\,\d t,
\end{align}
and hence
\begin{align}
\|c_r^{(x)}(\bs z)-c_r^{(x)}(\bs w)\|_2
  &\le
  \int_0^1
  \left\|
D c_r^{(x)}(\bs z_t)
  \right\|_\infty
  \|\bs z-\bs w\|_2\,\d t  \\
  &\le
  \sqrt{2x}\,(\sqrt r+R+1)\|\bs z-\bs w\|_2 .
\end{align}
This proves the claimed Lipschitz estimate with $L_C^{(x)}
  =
  \sqrt{2x}\,(\sqrt r+R+1)$.

We now prove the matrix perturbation bound.  Assume the conditions $\max_k\|\bs {\widehat\beta}_k-\beta_k\|_2
  \le
  \Delta_{\bs \beta}$, $\|\bs \beta_k\|_2\le R$, and $\Delta_{\bs \beta}\le1 $, we have 
\begin{align}
\|\bs {\widehat\beta}_k\|_2
  \le
  \|\bs \beta_k\|_2+\|\bs {\widehat\beta}_k-\bs \beta_k\|_2
  \le
  R+1 .
\end{align}
Thus the Lipschitz estimate applies to each pair
\((\bs {\widehat\beta}_k,\bs \beta_k)\), giving
\begin{align}
\left\|
  c_r^{(x)}(\bs {\widehat\beta}_k)
  -
  c_r^{(x)}(\bs \beta_k)
\right\|_2
  \le
  L_C^{(x)}\Delta_{\bs \beta} .
\end{align}
Therefore, we have 
\begin{align}
\|\widehat C-C\|_\infty
  &\le
  \|\widehat C-C\|_F \\
  &=
  \left(
\sum_{k=1}^d
\left\|
  c_r^{(x)}(\bs {\widehat\beta}_k)
  -
  c_r^{(x)}(\bs \beta_k)
\right\|_2^2
  \right)^{1/2}   \\
  &\le
  \sqrt d\,L_C^{(x)}\Delta_{\bs \beta} .
\end{align}

Finally, assume the conditions  $\sqrt d\,L_C^{(x)}\Delta_{\bs \beta}
  \le
  \frac1{2\kappa_C}$ and $\kappa_C
  =
  \|C^{-1}\|_\infty $, we have 
\begin{align}
\left\|
  C^{-1}(\widehat C-C)
\right\|_\infty
  \le
  \|C^{-1}\|_\infty
  \|\widehat C-C\|_\infty
  \le
  \frac12 .
\end{align}
Given the conditions $\widehat C
  =
  C+
  (\widehat C-C)
  =
  C
  \left(
I+C^{-1}(\widehat C-C)
  \right)$, and $\left\|
  C^{-1}(\widehat C-C)
\right\|_\infty
  \le
  \frac12$, the Neumann series implies that $I+C^{-1}(\widehat C-C)$ is invertible and that 
\begin{align}
\left\|
  \left(
I+C^{-1}(\widehat C-C)
  \right)^{-1}
\right\|_\infty
  \le
  \frac1{
1-\|C^{-1}(\widehat C-C)\|_\infty
  }
  \le
  2 .
\end{align}
Consequently, \(\widehat C\) is invertible and
\begin{align}
\|\widehat C^{-1}\|_\infty
  &=
  \left\|
\left(
  I+C^{-1}(\widehat C-C)
\right)^{-1}
C^{-1}
  \right\|_\infty\\
  &\le
  2\|C^{-1}\|_\infty \\
  &=
  2\kappa_C .
\end{align}
\end{proof}
The next lemma aligns the arbitrary phases of the reconstructed pure output
vectors.  

\begin{lemma}[Phase synchronization]
\label{lem:xmode-phase-sync}
Assume the exact gauge-fixed relation $|\phi_k\rangle
  =
  W_g|\underline{\bs \beta_k}\rangle$ and $\bs \beta_1=0$. Let $|u_k\rangle$ 
be arbitrary normalized representatives of the reconstructed pure states
\(\widetilde\rho_k\). Suppose the condition $D_{\rm tr}
\left(
  |u_k\rangle\langle u_k|,
  |\phi_k\rangle\langle\phi_k|
\right)
  \le
  \eta$ for every \(k\), and denote $D_\star
  =
  \max_{i<j}\|\bs \beta_i-\bs \beta_j\|_2$.

If we have the condition  $8\sqrt2\,\eta
  <
  e^{-D_\star^2/2}$, then one can choose phases of the \(u_k\)'s so as to obtain vectors
\(|\widehat\phi_k\rangle\) and a common phase \(\omega\in\mathbb R\) satisfying
\begin{align}
\left\|
  |\widehat\phi_k\rangle
  -
  e^{i\omega}|\phi_k\rangle
\right\|_2
  \le
  L_{\rm ph}\eta,
  \qquad
L_{\rm ph}
  =
  16e^{D_\star^2/2}.
\end{align}
Consequently, for any projection \(P\), we have $\left\|
  P|\widehat\phi_k\rangle
  -
  e^{i\omega}P|\phi_k\rangle
\right\|_2
  \le
  L_{\rm ph}\eta $.
\end{lemma}

\begin{proof}
We first convert trace-distance closeness of pure states into vector-norm
closeness after choosing phases.  For each \(k\), the pure-state trace-distance
identity gives $\frac 1 2 \left\|
  |u_k\rangle\langle u_k|-
  |\phi_k\rangle\langle\phi_k|
\right\|_1
  =
  \sqrt{1-|\langle u_k|\phi_k\rangle|^2}
  \le
  \eta$. Choose \(\xi_k\in\mathbb R\) so that 
$e^{-i\xi_k}\langle \phi_k|u_k\rangle
  =
  |\langle \phi_k|u_k\rangle|$. Then we have 
\begin{align}
\left\|
  |u_k\rangle-e^{i\xi_k}|\phi_k\rangle
\right\|_2^2
  &=
  2-2|\langle u_k|\phi_k\rangle|  \\
  &\le
  2\left(1-|\langle u_k|\phi_k\rangle|^2\right) \\
  &\le
  2\eta^2.
\end{align}
Thus, defining $|e_k\rangle
  :=
  |u_k\rangle-e^{i\xi_k}|\phi_k\rangle$, we have
\begin{align}
\|e_k\|_2
  \le
  \sqrt2\,\eta .
\label{eq:phase-sync-vector-error}
\end{align}

We now use \(\bs \beta_1=0\) as the phase anchor.  For \(k\ge2\), define $g_k
  :=
  \langle\phi_1|\phi_k\rangle$. Since \(|\phi_k\rangle=W_g|\underline{\bs \beta_k}\rangle\) and \(W_g\) is unitary, we have $g_k
  =
  \langle\underline{\bs \beta_1}|\underline{\bs \beta_k}\rangle$. Because \(\bs \beta_1=0\), the coherent overlap is real and positive: $g_k
  =
  e^{-\|\bs \beta_k\|_2^2/2}$. Moreover, the condition $\|\bs \beta_k\|_2
  =
  \|\bs \beta_k-\bs \beta_1\|_2
  \le
  D_\star$ gives 
\begin{align}
g_k
  \ge
  e^{-D_\star^2/2}.
\label{eq:gk-lower-bound}
\end{align}

Let us denote $h_k
  :=
  \langle u_1|u_k\rangle$. Using
the relation $|u_j\rangle
  =
  e^{i\xi_j}|\phi_j\rangle+|e_j\rangle$, we have an expansion 
\begin{align}
h_k
  &=
  \left(
e^{-i\xi_1}\langle\phi_1|+\langle e_1|
  \right)
  \left(
e^{i\xi_k}|\phi_k\rangle+|e_k\rangle
  \right)\\
  &=
  e^{i(\xi_k-\xi_1)}g_k
  +
  e^{-i\xi_1}\langle\phi_1|e_k\rangle
  +
  e^{i\xi_k}\langle e_1|\phi_k\rangle
  +
  \langle e_1|e_k\rangle .
\end{align}
Therefore, by Cauchy--Schwarz and Eq.~\eqref{eq:phase-sync-vector-error}, we have 
\begin{align}
\left|
  h_k-e^{i(\xi_k-\xi_1)}g_k
\right|
  &\le
  \|e_k\|_2+\|e_1\|_2+\|e_1\|_2\|e_k\|_2 \\
  &\le
  2\sqrt2\,\eta+2\eta^2.
\end{align}
The assumption $8\sqrt2\,\eta
  <
  e^{-D_\star^2/2}
  \le
  1$ implies \(\eta<1\). In addition, we have $2\sqrt2\,\eta+2\eta^2
  \le
  4\sqrt2\,\eta$. Thus we have 
\begin{align}
\left|
  h_k-e^{i(\xi_k-\xi_1)}g_k
\right|
  \le
  4\sqrt2\,\eta.
\label{eq:hk-ak-error}
\end{align}
Combining Eq.~\eqref{eq:hk-ak-error} with Eq.~\eqref{eq:gk-lower-bound}, we get
\begin{align}
|h_k|
  &\ge
  g_k
  -
  \left|
h_k-e^{i(\xi_k-\xi_1)}g_k
  \right|\\
  &\ge
  e^{-D_\star^2/2}
  -
  4\sqrt2\,\eta   \\
  &>
  \frac12 e^{-D_\star^2/2}.
\end{align}
In particular, \(h_k\neq0\), so \(\arg h_k\) is well defined.

We now define the synchronized representatives by $|\widehat\phi_1\rangle
  :=
  |u_1\rangle$ and $|\widehat\phi_k\rangle
  :=
  e^{-i\arg h_k}|u_k\rangle$ for $k\ge2$. This choice makes the anchor overlaps
\(\langle\widehat\phi_1|\widehat\phi_k\rangle\) real and positive, matching the
exact fact that \(g_k>0\). It remains to bound the error relative to one common phase.  Let us set $a_k
  :=
  e^{i(\xi_k-\xi_1)}g_k$. Then we have $\frac{a_k}{|a_k|}
  =
  e^{i(\xi_k-\xi_1)}$ and $\frac{h_k}{|h_k|}
  =
  e^{i\arg h_k}$. We further use the following inequality
\begin{align}
\left|
  \frac{a}{|a|}
  -
  \frac{b}{|b|}
\right|
  &\le
  \left|
\frac{a}{|a|}
-
\frac{b}{|a|}
  \right|
  +
  \left|
\frac{b}{|a|}
-
\frac{b}{|b|}
  \right|\\
  &=
  \frac{|a-b|}{|a|}
  +
  \frac{\bigl||b|-|a|\bigr|}{|a|}\\
  &\le
  \frac{2|a-b|}{|a|}.
\end{align}
Applying this with \(a=a_k\) and \(b=h_k\), and using
\(|a_k|=g_k\ge e^{-D_\star^2/2}\), gives
\begin{align}
\left|
  e^{i(\xi_k-\xi_1)}
  -
  e^{i\arg h_k}
\right|
  &\le
  2e^{D_\star^2/2}
  \left|
h_k-e^{i(\xi_k-\xi_1)}g_k
  \right|\\
  &\le
  8\sqrt2\,e^{D_\star^2/2}\eta .
\end{align}
Equivalently, we have 
\begin{align}
\left|
  e^{-i\arg h_k+i\xi_k-i\xi_1}
  -
  1
\right|
  \le
  8\sqrt2\,e^{D_\star^2/2}\eta .
\label{eq:relative-phase-error}
\end{align}

Take the common phase $\omega
  :=
  \xi_1$. For \(k\ge2\), we have
\begin{align}
|\widehat\phi_k\rangle-e^{i\omega}|\phi_k\rangle
  &=
  e^{-i\arg h_k}|u_k\rangle-e^{i\xi_1}|\phi_k\rangle \\
  &=
  e^{-i\arg h_k}
  \left(
e^{i\xi_k}|\phi_k\rangle+|e_k\rangle
  \right)
  -
  e^{i\xi_1}|\phi_k\rangle  \\
  &=
  e^{i\xi_1}
  \left(
e^{-i\arg h_k+i\xi_k-i\xi_1}
-
1
  \right)
  |\phi_k\rangle
  +
  e^{-i\arg h_k}|e_k\rangle .
\end{align}
Taking norms and using Eq.~\eqref{eq:relative-phase-error} and
Eq.~\eqref{eq:phase-sync-vector-error}, we obtain
\begin{align}
\left\|
  |\widehat\phi_k\rangle-e^{i\omega}|\phi_k\rangle
\right\|_2
  &\le
  8\sqrt2\,e^{D_\star^2/2}\eta
  +
  \sqrt2\,\eta   \\
  &\le
  16e^{D_\star^2/2}\eta .
\end{align}
For \(k=1\), the same bound holds because
\begin{align}
\left\|
  |\widehat\phi_1\rangle-e^{i\omega}|\phi_1\rangle
\right\|_2
  =
  \left\|
|u_1\rangle-e^{i\xi_1}|\phi_1\rangle
  \right\|_2
  \le
  \sqrt2\,\eta
  \le
  16e^{D_\star^2/2}\eta .
\end{align}
Thus, for every \(1\le k\le d\), we have 
\begin{align}
\left\|
  |\widehat\phi_k\rangle-e^{i\omega}|\phi_k\rangle
\right\|_2
  \le
  L_{\rm ph}\eta,
  \qquad
L_{\rm ph}
  =
  16e^{D_\star^2/2}.
\end{align}

Finally, if \(P\) is any projection, then \(\|P\|_\infty\le1\).  Therefore we have 
\begin{align}
\left\|
  P|\widehat\phi_k\rangle-e^{i\omega}P|\phi_k\rangle
\right\|_2
  &=
  \left\|
P
\left(
  |\widehat\phi_k\rangle-e^{i\omega}|\phi_k\rangle
\right)
  \right\|_2 \\
  &\le
  \left\|
|\widehat\phi_k\rangle-e^{i\omega}|\phi_k\rangle
  \right\|_2 \\
  &\le
  L_{\rm ph}\eta .
\end{align}
This proves the lemma.
\end{proof}
We now combine amplitude reconstruction, coherent Vandermonde inversion, and
phase synchronization to reconstruct the truncated operator.

\begin{lemma}[Truncated operator reconstruction]
\label{lem:xmode-truncated-operator}
Let us set integers \(r\in\mathbb N\), \(d=d_x(r)\), and suppose that the matrix $C
  =
  \left[
c_r^{(x)}(\beta_1)\ \cdots\ c_r^{(x)}(\beta_d)
  \right]$ is invertible.  Let us denote $W_{g,r}
  =
  \Pi_{\le r}^{(x)}W_g\Pi_{\le r}^{(x)}$. Define matrices $Y
  =
  \left[
\Pi_{\le r}^{(x)}|\phi_1\rangle\
\cdots\
\Pi_{\le r}^{(x)}|\phi_d\rangle
  \right]$ and $\widehat Y
  =
  \left[
\Pi_{\le r}^{(x)}|\widehat\phi_1\rangle\
\cdots\
\Pi_{\le r}^{(x)}|\widehat\phi_d\rangle
  \right]$. Let us additionally define $\widehat C
  =
  \left[
c_r^{(x)}(\widehat\beta_1)\
\cdots\
c_r^{(x)}(\widehat\beta_d)
  \right]$ and $\widehat W_r
  =
  \widehat Y\widehat C^{-1}$. Assume the hypotheses of Lemmas~\ref{lem:xmode-stable-gauge},
\ref{lem:xmode-coherent-vandermonde}, and \ref{lem:xmode-phase-sync} hold, and
assume the conditions $L_{\rm geom}^{(x)}\Delta_H(\eta)
  \le
  1$ and $\sqrt d\,L_C^{(x)}L_{\rm geom}^{(x)}\Delta_H(\eta)
  \le
  \frac{1}{2\kappa_C}$. Define the coherent tail 
$\tau_r^{(x)}(D_\star)
  =
  \left(
\sup_{\|z\|_2\le D_\star}
\sum_{|\boldsymbol n|>r}
e^{-\|z\|_2^2}
\frac{|z^{\boldsymbol n}|^2}{\boldsymbol n!}
  \right)^{1/2}$.

Then there exists a phase \(\omega\in\mathbb R\) such that
\begin{align}
\left\|
  e^{-i\omega}\widehat W_r-W_{g,r}
\right\|_{\infty}
  \le
  B_x(\eta),
\end{align}
where
\begin{align}
B_x(\eta)
  :=
  2\kappa_C
  \left[
\sqrt d\,L_{\rm ph}\eta
+
\sqrt d\,\tau_r^{(x)}(D_\star)
+
\sqrt d\,L_C^{(x)}L_{\rm geom}^{(x)}\Delta_H(\eta)
  \right].
\end{align}
\end{lemma}

\begin{proof}
For
brevity, let us write $P_r
  :=
  \Pi_{\le r}^{(x)}$. We first derive the exact matrix identity relating the output matrix \(Y\), the
truncated operator \(W_{g,r}\), and the coherent coefficient matrix \(C\).  Given the condition $|\phi_k\rangle
  =
  W_g|\underline{\bs \beta_k}\rangle$, we have $P_r|\phi_k\rangle
  =
  P_rW_g|\underline{\bs \beta_k}\rangle$. Now decompose the coherent input into its \(r\)-photon truncation and tail: $|\underline{\bs \beta_k}\rangle
  =
  P_r|\underline{\bs \beta_k}\rangle
  +
  (I-P_r)|\underline{\bs \beta_k}\rangle$. Therefore we have 
\begin{align}
P_r|\phi_k\rangle
  =
  P_rW_gP_r|\underline{\bs \beta_k}\rangle
  +
  P_rW_g(I-P_r)|\underline{\bs \beta_k}\rangle .
\end{align}
By definition, we have $W_{g,r}
  =
  P_rW_gP_r$. Moreover, the coordinate vector of \(P_r|\underline{\bs \beta_k}\rangle\) in the Fock basis
\(\{|\boldsymbol n\rangle:|\boldsymbol n|\le r\}\) is exactly $c_r^{(x)}(\bs \beta_k)$. Thus the \(k\)-th column of \(Y\) is $Y_k
  =
  W_{g,r}c_r^{(x)}(\bs \beta_k)
  +
  t_k$, with $t_k
  :=
  P_rW_g(I-P_r)|\underline{\bs \beta_k}\rangle$. Putting the columns together gives
\begin{align}
Y
  =
  W_{g,r}C+T,
\end{align}
where \(T=[t_1\ \cdots\ t_d]\).

We next bound \(T\).  Since \(W_g\) is unitary and \(P_r\) is a projection,
both are contractions.  Hence we have 
\begin{align}
\|t_k\|_2
  &=
  \left\|
P_rW_g(I-P_r)|\underline{\bs \beta_k}\rangle
  \right\|_2 \\
  &\le
  \left\|
(I-P_r)|\underline{\bs \beta_k}\rangle
  \right\|_2 .
\end{align}
Because \(\bs \beta_1=0\) in the gauge-fixed construction, we have
\begin{align}
\|\bs \beta_k\|_2
  =
  \|\bs \beta_k-\bs \beta_1\|_2
  \le
  D_\star .
\end{align}
Therefore, by the definition of \(\tau_r^{(x)}(D_\star)\), we have $\|t_k\|_2
  \le
  \tau_r^{(x)}(D_\star)$. Consequently, we have 
\begin{align}
\|T\|_\infty
  \le
  \|T\|_F
  =
  \left(
\sum_{k=1}^d\|t_k\|_2^2
  \right)^{1/2}
  \le
  \sqrt d\,\tau_r^{(x)}(D_\star).
\label{eq:xmode-T-bound}
\end{align}

We now control the error in the reconstructed output matrix.  By the phase
synchronization lemma, there exists a common phase \(\omega\in\mathbb R\) such
that, for every \(k\), we have
\begin{align}
\left\|
  P_r|\widehat\phi_k\rangle
  -
  e^{i\omega}P_r|\phi_k\rangle
\right\|_2
  \le
  L_{\rm ph}\eta .
\end{align}
Equivalently, we have $\left\|
  e^{-i\omega}P_r|\widehat\phi_k\rangle
  -
  P_r|\phi_k\rangle
\right\|_2
  \le
  L_{\rm ph}\eta$.

Since \(e^{-i\omega}\widehat Y-Y\) has these vectors as columns, we get
\begin{align}
\|e^{-i\omega}\widehat Y-Y\|_\infty
  &\le
  \|e^{-i\omega}\widehat Y-Y\|_F  \\
  &=
  \left(
\sum_{k=1}^d
\left\|
  e^{-i\omega}P_r|\widehat\phi_k\rangle
  -
  P_r|\phi_k\rangle
\right\|_2^2
  \right)^{1/2}\\
  &\le
  \sqrt d\,L_{\rm ph}\eta .
\label{eq:xmode-Y-bound}
\end{align}

Next, we control the input coefficient matrix.  By the stable gauge recovery
lemma, we have 
\begin{align}
\max_k\|\bs{\widehat\beta}_k-\bs \beta_k\|_2
  \le
  L_{\rm geom}^{(x)}\Delta_H(\eta).
\end{align}
The additional assumption $L_{\rm geom}^{(x)}\Delta_H(\eta)
  \le
  1$ implies
\begin{align}
\|\bs {\widehat\beta}_k\|_2
  \le
  \|\bs \beta_k\|_2+\|\bs {\widehat\beta}_k-\bs \beta_k\|_2
  \le
  D_\star+1.
\end{align}
Therefore the Lipschitz estimate from the coherent coefficient lemma applies
with \(R=D_\star\), giving
\begin{align}
\left\|
  c_r^{(x)}(\bs {\widehat\beta}_k)
  -
  c_r^{(x)}(\bs \beta_k)
\right\|_2
  \le
  L_C^{(x)}L_{\rm geom}^{(x)}\Delta_H(\eta).
\end{align}
Thus we have 
\begin{align}
\|\widehat C-C\|_\infty
  &\le
  \|\widehat C-C\|_F   \\
  &=
  \left(
\sum_{k=1}^d
\left\|
  c_r^{(x)}(\bs {\widehat\beta}_k)
  -
  c_r^{(x)}(\bs \beta_k)
\right\|_2^2
  \right)^{1/2} \\
  &\le
  \sqrt d\,L_C^{(x)}L_{\rm geom}^{(x)}\Delta_H(\eta).
\label{eq:xmode-C-bound}
\end{align}
By the assumed invertibility condition, we have 
\begin{align}
\sqrt d\,L_C^{(x)}L_{\rm geom}^{(x)}\Delta_H(\eta)
  \le
  \frac1{2\kappa_C},
  \qquad
\kappa_C=\|C^{-1}\|_\infty,
\end{align}
the Neumann-series estimate from Lemma~\ref{lem:xmode-coherent-vandermonde}
implies that \(\widehat C\) is invertible and the condition 
\begin{align}
\|\widehat C^{-1}\|_\infty
  \le
  2\kappa_C .
\label{eq:xmode-C-inverse-bound}
\end{align}

We now compare the reconstructed truncated operator $\widehat W_r
  =
  \widehat Y\widehat C^{-1}$ with the exact truncated operator \(W_{g,r}\).  Using the exact relation
\(Y=W_{g,r}C+T\), we compute
\begin{align}
e^{-i\omega}\widehat W_r-W_{g,r}
  &=
  e^{-i\omega}\widehat Y\widehat C^{-1}-W_{g,r}\\
  &=
  (e^{-i\omega}\widehat Y-Y)\widehat C^{-1}
  +
  Y\widehat C^{-1}
  -
  W_{g,r}  \\
  &=
  (e^{-i\omega}\widehat Y-Y)\widehat C^{-1}
  +
  (W_{g,r}C+T)\widehat C^{-1}
  -
  W_{g,r}  \\
  &=
  (e^{-i\omega}\widehat Y-Y)\widehat C^{-1}
  +
  T\widehat C^{-1}
  +
  W_{g,r}\left(C\widehat C^{-1}-I\right).
\end{align}
Since we have $C\widehat C^{-1}-I
  =
  (C-\widehat C)\widehat C^{-1}$, we obtain
\begin{align}
e^{-i\omega}\widehat W_r-W_{g,r}
  =
  (e^{-i\omega}\widehat Y-Y)\widehat C^{-1}
  +
  T\widehat C^{-1}
  +
  W_{g,r}(C-\widehat C)\widehat C^{-1}.
\end{align}

Taking operator norms and using submultiplicativity gives
\begin{align}
\left\|
  e^{-i\omega}\widehat W_r-W_{g,r}
\right\|_\infty
&\le
\|e^{-i\omega}\widehat Y-Y\|_\infty
\|\widehat C^{-1}\|_\infty  \\
&\quad+
\|T\|_\infty
\|\widehat C^{-1}\|_\infty  \\
&\quad+
\|W_{g,r}\|_\infty
\|C-\widehat C\|_\infty
\|\widehat C^{-1}\|_\infty .
\end{align}
Now we have $\|W_{g,r}\|_\infty
  =
  \|P_rW_gP_r\|_\infty
  \le
  1$, because \(P_r\) is a projection and \(W_g\) is unitary.  Substituting
Eqs.~\eqref{eq:xmode-T-bound}, \eqref{eq:xmode-Y-bound},
\eqref{eq:xmode-C-bound}, and \eqref{eq:xmode-C-inverse-bound}, we get
\begin{align}
\left\|
  e^{-i\omega}\widehat W_r-W_{g,r}
\right\|_\infty
&\le
2\kappa_C
\left[
  \sqrt d\,L_{\rm ph}\eta
  +
  \sqrt d\,\tau_r^{(x)}(D_\star)
  +
  \sqrt d\,L_C^{(x)}L_{\rm geom}^{(x)}\Delta_H(\eta)
\right] \\
&=
B_x(\eta).
\end{align}
This proves the claim.
\end{proof}
The final deterministic step completes the reconstructed truncated operator to
a finite-dimensional unitary.

\begin{lemma}[Stable unitary completion]
\label{lem:xmode-unitary-completion}
Let us denote projectors $P_q
  :=
  \Pi_{\le q}^{(x)}$ and $P_r
  :=
  \Pi_{\le r}^{(x)}$ for $q\le r$. Define the finite-dimensional subspaces $\mathcal K_q
  :=
  P_q\mathcal H$ and $\mathcal K_r
  :=
  P_r\mathcal H$. Let us denote maps $A_r
  :=
  P_rW_gP_q:
  \mathcal K_q\to\mathcal K_r$ and $\widehat A_r
  :=
  \widehat W_rP_q:
  \mathcal K_q\to\mathcal K_r$. Assume that \(A_r\) has full column rank.  Fix a deterministic local completion
map $\mathcal Q$ defined in a neighborhood of the phase orbit $\{e^{i\theta}A_r:\theta\in\mathbb R\}$, such that \(\mathcal Q(B)\) is a unitary on \(\mathcal K_r\), depends smoothly
on an operator \(B\), and is phase-equivariant: $\mathcal Q(e^{i\theta}B)
  =
  e^{i\theta}\mathcal Q(B)$. For example, \(\mathcal Q\) may be taken to be a fixed local Gram--Schmidt or
polar-completion map. Set $V_r
  :=
  \mathcal Q(A_r)$ and $\widehat V_r
  :=
  \mathcal Q(\widehat A_r)$.

Then there exist constants $\rho_{\rm QR}^{(x)}>0$ and $L_{\rm QR}^{(x)}<\infty$, depending only on the exact frame \(A_r\) and on the chosen completion chart,
such that whenever
\begin{align}
\|e^{-i\omega}\widehat A_r-A_r\|_{\infty}
  \le
  \rho_{\rm QR}^{(x)}
\end{align}
for some \(\omega\in\mathbb R\), we have
\begin{align}
\inf_{\vartheta\in\mathbb R}
\|\widehat V_r-e^{i\vartheta}V_r\|_{\infty}
  \le
  L_{\rm QR}^{(x)}
  \|e^{-i\omega}\widehat A_r-A_r\|_{\infty}.
\end{align}
In particular, if Lemma~\ref{lem:xmode-truncated-operator} gives $\|e^{-i\omega}\widehat W_r-W_{g,r}\|_{\infty}
  \le
  B_x(\eta)$, and if \(B_x(\eta)\le \rho_{\rm QR}^{(x)}\), then $\inf_{\vartheta\in\mathbb R}
\|\widehat V_r-e^{i\vartheta}V_r\|_{\infty}
  \le
  L_{\rm QR}^{(x)}B_x(\eta)$.
\end{lemma}

\begin{proof}
Since \(A_r\) has full column rank, it lies in the open set of full-rank
operators from \(\mathcal K_q\) to \(\mathcal K_r\).  On this open set, any
fixed Gram--Schmidt or polar-completion chart is smooth.  Hence, after possibly
shrinking the neighborhood of the compact phase orbit $\{e^{i\theta}A_r:\theta\in\mathbb R\}$, there exist constants \(\rho_{\rm QR}^{(x)}>0\) and
\(L_{\rm QR}^{(x)}<\infty\) such that
\begin{align}
\|\mathcal Q(B)-\mathcal Q(A_r)\|_{\infty}
  \le
  L_{\rm QR}^{(x)}\|B-A_r\|_{\infty}
\end{align}
whenever \(\|B-A_r\|_{\infty}\le \rho_{\rm QR}^{(x)}\).

Now set $B
  :=
  e^{-i\omega}\widehat A_r $. If we have $\|B-A_r\|_{\infty}
  \le
  \rho_{\rm QR}^{(x)}$, then the local Lipschitz bound gives
\begin{align}
\|\mathcal Q(B)-\mathcal Q(A_r)\|_{\infty}
  \le
  L_{\rm QR}^{(x)}
  \|e^{-i\omega}\widehat A_r-A_r\|_{\infty}.
\end{align}
By phase equivariance, we have 
\begin{align}
\mathcal Q(B)
  =
  \mathcal Q(e^{-i\omega}\widehat A_r)
  =
  e^{-i\omega}\mathcal Q(\widehat A_r)
  =
  e^{-i\omega}\widehat V_r,
\end{align}
and \(\mathcal Q(A_r)=V_r\).  Therefore, we have 
\begin{align}
\|e^{-i\omega}\widehat V_r-V_r\|_{\infty}
  \le
  L_{\rm QR}^{(x)}
  \|e^{-i\omega}\widehat A_r-A_r\|_{\infty}.
\end{align}
Multiplying by \(e^{i\omega}\) gives $\|\widehat V_r-e^{i\omega}V_r\|_{\infty}
  \le
  L_{\rm QR}^{(x)}
  \|e^{-i\omega}\widehat A_r-A_r\|_{\infty}$. Taking the infimum over \(\vartheta\in\mathbb R\) proves the first claim.

For the final claim, note that \(q\le r\), so we have 
\begin{align}
W_{g,r}P_q
  =
  P_rW_gP_rP_q
  =
  P_rW_gP_q
  =
  A_r.
\end{align}
In addition, we have $\widehat A_r
  =
  \widehat W_rP_q$. Hence we have 
\begin{align}
\|e^{-i\omega}\widehat A_r-A_r\|_{\infty}
  &=
  \|(e^{-i\omega}\widehat W_r-W_{g,r})P_q\|_{\infty} \\
  &\le
  \|e^{-i\omega}\widehat W_r-W_{g,r}\|_{\infty}  \\
  &\le
  B_x(\eta).
\end{align}
If \(B_x(\eta)\le \rho_{\rm QR}^{(x)}\), the first part gives
\begin{align}
\inf_{\vartheta\in\mathbb R}
\|\widehat V_r-e^{i\vartheta}V_r\|_{\infty}
  \le
  L_{\rm QR}^{(x)}B_x(\eta).
\end{align}
\end{proof}

We can now state the main theorem.  The theorem is deterministic conditional on
the tomography-successful event.

\begin{theorem}[Stable \(x\)-mode coherent-probe reconstruction]
\label{thm:xmode-main}
Fix the number of modes \(x\), a photon-number bound \(N\), and an accuracy
\(0<\epsilon<1\).  Let \(W_\star\) be an unknown \(x\)-mode unitary, and define
\(q:=\lceil 256N/\epsilon^2\rceil\).  Choose \(r\ge q\) and set
\(d:=d_x(r)=\binom{x+r}{x}\).  Let $|\underline{\bs \alpha_1}\rangle,\ldots,|\underline{\bs \alpha_d}\rangle$ be the coherent probes, and let $|\psi_k\rangle:=W_\star|\underline{\bs \alpha_k}\rangle$ and $\rho_k:=|\psi_k\rangle\langle\psi_k|$ denote the output state. Let \(\bs \beta_1,\ldots,\bs \beta_d\) be the exact gauge-fixed amplitudes from
Lemma~\ref{lem:xmode-exact-gauge}.  Assume the following conditions.

\begin{enumerate}
\item The centered amplitudes span \(\mathbb C^x\), the minimum distance
\(d_{\min}:=\min_{i<j}\|\bs \beta_i-\bs \beta_j\|_2\) is positive, and the Bargmann
phase unwrapping condition holds with margin \(\zeta>0\).

\item The points \(\beta_1,\ldots,\beta_d\) are unisolvent for holomorphic
polynomials of total degree at most \(r\).

\item The coherent coefficient matrix $C=\left[c_r^{(x)}(\bs \beta_1)\ \cdots\ c_r^{(x)}(\bs \beta_d)\right]$ is invertible, and \(\kappa_C:=\|C^{-1}\|_\infty\).

\item Let \(G=D(\bs \gamma)\Gamma(R)\) be the physical coherent gauge from
Lemma~\ref{lem:xmode-exact-gauge}, and define \(W_g:=W_\star G^\dagger\).
There exists \(N_{\rm dyn}\) such that $\mathsf N_{W_g}^{(x)}(q)\le N_{\rm dyn}$, and the cutoff \(r\) satisfies $r\ge \left\lceil \frac{256N_{\rm dyn}}{\epsilon^2}\right\rceil$.
\end{enumerate}

Let $D_\star:=\max_{i<j}\|\bs \beta_i-\bs \beta_j\|_2$. Let \(\Delta_{\rm dist}\), \(\Delta_H\), \(L_{\rm geom}^{(x)}\), \(\eta_0\),
\(L_C^{(x)}\), \(L_{\rm ph}\), \(B_x\), \(L_{\rm QR}^{(x)}\), and
\(\rho_{\rm QR}^{(x)}\) be the constants from the preceding lemmas. Choose \(\eta_{\rm stab}>0\) such that
\begin{align}
\eta_{\rm stab} &\le \eta_0, \\
4\eta_{\rm stab}
&<
\min\left\{e^{-D_\star^2},\,1-e^{-d_{\min}^2}\right\}, \\
12\eta_{\rm stab}
&<
e^{-3D_\star^2/2}, \\
12e^{3D_\star^2/2}\eta_{\rm stab}
&<
\frac{\zeta}{2}, \\
8\sqrt2\,\eta_{\rm stab}
&<
e^{-D_\star^2/2}, \\
L_{\rm geom}^{(x)}\Delta_H(\eta_{\rm stab})
&\le 1, \\
\sqrt d\,L_C^{(x)}L_{\rm geom}^{(x)}\Delta_H(\eta_{\rm stab})
&\le \frac1{2\kappa_C}, \\
B_x(\eta_{\rm stab})
&\le \rho_{\rm QR}^{(x)}, \\
2L_{\rm QR}^{(x)}B_x(\eta_{\rm stab})
&\le \frac{\epsilon}{2}.
\end{align}
Suppose the tomography-successful event holds: $\frac 1 2 \left\|\widetilde\rho_k-\rho_k\right\|_1\le \eta_{\rm stab}$ for $1\le k\le d$.

Then the reconstruction procedure described in
Lemmas~\ref{lem:xmode-stable-gauge}--\ref{lem:xmode-unitary-completion}
returns a unitary-channel estimator \(\widehat{\mathcal W}\) satisfying
\begin{align}
d_{\diamond,N}^{\rm coh,x}(\widehat{\mathcal W},\mathcal W_\star)
\le
\epsilon,
\end{align}
where $d_{\diamond,N}^{\rm coh,x}(\widehat{\mathcal W},\mathcal W_\star)
:=
\inf_{\bs \gamma\in\mathbb C^x,\ R\in U(x)}
\left\|
\widehat{\mathcal W}
-
\mathcal W_\star\circ\mathcal G_{\bs \gamma,R}^\dagger
\right\|_{\diamond}^{(x,N)}$ is the coherent-gauge quotient distance, and $\mathcal G_{\bs \gamma,R}
=
G_{\bs \gamma,R}\,\boldsymbol{\cdot}\,G_{\bs \gamma,R}^\dagger, 
G_{\bs \gamma,R}=D(\bs \gamma)\Gamma(R)$. Consequently, if the tomography-successful event holds with probability at least
\(1-\delta\), then the same reconstruction guarantee holds with probability at
least \(1-\delta\).
\end{theorem}

\begin{proof}
By Lemma~\ref{lem:xmode-exact-gauge}, there is a physical coherent gauge operator 
\(G=D(\bs \gamma)\Gamma(R)\) such that, after choosing suitable output phases, we have $|\phi_k\rangle=W_g|\underline{\bs \beta_k}\rangle$ with $W_g=W_\star G^\dagger$. Therefore the quotient distance can only be smaller than the distance to this
particular gauge-fixed channel:
\begin{align}
d_{\diamond,N}^{\rm coh,x}(\widehat{\mathcal W},\mathcal W_\star)
\le
\|\widehat{\mathcal W}-\mathcal W_g\|_{\diamond}^{(x,N)}.
\end{align}

We first approximate the gauge-fixed unitary \(W_g\) by a finite-dimensional
unitary.  Since \(\mathsf N_{W_g}^{(x)}(q)\le N_{\rm dyn}\) and
\(r\ge \lceil256N_{\rm dyn}/\epsilon^2\rceil\), Lemma~\ref{lem:xmode-cutoff}
gives a unitary \(V_r\) on \(\mathcal H_{\le r}^{(x)}\) such that, for every
unitary \(V_{>r}\) on the complement,
\begin{align}
\left\|
\mathcal W_g
-
(V_r\oplus V_{>r})\,\boldsymbol{\cdot}\,(V_r\oplus V_{>r})^\dagger
\right\|_{\diamond}^{(x,N)}
\le
\frac{\epsilon}{2}.
\label{eqs800}
\end{align}

Now condition on the tomography-good event.  Lemma~\ref{lem:xmode-stable-gauge}
recovers the gauge-fixed amplitudes with error controlled by
\(\Delta_H(\eta_{\rm stab})\).  The assumptions $L_{\rm geom}^{(x)}\Delta_H(\eta_{\rm stab})\le1$ and $\sqrt d\,L_C^{(x)}L_{\rm geom}^{(x)}\Delta_H(\eta_{\rm stab})
\le
\frac1{2\kappa_C}$ allow Lemma~\ref{lem:xmode-coherent-vandermonde} to be applied.  Hence
\(\widehat C\) is invertible and \(\|\widehat C^{-1}\|_\infty\le2\kappa_C\).

Lemma~\ref{lem:xmode-phase-sync} then aligns the reconstructed output vectors
up to a single common phase.  Combining this phase alignment with the
coherent-coefficient inversion, Lemma~\ref{lem:xmode-truncated-operator} gives
a phase \(\omega\in\mathbb R\) such that
\begin{align}
\|e^{-i\omega}\widehat W_r-W_{g,r}\|_\infty
\le
B_x(\eta_{\rm stab}).
\label{eqs801}
\end{align}

Let \(P_q:=\Pi_{\le q}^{(x)}\) and $P_r:=\Pi_{\le r}^{(x)}$.  Define $A_r:=P_rW_gP_q$ and $\widehat A_r:=\widehat W_rP_q$. Since \(q\le r\), we have \(W_{g,r}P_q=A_r\).  Therefore Eq. (\ref{eqs801}) implies
\begin{align}
\|e^{-i\omega}\widehat A_r-A_r\|_\infty
=
\|(e^{-i\omega}\widehat W_r-W_{g,r})P_q\|_\infty
\le
B_x(\eta_{\rm stab}).
\end{align}
Because \(B_x(\eta_{\rm stab})\le \rho_{\rm QR}^{(x)}\),
Lemma~\ref{lem:xmode-unitary-completion} gives
\begin{align}
\inf_{\vartheta\in\mathbb R}
\|\widehat V_r-e^{i\vartheta}V_r\|_\infty
\le
L_{\rm QR}^{(x)}B_x(\eta_{\rm stab}).
\label{eqs803}
\end{align}

Choose \(\vartheta\) satisfying Eq. (\ref{eqs803}), and define the final estimator $\widehat U:=\widehat V_r\oplus I_{>r}$ and $\widehat{\mathcal W}
=
\widehat U\,\boldsymbol{\cdot}\,\widehat U^\dagger$.
Also define the unitary channel  $U_\vartheta:=e^{i\vartheta}V_r\oplus I_{>r}$ and $\mathcal U_\vartheta
=
U_\vartheta\,\boldsymbol{\cdot}\,U_\vartheta^\dagger$.

For unitary channels, we use the elementary bound
\begin{align}
\|U\,\boldsymbol{\cdot}\,U^\dagger
-
V\,\boldsymbol{\cdot}\,V^\dagger\|_\diamond
\le
2\|U-V\|_\infty .
\end{align}
Indeed,
\begin{align}
U\rho U^\dagger-V\rho V^\dagger
=
(U-V)\rho U^\dagger
+
V\rho(U^\dagger-V^\dagger),
\end{align}
and the trace norm of the right-hand side is at most \(2\|U-V\|_\infty\).
Therefore, by Eq. (\ref{eqs803}), we have 
\begin{align}
\|\widehat{\mathcal W}-\mathcal U_\vartheta\|_{\diamond}^{(x,N)}
\le
\|\widehat{\mathcal W}-\mathcal U_\vartheta\|_{\diamond}
\le
2L_{\rm QR}^{(x)}B_x(\eta_{\rm stab}).
\label{eqs806}
\end{align}

It remains to compare \(\mathcal U_\vartheta\) with \(\mathcal W_g\).  The unitary
\(U_\vartheta=e^{i\vartheta}V_r\oplus I_{>r}\) induces the same channel as $V_r\oplus e^{-i\vartheta}I_{>r}$, because the two differ only by a global phase.  Since Lemma~\ref{lem:xmode-cutoff}
allows an arbitrary complement unitary \(V_{>r}\), estimate Eq. (\ref{eqs800}) applies to
\(\mathcal U_\vartheta\).  Hence we have 
\begin{align}
\|\mathcal U_\vartheta-\mathcal W_g\|_{\diamond}^{(x,N)}
\le
\frac{\epsilon}{2}.
\label{eqs807}
\end{align}

Combining Eqs. (\ref{eqs806}) and (\ref{eqs807}), we have 
\begin{align}
\|\widehat{\mathcal W}-\mathcal W_g\|_{\diamond}^{(x,N)}
\le
2L_{\rm QR}^{(x)}B_x(\eta_{\rm stab})
+
\frac{\epsilon}{2}.
\end{align}
By the final stability assumption, we have  $2L_{\rm QR}^{(x)}B_x(\eta_{\rm stab})
\le
\frac{\epsilon}{2}$. Therefore, we have 
\begin{align}
\|\widehat{\mathcal W}-\mathcal W_g\|_{\diamond}^{(x,N)}
\le
\epsilon.
\end{align}
Since the coherent-gauge quotient distance is bounded by this particular
gauge-fixed distance, we obtain
\begin{align}
d_{\diamond,N}^{\rm coh,x}(\widehat{\mathcal W},\mathcal W_\star)
\le
\epsilon.
\end{align}

If the tomography-good event holds with probability at least \(1-\delta\), the
entire deterministic argument above holds on that event, so the final guarantee
also holds with probability at least \(1-\delta\).
\end{proof}

The sample complexity of the proposed protocol is given in the following corollary. 

\begin{corollary}[Query complexity after instantiating tomography]
\label{cor:xmode-copy-complexity}
Assume the hypotheses of Theorem~\ref{thm:xmode-main}.  In addition, assume that
the training outputs satisfy the total photon-number bound $\langle\psi_k|\hat n_x |\psi_k\rangle
  \le
  N'$ for $1\le k\le d$, where \(d=d_x(r)=\binom{x+r}{x}\).  Let \(\eta_{\rm stab}\) be the stability
threshold appearing in Theorem~\ref{thm:xmode-main}.  Then the tomography-successful event $\frac 1 2\|\widetilde\rho_k-\rho_k\|_1
  \le
  \eta_{\rm stab}$ for $1\le k\le d$, holds simultaneously with probability at least \(1-\delta\) using at most
\begin{align}
M_{\rm tot}
  =
  d
  \left\lceil
  2^{21}
  \left(
\frac{eN'}{x\eta_{\rm stab}^{2}}
+
2e
  \right)^x
  \eta_{\rm stab}^{-2}
  \log\frac{4d}{\delta}
  \right\rceil
\end{align}
uses of \(W_\star\).  Consequently, with probability at least \(1-\delta\), the
estimator returned by the protocol satisfies
\begin{align}
d_{\diamond,N}^{\rm coh,x}
(\widehat{\mathcal W},\mathcal W_\star)
  \le
  \epsilon .
\end{align}

If, in addition, the chosen probe design satisfies the certified stability bound
\begin{align}
\eta_{\rm stab}^{-1}
  \le
  \textbf{\em exp}
  \left[
\textbf{\em poly}
\left(
  x,
  N,
  N_{\rm dyn},
  \frac1\epsilon
\right)
  \right],
\end{align}
then the resulting conservative unconditional bound is
\begin{align}
M_{\rm tot}
  =
  \textbf{\em exp}
  \left[
\textbf{\em poly}
\left(
  x,
  N,
  N_{\rm dyn},
  \frac1\epsilon,\log (N'+1)
\right)
  \right]
  \textbf{\em poly}
  \left(\log\frac1\delta
  \right),
\end{align}
for fixed \(x\). 
\end{corollary}

\begin{proof}
We instantiate the tomography-good event required by
Theorem~\ref{thm:xmode-main}.  For each \(k\), the output state
\(\rho_k=|\psi_k\rangle\langle\psi_k|\) is an \(x\)-mode pure state satisfying
\begin{align}
\operatorname{Tr}(\hat n_x \rho_k)
  =
  \langle\psi_k|\hat n_x |\psi_k\rangle
  \le
  N' .
\end{align}
Theorem~S36 of \cite{mele2025learning}, applied with moment order \(1\),
accuracy \(\eta_{\rm stab}\), and failure probability \(\delta/d\), gives a
tomography procedure using
\begin{align}
M_k
  =
  \left\lceil
  2^{21}
  d_{\rm eff}^{(x)}
  \eta_{\rm stab}^{-2}
  \log\frac{4d}{\delta}
  \right\rceil
\end{align}
copies of \(\rho_k\), where the effective dimension satisfies
\begin{align}
d_{\rm eff}^{(x)}
  \le
  \left(
\frac{eN'}{x\eta_{\rm stab}^{2}}
+
2e
  \right)^x .
\end{align}
The factor \(N'/x\) appears because Theorem~S36 is stated for an \(x\)-mode
state with first moment bounded by \(xN_{\rm phot}\).  Our assumption gives
\(\operatorname{Tr}(\hat n_x \rho_k)\le N'\), so we may take
\(N_{\rm phot}=N'/x\).

Therefore, for each fixed \(k\), tomography returns
\(\widetilde\rho_k\) satisfying
\begin{align}
D_{\rm tr}(\widetilde\rho_k,\rho_k)
  \le
  \eta_{\rm stab}
\end{align}
with probability at least \(1-\delta/d\).  By the union bound,
all \(d\) tomography calls succeed simultaneously with probability at least $1-d\cdot\frac{\delta}{d}
  =
  1-\delta$. On this simultaneous-success event, the tomography-good event required by
Theorem~\ref{thm:xmode-main} holds.  Hence Theorem~\ref{thm:xmode-main} implies
\begin{align}
d_{\diamond,N}^{\rm coh,x}
(\widehat{\mathcal W},\mathcal W_\star)
  \le
  \epsilon .
\end{align}

It remains only to count uses of \(W_\star\).  Each copy of \(\rho_k\) is
prepared by one use of \(W_\star\) on the coherent input \(|\underline{\alpha_k}\rangle\).
Thus the total number of uses is
\begin{align}
\sum_{k=1}^d M_k
  \le
  d
  \left\lceil
  2^{21}
  \left(
\frac{eN'}{x\eta_{\rm stab}^{2}}
+
2e
  \right)^x
  \eta_{\rm stab}^{-2}
  \log\frac{4d}{\delta}
  \right\rceil .
\end{align}

Finally, we have 
\begin{align}
d_x(r)
  =
  \binom{x+r}{x}.
\end{align}
For fixed \(x\), this satisfies
\begin{align}
d_x(r)
  =
  \mathcal O(r^x).
\end{align}
If \(r=O((N+N_{\rm dyn})/\epsilon^2)\), then \(d_x(r)\) is polynomial in
\(N\), \(N_{\rm dyn}\), and \(1/\epsilon\) for fixed \(x\).  Substituting this
into the explicit expression for \(M_{\rm tot}\) gives the polynomial dependence
on $N, 
N_{\rm dyn}, 
\frac1\epsilon, 
N',
\frac1{\eta_{\rm stab}}, 
\log\frac1\delta$ for fixed \(x\).  If the probe design further certifies
\begin{align}
\eta_{\rm stab}^{-1}
  \le
  \textbf{exp}
  \left[
\textbf{poly}
\left(
  x,
  N,
  N_{\rm dyn},
  \frac1\epsilon
\right)
  \right],
\end{align}
then substituting this bound yields
\begin{align}
M_{\rm tot}
  =
  \textbf{exp}
  \left[
\textbf{poly}
\left(
  x,
  N,
  N_{\rm dyn},
  \frac1\epsilon,\log (N'+1)
\right)
  \right]
  \textbf{poly}
  \left(
\log\frac1\delta
  \right),
\end{align}
again for fixed \(x\).  When \(x\) is not fixed, the factor
\begin{align}
\left(
  \frac{eN'}{x\eta_{\rm stab}^{2}}
  +
  2e
\right)^x
\end{align}
must remain explicit.  This completes the proof.
\end{proof}

In summary, the unitary learning protocol with unknown coherent probe state consists of the following steps.

\begin{enumerate}
\item[(STU1')] \emph{Probe design and cutoffs.}
Fix the number of modes \(x\), the photon-number budget \(N\), the target
accuracy \(0<\epsilon<1\), and the failure probability \(0<\delta<1\).  Define
the input cutoff $q
  :=
  \left\lceil
\frac{256N}{\epsilon^2}
  \right\rceil$. Choose an output cutoff \(r\ge q\) satisfying $r
  \ge
  \left\lceil
\frac{256N_{\rm dyn}}{\epsilon^2}
  \right\rceil$, where \(N_{\rm dyn}\) is a dynamical photon-number bound for the gauge-fixed
unitary, as in Theorem~\ref{thm:xmode-main}.  Set $d
  :=
  d_x(r)
  =
  \binom{x+r}{x}$. Choose \(d\) coherent probe amplitudes $\bs \alpha_1,\ldots,\bs \alpha_d\in\mathbb C^x$ such that the centered amplitudes $\bs v_k
  :=
  \bs \alpha_k-\bs \alpha_1$ span \(\mathbb C^x\), satisfy the non-aliasing condition $\operatorname{Im}(\bs v_i^\dagger\bs v_j)
  \in
  (-\pi+\zeta,\pi-\zeta)$ for $2\le i,j\le d$, and form a unisolvent set for holomorphic polynomials of total degree at most
\(r\), after fixing the coherent gauge.  A convenient way to guarantee this is
to use a generic total-degree interpolation design, scaled small enough so that
the Bargmann phases do not cross the \(\pm\pi\) branch cut.

The purpose of this step is to ensure that the multi-mode coherent
Vandermonde matrix is invertible and that the Bargmann phases can be unwrapped
unambiguously.

\item[(STU2')] \emph{State tomography of the output probes.}
For each \(k=1,\ldots,d\), prepare \(M_k\) independent copies of the coherent
probe state \(|\underline{\bs \alpha_k}\rangle\), send them through the unknown unitary
\(W_\star\), and obtain $|\psi_k\rangle
  :=
  W_\star|\underline{\bs \alpha_k}\rangle$ and $\rho_k
  :=
  |\psi_k\rangle\langle\psi_k|$.

Perform \(x\)-mode continuous-variable pure-state tomography on each output
state.  The tomography procedure returns estimates $\widetilde\rho_k
  =
  |u_k\rangle\langle u_k|$ for $1\le k\le d$, where \(|u_k\rangle\) is an arbitrary normalized representative, and the desired
tomography-good event is
\begin{align}
\frac12
\|\widetilde\rho_k-\rho_k\|_1
  \le
  \eta_{\rm stab},
  \qquad
1\le k\le d.
\end{align}

If the training outputs satisfy the photon-number bound $\langle \psi_k|\hat n_x |\psi_k\rangle
  \le
  N'$ with $1\le k\le d$, then one may choose $M_k
  =
  \left\lceil
  2^{21}
  \left(
\frac{eN'}{x\eta_{\rm stab}^{2}}
+
2e
  \right)^x
  \eta_{\rm stab}^{-2}
  \log\frac{4d}{\delta}
  \right\rceil$. By a union bound, all \(d\) tomography calls succeed simultaneously with
probability at least \(1-\delta\).

\item[(STU3')] \emph{Estimate pairwise distances and Bargmann phases.}
From the reconstructed density matrices, estimate the pairwise transition
probabilities by
\begin{align}
\widehat F_{ij}
  :=
  \operatorname{Tr}(\widetilde\rho_i\widetilde\rho_j),
  \qquad
1\le i<j\le d.
\end{align}
Then define the estimated Euclidean distances $\widehat d_{ij}
  :=
  \sqrt{-\log \widehat F_{ij}}$. These distances alone would reconstruct the probe amplitudes only up to $\mathbb R^{2x}\rtimes O(2x)$, which is not a physical coherent-unitary gauge. Therefore, in addition to the pairwise overlaps, estimate the Bargmann triple
products
\begin{align}
\widetilde B_{ij1}
  :=
  \operatorname{Tr}(\widetilde\rho_i\widetilde\rho_j\widetilde\rho_1),
  \qquad
2\le i<j\le d.
\end{align}
In the noiseless case, we have  $\arg B_{ij1}
  =
  \operatorname{Im}
  \left(
(\bs \alpha_i-\bs \alpha_1)^\dagger
(\bs \alpha_j-\bs \alpha_1)
  \right)
  \quad
  \operatorname{mod} 2\pi$. Using the non-aliasing condition, unwrap the phase by choosing the unique branch
in $(-\pi+\zeta,\pi-\zeta)$.

Construct the estimated Hermitian Gram matrix $\widehat H
  =
  (\widehat H_{ij})_{2\le i,j\le d}$ as follows.  For \(2\le i<j\le d\), set $\operatorname{Re}\widehat H_{ij}
  =
  \frac12
  \left(
\widehat d_{i1}^2
+
\widehat d_{j1}^2
-
\widehat d_{ij}^2
  \right)$, and $\operatorname{Im}\widehat H_{ij}
  =
  \operatorname{unwrap}_{(-\pi+\zeta,\pi-\zeta)}
  \left(
\arg \widetilde B_{ij1}
  \right)$. For the diagonal, set $\widehat H_{ii}
  :=
  \widehat d_{i1}^2$ with $2\le i\le d$, and for \(j<i\), set $\widehat H_{ji}
  :=
  \overline{\widehat H_{ij}}$.

This is the key multi-mode replacement of the single-mode distance-only
trilateration step.  The imaginary parts from the Bargmann phases remove the
conjugation/reflection branch.

\item[(STU4')] \emph{Gauge-fixed input amplitude reconstruction.}
Factor the estimated Hermitian Gram matrix in the fixed local chart used in
Lemma~\ref{lem:xmode-stable-gauge}.  That is, choose a rank-\(x\) factorization $\widehat H
  \approx
  \widehat Z^\dagger\widehat Z$ and $\widehat Z\in\mathbb C^{x\times(d-1)}$. Let the columns of \(\widehat Z\) be $\bs{\widehat\beta}_2,\ldots,\bs{\widehat\beta}_d$, and set $\bs{\widehat\beta}_1
  :=
  0$.

In the noiseless case, this produces amplitudes satisfying $\bs \beta_k
  =
  R(\bs \alpha_k-\bs \alpha_1)$ with $1\le k\le d$, for some \(R\in {\rm U}(x)\).  Equivalently, there is a physical coherent gauge $G
  =
  D(\bs\gamma)\Gamma(R)$ with $\bs\gamma
  =
  -R\bs \alpha_1$, such that
\begin{align}
G|\underline{\bs \alpha_k}\rangle
  =
  e^{i\chi_k}|\underline{\bs \beta_k}\rangle.
\end{align}
Thus the remaining gauge is physically implementable: $\mathbb C^x\rtimes {\rm U}(x)$, namely displacement plus passive linear optics.  No extra reflection branch has
to be run, because the Bargmann phases have already selected the complex-linear
branch.

\item[(STU5')] \emph{Phase synchronization of output representatives.}
The vectors \(|u_k\rangle\) obtained from pure-state tomography are defined only
up to individual phases.  We synchronize them using the vacuum anchor
\(\bs{\widehat\beta}_1=0\).

Set $|\widehat\phi_1\rangle
  :=
  |u_1\rangle$. For \(k\ge2\), compute $h_k
  :=
  \langle u_1|u_k\rangle$. Whenever \(h_k\neq0\), define $|\widehat\phi_k\rangle
  :=
  e^{-i\arg h_k}|u_k\rangle$. This makes the anchor overlaps $\langle \widehat\phi_1|\widehat\phi_k\rangle$ real and positive, matching the exact relation $\langle \phi_1|\phi_k\rangle
  =
  \langle \underline{\bs \beta_1}|\underline{\bs \beta_k}\rangle
  =
  e^{-\|\bs \beta_k\|_2^2/2}
  >
  0$. Under the stability condition in Lemma~\ref{lem:xmode-phase-sync}, there exists
a single common phase \(\omega\) such that $\left\|
|\widehat\phi_k\rangle
-
e^{i\omega}|\phi_k\rangle
\right\|_2
  \le
  L_{\rm ph}\eta_{\rm stab}$ for $1\le k\le d$.

\item[(STU6')] \emph{Coherent coefficient inversion and truncated operator reconstruction.}
Let $\mathcal I_r^{(x)}
  =
  \left\{
\boldsymbol n\in\mathbb N^x:
|\boldsymbol n|\le r
  \right\}$. For \(\bs z\in\mathbb C^x\), define the truncated coherent coefficient vector $c_r^{(x)}(\bs z)
  =
  \left(
e^{-\|\bs z\|_2^2/2}
\frac{\bs z^{\boldsymbol n}}{\sqrt{\boldsymbol n!}}
  \right)_{\boldsymbol n\in\mathcal I_r^{(x)}}$. Construct the estimated coherent coefficient matrix $\widehat C
  =
  \left[
c_r^{(x)}(\bs{\widehat\beta}_1)\
\cdots\
c_r^{(x)}(\bs{\widehat\beta}_d)
  \right]$.

By the unisolvency condition and the perturbation bound in
Lemma~\ref{lem:xmode-coherent-vandermonde}, \(\widehat C\) is invertible on the
tomography-good event.

Then, let $P_r
  :=
  \Pi_{\le r}^{(x)}$. Construct the output matrix $\widehat Y
  =
  \left[
P_r|\widehat\phi_1\rangle\
\cdots\
P_r|\widehat\phi_d\rangle
  \right]$. Define the reconstructed truncated operator $\widehat W_r
  :=
  \widehat Y\widehat C^{-1}$. In the noiseless case this is exactly the matrix relation $Y
  =
  W_{g,r}C+T$, where $W_{g,r}
  =
  P_rW_gP_r$ and \(T\) is the coherent tail beyond the \(r\)-photon cutoff.  In the noisy
case, Lemma~\ref{lem:xmode-truncated-operator} gives $\|e^{-i\omega}\widehat W_r-W_{g,r}\|_\infty
  \le
  B_x(\eta_{\rm stab})$. If more than \(d=d_x(r)\) probes are used, replace the inverse
\(\widehat C^{-1}\) by the Moore--Penrose pseudoinverse and use the
corresponding least-squares estimator $\widehat W_r
  =
  \widehat Y\widehat C^{+}$.

\item[(STU7')] \emph{Completion to a finite-dimensional unitary.}
Let $P_q
  :=
  \Pi_{\le q}^{(x)}$. Define the reconstructed truncated frame $\widehat A_r
  :=
  \widehat W_rP_q$. Apply the fixed local unitary-completion map from
Lemma~\ref{lem:xmode-unitary-completion}: $\widehat V_r
  :=
  \mathcal Q(\widehat A_r)$. Here \(\widehat V_r\) is a unitary on
\(\mathcal H_{\le r}^{(x)}\).  Finally, extend by the identity on the orthogonal
complement: $\widehat U
  :=
  \widehat V_r\oplus I_{>r}$ and $\widehat{\mathcal W}
  :=
  \widehat U\,\boldsymbol{\cdot}\,\widehat U^\dagger$.

Under the stability assumptions in Theorem~\ref{thm:xmode-main}, the estimator
satisfies
\begin{align}\label{eqs803a}
d_{\diamond,N}^{\rm coh,x}
(\widehat{\mathcal W},\mathcal W_\star)
  \le
  \epsilon,
\end{align}
where $d_{\diamond,N}^{\rm coh,x}(\widehat{\mathcal W},\mathcal W_\star)
:=
\inf_{\bs \gamma\in\mathbb C^x,\ R\in U(x)}
\left\|
\widehat{\mathcal W}
-
\mathcal W_\star\circ\mathcal G_{\bs \gamma,R}^\dagger
\right\|_{\diamond}^{(x,N)}$ is the coherent-gauge quotient distance, and $\mathcal G_{\bs \gamma,R}
=
G_{\bs \gamma,R}\,\boldsymbol{\cdot}\,G_{\bs \gamma,R}^\dagger, 
G_{\bs \gamma,R}=D(\bs \gamma)\Gamma(R)$ is a Gaussian unitary channel with $D(\bs \gamma)$ being a displacement operator, $\Gamma(R)$ being a passive-Gaussian unitary. Equivalently, the protocol learns the unknown unitary channel up to the
physically implementable coherent gauge $G_{\bs\gamma,R}
  =
  D(\bs\gamma)\Gamma(R)$ with $R\in U(x)$.

\item[(STU8')] \emph{Total copy complexity.}
If the output photon-number bound $\langle\psi_k|\hat n_x |\psi_k\rangle
  \le
  N'$ with $1\le k\le d$, holds, then the total number of uses of \(W_\star\) sufficient to make the
tomography-good event hold with probability at least \(1-\delta\) is
\begin{align}
M_{\rm tot}
  =
  d
  \left\lceil
  2^{21}
  \left(
\frac{eN'}{x\eta_{\rm stab}^{2}}
+
2e
  \right)^x
  \eta_{\rm stab}^{-2}
  \log\frac{4d}{\delta}
  \right\rceil .
\end{align}
Since $d
  =
  d_x(r)
  =
  \binom{x+r}{x}$, for fixed \(x\) and $r
  =
  \mathcal O\left(
\frac{N+N_{\rm dyn}}{\epsilon^2}
  \right)$, the explicit cutoff-dimensional part is polynomial in $N,$ $
N_{\rm dyn},$ $
\frac1\epsilon,$ $
N',$ $
\frac1{\eta_{\rm stab}},$ $
\log\frac1\delta$.

If the chosen probe design certifies $\eta_{\rm stab}^{-1}
  \le
  \exp
  \left[
\operatorname{poly}
\left(
  x,
  N,
  N_{\rm dyn},
  \frac1\epsilon
\right)
  \right]$, then the conservative unconditional scaling becomes
\begin{align}
M_{\rm tot}
  =
  \textbf{exp}
  \left[
\textbf{poly}
\left(
  x,
  N,
  N_{\rm dyn},
  \frac1\epsilon,\log (N'+1)
\right)
  \right]
  \textbf{poly}
  \left(
\log\frac1\delta
  \right),\label{eqs805}
\end{align}
for fixed \(x\).  If \(x\) is allowed to grow, the factor $\left(
  \frac{eN'}{x\eta_{\rm stab}^{2}}
  +
  2e
\right)^x$ should be kept explicitly.
\end{enumerate}


Finally, we apply the proposed learning protocol to learn the unitary \(U_{(x)}^\star\) in Eq.~(\ref{U_rest_t_doped}). Nevertheless, even under the assumption \(x=\kappa t=\mathcal O(1)\), the first passive-Gaussian layer \(U_{O^\star}\) may concentrate the input energy, causing both \(N\) and \(N_{\rm dyn}\) to scale linearly with the total mode number \(m\), even though the target accuracy \(\epsilon\) can be defined as half of the total error, remains independent of \(m\). Here, we address this issue with the following theorem. The theorem shows that,
under the i.i.d. random coherent-amplitude assumption, energy concentration into
any prescribed set of output modes is not a typical event.

\begin{theorem}[Vanishing probability of energy concentration]
\label{lem:selected-mode-energy-bound}
Let \(U\in \mathrm{U}(m)\) be the interferometric matrix of an \(m\)-mode passive-Gaussian unitary. 
Assume that the input coherent amplitudes \(\alpha_1,\ldots,\alpha_m\) are independent and identically distributed according to the uniform probability distribution on
\begin{align}
D_{\sqrt N}
=
\{\alpha\in\mathbb C:\ |\alpha|^2\le N\}.
\end{align}
For each output mode \(k\in[m]\), let us denote the output amplitudes $\beta_k
=
\sum_{j=1}^m U_{kj}\alpha_j$ at the $k$-th mode. Let \(\mathcal S\subseteq [m]\) be any nonempty set of selected output modes. Then, for every \(t>0\), we have 
\begin{align}
\Pr\!\left[
\max_{k\in\mathcal S} |\beta_k|^2 \ge 4N\log\!\left(\frac{4|\mathcal S|}{\delta}\right)
\right]
\le \delta .
\label{eq:selected-mode-tail}
\end{align}
\end{theorem}

\begin{proof}
Let us fix the output mode to \(k\in[m]\) and denote the real and imaginary components of its output amplitude  $\beta_k
:=
\sum_{j=1}^m U_{kj}\alpha_j$ as follows
\begin{align}
X_{k,j}:=\Re(U_{kj}\alpha_j),
\qquad
Y_{k,j}:=\Im(U_{kj}\alpha_j),
\qquad j=1,\dots,m.
\end{align}
Since \(\alpha_1,\dots,\alpha_m\) are independent, the families
\(\{X_{k,j}\}_{j=1}^m\) and \(\{Y_{k,j}\}_{j=1}^m\) are independent.

Moreover, by rotational symmetry of the uniform probability measure on \(D_{\sqrt N}\), we have $\mathbb E[\alpha_j]=0$ for $j=1,\dots,m$. Hence, we have 
\begin{align}
\mathbb E[X_{k,j}]=\mathbb E[Y_{k,j}]=0,
\qquad
j=1,\dots,m.
\end{align}
Also, for every realization of \(\alpha_j\), we have upper bounds
\begin{align}
|X_{k,j}|
\le
|U_{kj}|\,|\alpha_j|
\le
|U_{kj}|\sqrt N,
\qquad
|Y_{k,j}|
\le
|U_{kj}|\,|\alpha_j|
\le
|U_{kj}|\sqrt N.
\end{align}

Applying Hoeffding's inequality \cite{hoeffding1963probability} to the centered independent sum
\(\Re\beta_k=\sum_{j=1}^m X_{k,j}\), we obtain the following for every \(s>0\)
\begin{align}
\Pr\!\left(
|\Re\beta_k|\ge s
\right)
&=
\Pr\!\left(
\left|\sum_{j=1}^m X_{k,j}\right|\ge s
\right)
\nonumber\\
&\le
2\exp\!\left(
-\frac{2s^2}{\sum_{j=1}^m (2|U_{kj}|\sqrt N)^2}
\right)
\label{eq:real-tail-step}
\\
&=
2\exp\!\left(
-\frac{s^2}{2N}
\right).\label{eq:real-tail}
\end{align}
where the last equation follows from the fact that the \(k\)-th row of \(U\) is normalized.  Exactly the same argument gives
$\Pr\!\left(
|\Im\beta_k|\ge s
\right)
\le
2\exp\!\left(-\frac{s^2}{2N}\right)$. Further, if \(|\beta_k|\ge r\), then at least one of the two inequalities
$|\Re\beta_k|\ge \frac{r}{\sqrt 2}$ and $|\Im\beta_k|\ge \frac{r}{\sqrt 2}$ must hold. Hence, by the union bound shown in Lemma \ref{lem:union bound}, we have 
\begin{align}
\Pr\!\left(
|\beta_k|\ge r
\right)
&\le
\Pr\!\left(
|\Re\beta_k|\ge \frac{r}{\sqrt 2}
\right)
+
\Pr\!\left(
|\Im\beta_k|\ge \frac{r}{\sqrt 2}
\right)
\nonumber\\
&\le
4\exp\!\left(-\frac{r^2}{4N}\right).
\end{align}
Equivalently, we have 
\begin{align}
\Pr\!\left(
|\beta_k|^2\ge t
\right)
\le
4\exp\!\left(-\frac{t}{4N}\right),\ \ \ \forall t>0.
\label{eq:single-mode-tail}
\end{align}

Finally, applying the union bound over all \(k\in\mathcal S\) yields
\begin{align}
\Pr\!\left[
\max_{k\in\mathcal S} |\beta_k|^2 \ge t
\right]
&\le
\sum_{k\in\mathcal S}
\Pr\!\left(
|\beta_k|^2\ge t
\right)
\nonumber\\
&\le
4|\mathcal S|\exp\!\left(-\frac{t}{4N}\right).
\end{align}
To obtain Eq.~\eqref{eq:selected-mode-tail}, we set $t
=
4N\log\!\left(\frac{4|\mathcal S|}{\delta}\right)$ and the claim follows.
\end{proof}

Here, the key point is that
each selected output amplitude is a weighted sum of independent zero-mean
random variables, and the weights are normalized by unitarity. Hence the
sub-Gaussian scale of each output amplitude is controlled by $N\sum_{j=1}^m |U_{kj}|^2=N$, rather than by \(mN\). Therefore the resulting tail probability depends on the
number of selected modes \(|\mathcal S|\), but not on the total number of modes
\(m\). In particular, for fixed \(|\mathcal S|\), the selected-mode energies
remain \(\mathcal O(E)\) with high probability, even though the total input energy is
typically \(\mathcal O(mE)\).

\subsubsection{Learning of the remaining Gaussian unitary}\label{supp:Learning of the remaining Gaussian unitary_doped}

Recall that we proposed Steps (RE1') and (RE2') in \ref{supp:Identification of output mode types_doped} to  probe the unitary $U_{\rm t,rest}$ with coherent state $|\underline{\bs\alpha^{(k)}}\>$ and  identify the modes that stably output coherent states. The identified mode set $\chi_{\rm coh}$ provide output state with a trace distance error $\sqrt{\epsilon_{\rm coh}/2}$. In Steps (STU1')-(STU8') of \ref{supp:Learning of the local unitary that might not preserve coherent states_doped}, we propose to probe the unitary $U_{\rm t,rest}$ with i.i.d. coherent probe states $\{|\underline{\bs\alpha^{(k')}}\rangle\}$. Given the fact that these steps depend on which mode to implement and that steps (RE1') and (RE2') only require to estimate $\mathcal O(1)$ parameter, we propose to repeat steps (RE1') and (RE2') when learning $U_{(x)}^\star$. For the probe-level reconstruction, set  $N^\star_{\rm eff}
\coloneqq
\left(\sqrt{N^\star}+\|\bs\gamma^\star\|_2\right)^2$. For the final energy-constrained diamond-norm accounting, the protocol has a photon-number bound $N_{\rm learn}\ge xE_{\rm ng}-\frac x2$, and we may take \(N_{\rm learn}\ge N^\star_{\rm eff}\) for the selected coherent probes. As a result, we can reconstruct $\widetilde U_{(x)}^\star$ using Steps (STU1')-(STU8') such that

\begin{align}\label{eqs816}
d_{\diamond,N^\star_{\rm learn}}^{\rm coh,x}
\left({\widetilde{\mathcal U}_{(x)}^\star},\mathcal U_{(x)}^\star\right)
  \le
  \epsilon_{\rm ng},
\end{align}
where $d_{\diamond,N^\star_{\rm learn}}^{\rm coh,x}(\widetilde{\mathcal W},\mathcal W_\star)
:=
\inf_{\bs \gamma\in\mathbb C^x,\ R\in U(x)}
\left\|
\widetilde{\mathcal W}
-
\mathcal W_\star\circ\mathcal G_{\bs \gamma,R}^\dagger
\right\|_{\diamond}^{(x,N_{\rm eff})}$ denotes the coherent-gauge quotient distance, \(N^\star_{\rm eff}\) denotes the typical total photon-number bound of the selected \(x\)-mode subsystem, taking into consideration of the vanishing possibility of energy concentration (see Theorem \ref{lem:selected-mode-energy-bound}), $\mathcal G_{\bs \gamma,R}
=
G_{\bs \gamma,R}\,\boldsymbol{\cdot}\,G_{\bs \gamma,R}^\dagger, 
G_{\bs \gamma,R}=D(\bs \gamma)\Gamma(R)$ is a Gaussian unitary channel with $D(\bs \gamma)$ being a displacement operator, $\Gamma(R)$ being a passive-Gaussian unitary. 

By classically implementing a counter-rotation ${{\widetilde  U_{(x)}^{\star\dag}}}$ to the output modes of ${ U_{(x)}^\star}$ one can reconstruct $x$-mode output states 
\begin{align}
|\widetilde \Psi_{\rm final}^{(k')}\>=\widetilde U_{(x)}^{\star\dag} U_{(x)}^\star|\underline{\overline{\bs\alpha}_{(x)}^{(k')}}\>
\end{align}
up to an error 
\begin{align}
&\frac 12\left\||\widetilde \Psi_{\rm final}^{(k')}\>\<\widetilde \Psi_{\rm final}^{(k')}|-| \Psi_{\rm final}^{(k')}\>\< \Psi_{\rm final}^{(k')}|\right\|_1\nonumber \\
= &\frac 1 2 \left\|\widetilde U_{(x)}^{\star\dag} U_{(x)}^\star|\underline{\overline{\bs\alpha}_{(x)}^{(k')}}\>\<\underline{\overline{\bs\alpha}_{(x)}^{(k')}}|U_{(x)}^{\star\dag}\widetilde U_{(x)}^{\star} -D(\bs \gamma^\star)\Gamma (R^\star)|\underline{\overline{\bs\alpha}_{(x)}^{(k')}}\>\<\underline{\overline{\bs\alpha}_{(x)}^{(k')}}|\Gamma^\dag  (R^\star)D^\dag (\bs \gamma^\star)\right\|_1\\
= &\frac 1 2 \left\|U_{(x)}^\star|\underline{\overline{\bs\alpha}_{(x)}^{(k')}}\>\<\underline{\overline{\bs\alpha}_{(x)}^{(k')}}|U_{(x)}^{\star\dag} -\widetilde U_{(x)}^{\star} D(\bs \gamma^\star)\Gamma (R^\star)|\underline{\overline{\bs\alpha}_{(x)}^{(k')}}\>\<\underline{\overline{\bs\alpha}_{(x)}^{(k')}}|\Gamma^\dag  (R^\star)D^\dag (\bs \gamma^\star)\widetilde U_{(x)}^{\star\dag}\right\|_1\\
\le &\,d_{\diamond,N_{\rm learn}}^{\rm coh,x}
\left({\widetilde{\mathcal U}_{(x)}^\star},\mathcal U_{(x)}^\star\right)\\
:=&\,x\cdot\epsilon_{\rm ng}',
\end{align}
where $| \Psi_{\rm final}^{(k')}\>=D(\bs \gamma^\star)\Gamma (R^\star)|\underline{\overline{\bs\alpha}_{(x)}^{(k')}}\>$ corresponds to the output state if one can estimate and counter-rotate $\widetilde U^\star_{(x)}$ with no error, $|\underline{\overline{\bs\alpha}_{(x)}^{(k')}}\>$ refers to the input coherent state for $U^\star_{(x)}$. In the meantime, each of the remaining \(m-x\) output modes stably produce approximate coherent states with trace-distance error at most \(\sqrt{\epsilon_{\rm cov}/2}\) (see Eq.~(\ref{eqs593})). The remaining unitary takes the following form: 
\begin{align}
\widetilde U_{\rm t,final}&:= \left[\widetilde U_{(x)}^{\star\dag}\otimes I_{(m-x)}\right] U_{\rm t,rest}\\
&\simeq \left(D(\bs \gamma^\star)\Gamma (R^\star)\otimes D_{(m-x)}^\star \right)U_{O^\star}
\end{align}
which is a special case of the structured unitary $U_{\rm final}$ in Eq. (\ref{eqs251}). Therefore, we can apply the efficient learning protocol introduced in the proof of Proposition \ref{thm:structured-product-output-learning}. The corresponding query complexity is given in Eq. (\ref{eq:corrected-query-bound-factorized}).

\subsection{Summary of the learning protocol}\label{supp:Summary of the learning protocol-doped}

The learning protocol for $t$-doped unitaries (see Eq. (\ref{supp_def:U_doped}) and its equivalent form in Eq. (\ref{supp_def:U_doped_three_leyer})) is summarized as follows.  

\begin{enumerate}
\item [Step 1] \emph{Learning of the Gaussian layer.} Apply Steps (GD1')-(GD3') to learn and physically counter-rotate the final Gaussian layer with  $U_{\widetilde S_{\bs\alpha^{(\ell)}} }^\dag$ in the follow-up steps. By Lemma~\ref{lem:Gaussian-disentangling for different input}, the output state remains a generalized passive-separable state even when the probe state is changed.
\item[Step 2] \emph{Learning of the second passive-Gaussian layer.} Apply the shadow tomography protocol of Ref.~\cite{zhao2025complexity} to the generalized passive-separable state. In this step, we learn a passive-Gaussian unitary $U_{\widetilde O_{2,\bs\alpha^{(\ell),\bs\alpha^{(h)}}}}$. In the following steps, we always rotate the heterodyne measurement outcomes to effectively implement $U_{\widetilde O_{2,\bs\alpha^{(\ell),\bs\alpha^{(h)}}}}^\dag$. After the counter-rotation, the resulting state stably factorizes into an $x$-mode state and $(m-x)$ coherent states, even when the probe state is changed. The learning problem therefore reduces to learning the unitary $U_{\rm t,rest}:=\left[U_{(x)}^\star\otimes D_{(m-x)}^\star\right]U_{O^\star}$.

\item[Step 3] \emph{Learning of the first passive-Gaussian layer.} Apply Steps~(STU1')--(STU8'), while repeating Steps~(RE1')--(RE2') for each probe, to reconstruct an $x$-mode unitary $\widetilde U_{(x)}^\star$ such that $\widetilde U_{\rm t,final}:=\left[\widetilde U_{(x)}^{\star\dag}\otimes I_{(m-x)}\right]U_{\rm t,rest}$ converges to a Gaussian unitary.

\item[Step 4] \emph{Learning of the remaining Gaussian unitary.} Apply the efficient learning protocol introduced in the proof of Proposition \ref{thm:structured-product-output-learning}.

\item[Step 5] \emph{Assemble.} 
Return the reconstructed overall unitary 
$\widetilde{U}_{\rm doped}=U_{\widetilde S_{\bs\alpha^{(\ell)}} }U_{\widetilde O_{2,\bs\alpha^{(\ell),\bs\alpha^{(h)}}}}\left[\widetilde U_{(x)}^{\star}\otimes I_{(m-x)}\right]\widetilde U_{\rm t,final}$.
\end{enumerate}

\subsection{A brief proof that the query complexity grows polynomially}
\label{supp:query complexity of t doped unitary}

We now analyze the query complexity of the learning protocol for the
$t$-doped Gaussian unitary, under the energy constraints in Eq.~(\ref{supp:three energy constraints t doped}). We emphasize that the dependence on \(\kappa t=\mathcal O(1)\) is separated from the dependence on
the number of modes \(m\). Thus, for fixed doped width \(\kappa t\), the final query
complexity is polynomial in \(m\), the energy parameters, reconstruction error \(\epsilon\), and the function 
\(\log(1/\delta)\) of the failure probability.

\begin{theorem}[Query complexity of learning \(t\)-doped Gaussian unitaries]
\label{thm:query-complexity-t-doped-final}
Consider a $t$-doped unitary $U_{\rm doped}
=
U_S
\left(
U_{(\kappa t)}\otimes D_{(m-\kappa t)}
\right)
U_O$ with $\kappa t=\mathcal O(1)$.
Assume the input and output energy constraints in
Eq.~(\ref{supp:three energy constraints t doped}). Then there exists a learning protocol that returns a unitary \(\widetilde U_{\rm doped}\) such that
\begin{align}
\left\|
\mathcal U_{\rm doped}
-
\widetilde{\mathcal U}_{\rm doped}
\right\|_{\diamond}^{mE}
\le
\epsilon
\end{align}
with probability at least \(1-\delta\). For fixed \(\kappa t\), fixed global energy parameters
\(E,E_{\rm probe},E_{\rm II}\), and fixed block parameters
\(E_\star,E_\star',E_{\rm dyn},\epsilon,\delta\), the total number of
queries satisfies $M_{\rm tot}
\le\textbf{\em poly}\!\left(
m\right)$.
\end{theorem}

\begin{proof} Here we employ the learning protocol described in Steps 1-6. We use the perturbative error-propagation assumption stated in the learning
task: when estimating the error contributed by one step, all previous steps are
treated as exact, and the total error is bounded by the sum of the stepwise
errors. On this account, let us choose error and failure-probability bound
\begin{align}
\begin{cases}
\epsilon
&=
\epsilon_{\rm gd}
+
\epsilon_{\rm ps}
+
\epsilon_{\rm ng}
+
\epsilon_{\rm G},
\\
\delta
&=
\delta_{\rm gd}
+
\delta_{\rm ps}
+
\delta_{\rm ng}
+
\delta_{\rm G},
\end{cases}
\label{eq:t-doped-failure-budget-final}
\end{align}
where $\epsilon_{\rm gd}(/
\epsilon_{\rm ps}/
\epsilon_{\rm ng}/
\epsilon_{\rm G})$ denotes the reconstruction error, $\delta_{\rm gd}(/\delta_{\rm ps}/\delta_{\rm ng}/\delta_{\rm G})$ refers to the failure probability for the unitary learning step 1(/2/3/4), respectively. Here, the parameters $\epsilon_{\rm ng}, \epsilon_{\rm G}$, $\delta_{\rm ng}$, and $\delta_{\rm G}$ already account for the vanishing failure probability associated with the activation argument in Corollary~\ref{cor:random-coherent-block-factorization} in Step~2. The quantities $\epsilon_{\rm ng}$ and $\delta_{\rm ng}=\delta_{\rm ng}'+md\,\delta_{\rm cov}$ also take into consideration the repeatedly implementing Steps~(RE1')--(RE2') to identify the output location and scale of $U_{(x)}^\star$. In particular, the failure probability $\delta_{\rm ng}'$ is from unitary learning steps (STU1')-(STU8') with unknown coherent probe states. The repeated location identification procedure contributes a covariance-estimation failure probability $\delta_{\rm cov}$ and an identification error $\sqrt{\epsilon_{\rm cov}/2}$ for each of the $m$ modes. In total, we need to repeat this identification for $d$ rounds, where $d$ denotes the number of distinct probe states. Then, the $(m-x)$ modes that stably output approximated coherent states are used as the input to Step~4, and hence $\epsilon_{\rm cov}$ enters the Gaussian-learning error $\epsilon_{\rm G}$ through its parameter dependence. In addition to the activation theorem and the locating procedure, Theorem~\ref{lem:selected-mode-energy-bound} implies that, since the probe states for $U_{\rm doped}$ are always prepared in an i.i.d. form, the input energy seen by $U_{(x)}^\star$ remains independent of the total number of modes $m$ with constant success probability. Since this constant-probability event does not affect the asymptotic dependence on $m$, we suppress the corresponding failure probability in Eq.~\eqref{eq:t-doped-failure-budget-final}.

Now, let us prove that the reconstruction error for $U_{\rm doped}$ is bounded by $\epsilon$. In the ideal limit of Step~1, the covariance-matrix learning and Gaussian
counter-rotation produce a Gaussian unitary \(U_{S,\bs\alpha^{(\ell)}}\) such
that $U_{S,\bs\alpha^{(\ell)}}^\dagger U_{\rm doped}$ maps every later coherent probe to a generalized passive-separable state.
In the ideal limit of Step~2, the generalized passive-separable shadow
tomography reconstructs a passive-Gaussian unitary
\(U_{O_2,\bs\alpha^{(\ell)},\bs\alpha^{(h)}}\) such that $U_{\rm t,rest}
:=
U_{O_2,\bs\alpha^{(\ell)},\bs\alpha^{(h)}}^\dagger
U_{S,\bs\alpha^{(\ell)}}^\dagger
U_{\rm doped}$ has the form $U_{\rm t,rest}
=
\left[
U_{(x)}^\star
\otimes
D_{(m-x)}^\star
\right]
U_{O^\star}$ with $x\le \kappa t$. The activation and coherent-block factorization results imply that this value
of \(x\), and the corresponding coherent versus non-coherent output
partition, are stable under changing the coherent probe, except on a
Lebesgue-null set of probe amplitudes. Hence, for probes drawn from an
absolutely continuous distribution, this structural reduction holds with
probability one. Step~3 learns the \(x\)-mode unitary \(U_{(x)}^\star\), up to the natural
coherent-input gauge, and returns \(\widetilde U_{(x)}^\star\). After this
counter-rotation, the remaining unitary is $\widetilde U_{\rm t,final}
:=
\left[
\widetilde U_{(x)}^{\star\dagger}
\otimes
I_{(m-x)}
\right]
U_{\rm t,rest}$. In the noiseless limit this is a structured Gaussian unitary $U_{\rm t,final}$. The final estimator is assembled as
\begin{align}
\widetilde U_{\rm doped}
=
U_{\widetilde S_{\bs\alpha^{(\ell)}}}
U_{\widetilde O_{2,\bs\alpha^{(\ell)},\bs\alpha^{(h)}}}
\left[
\widetilde U_{(x)}^\star
\otimes
I_{(m-x)}
\right]
\widetilde U_{\rm t,final}.
\label{eq:t-doped-final-estimator}
\end{align}
To make the error propagation explicit, let us write the ideal decomposition as $\mathcal U_{\rm doped}
=
\mathcal S_{\ell}
\circ
\mathcal O_{\ell,h}
\circ
\mathcal N
\circ
\mathcal G$, with $\mathcal S_{\ell}\equiv
\mathcal U_{S_{\bs\alpha^{(\ell)}}}$, $\mathcal O_{\ell,h}\equiv
\mathcal U_{O_{2,\bs\alpha^{(\ell)},\bs\alpha^{(h)}}}$, $\mathcal N\equiv
\mathcal U_{(x)}^\star\otimes \mathcal I_{(m-x)}$, $\mathcal G
\equiv
\mathcal U_{\rm t,final}$. Similarly, the learned channel is $\widetilde{\mathcal U}_{\rm doped}
=
\widetilde{\mathcal S}_{\ell}
\circ
\widetilde{\mathcal O}_{\ell,h}
\circ
\widetilde{\mathcal N}
\circ
\widetilde{\mathcal G}$, with the evident definitions $\widetilde{\mathcal S}_{\ell}\equiv
\mathcal U_{\widetilde S_{\bs\alpha^{(\ell)}}}$, $\widetilde{\mathcal O}_{\ell,h}
\equiv
\mathcal U_{\widetilde O_{2,\bs\alpha^{(\ell)},\bs\alpha^{(h)}}}$, $\widetilde{\mathcal N}\equiv
\widetilde{\mathcal U}_{(x)}^\star\otimes \mathcal I_{(m-x)}$, $\widetilde{\mathcal G}\equiv
\widetilde{\mathcal U}_{\rm t,final}$. Let us denote the Hamiltonians as $H_m
:=
\sum_{j=1}^m a_j^\dagger a_j+\frac m2$ and $H_x
:=
\sum_{j=1}^x a_j^\dagger a_j+\frac x2$. Since the energy can change after each non-passive step, we define the exact intermediate energy bound by
\begin{align}
\begin{cases}
mE_{\rm G}
&:=
mE,
\\
xE_{\rm ng}
&:=
\sup_{\Tr(H_m\rho)\le mE}
\Tr\!\left[
H_x\,
\Tr_{m-x}
\!\left(
\mathcal G(\rho)
\right)
\right],
\\
mE_{\rm ps}
&:=
\sup_{\Tr(H_m\rho)\le mE}
\Tr\!\left[
H_m\,
(\mathcal N\circ \mathcal G)(\rho)
\right],
\\
mE_{\rm gd}
&:=
\sup_{\Tr(H_m\rho)\le mE}
\Tr\!\left[
H_m\,
(\mathcal O_{\ell,h}\circ \mathcal N\circ \mathcal G)(\rho)
\right].
\end{cases}
\end{align}
Equivalently, the four learning steps are required to satisfy
\begin{align}
\begin{cases}
\left\|
\mathcal G-\widetilde{\mathcal G}
\right\|_{\diamond}^{mE_{\rm G}}
&\le
\epsilon_{\rm G},
\\
\left\|
\mathcal U_{(x)}^\star-\widetilde{\mathcal U}_{(x)}^\star
\right\|_{\diamond}^{xE_{\rm ng}}
&\le
\epsilon_{\rm ng},
\\
\left\|
\mathcal O_{\ell,h}-\widetilde{\mathcal O}_{\ell,h}
\right\|_{\diamond}^{mE_{\rm ps}}
&\le
\epsilon_{\rm ps},
\\
\left\|
\mathcal S_{\ell}-\widetilde{\mathcal S}_{\ell}
\right\|_{\diamond}^{mE_{\rm gd}}
&\le
\epsilon_{\rm gd}.
\end{cases}
\end{align}
Here the \((m-x)\)-mode system is treated as a reference system in the
\(x\)-mode bound for \(\mathcal U_{(x)}^\star\).

Now take an arbitrary input-reference state \(\rho_{AR}\) satisfying $\Tr\!\left[(H_m\otimes I_R)\rho_{AR}\right]\le mE$. Using the exact telescoping identity for the difference of composed channels,
we obtain
\begin{align}
&
\left\|
\left[
(\mathcal U_{\rm doped}-\widetilde{\mathcal U}_{\rm doped})
\otimes \mathcal I_R
\right](\rho_{AR})
\right\|_1
\nonumber\\
\le\;&
\left\|
\left[
(\mathcal S_{\ell}-\widetilde{\mathcal S}_{\ell})
\circ
\mathcal O_{\ell,h}
\circ
\mathcal N
\circ
\mathcal G
\otimes \mathcal I_R
\right](\rho_{AR})
\right\|_1
\nonumber\\
&+
\left\|
\left[
\widetilde{\mathcal S}_{\ell}
\circ
(\mathcal O_{\ell,h}-\widetilde{\mathcal O}_{\ell,h})
\circ
\mathcal N
\circ
\mathcal G
\otimes \mathcal I_R
\right](\rho_{AR})
\right\|_1
\nonumber\\
&+
\left\|
\left[
\widetilde{\mathcal S}_{\ell}
\circ
\widetilde{\mathcal O}_{\ell,h}
\circ
(\mathcal N-\widetilde{\mathcal N})
\circ
\mathcal G
\otimes \mathcal I_R
\right](\rho_{AR})
\right\|_1
\nonumber\\
&+
\left\|
\left[
\widetilde{\mathcal S}_{\ell}
\circ
\widetilde{\mathcal O}_{\ell,h}
\circ
\widetilde{\mathcal N}
\circ
(\mathcal G-\widetilde{\mathcal G})
\otimes \mathcal I_R
\right](\rho_{AR})
\right\|_1 .
\label{eq:t-doped-telescoping-error}
\end{align}
All channels appearing to the left of a difference term are unitary channels and
therefore preserve trace norm. Hence, using the corresponding intermediate
energy bounds defined above, Eq.~(\ref{eq:t-doped-telescoping-error}) gives
\begin{align}
&
\left\|
\left[
(\mathcal U_{\rm doped}-\widetilde{\mathcal U}_{\rm doped})
\otimes \mathcal I_R
\right](\rho_{AR})
\right\|_1
\nonumber\\
\le\;&
\left\|
\mathcal S_{\ell}
-
\widetilde{\mathcal S}_{\ell}
\right\|_{\diamond}^{mE_{\rm gd}}
+
\left\|
\mathcal O_{\ell,h}
-
\widetilde{\mathcal O}_{\ell,h}
\right\|_{\diamond}^{mE_{\rm ps}}
\nonumber\\
&+
\left\|
\mathcal U_{(x)}^\star
-
\widetilde{\mathcal U}_{(x)}^\star
\right\|_{\diamond}^{xE_{\rm ng}}
+
\left\|
\mathcal G
-
\widetilde{\mathcal G}
\right\|_{\diamond}^{mE_{\rm G}}
\nonumber\\
\le\;&
\epsilon_{\rm gd}
+
\epsilon_{\rm ps}
+
\epsilon_{\rm ng}
+
\epsilon_{\rm G}.
\end{align}
Here, the parameter $\epsilon_{\rm gd}$ is achieved by estimating the covariance matrix and applying the Gaussian
Solovay–Kitaev theorem \cite{becker2021energy}, $\epsilon_{\rm ps}$ is achieved by the shadow tomography protocol introduced in Ref. \cite{zhao2025complexity}, $\epsilon_{\rm ng}$ is given by Eq. \ref{eqs816} after Steps (STU1')-(STU8'), $\epsilon_{\rm G}$ is given by the efficient learning protocol introduced in the proof of Proposition \ref{thm:structured-product-output-learning}.

Taking the supremum over all reference systems \(R\) and all admissible
\(\rho_{AR}\) with input energy at most \(mE\), we conclude that
\begin{align}
\left\|
\mathcal U_{\rm doped}
-
\widetilde{\mathcal U}_{\rm doped}
\right\|_{\diamond}^{mE}
\le
\epsilon_{\rm gd}
+
\epsilon_{\rm ps}
+
\epsilon_{\rm ng}
+
\epsilon_{\rm G}
=
\epsilon .
\label{eq:t-doped-error-sum}
\end{align}
Here \(\epsilon_{\rm coh}\) denotes the contribution of the approximate
coherent-mode identification. 

It remains to bound the success probability. The exceptional probe choices in
the activation form sets of Lebesgue
measure zero, and hence contribute no failure probability under absolutely
continuous coherent-probe sampling. The finite-sample failures arise only from
the four reconstruction steps and the covariance-based coherent-mode test.
Therefore, by the union bound, we have 
\begin{align}
\Pr\!\left[
\left\|
\mathcal U_{\rm doped}
-
\widetilde{\mathcal U}_{\rm doped}
\right\|_{\diamond}^{mE}
\le
\epsilon
\right]
\ge
1
-
\delta_{\rm gd}
-
\delta_{\rm ps}
-
\delta_{\rm ng}
-
\delta_{\rm G}
=
1-\delta .
\label{eq:t-doped-success-probability}
\end{align}

We now count queries. Step~1 is covariance-matrix learning followed by
Gaussian counter-rotation. As in the Gaussian-entanglable case, the required
number of queries is
\begin{align}
M_{\rm gd}
=
\textbf{poly}\!\left(
m,\,
E_{\rm II},\,
\frac{1}{\epsilon_{\rm gd}},\,
\log\frac{1}{\delta_{\rm gd}}
\right),
\label{eq:t-doped-Mgd}
\end{align}
where 
\begin{align}
mE_{\rm II}\ge \max_{\substack{|\underline{\bs\alpha}\>,\bs\alpha\in \C^m \\
\|\bs\alpha\|_2^2\le m(E_{\rm probe}-\frac12)}}\sqrt{\Tr\left[\left(\sum_{j=1}^m a^\dag_j a_j +\frac m 2 \right)^2 U_{\rm doped}|\underline{\bs\alpha}\>\<\underline{\bs\alpha}| U_{\rm doped}^\dag \right]}
\end{align}
denotes the output square energy of $U_{\rm doped}$ (see \cite{mele2025learning}). Step~2 is generalized passive-separable shadow tomography. The input
non-coherent block has size at most \(\kappa t\), so its cost is (see \cite{zhao2025complexity})
\begin{align}
M_{\rm ps}
=\textbf{exp}\!\left\{\mathcal O\left[\kappa t\left(1+ \log m +\log (E_{\rm II}+1)+\log 
\frac{1}{\epsilon_{\rm ps}}\right) \right]\right\}\textbf{poly}\!\left(\log\frac{1}{\delta_{\rm ps}}
\right).
\label{eq:t-doped-Mps}
\end{align}
The coherent-mode identification steps (RE1') and (RE2') estimate only second moments
for each output mode. Using Eq.~(\ref{eqs595}), its cost is
\begin{align}
M_{\rm cov}
=
\textbf{poly}\!\left(E_{\rm II},\,
\frac{1}{\epsilon_{\rm coh}},\,
\log\frac{1}{\delta_{\rm cov}}
\right),
\label{eq:t-doped-Mcov}
\end{align}
where $\epsilon_{\rm cov}=\mathcal O(\epsilon_{\rm ng})$ and $\delta_{\rm cov}=\mathcal O(\delta_{\rm ng})$ refers to the error and failure probability of single-mode coherent state identification procedure, respectively. Then, Step~3 learns the \(x\)-mode unitary \(U_{(x)}^\star\) from unknown coherent
probes. The number of queries for this step has the form
\begin{align}
M_{\rm ng}&=M_{\rm ng}'+ md \cdot M_{\rm cov}\\
&=\textbf{exp}
  \left[
\textbf{poly}
\left(
  x,
  E_{\rm *},
  E_{\rm dyn},
  \frac1{\epsilon_{\rm ng}},\log (E_\star'+1)
\right)
  \right]
  \textbf{poly}
  \left(m,
\log\frac1{\delta_{\rm ng}}
  \right),
\label{eq:t-doped-Mng}
\end{align}
where $M_{\rm ng}'$ is the sample complexity for Steps (STU1')-(STU8') (see Corollary \ref{cor:xmode-copy-complexity}), $d=\textbf{exp}(\mathcal O(x))=\textbf{poly}(E_\star,E_{\rm dyn},1/\epsilon^2_{\rm ng})$ denotes the number of distinct probe state used in these steps, $E_\star(E_\star')$ denotes the typical input(/output) energy per mode for $U_{(x)}^\star$ (see Theorem~\ref{lem:selected-mode-energy-bound}), $E_{\rm dyn}$ denotes the total output energy of $U_{(x)}^\star$ for inputs supported on the truncated subspace with maximal photon number
$256x(E_\star-1/2)/\epsilon_{\rm ng}^2$ (see Lemma \ref{lem:xmode-cutoff}).

Finally, Step~4 learns the remaining structured Gaussian unitary, giving (see Eq. (\ref{eq:corrected-query-bound-factorized}))
\begin{align}
M_{\rm final}
=
\textbf{poly}\!\left(
m,\,
E,E_\star,\,
\frac{1}{\epsilon_{\rm G}},\,
\log\frac{1}{\delta_{\rm G}}
\right).
\label{eq:t-doped-Mfinal}
\end{align}

Combining Eqs.~(\ref{eq:t-doped-Mgd})--(\ref{eq:t-doped-Mfinal}) and selecting all stepwise errors and failure probabilities to
be constant fractions of \(\epsilon\) and \(\delta\): $\epsilon/4
=
\epsilon_{\rm gd}
=
\epsilon_{\rm ps}
=
\epsilon_{\rm ng}
=
\epsilon_{\rm G}, \, 
\delta/4=
\delta_{\rm gd}
=
\delta_{\rm ps}
=
\delta_{\rm ng}
=
\delta_{\rm G}$, we obtain
\begin{align}
M_{\rm tot}
=&
M_{\rm gd}
+
M_{\rm ps}
+
M_{\rm ng}
+
M_{\rm final}
\nonumber\\
\le &
\textbf{poly}\!\left(
m,\,
E_{\rm II},\,
\frac{1}{\epsilon},\,
\log\frac{1}{\delta}
\right)
+
\textbf{exp}\!\left\{\mathcal O\left[\kappa t\left(1+ \log m +\log (E_{\rm II}+1)+\log 
\frac{1}{\epsilon}\right) \right]\right\}\textbf{poly}\!\left(\log\frac{1}{\delta}
\right)\nonumber \\
&+ \textbf{exp}
  \left[
\textbf{poly}
\left(
  x,
  E_{\rm *},
  E_{\rm dyn},
  \frac1{\epsilon},\log (E_\star'+1)
\right)
  \right]
  \textbf{poly}
  \left(m,
\log\frac1{\delta}
  \right)\nonumber \\
  &+ \textbf{poly}\!\left(
m,\,
E,E_\star,\,
\frac{1}{\epsilon},\,
\log\frac{1}{\delta}
\right).
\end{align}
We can further simplify the sample complexity using the relations  
\begin{align}
\begin{cases}
x\le \kappa t=\mathcal O(1)\\
E_\star' \le E_{\rm II}
\end{cases}
\end{align}
where the second inequality  follows from Lemma S22 of \cite{zhao2025complexity}, the third line is obtained from Theorem \ref{lem:selected-mode-energy-bound}. In addition, we know that $E_\star$ is independent of $m$. In terms of the scaling of mode number $m$, we have 
\begin{align}
M_{\rm tot}
=
\textbf{poly}\!\left(
m\right),
\end{align}
which proves the claimed polynomial query complexity for fixed doped width.
\end{proof}

\section{Learning of a general unitary}\label{sec:coh-het-unitary}

In this Supplementary Note, we introduce a unitary learning protocol based on coherent state probing and heterodyne measurement. In particular, we first describe the learning task. Then, we demonstrate the estimator built from heterodyne measurement result. Further, we analyze the reconstruction error led by our proposed protocol. 

\subsection{Set-up}\label{subsec:setup-why-many}

\subsubsection{Task}

Let $\mathcal U(\rho)\coloneqq U\rho\,U^\dagger$ be an arbitrary unknown $m$-mode unitary channel on the bosonic Hilbert space $\mathcal H$. Without loss of generality, we evaluate the learning error in the energy-constrained diamond norm
\begin{equation}
\label{eq:ecd-setup}
\|\mathcal V\|_{\diamond}^{mE}
\coloneqq
\sup_{\Tr[(\widehat H_m\otimes I)\rho]\le mE}
\big\|(\mathcal V\otimes {\rm id})(\rho)\big\|_1,
\qquad
H_m=\sum_{s=1}^m a_s^\dagger a_s+\frac m2,
\end{equation}
where \(E\) is the per-mode physical input-energy upper bound.

The learning task is naturally split into two steps.
First, under the energy constraint \eqref{eq:ecd-setup}, one replaces the infinite-dimensional channel $\mathcal U$ by an effective finite-dimensional model.
Fix a total-photon cutoff $L\in\mathbb N$, and denote the new Hilbert space by
\begin{equation}
\label{eq:cutoff-space-setup}
\mathcal H_{\le L}
\coloneqq
{\rm span}\bigl\{\ket{\bm n}\,:\,\bm n\in\mathbb N^m,\ \|\bm n\|_1\le L\bigr\},
\qquad
\mathcal J_L
\coloneqq
\bigl\{\bm n\in\mathbb N^m\,:\,\|\bm n\|_1\le L\bigr\},
\end{equation}
with projector $\Pi_L$ onto $\mathcal H_{\le L}$.
Later, using the effective-dimension theorem \cite{arzani2025effective}, we will choose $L$ and a unitary $V_L$ acting on $\mathcal H_{\le L}$ such that some full-space unitary extension
\begin{equation}
\label{eq:V-extension-setup}
V = V_L \oplus V_L^\perp
\end{equation}
satisfies the following condition 
\begin{equation}
\label{eq:ecd-trunc-setup}
\|\mathcal U-\mathcal V\|_{\diamond}^{mE}\le \epsilon_{\rm trunc},
\qquad
\mathcal V(\rho)\coloneqq V\rho V^\dagger .
\end{equation}

Once this reduction is made, the remaining task is to learn the finite-dimensional unitary matrix
\begin{equation}
\label{eq:target-matrix-elements-setup}
(V_L)_{{\bm j}{\bm n}}
\coloneqq
\langle {\bm j}|V_L|{\bm n}\rangle,
\qquad
{\bm j},{\bm n}\in\mathcal J_L,
\end{equation}
up to a global phase.
Its dimension is
\begin{equation}
\label{eq:D-setup}
D\coloneqq \dim(\mathcal H_{\le L}) = |\mathcal J_L| = \binom{L+m}{m}.
\end{equation}
Accordingly, the truncated learning problem is a $D$-dimensional unitary-learning problem.

\subsubsection{Probe states and measurement}

We use multimode coherent probe states
\begin{equation}
\label{eq:coh-probe-setup}
\ket{\underline{\bm\alpha}}
\coloneqq
\bigotimes_{s=1}^m \ket{\underline{\alpha_s}},
\qquad
\bm\alpha=(\alpha_1,\ldots,\alpha_m)\in\mathbb C^m.
\end{equation}
Their total photon number is $\langle\underline{ \bm\alpha}|\widehat N|\underline{\bm\alpha}\rangle = \|\bm\alpha\|_2^2$. Thus the corresponding physical energy is
\(\|\bm\alpha\|_2^2+\frac m2\), and the probe respects the input-energy constraint whenever $\|\bm\alpha\|_2^2\le m\left(E-\frac12\right)$.

On the output, we perform heterodyne detection with POVM
\begin{equation}
\label{eq:heterodyne-povm-setup}
\Pi(\bm\beta)=\pi^{-m}\ket{\underline{\bm\beta}}\!\bra{\underline{\bm\beta}},
\qquad
\bm\beta=(\beta_1,\ldots,\beta_m)\in\mathbb C^m.
\end{equation}
Equivalently, a single shot produces the real vector
\begin{equation}
\label{eq:heterodyne-data-setup}
\bm y_{\bm\alpha}^{(\ell)}
=
\bigl(
2\Re\beta_{\bm\alpha,1}^{(\ell)},
2\Im\beta_{\bm\alpha,1}^{(\ell)},
\ldots,
2\Re\beta_{\bm\alpha,m}^{(\ell)},
2\Im\beta_{\bm\alpha,m}^{(\ell)}
\bigr)^{\mathsf T}\in\mathbb R^{2m},
\end{equation}
where $\ell$ labels the measurement round.

For the finite-dimensional reconstruction step, it is convenient to introduce, for each ${\bm j},{\bm k}\in\mathcal J_L$, the cutoff matrix element
\begin{equation}
\label{eq:rho-jk-setup}
\rho_{{\bm j}{\bm k}}(\bm\alpha)
\coloneqq
\langle {\bm j}|
\mathcal V(\ket{\underline{\bm\alpha}}\!\bra{\underline{\bm\alpha}})
|{\bm k}\rangle .
\end{equation}
Following the standard coherent-state expansion, define also the rescaled quantity
\begin{equation}
\label{eq:g-jk-setup}
g_{{\bm j}{\bm k}}(\bm\alpha,\bar{\bm\alpha})
\coloneqq
e^{\|\bm\alpha\|_2^2}\rho_{{\bm j}{\bm k}}(\bm\alpha).
\end{equation}
Applying the definitions 
\begin{equation}
\label{eq:coh-fock-expansion-setup}
\ket{\underline{\bm\alpha}}
=
e^{-\|\bm\alpha\|_2^2/2}
\sum_{{\bm n}\in\mathbb N^m}
\frac{\bm\alpha^{\bm n}}{\sqrt{{\bm n}!}}
\ket{\bm n},
\qquad
\bm\alpha^{\bm n}\coloneqq \prod_{s=1}^m \alpha_s^{n_s},
\qquad
{\bm n}!\coloneqq \prod_{s=1}^m n_s!,
\end{equation}
one obtains the expression 
\begin{equation}
\label{eq:g-expansion-setup}
g_{{\bm j}{\bm k}}(\bm\alpha,\bar{\bm\alpha})
=
\sum_{{\bm n},{\bm n}'\in\mathcal J_L}
E_{{\bm j}{\bm k}}^{{\bm n}{\bm n}'}
\frac{\bm\alpha^{\bm n}\bar{\bm\alpha}^{{\bm n}'}}{\sqrt{{\bm n}!\,{\bm n}'!}},
\end{equation}
where
\begin{equation}
\label{eq:process-tensor-full-setup}
E_{{\bm j}{\bm k}}^{{\bm n}{\bm n}'}
\coloneqq
\langle {\bm j}|V_L|{\bm n}\rangle
\langle {\bm n}'|V_L^\dagger|{\bm k}\rangle .
\end{equation}
In particular, the slice with ${\bm n}'={\bm 0}$ factorizes as
\begin{equation}
\label{eq:process-slice-factorization-setup}
E_{{\bm j}{\bm k}}^{{\bm n}{\bm 0}}
=
(V_L)_{{\bm j}{\bm n}}\,(V_L)_{{\bm k}{\bm 0}}^{*},
\end{equation}
which will later allow us to reconstruct the whole unitary matrix from one reference column.

\subsubsection{Why multiple coherent probes are necessary}

A single probe state cannot determine an arbitrary $D$-dimensional unitary.
Indeed, up to a global phase, the target unitary channel on $\mathcal H_{\le L}$ has $D^2-1$ real degrees of freedom, whereas the output generated by one fixed coherent probe is a single pure state, which carries only $2D-2$ real degrees of freedom.
Therefore one must use a family of different probe amplitudes.

More importantly, \eqref{eq:g-expansion-setup} shows that the relevant quantity
$g_{{\bm j}{\bm k}}(\bm\alpha,\bar{\bm\alpha})$
depends on both $\bm\alpha$ and $\bar{\bm\alpha}$.
Hence it is, in general, not holomorphic as a function of $\bm\alpha$ alone.
For this reason, sampling only along a real line or only on a purely real grid is not sufficient to isolate the coefficients
$E_{{\bm j}{\bm k}}^{{\bm n}{\bm 0}}$.
To separate the monomials
$\bm\alpha^{\bm n}\bar{\bm\alpha}^{{\bm n}'}$,
one must vary both quadratures of each mode, equivalently, use coherent amplitudes with different complex phases in a neighborhood of the origin.

Accordingly, in the reconstruction stage we will choose a finite probe set
\begin{equation}
\label{eq:probe-set-setup}
\mathcal A_L\subset \mathbb C^m
\end{equation}
such that the associated interpolation matrix for the monomials
\begin{equation}
\label{eq:monomial-family-setup}
\Bigl\{
\bm\alpha^{\bm n}\bar{\bm\alpha}^{{\bm n}'}
\,\Big|\,
{\bm n},{\bm n}'\in\mathcal J_L
\Bigr\}
\end{equation}
has full column rank on $\mathcal A_L$.
This guarantees that finitely many phase-resolved coherent probes suffice to recover the required coefficients of \eqref{eq:g-expansion-setup}.
The explicit construction of such a probe set and the resulting estimators will be given in the next subsection.

\subsection{Reconstruction of the truncated unitary}\label{subsec:reconstruction}

In this subsection, we reconstruct the finite-dimensional unitary $V_L$ on the cutoff space $\mathcal H_{\le L}$ from coherent probes and heterodyne data.
As explained in the previous subsection, the key quantity is
\begin{equation}
\label{eq:g-expansion-recon}
g_{{\bm j}{\bm k}}(\bm\alpha,\bar{\bm\alpha})
=
\sum_{{\bm n},{\bm n}'\in\mathcal J_L}
E_{{\bm j}{\bm k}}^{{\bm n}{\bm n}'}
\frac{\bm\alpha^{\bm n}\bar{\bm\alpha}^{{\bm n}'}}{\sqrt{{\bm n}!\,{\bm n}'!}},
\qquad
{\bm j},{\bm k}\in\mathcal J_L,
\end{equation}
where
\begin{equation}
\label{eq:E-def-recon}
E_{{\bm j}{\bm k}}^{{\bm n}{\bm n}'}
\coloneqq
\langle {\bm j}|V_L|{\bm n}\rangle
\langle {\bm n}'|V_L^\dagger|{\bm k}\rangle.
\end{equation}
Since \eqref{eq:g-expansion-recon} depends jointly on $\bm\alpha$ and $\bar{\bm\alpha}$, the reconstruction problem is naturally a finite-dimensional linear inversion problem for the coefficients
$\{E_{{\bm j}{\bm k}}^{{\bm n}{\bm n}'}\}$.

\subsubsection{Estimator of the process tensor by interpolation}

Fix a finite probe set
\begin{equation}
\label{eq:probe-set-recon}
\mathcal A_L=\bigl\{\bm\alpha^{(r)}\in\mathbb C^m\bigr\}_{r=1}^{M},
\qquad
M\ge D^2,
\qquad
D=|\mathcal J_L|=\binom{L+m}{m},
\end{equation}
and assume that the associated interpolation matrix has full column rank.
More explicitly, define
\begin{equation}
\label{eq:design-matrix-recon}
\Phi\in\mathbb C^{M\times D^2},
\qquad
\Phi_{r,({\bm n},{\bm n}')}
\coloneqq
\frac{\bigl(\bm\alpha^{(r)}\bigr)^{\bm n}\,
\overline{\bigl(\bm\alpha^{(r)}\bigr)}^{{\bm n}'}}{\sqrt{{\bm n}!\,{\bm n}'!}},
\qquad
{\bm n},{\bm n}'\in\mathcal J_L .
\end{equation}
Here the column index $({\bm n},{\bm n}')$ ranges over $\mathcal J_L\times\mathcal J_L$ in any fixed order.
The full-column-rank assumption on $\Phi$ is equivalent to saying that the probe set $\mathcal A_L$ separates all monomials in
\eqref{eq:g-expansion-recon} up to total cutoff $L$.

For each probe amplitude $\bm\alpha^{(r)}$, perform heterodyne measurement on the output state $\rho^{(r)}
\coloneqq
\mathcal V\!\left(\ket{\underline{\bm\alpha^{(r)}}}\!\bra{\underline{\bm\alpha^{(r)}}}\right)$. For every ${\bm j},{\bm k}\in\mathcal J_L$, let us denote $\rho_{{\bm j}{\bm k}}^{(r)}
\coloneqq
\langle {\bm j}|\rho^{(r)}|{\bm k}\rangle$. Choose any unbiased heterodyne estimator for these matrix elements.
That is, for each pair $({\bm j},{\bm k})$ and each probe $\bm\alpha^{(r)}$, let
$F_{{\bm j}{\bm k}}(\bm y)$ be a single-shot estimator such that, under the heterodyne distribution of $\rho^{(r)}$, $\mathbb E\!\left[F_{{\bm j}{\bm k}}(\bm y)\right]
=
\rho_{{\bm j}{\bm k}}^{(r)}$. Given $N_r$ heterodyne samples
$\{\bm y_r^{(\ell)}\}_{\ell=1}^{N_r}$ at probe $\bm\alpha^{(r)}$, define
\begin{equation}
\label{eq:rho-hat-recon}
\widehat\rho_{{\bm j}{\bm k}}^{(r)}
\coloneqq
\frac{1}{N_r}\sum_{\ell=1}^{N_r}
F_{{\bm j}{\bm k}}\!\bigl(\bm y_r^{(\ell)}\bigr),
\qquad
\widehat g_{{\bm j}{\bm k}}^{(r)}
\coloneqq
e^{\|\bm\alpha^{(r)}\|_2^2}\,
\widehat\rho_{{\bm j}{\bm k}}^{(r)} .
\end{equation}
Their expectations satisfy
\begin{equation}
\label{eq:g-hat-mean-recon}
\mathbb E\!\left[\widehat g_{{\bm j}{\bm k}}^{(r)}\right]
=
g_{{\bm j}{\bm k}}\!\bigl(\bm\alpha^{(r)},\overline{\bm\alpha^{(r)}}\bigr).
\end{equation}

Now fix ${\bm j},{\bm k}\in\mathcal J_L$ and collect the sampled values into the vector
\begin{equation}
\label{eq:g-vector-recon}
\widehat{\bm g}_{{\bm j}{\bm k}}
\coloneqq
\bigl(
\widehat g_{{\bm j}{\bm k}}^{(1)},
\ldots,
\widehat g_{{\bm j}{\bm k}}^{(M)}
\bigr)^{\mathsf T}\in\mathbb C^{M}.
\end{equation}
Likewise, collect the unknown coefficients into
\begin{equation}
\label{eq:e-vector-recon}
{\bm e}_{{\bm j}{\bm k}}
\coloneqq
\bigl(
E_{{\bm j}{\bm k}}^{{\bm n}{\bm n}'}
\bigr)_{({\bm n},{\bm n}')\in\mathcal J_L^2}
\in\mathbb C^{D^2}.
\end{equation}
By \eqref{eq:g-expansion-recon}, these quantities obey the linear relation
\begin{equation}
\label{eq:linear-system-recon}
{\bm g}_{{\bm j}{\bm k}}=\Phi\,{\bm e}_{{\bm j}{\bm k}} .
\end{equation}
Since $\Phi$ has full column rank, the coefficient vector is uniquely determined by
\begin{equation}
\label{eq:e-recover-recon}
{\bm e}_{{\bm j}{\bm k}}
=
\Phi^\dagger {\bm g}_{{\bm j}{\bm k}},
\end{equation}
where $\Phi^\dagger=(\Phi^*\Phi)^{-1}\Phi^*$ is the Moore--Penrose left pseudoinverse \cite{ben2003generalized} as $\Phi$ is not a square matrix.
This leads to the estimator
\begin{equation}
\label{eq:e-hat-recon}
\widehat{\bm e}_{{\bm j}{\bm k}}
\coloneqq
\Phi^\dagger \widehat{\bm g}_{{\bm j}{\bm k}} .
\end{equation}
Equivalently, for each ${\bm n},{\bm n}'\in\mathcal J_L$, we have 
\begin{equation}
\label{eq:E-hat-entry-recon}
\widehat E_{{\bm j}{\bm k}}^{{\bm n}{\bm n}'}
=
\sum_{r=1}^{M}
(\Phi^\dagger)_{({\bm n},{\bm n}'),r}\,
\widehat g_{{\bm j}{\bm k}}^{(r)} .
\end{equation}

This interpolation-based estimator replaces the derivative-based reconstruction used in earlier coherent-state tomography arguments.
Its advantage here is that it is exactly matched to the non-holomorphic dependence on $(\bm\alpha,\bar{\bm\alpha})$ in \eqref{eq:g-expansion-recon}.
In particular, no finite-difference approximation in the complex amplitudes is needed at this stage.

\subsubsection{Reconstruction of the unitary matrix}

We now recover the matrix of $V_L$ from the estimated process tensor.
For each ${\bm n}\in\mathcal J_L$, define the ${\bm n}$-th column vector of $V_L$ by
\begin{equation}
\label{eq:column-vector-recon}
u^{({\bm n})}
\coloneqq
\bigl((V_L)_{{\bm j}{\bm n}}\bigr)_{{\bm j}\in\mathcal J_L}
\in\mathbb C^{D}.
\end{equation}
Then \eqref{eq:E-def-recon} can be written compactly as
\begin{equation}
\label{eq:E-outer-recon}
E^{{\bm n}{\bm n}'}
=
u^{({\bm n})}u^{({\bm n}')\dagger},
\qquad
{\bm n},{\bm n}'\in\mathcal J_L,
\end{equation}
where $E^{{\bm n}{\bm n}'}\in\mathbb C^{D\times D}$ denotes the matrix with entries
$E_{{\bm j}{\bm k}}^{{\bm n}{\bm n}'}$.

In particular, the vacuum block satisfies
\begin{equation}
\label{eq:E00-rankone-recon}
E^{{\bm 0}{\bm 0}}
=
u^{({\bm 0})}u^{({\bm 0})\dagger}.
\end{equation}
Since $V_L$ is unitary, the column $u^{({\bm 0})}$ has unit norm, and therefore
$E^{{\bm 0}{\bm 0}}$ is a rank-one positive semidefinite matrix with unique nonzero eigenvalue equal to $1$.
Thus, in the noiseless case, $u^{({\bm 0})}$ is determined by the top eigenspace of $E^{{\bm 0}{\bm 0}}$, up to a global phase.

This observation yields a clean algebraic recovery rule.
Indeed, for every ${\bm n}\in\mathcal J_L$,
\begin{equation}
\label{eq:column-recovery-recon}
E^{{\bm n}{\bm 0}}u^{({\bm 0})}
=
u^{({\bm n})}u^{({\bm 0})\dagger}u^{({\bm 0})}
=
u^{({\bm n})},
\end{equation}
because $\|u^{({\bm 0})}\|_2=1$.
Hence, once the reference column $u^{({\bm 0})}$ is fixed, all other columns follow linearly from the slice
$\{E^{{\bm n}{\bm 0}}\}_{{\bm n}\in\mathcal J_L}$.

In the empirical setting, we proceed as follows. First, form the estimated vacuum block
\begin{equation}
\label{eq:E00-hat-recon}
\widehat E^{{\bm 0}{\bm 0}}
\coloneqq
\bigl(\widehat E_{{\bm j}{\bm k}}^{{\bm 0}{\bm 0}}\bigr)_{{\bm j},{\bm k}\in\mathcal J_L}.
\end{equation}
Let $\widehat u^{({\bm 0})}$ be a unit-norm eigenvector associated with the largest eigenvalue of
$\widehat E^{{\bm 0}{\bm 0}}$.
Its phase is irrelevant at the channel level.
If one wishes to fix a convention, one may require the first nonzero entry of
$\widehat u^{({\bm 0})}$ to be real and positive.

Next, for every ${\bm n}\in\mathcal J_L$, define
\begin{equation}
\label{eq:column-hat-recon}
\widehat u^{({\bm n})}
\coloneqq
\widehat E^{{\bm n}{\bm 0}}\widehat u^{({\bm 0})},
\qquad
\widehat E^{{\bm n}{\bm 0}}
\coloneqq
\bigl(\widehat E_{{\bm j}{\bm k}}^{{\bm n}{\bm 0}}\bigr)_{{\bm j},{\bm k}\in\mathcal J_L}.
\end{equation}
Assemble these columns into the matrix
\begin{equation}
\label{eq:U-tilde-recon}
\widetilde V_L
\coloneqq
\bigl[\widehat u^{({\bm n})}\bigr]_{{\bm n}\in\mathcal J_L}
\in\mathbb C^{D\times D}.
\end{equation}

Because of sampling noise, $\widetilde V_L$ need not be exactly unitary.
We therefore impose exact unitarity by polar projection.
Let
\begin{equation}
\label{eq:polar-recon}
\widetilde V_L = QH
\end{equation}
be the polar decomposition, where $Q$ is unitary and $H$ is positive semidefinite.
We define the final estimator of the truncated unitary by
\begin{equation}
\label{eq:Vhat-final-recon}
\widehat V_L \coloneqq Q .
\end{equation}
The corresponding channel estimator is
\begin{equation}
\label{eq:channel-final-recon}
\widehat{\mathcal V}_L(\rho)
\coloneqq
\widehat V_L\,\rho\,\widehat V_L^\dagger
\qquad
\text{on } \mathcal H_{\le L}.
\end{equation}
The corresponding full-space estimator is obtained by extending $\widehat V_L$ with the same complementary unitary as in the effective model $V = V_L \oplus V_L^\perp $, namely, $\widehat V \coloneqq \widehat V_L \oplus V_L^\perp $. We then define $\widehat{\mathcal V}(\rho)\coloneqq \widehat V \rho \widehat V^\dagger $. Since quantum channels are insensitive to a global phase, the reconstruction is unique at the channel level.

To summarize, the reconstruction consists of two linear-algebraic steps.
First, interpolate the coefficient tensor
$\{E_{{\bm j}{\bm k}}^{{\bm n}{\bm n}'}\}$
from the probe values
$\{g_{{\bm j}{\bm k}}(\bm\alpha^{(r)},\overline{\bm\alpha^{(r)}})\}$.
Second, recover the unitary columns from the rank-one structure
$E^{{\bm n}{\bm n}'}=u^{({\bm n})}u^{({\bm n}')\dagger}$,
using the top eigenvector of $E^{{\bm 0}{\bm 0}}$ as the reference column.

\subsection{Statistical error and dependence on the probe design}\label{subsec:probe-design-stability}

We now quantify the estimation error of the interpolation-based reconstruction introduced in the previous subsection. Once we reconstruct the process tensor by solving a linear equation, the error is controlled by how well conditioned the interpolation matrix $\Phi$ is.

\subsubsection{Uniform estimation error for the sampled values}

Recall the probe set defined by $\mathcal A_L=\{\bm\alpha^{(r)}\}_{r=1}^{M}\subset \mathbb C^m,$ $M\ge D^2,$  $D=\binom{L+m}{m}$, and the associated design matrix $\Phi_{r,({\bm n},{\bm n}')}
=
\frac{\bigl(\bm\alpha^{(r)}\bigr)^{\bm n}\,
\overline{\bigl(\bm\alpha^{(r)}\bigr)}^{{\bm n}'}}{\sqrt{{\bm n}!\,{\bm n}'!}},$ for ${\bm n},{\bm n}'\in\mathcal J_L$. We assume throughout that $\Phi$ has full column rank. For each probe index $r\in[M]$ and each pair ${\bm j},{\bm k}\in\mathcal J_L$, suppose we perform $N$ independent heterodyne measurements on the output state associated with the coherent probe $\bm\alpha^{(r)}$.
Let $Z_{{\bm j}{\bm k}}^{(r,1)},\ldots,Z_{{\bm j}{\bm k}}^{(r,N)}$ denote the corresponding single-shot random variables used to estimate $g_{{\bm j}{\bm k}}^{(r)}
\coloneqq
g_{{\bm j}{\bm k}}\!\bigl(\bm\alpha^{(r)},\overline{\bm\alpha^{(r)}}\bigr)$. Assume that these single-shot estimators are unbiased and have uniformly bounded variance, namely
\begin{equation}
\mathbb E\!\left[Z_{{\bm j}{\bm k}}^{(r,\ell)}\right]
=
g_{{\bm j}{\bm k}}^{(r)},
\qquad
\mathbb E\!\left[
\left|Z_{{\bm j}{\bm k}}^{(r,\ell)}-g_{{\bm j}{\bm k}}^{(r)}\right|^2
\right]
\le
\nu_{\max}
\end{equation}
for all $r\in[M]$, ${\bm j},{\bm k}\in\mathcal J_L$, and $\ell\in[N]$.
Here $\nu_{\max}$ is a uniform upper bound on the single-shot second moment over all probe settings and all matrix elements.

Now let $\widehat g_{{\bm j}{\bm k}}^{(r)}$ be a robust empirical estimator constructed from
$\{Z_{{\bm j}{\bm k}}^{(r,\ell)}\}_{\ell=1}^N$,
for example by the median-of-means procedure.
Then there exists a universal constant $C_{\rm rob}>0$, independent of $r,{\bm j},{\bm k},N$, such that
\begin{equation}
\label{eq:single-robust-bound-third}
\Pr\!\left(
\left|
\widehat g_{{\bm j}{\bm k}}^{(r)}-g_{{\bm j}{\bm k}}^{(r)}
\right|
>
C_{\rm rob}\sqrt{\frac{\nu_{\max}}{N}\,t}
\right)
\le
2e^{-t},
\qquad
t\ge 0 .
\end{equation}
This is a standard consequence of robust mean estimation under a bounded second moment
\cite{lugosi2019mean,minsker2023efficient}.

There are in total $M\,D^2$ quantities of the form
$\widehat g_{{\bm j}{\bm k}}^{(r)}$
to control uniformly, since there are $M$ probe amplitudes and $D^2$ pairs
$({\bm j},{\bm k})\in\mathcal J_L^2$. Setting
\begin{equation}
t=\log\!\Bigl(\frac{2MD^2}{\delta}\Bigr)
\end{equation}
in Eq. \eqref{eq:single-robust-bound-third} and applying a union bound, we obtain that, with probability at least $1-\delta$,
\begin{equation}
\label{eq:uniform-g-bound-third}
\max_{1\le r\le M}\max_{{\bm j},{\bm k}\in\mathcal J_L}
\left|
\widehat g_{{\bm j}{\bm k}}^{(r)}-g_{{\bm j}{\bm k}}^{(r)}
\right|
\le
\varepsilon_g ,
\end{equation}
where
\begin{equation}
\label{eq:eps-g-third}
\varepsilon_g
\coloneqq
C_{\rm rob}
\sqrt{
\frac{\nu_{\max}}{N}
\log\!\Bigl(\frac{2MD^2}{\delta}\Bigr)
}.
\end{equation}

\subsubsection{Propagation through the interpolation map}

Let us fix ${\bm j},{\bm k}\in\mathcal J_L$. Recall the vector form of the reconstruction problem ${\bm g}_{{\bm j}{\bm k}}=\Phi\,{\bm e}_{{\bm j}{\bm k}},$ and $\widehat{\bm e}_{{\bm j}{\bm k}}=\Phi^\dagger \widehat{\bm g}_{{\bm j}{\bm k}}$. Hence, we have 
\begin{equation}
\label{eq:e-error-third}
\widehat{\bm e}_{{\bm j}{\bm k}}-{\bm e}_{{\bm j}{\bm k}}
=
\Phi^\dagger
\bigl(
\widehat{\bm g}_{{\bm j}{\bm k}}-{\bm g}_{{\bm j}{\bm k}}
\bigr).
\end{equation}

To quantify the worst-case entrywise error, define the stability constant
\begin{equation}
\label{eq:mu-infty-third}
\mu_{\infty}(\Phi)
\coloneqq
\max_{({\bm n},{\bm n}')\in\mathcal J_L^2}
\sum_{r=1}^{M}
\bigl|
(\Phi^\dagger)_{({\bm n},{\bm n}'),r}
\bigr| .
\end{equation}

Then, on the event in Eq. \eqref{eq:uniform-g-bound-third}, for every
${\bm j},{\bm k},{\bm n},{\bm n}'\in\mathcal J_L$, we have 
\begin{align}
\left|
\widehat E_{{\bm j}{\bm k}}^{{\bm n}{\bm n}'}
-
E_{{\bm j}{\bm k}}^{{\bm n}{\bm n}'}
\right|
&=
\left|
\sum_{r=1}^{M}
(\Phi^\dagger)_{({\bm n},{\bm n}'),r}
\bigl(
\widehat g_{{\bm j}{\bm k}}^{(r)}-g_{{\bm j}{\bm k}}^{(r)}
\bigr)
\right| \nonumber\\
&\le
\left(
\sum_{r=1}^{M}
\bigl|
(\Phi^\dagger)_{({\bm n},{\bm n}'),r}
\bigr|
\right)
\max_{1\le r\le M}
\left|
\widehat g_{{\bm j}{\bm k}}^{(r)}-g_{{\bm j}{\bm k}}^{(r)}
\right| \nonumber\\
&\le
\mu_{\infty}(\Phi)\,\varepsilon_g ,
\end{align}
where we applied the Cauchy-Schwarz inequality.

Therefore, with probability at least $1-\delta$, we have 
\begin{equation}
\label{eq:uniform-E-bound-third}
\Delta_E
\coloneqq
\max_{{\bm j},{\bm k},{\bm n},{\bm n}'\in\mathcal J_L}
\left|
\widehat E_{{\bm j}{\bm k}}^{{\bm n}{\bm n}'}
-
E_{{\bm j}{\bm k}}^{{\bm n}{\bm n}'}
\right|
\le
\mu_{\infty}(\Phi)\,\varepsilon_g .
\end{equation}
Substituting Eq. \eqref{eq:eps-g-third} gives the upper bound of the error
\begin{equation}
\label{eq:uniform-E-bound-explicit-third}
\Delta_E
\le
C_{\rm rob}\,
\mu_{\infty}(\Phi)\,
\sqrt{
\frac{\nu_{\max}}{N}
\log\!\Bigl(\frac{M D^2}{\delta}\Bigr)
},
\end{equation}
where the quantity $C_{\rm rob}$ is the universal constant associated with the chosen robust mean estimator, the parameter $\mu_\infty(\Phi)$ measures the stability of the interpolation step, namely, how much the linear inversion can amplify entrywise errors in the sampled data, the parameter $\nu_{\max}$ is a uniform upper bound on the single-shot second moment of the heterodyne-based estimators over all probe settings and all matrix elements, the number $N$ is the number of heterodyne shots used for each probe amplitude, the factor $M=|\mathcal A_L|$ is the number of probe amplitudes, while $D$ is the cutoff dimension, $\delta$ is the overall failure probability, the logarithmic factor $\log(2MD^2/\delta)$ comes from applying a union bound over all $MD^2$ sampled quantities. 

This is the basic uniform recovery bound for the process tensor.
It shows that the interpolation error depends on the probe design only through the stability parameter $\mu_\infty(\Phi)$.

\subsubsection{RMSE and total query complexity}

Eq. \eqref{eq:uniform-E-bound-explicit-third} immediately yields the shot requirement needed to achieve a target entrywise process-tensor accuracy.
Indeed, if one requires $\Delta_E\le \tau$ with probability at least $1-\delta$, it suffices to choose
\begin{equation}
\label{eq:N-sufficient-third}
N
\ge
C_{\rm rob}^2\,
\mu_{\infty}(\Phi)^2\,
\nu_{\max}\,
\tau^{-2}
\log\!\Bigl(\frac{M D^2}{\delta}\Bigr).
\end{equation}
Since there are $M$ probe settings and $N$ heterodyne shots per setting, the total number of channel uses is
\begin{equation}
\label{eq:Ntot-third}
N_{\rm tot}=MN .
\end{equation}
Hence a sufficient total query budget is
\begin{equation}
\label{eq:Ntot-sufficient-third}
N_{\rm tot}
\ge
C_{\rm rob}^2\,
M\,
\mu_{\infty}(\Phi)^2\,
\nu_{\max}\,
\tau^{-2}
\log\!\Bigl(\frac{M D^2}{\delta}\Bigr).
\end{equation}

\subsubsection{Consequences for probe design}

The energy constraint restricts the admissible probe set to $\mathcal A_L
\subseteq
\bigl\{
\bm\alpha\in\mathbb C^m
\,\big|\,
\|\bm\alpha\|_2^2\le N_{\rm probe}
\bigr\},$ for $N_{\rm probe}\le m(E-1/2)$. Within the admissible energy range, successful reconstruction requires two conditions. First, the interpolation matrix $\Phi$ must have full column rank. Otherwise, different coefficient tensors can produce the same sampled data, so the process tensor is not identifiable. Second, $\Phi$ should be well conditioned, or equivalently, $\mu_\infty(\Phi)$ should be small. If $\Phi$ is poorly conditioned, small statistical errors in the sampled values are greatly amplified by the linear inversion. This does not require unbounded probe energy. Rather, one must choose finitely many coherent probes with sufficiently diverse amplitudes and phases inside the allowed energy region. 

In practice, using phase-resolved complex probes with several radii and allowing moderate oversampling, namely $M>D^2$, improves conditioning and hence the robustness of the reconstruction. By contrast, if all probe amplitudes lie on a real line, then $\bar{\bm\alpha}=\bm\alpha$, different monomials $\bm\alpha^{\bm n}\bar{\bm\alpha}^{{\bm n}'}$ become indistinguishable, and the full process tensor is generally not identifiable. Therefore, the main trade-off is between respecting the probe-energy budget and choosing a probe set that yields a stable interpolation problem.

\subsection{Effective dimension}\label{subsec:effective-dimension}

The previous subsections described how to reconstruct a finite-dimensional unitary from coherent probes and heterodyne data.
We now justify why such a finite-dimensional reduction is valid under an input-energy constraint.

\begin{theorem}[Multimode extension of Theorem 1 in \cite{arzani2025effective}]
\label{thm:effective-dimension-corrected}
Let $U$ be an $m$-mode bosonic unitary on $\mathcal H$, and let $\widehat N\coloneqq \sum_{s=1}^m a_s^\dagger a_s$ be the total photon-number operator.
For $L\in\mathbb N$, define $\mathcal H_{\le L}
\coloneqq
{\rm span}\bigl\{\ket{\bm n}:\bm n\in\mathbb N^m,\ \|\bm n\|_1\le L\bigr\},$ $\Pi_L \text{ the projector onto } \mathcal H_{\le L}$, and the energy blow-up function
\begin{equation}
\label{eq:EU-def-corrected}
E_U(L)
\coloneqq
\sup_{\ket{\psi}\in\mathcal H_{\le L},\,\|\psi\|_2=1}
\bra{\psi}U^\dag \widehat N U\ket{\psi}
=
\bigl\|\Pi_L U^\dag \widehat N U \Pi_L\bigr\|_\infty .
\end{equation}

Fix an input-energy bound $E\ge 1/2$ and an accuracy $\epsilon\in(0,1)$. Assume that \(E_U(M)<\infty\) for the cutoff \(M\) below. 
Set $N_{\rm in}\coloneqq m\left(E-\frac12\right)$ and 
\begin{equation}
\label{eq:MN-def-corrected}
M\coloneqq \left\lceil \frac{256 N_{\rm in}}{\epsilon^2}\right\rceil,
\qquad
N_{\rm out}\coloneqq \left\lceil \frac{256\,E_U(M)}{\epsilon^2}\right\rceil,
\qquad
L\coloneqq \max\{M,N_{\rm out}\}.
\end{equation}
Then there exists a unitary $V_L$ on $\mathcal H_{\le L}$ such that, for any full-space unitary extension
\begin{equation}
V = V_L \oplus V_L^\perp
\qquad\text{on}\qquad
\mathcal H = \mathcal H_{\le L}\oplus \mathcal H_{\le L}^\perp,
\end{equation}
the corresponding unitary channel $\mathcal V(\rho)=V\rho V^\dagger$ satisfies
\begin{equation}
\label{eq:effective-main-corrected}
\|\mathcal U-\mathcal V\|_\diamond^{mE} \le \epsilon,
\qquad
\mathcal U(\rho)\coloneqq U\rho U^\dagger.
\end{equation}
In particular, under the energy constraint $mE$, the channel $\mathcal U$ admits an $\epsilon$-accurate effective finite-dimensional description supported on $\mathcal H_{\le L}$. In addition, we have an effective dimension $\dim(\mathcal H_{\le L})
=
\#\bigl\{\bm n\in\mathbb N^m:\|\bm n\|_1\le L\bigr\}
=
\binom{L+m}{m}$.
\end{theorem}

\begin{proof}
Fix an arbitrary bipartite input state $\rho_{AR}$ satisfying $\Tr\!\bigl[(\widehat H_m\otimes I_R)\rho_{AR}\bigr]\le mE$. We will bound
\begin{equation}
\bigl\|(\mathcal U\otimes{\rm id}_R)(\rho_{AR})-(\mathcal V\otimes{\rm id}_R)(\rho_{AR})\bigr\|_1
\end{equation}
uniformly over all such inputs. We will repetitively use the standard operator inequality
\begin{equation}
\label{eq:projector-tail-corrected}
\widehat N\ge (K+1)(I-\Pi_K),
\qquad K\in\mathbb N,
\end{equation}
which implies the inequality  
\begin{equation}
\label{eq:tail-prob-corrected}
\Tr\!\bigl[(I-\Pi_K)\omega\bigr]
\le
\frac{\Tr(\widehat N\omega)}{K+1}
\end{equation}
for every state $\omega$ with finite energy.
We will also repetitively use the gentle-measurement lemma \cite{wilde2017qit}
\begin{equation}
\label{eq:gentle-subnorm-corrected}
\|\omega-(\Pi_K\otimes I)\omega(\Pi_K\otimes I)\|_1
\le
2\sqrt{\Tr\!\bigl[((I-\Pi_K)\otimes I)\omega\bigr]}
\end{equation}
for every subnormalized positive operator $\omega$.

Now, let us look at the input truncation. Define
\begin{equation}
\rho_{AR}^{(M)}
\coloneqq
(\Pi_M\otimes I_R)\rho_{AR}(\Pi_M\otimes I_R).
\end{equation}
By Eq. \eqref{eq:tail-prob-corrected} with $K=M$, we have 
\begin{equation}
\Tr\!\bigl[(I-\Pi_M)\rho_A\bigr]
\le
\frac{N_{\rm in}}{M+1},
\end{equation} with $N_{\rm in}=m(E-1/2)$.

Therefore, by Eq. \eqref{eq:gentle-subnorm-corrected}, we have 
\begin{equation}
\label{eq:input-trunc-bound-corrected}
\|\rho_{AR}-\rho_{AR}^{(M)}\|_1
\le
2\sqrt{\frac{N_{\rm in}}{M+1}}.
\end{equation}
Then, using the trace-norm contractivity of channels, we have
\begin{equation}
\label{eq:input-trunc-channel-corrected}
\bigl\|(\mathcal U\otimes{\rm id}_R)(\rho_{AR})-(\mathcal U\otimes{\rm id}_R)(\rho_{AR}^{(M)})\bigr\|_1
\le
2\sqrt{\frac{N_{\rm in}}{M+1}},
\end{equation}
and the same bound holds with $\mathcal U$ replaced by $\mathcal V$.

Next, let us look at the output truncation. Since $\rho_{AR}^{(M)}$ is supported on $\mathcal H_{\le M}$ in system $A$, its reduced state $\rho_A^{(M)}$ obeys
\begin{equation}
\Tr\!\bigl[\widehat N\,U\rho_A^{(M)}U^\dagger\bigr]
\le
E_U(M)\,\Tr(\rho_A^{(M)})\le E_U(M).
\end{equation}
Applying Eq. \eqref{eq:tail-prob-corrected} with $K=L$ gives
\begin{equation}
\Tr\!\bigl[(I-\Pi_L)\,U\rho_A^{(M)}U^\dagger\bigr]
\le
\frac{E_U(M)}{L+1}
\le
\frac{E_U(M)}{N_{\rm out}+1}.
\end{equation}
Therefore, we have 
\begin{equation}
\label{eq:output-trunc-bound-corrected}
\bigl\|
(\mathcal U\otimes{\rm id}_R)(\rho_{AR}^{(M)})
-
(\Pi_L\otimes I_R)(\mathcal U\otimes{\rm id}_R)(\rho_{AR}^{(M)})(\Pi_L\otimes I_R)
\bigr\|_1
\le
2\sqrt{\frac{E_U(M)}{L+1}}.
\end{equation}

Further, let us consider the  approximate the truncated action by an isometry. Define
\begin{equation}
\label{eq:A-def-corrected}
A\coloneqq \Pi_L U \Pi_M:\mathcal H_{\le M}\to\mathcal H_{\le L}.
\end{equation}
Then, we have 
\begin{align}
A^\dagger A
&=
\Pi_M U^\dagger \Pi_L U \Pi_M \nonumber\\
&=
\Pi_M-\Pi_M U^\dagger(I-\Pi_L)U\Pi_M \nonumber\\
&\ge
\Pi_M-\frac{1}{L+1}\Pi_M U^\dagger \widehat N U \Pi_M \nonumber\\
&\ge
\left(1-\frac{E_U(M)}{L+1}\right)\Pi_M.
\label{eq:AdA-lower-corrected}
\end{align}
Set $\delta_L\coloneqq \frac{E_U(M)}{L+1}$. Eq.  \eqref{eq:AdA-lower-corrected} gives
\begin{equation}
\label{eq:AdA-gap-corrected}
(1-\delta_L)\Pi_M \le A^\dagger A \le \Pi_M .
\end{equation}
Since $L\ge M$, one has
$\dim(\mathcal H_{\le L})\ge \dim(\mathcal H_{\le M})$,
so an isometric embedding from $\mathcal H_{\le M}$ into $\mathcal H_{\le L}$ is possible. Let $A=W\sqrt{A^\dagger A}$ be the polar decomposition of $A$, where
$W:\mathcal H_{\le M}\to \mathcal H_{\le L}$ is an isometry.
Since the spectrum of $A^\dagger A$ on $\mathcal H_{\le M}$ lies in $[1-\delta_L,1]$, we have 
\begin{equation}
\bigl\|\sqrt{A^\dagger A}-\Pi_M\bigr\|_\infty
\le
1-\sqrt{1-\delta_L}
\le
\delta_L.
\end{equation}
Hence, we have 
\begin{equation}
\label{eq:A-W-bound-corrected}
\|A-W\|_\infty
=
\bigl\|W(\sqrt{A^\dagger A}-\Pi_M)\bigr\|_\infty
\le
\delta_L.
\end{equation}

Extend $W$ to a unitary $V_L$ on $\mathcal H_{\le L}$, and let
$V=V_L\oplus V_L^\perp$ be any full-space unitary extension.
For every subnormalized state $\tau_{AR}$ supported on $\mathcal H_{\le M}$ in register $A$, we have 
\begin{align}
&\bigl\|
(A\otimes I_R)\tau_{AR}(A^\dagger\otimes I_R)
-
(W\otimes I_R)\tau_{AR}(W^\dagger\otimes I_R)
\bigr\|_1 \nonumber\\
&\le
\bigl\|((A-W)\otimes I_R)\tau_{AR}(A^\dagger\otimes I_R)\bigr\|_1
+
\bigl\|(W\otimes I_R)\tau_{AR}((A^\dagger-W^\dagger)\otimes I_R)\bigr\|_1 \nonumber\\
&\le
2\|A-W\|_\infty
\le
2\delta_L .
\label{eq:AW-state-bound-corrected}
\end{align}

Since $\rho_{AR}^{(M)}$ is supported on $\mathcal H_{\le M}$ in system $A$, we have 
\begin{equation}
\label{eq:PiLUrho-corrected}
(\Pi_L\otimes I_R)(\mathcal U\otimes{\rm id}_R)(\rho_{AR}^{(M)})(\Pi_L\otimes I_R)
=
(A\otimes I_R)\rho_{AR}^{(M)}(A^\dagger\otimes I_R),
\end{equation}
with
\begin{equation}
\label{eq:Vrhom-corrected}
(\mathcal V\otimes{\rm id}_R)(\rho_{AR}^{(M)})
=
(W\otimes I_R)\rho_{AR}^{(M)}(W^\dagger\otimes I_R).
\end{equation}
Combining Eqs. \eqref{eq:input-trunc-channel-corrected}, \eqref{eq:output-trunc-bound-corrected}, and \eqref{eq:AW-state-bound-corrected}, we obtain
\begin{align}
&\bigl\|(\mathcal U\otimes{\rm id}_R)(\rho_{AR})-(\mathcal V\otimes{\rm id}_R)(\rho_{AR})\bigr\|_1 \nonumber\\
&\le
2\sqrt{\frac{N_{\rm in}}{M+1}}
+
2\sqrt{\frac{N_{\rm in}}{M+1}}
+
2\sqrt{\frac{E_U(M)}{L+1}}
+
2\frac{E_U(M)}{L+1}.
\label{eq:pre-final-effective-corrected}
\end{align}
By the definitions of $M$ and $L$ in Eq. \eqref{eq:MN-def-corrected}, we have 
\begin{equation}
\frac{N_{\rm in}}{M+1}\le \frac{\epsilon^2}{256},
\qquad
\frac{E_U(M)}{L+1}\le \frac{\epsilon^2}{256}.
\end{equation}
In addition, we have 
\begin{equation}
4\sqrt{\frac{N_{\rm in}}{M+1}}\le \frac{\epsilon}{4},
\qquad
2\sqrt{\frac{E_U(M)}{L+1}}\le \frac{\epsilon}{8},
\qquad
2\frac{E_U(M)}{L+1}\le \frac{\epsilon^2}{128}\le \frac{\epsilon}{128},
\end{equation}
Therefore, the right-hand side of Eq. \eqref{eq:pre-final-effective-corrected} is at most $\epsilon$.
Taking the supremum over all admissible $\rho_{AR}$ proves Eq. \eqref{eq:effective-main-corrected}.
\end{proof}

\subsection{Query complexity under an energy constraint}\label{subsec:query-complexity-corrected}

We now combine the previous results into a quantitative statement on the number of channel uses. Here, the key point is that there are two logically distinct sources of error. First, one must choose a cutoff $L$ so that the effective finite-dimensional model is accurate under the energy-constrained diamond norm.
Second, one must estimate the corresponding finite-dimensional unitary from coherent probes and heterodyne data.
These two steps should be separated cleanly. Here we denote 
$\epsilon = \epsilon_{\rm model}+\epsilon_{\rm recon},$ with $\epsilon_{\rm model},\epsilon_{\rm recon}\in(0,1)$.

\subsubsection{A rigorous lower bound from an energy-accessible code subspace}

The lower bound should depend on the dimension of a subspace that is genuinely accessible under the input-energy bound $E$, not on the truncation dimension chosen later for the upper bound. Let $q\coloneqq \lfloor m(E-1/2)\rfloor,$ and $d_E\coloneqq \dim(\mathcal H_{\le q})=\binom{q+m}{m}$. Every state supported on $\mathcal H_{\le q}$ has photon number at most $q\le mE-m/2$, and is admissible in the energy-constrained diamond norm. 
Therefore, any protocol that learns arbitrary bosonic unitaries under $\|\cdot\|_\diamond^{mE}$ must in particular be able to learn the subclass of physical unitaries that act arbitrarily on $\mathcal H_{\le q}$ and as identity on $\mathcal H_{\le q}^\perp$.

Let $\mathcal K\coloneqq \mathcal H_{\le q}$. Consider the subclass of bosonic unitaries
\begin{align}
\mathfrak U_E
\coloneqq
\left\{
U_W = W\oplus I_{\mathcal K^\perp}
\;\middle|\;
W\in U(\mathcal K)
\right\}.
\end{align}
For every state \(\rho\) supported on \(\mathcal K\), one has
\begin{align}
\Tr(\widehat N \rho)\le q\le mE-\frac m2,
\end{align}
so \(\rho\) is admissible in the energy-constrained diamond norm. Moreover, since \(U_W\mathcal K=\mathcal K\), the output of \(U_W\) also remains in \(\mathcal K\).

Now let \(W_1,W_2\in U(\mathcal K)\), and denote by \(\mathcal U_{W_1},\mathcal U_{W_2}\) the corresponding bosonic unitary channels on the full Hilbert space. Then
\begin{align}
\|\mathcal U_{W_1}-\mathcal U_{W_2}\|_\diamond^{mE}
&=
\sup_{\Tr[(\widehat H_m\otimes I)\rho]\le mE}
\left\|
\bigl((\mathcal U_{W_1}-\mathcal U_{W_2})\otimes{\rm id}\bigr)(\rho)
\right\|_1 \\
&\ge
\sup_{\substack{\rho\ \text{supported on }\mathcal K\otimes\mathcal H_R}}
\left\|
\bigl((\mathcal U_{W_1}-\mathcal U_{W_2})\otimes{\rm id}\bigr)(\rho)
\right\|_1 \\
&=
\|\mathcal W_1-\mathcal W_2\|_\diamond,
\end{align}
where \(\mathcal W_j(\rho)\coloneqq W_j\rho W_j^\dagger\) denotes the corresponding \(d_E\)-dimensional unitary channel on \(\mathcal K\). Since any \(d_E\)-dimensional unitary channel has an extended bosonic unitary version, the above indicates that learning arbitrary bosonic unitaries under \(\|\cdot\|_\diamond^{mE}\) is at least as hard as learning arbitrary \(d_E\)-dimensional unitary channels. Therefore, the learning problem reduces exactly to learning a $d_E$-dimensional unitary channel in ordinary diamond distance.
Thus, by the finite-dimensional minimax lower bound for unitary-channel learning \cite{haah2023query}, the query complexity takes the following form: 
\begin{equation}
\label{eq:lower-bound-corrected}
n = \Omega\!\left(\frac{d_E^2}{\epsilon}\right).
\end{equation}
More generally, if $\mathcal K\subseteq \mathcal H$ is any finite-dimensional subspace such that $\sup_{\ket{\psi}\in\mathcal K,\ \|\psi\|_2=1}\bra{\psi}\widehat N\ket{\psi}\le N_{\rm in},$ then the same argument yields
\begin{equation}
n = \Omega\!\left(\frac{\dim(\mathcal K)^2}{\epsilon}\right).
\end{equation}
Here, Eq. \eqref{eq:lower-bound-corrected} is the canonical choice obtained by taking $\mathcal K=\mathcal H_{\le q}$.

\subsubsection{Stability of the algebraic reconstruction}

We next quantify how errors in the reconstructed process tensor propagate to the estimated finite-dimensional unitary.

\begin{proposition}[Local stability of the reconstruction map]
\label{prop:reconstruction-stability-corrected}
Let $V_L$ be a unitary on $\mathcal H_{\le L}$ with $D=\dim(\mathcal H_{\le L})$. Let
$E^{{\bm n}{\bm n}'}=u^{({\bm n})}u^{({\bm n}')\dagger},$ with ${\bm n},{\bm n}'\in\mathcal J_L$, 
be its exact process-tensor blocks, where $u^{({\bm n})}$ denotes the ${\bm n}$th column of $V_L$.
Let $\widehat E^{{\bm n}{\bm n}'}$ be estimated blocks satisfying
\begin{equation}
\Delta_E
\coloneqq
\max_{{\bm j},{\bm k},{\bm n},{\bm n}'\in\mathcal J_L}
\left|
\widehat E_{{\bm j}{\bm k}}^{{\bm n}{\bm n}'}
-
E_{{\bm j}{\bm k}}^{{\bm n}{\bm n}'}
\right|.
\end{equation}
Construct $\widehat V_L$ from $\widehat E$ exactly as in \ref{subsec:reconstruction}.
Then there exist universal constants $c_{\rm rec},C_{\rm rec}>0$ such that, whenever
\begin{equation}
\label{eq:Delta-small-corrected}
\Delta_E \le c_{\rm rec} D^{-3/2},
\end{equation}
one has
\begin{equation}
\label{eq:VL-stability-corrected}
\min_{\theta\in\mathbb R}
\bigl\|
\widehat V_L-e^{i\theta}V_L
\bigr\|_\infty
\le
C_{\rm rec} D^{3/2}\Delta_E .
\end{equation}
Consequently,
\begin{equation}
\label{eq:channel-stability-corrected}
\|\widehat{\mathcal V}_L-\mathcal V_L\|_\diamond
\le
2 C_{\rm rec} D^{3/2}\Delta_E ,
\end{equation}
where $\mathcal V_L(\rho)=V_L\rho V_L^\dagger$ and
$\widehat{\mathcal V}_L(\rho)=\widehat V_L\rho \widehat V_L^\dagger$ on $\mathcal H_{\le L}$.
\end{proposition}

\begin{proof}
Let $u^{({\bm 0})}$ denote the vacuum column of $V_L$.
Then $E^{{\bm 0}{\bm 0}}=u^{({\bm 0})}u^{({\bm 0})\dagger}$ is a rank-one projector with spectral gap $1$.
Since every entry of $\widehat E^{{\bm 0}{\bm 0}}-E^{{\bm 0}{\bm 0}}$ is bounded by $\Delta_E$, we have 
\begin{equation}
\bigl\|\widehat E^{{\bm 0}{\bm 0}}-E^{{\bm 0}{\bm 0}}\bigr\|_\infty
\le
D\Delta_E ,
\end{equation}
where we applied the inequality $\|A\|_\infty \le \|A\|_F\le r \max_{jk}|A_{jk}|$ for any $r\times r$ matrix $A$. If $D\Delta_E\le 1/4$, Davis--Kahan perturbation theory \cite{davis1970rotation,yu2015useful} yields a phase $\theta\in\mathbb R$ such that
\begin{equation}
\label{eq:u0-dk-corrected}
\bigl\|
\widehat u^{({\bm 0})}-e^{i\theta}u^{({\bm 0})}
\bigr\|_2
\le
4D\Delta_E .
\end{equation}

For any ${\bm n}\in\mathcal J_L$, the reconstruction rule gives
\begin{equation}
\widehat u^{({\bm n})}
=
\widehat E^{{\bm n}{\bm 0}}\widehat u^{({\bm 0})},
\qquad
u^{({\bm n})}
=
E^{{\bm n}{\bm 0}}u^{({\bm 0})}.
\end{equation}
Therefore, we have 
\begin{align}
\bigl\|
\widehat u^{({\bm n})}-e^{i\theta}u^{({\bm n})}
\bigr\|_2
&\le
\bigl\|
(\widehat E^{{\bm n}{\bm 0}}-E^{{\bm n}{\bm 0}})\widehat u^{({\bm 0})}
\bigr\|_2
+
\bigl\|
E^{{\bm n}{\bm 0}}
(\widehat u^{({\bm 0})}-e^{i\theta}u^{({\bm 0})})
\bigr\|_2 \nonumber\\
&\le
\bigl\|\widehat E^{{\bm n}{\bm 0}}-E^{{\bm n}{\bm 0}}\bigr\|_\infty
+
\bigl\|E^{{\bm n}{\bm 0}}\bigr\|_\infty
\bigl\|
\widehat u^{({\bm 0})}-e^{i\theta}u^{({\bm 0})}
\bigr\|_2 \nonumber\\
&\le
D\Delta_E + 4D\Delta_E
=
5D\Delta_E,
\label{eq:column-err-corrected}
\end{align}
where we applied the triangle inequality, the definition of the operator norm $\|Av\|_2\le \|A\|_\infty \|v\|_2$ for a matrix $A$ and a vector $v$, and the relation 
$\|E^{{\bm n}{\bm 0}}\|_\infty=\|u^{({\bm n})}u^{({\bm 0})\dagger}\|_\infty=1$.

Let $\widetilde V_L$ be the matrix whose columns are $\widehat u^{({\bm n})}$ before the final polar projection.
Using Eq. \eqref{eq:column-err-corrected}, we have 
\begin{equation}
\bigl\|
\widetilde V_L-e^{i\theta}V_L
\bigr\|_{\rm F}
=
\left(
\sum_{{\bm n}\in\mathcal J_L}
\bigl\|
\widehat u^{({\bm n})}-e^{i\theta}u^{({\bm n})}
\bigr\|_2^2
\right)^{1/2}
\le
\sqrt{D}\,(5D\Delta_E)
=
5D^{3/2}\Delta_E .,
\end{equation}
where we used that the operator norm is bounded by the Frobenius norm, namely $\|A\|_\infty\le \|A\|_{\rm F}$, and that $\widetilde V_L-e^{i\theta}V_L$ has $D$ columns, each with Euclidean norm at most $5D\Delta_E$ by Eq.~\eqref{eq:column-err-corrected}. 

If Eq. \eqref{eq:Delta-small-corrected} holds with $c_{\rm rec}$ chosen sufficiently small, then we have 
\begin{equation}
\bigl\|
\widetilde V_L-e^{i\theta}V_L
\bigr\|_\infty \le \frac12 .
\end{equation}
Let $A\coloneqq \widetilde V_L,$ $U\coloneqq e^{i\theta}V_L,$ and $\varepsilon\coloneqq \|A-U\|_\infty$. Since $U$ is unitary and $\varepsilon\le \tfrac12$, the matrix $A$ is invertible. Let $A=\widehat V_L H$ be the polar decomposition of $A$, where $\widehat V_L$ is unitary and $H=(A^\dagger A)^{1/2}$ is positive semidefinite. Then, we have 
\begin{align}
\bigl\|
\widehat V_L-U
\bigr\|_\infty
&\le
\bigl\|
\widehat V_L-A
\bigr\|_\infty
+
\bigl\|
A-U
\bigr\|_\infty \nonumber\\
&=
\bigl\|
\widehat V_L(I-H)
\bigr\|_\infty
+
\varepsilon \nonumber\\
&=
\bigl\|
H-I
\bigr\|_\infty
+
\varepsilon .
\end{align}
Now the eigenvalues of $H$ are the singular values of $A$. Since $U$ is unitary, all singular values of $U$ are equal to $1$, and Weyl's perturbation bound for singular values gives
\begin{align}
\|H-I\|_\infty
=
\max_i |\sigma_i(A)-1|
\le
\|A-U\|_\infty
=
\varepsilon .
\end{align}
Therefore, we have 
\begin{equation}
\bigl\|
\widehat V_L-e^{i\theta}V_L
\bigr\|_\infty
=
\|\widehat V_L-U\|_\infty
\le
2\varepsilon
=
2
\bigl\|
\widetilde V_L-e^{i\theta}V_L
\bigr\|_\infty
\le
10 D^{3/2}\Delta_E .
\end{equation}
This proves Eq.~\eqref{eq:VL-stability-corrected} with $C_{\rm rec}=10$ after adjusting constants.

Finally, for unitary channels, we have \cite{haah2023query} 
\begin{equation}
\|\widehat{\mathcal V}_L-\mathcal V_L\|_\diamond
\le
2\min_{\theta\in\mathbb R}
\bigl\|
\widehat V_L-e^{i\theta}V_L
\bigr\|_\infty,
\end{equation}
which yields Eq.~\eqref{eq:channel-stability-corrected}.
\end{proof}

\subsubsection{Upper bound from coherent probes, heterodyne data, and interpolation}

We now combine the effective-dimension reduction with the interpolation bound from \ref{subsec:probe-design-stability} and the reconstruction stability of Proposition~\ref{prop:reconstruction-stability-corrected}. Choose $L$ and a full-space unitary extension $V$ as in Theorem~\ref{thm:effective-dimension-corrected}, but with accuracy parameter $\epsilon_{\rm model}$ instead of $\epsilon$. Then, we have $\|\mathcal U-\mathcal V\|_\diamond^{mE} \le \epsilon_{\rm model}$. Let 
$D=\dim(\mathcal H_{\le L}),$ $M_{\rm probe}=|\mathcal A_L|$, and choose a probe set $\mathcal A_L=\{\bm\alpha^{(r)}\}_{r=1}^{M_{\rm probe}}
\subseteq
\bigl\{\bm\alpha\in\mathbb C^m:\|\bm\alpha\|_2^2\le N_{\rm probe}\bigr\},$ with $N_{\rm probe}$ being the input photon number, whose interpolation matrix $\Phi$ has full column rank.

For each probe $\bm\alpha^{(r)}$, heterodyne data are generated by the true channel $\mathcal U$, not by the approximate channel $\mathcal V$.
Accordingly, there are two contributions to the error in the recovered process tensor. First, the statistical fluctuation of the sampled values obeys the bound from \ref{subsec:probe-design-stability}.
Namely, if $\nu_{\max}$ is the corresponding uniform second-moment parameter for the observed heterodyne estimators, then with probability at least $1-\delta$, we have 
\begin{equation}
\label{eq:g-stat-bound-fifth}
\max_{r,{\bm j},{\bm k}}
\left|
\widehat g_{{\bm j}{\bm k}}^{(r)} - g_{{\bm j}{\bm k}}^{U,(r)}
\right|
\le
\varepsilon_g,
\qquad
\varepsilon_g
=
C_{\rm rob}
\sqrt{
\frac{\nu_{\max}}{N}
\log\!\Bigl(\frac{M_{\rm probe}D^2}{\delta}\Bigr)
},
\end{equation}
where $g_{{\bm j}{\bm k}}^{U,(r)}$ denotes the exact value associated with the true channel $\mathcal U$ at probe $\bm\alpha^{(r)}$.

Second, the model mismatch between $\mathcal U$ and $\mathcal V$ induces a deterministic bias.
Indeed, for every admissible probe $\bm\alpha^{(r)}$, we have 
\begin{equation}
\bigl\|
\mathcal U(\ket{\underline{\bm\alpha^{(r)}}}\!\bra{\underline{\bm\alpha^{(r)}}})
-
\mathcal V(\ket{\underline{\bm\alpha^{(r)}}}\!\bra{\underline{\bm\alpha^{(r)}}})
\bigr\|_1
\le
\epsilon_{\rm model}.
\end{equation}
Therefore, for every ${\bm j},{\bm k}\in\mathcal J_L$, we have 
\begin{align}
\left|
g_{{\bm j}{\bm k}}^{U,(r)}-g_{{\bm j}{\bm k}}^{V,(r)}
\right|
&=
e^{\|\bm\alpha^{(r)}\|_2^2}
\left|
\langle {\bm j}|
\bigl(
\mathcal U-\mathcal V
\bigr)
(\ket{\underline{\bm\alpha^{(r)}}}\!\bra{\underline{\bm\alpha^{(r)}}})
|{\bm k}\rangle
\right| \nonumber\\
&\le
e^{N_{\rm probe}}
\epsilon_{\rm model}.
\label{eq:model-bias-g-fifth}
\end{align}

Let $\mu_\infty(\Phi)\coloneqq \max_{({\bm n},{\bm n}')\in\mathcal J_L^2}
\sum_{r=1}^{M}
\bigl|
(\Phi^\dagger)_{({\bm n},{\bm n}'),r}
\bigr|$ be the interpolation stability constant from \ref{subsec:probe-design-stability}.
Combining Eqs. \eqref{eq:g-stat-bound-fifth}, \eqref{eq:model-bias-g-fifth}, and the linear inversion formula yields the uniform process-tensor bound
\begin{equation}
\label{eq:DeltaE-fifth}
\Delta_E
\le
\mu_\infty(\Phi)
\bigl(
\varepsilon_g + e^{N_{\rm probe}}\epsilon_{\rm model}
\bigr)
\end{equation}
with probability at least $1-\delta$. Now define the target process-tensor tolerance
\begin{equation}
\label{eq:etaE-fifth}
\eta_E
\coloneqq
\min\!\left\{
c_{\rm rec}D^{-3/2},
\frac{\epsilon_{\rm recon}}{2C_{\rm rec}D^{3/2}}
\right\}.
\end{equation}
If the following conditions 
\begin{equation}
\label{eq:model-bias-condition-fifth}
\mu_\infty(\Phi)e^{N_{\rm probe}}\epsilon_{\rm model}
\le
\frac{\eta_E}{2}
\end{equation}
and
\begin{equation}
\label{eq:stat-condition-fifth}
\mu_\infty(\Phi)\varepsilon_g
\le
\frac{\eta_E}{2},
\end{equation}
are satisfied, then Eq.  \eqref{eq:DeltaE-fifth} implies
\begin{equation}
\Delta_E\le \eta_E.
\end{equation}
By Proposition~\ref{prop:reconstruction-stability-corrected}, the reconstructed cutoff unitary $\widehat V_L$ satisfies
\begin{equation}
\label{eq:cutoff-recon-error-fifth}
\min_{\theta\in\mathbb R}
\bigl\|
\widehat V_L-e^{i\theta}V_L
\bigr\|_\infty
\le
\frac{\epsilon_{\rm recon}}{2}.
\end{equation}
We now extend $\widehat V_L$ to the full space using the same complementary unitary as in the effective model $V=V_L\oplus V_L^\perp$, namely, $\widehat V \coloneqq \widehat V_L\oplus V_L^\perp,$ and $\widehat{\mathcal V}(\rho)\coloneqq \widehat V\rho \widehat V^\dagger$. Since $\widehat V$ and $V$ coincide on $\mathcal H_{\le L}^\perp$, we have
\begin{equation}
\label{eq:fullspace-unitary-error-fifth}
\min_{\theta\in\mathbb R}
\bigl\|
\widehat V-e^{i\theta}V
\bigr\|_\infty
=
\min_{\theta\in\mathbb R}
\bigl\|
\widehat V_L-e^{i\theta}V_L
\bigr\|_\infty
\le
\frac{\epsilon_{\rm recon}}{2}.
\end{equation}
Therefore, using the standard bound for unitary channels,
\begin{equation}
\label{eq:recon-error-fifth}
\|\widehat{\mathcal V}-\mathcal V\|_\diamond
\le
2\min_{\theta\in\mathbb R}
\bigl\|
\widehat V-e^{i\theta}V
\bigr\|_\infty
\le
\epsilon_{\rm recon}.
\end{equation}
Since $\|\cdot\|_\diamond^{mE}\le \|\cdot\|_\diamond$, we also have
\begin{equation}
\|\widehat{\mathcal V}-\mathcal V\|_\diamond^{mE}
\le
\epsilon_{\rm recon}.
\end{equation}
Further , we have 
\begin{equation}
\label{eq:final-query-upper-fifth}
\|\widehat{\mathcal V}-\mathcal U\|_\diamond^{mE}
\le
\epsilon_{\rm model}+\epsilon_{\rm recon}
=
\epsilon.
\end{equation}

Condition in Eq. \eqref{eq:stat-condition-fifth} is ensured by choosing
\begin{equation}
\label{eq:N-upper-fifth}
N
\ge
4C_{\rm rob}^2\,
\mu_\infty(\Phi)^2\,
\nu_{\max}\,
\eta_E^{-2}
\log\!\Bigl(\frac{M_{\rm probe}D^2}{\delta}\Bigr).
\end{equation}
Since each of the $M_{\rm probe}$ probe settings is used $N$ times, the total number of channel uses is
\begin{equation}
\label{eq:total-query-fifth}
n = M_{\rm probe}N.
\end{equation}
Thus a sufficient query bound is
\begin{equation}
\label{eq:total-query-upper-fifth}
n
\ge
4C_{\rm rob}^2\,
M_{\rm probe}\,
\mu_\infty(\Phi)^2\,
\nu_{\max}\,
\eta_E^{-2}
\log\!\Bigl(\frac{M_{\rm probe}D^2}{\delta}\Bigr),
\end{equation}
provided the model-bias constraint \eqref{eq:model-bias-condition-fifth} is also satisfied.

\emph{Interpretation.}
The corrected upper bound differs conceptually from the earlier finite-difference analysis.
There is no small-step parameter and no $h^{-L}$ noise amplification.
Instead, the sample complexity is controlled by four quantities

\begin{equation}
\left\{ D,
\ 
M_{\rm probe},
\ 
\mu_\infty(\Phi),
\ 
\nu_{\max}\right\} ,
\end{equation}
together with the effective-dimension error $\epsilon_{\rm model}$.
Among these, the genuinely design-dependent quantity is the interpolation stability constant $\mu_\infty(\Phi)$.
This is the correct object replacing the earlier step-size analysis.

\subsubsection{A compact summary}

We now summarize the preceding analysis in a form that makes the lower and upper bounds directly comparable.
The lower bound is unconditional.
For the upper bound, we isolate the experimentally meaningful ingredients of the coherent-probe heterodyne protocol, namely the number of probe coherent states, the conditioning of the interpolation map, the single-shot heterodyne fluctuation, and the input-energy bound.

\begin{theorem}[Query complexity of general unitary learning]
\label{thm:query-summary-corrected}
Fix the number of modes \(m\in\mathbb N\), an input-energy bound \(E\ge 1/2\), a target accuracy \(\epsilon\in(0,1)\), and a confidence level \(\delta\in(0,1/2)\). Set \(N_{\rm in}\coloneqq m(E-\frac12)\). 
Define the function $d_x \coloneqq \binom{\lfloor x\rfloor+m}{m},$ for $x\ge 0$. Let \(q\coloneqq \lfloor N_{\rm in}\rfloor\), so that \(d_E=\dim(\mathcal H_{\le q})=\binom{q+m}{m}\).
Let \(L\) be the effective cutoff from Theorem~\ref{thm:effective-dimension-corrected}, and write
$d_L=\dim(\mathcal H_{\le L})=\binom{L+m}{m}$. Assume that at cutoff \(L\), one can implement a coherent probe state family \(\mathcal A_L\) with total photon number \(N_{\rm probe}\le N_{\rm in}\), and that there exist constants \(C_M,C_\mu,C_\nu>0\) and exponents \(b,c\ge 0\) such that
\begin{equation}
\label{eq:summary-assumptions-probes}
M_{\rm probe}\le C_M d_L^2,
\qquad
\mu_\infty(\Phi)\le C_\mu d_L^b,
\qquad
\nu_{\max}\le C_\nu d_L^c,
\end{equation}
where \(M_{\rm probe}=|\mathcal A_L|\) is the number of coherent probes, \(\mu_\infty(\Phi)\) is the stability constant of the interpolation map, and \(\nu_{\max}\) is a uniform upper bound on the single-shot second moment of the heterodyne estimators. The constants \(C_M,C_\mu,C_\nu>0\) are independent of \(d_L\), while \(b,c\ge 0\) describe the polynomial scaling of \(\mu_\infty(\Phi)\) and \(\nu_{\max}\) with \(d_L\). Assume also that the modeling error \(\epsilon_{\rm model}\) and reconstruction error \(\epsilon_{\rm recon}\) satisfy
\begin{equation}
\label{eq:summary-bias-condition}
\epsilon=\epsilon_{\rm model}+\epsilon_{\rm recon},
\qquad
\epsilon_{\rm model}
\le
\frac{e^{-N_{\rm probe}}}{2C_\mu}\,
d_L^{-(b+3/2)}
\min\!\left\{
c_{\rm rec},
\frac{\epsilon_{\rm recon}}{2C_{\rm rec}}
\right\},
\end{equation}
where \(c_{\rm rec},C_{\rm rec}>0\) are the constants from Proposition~\ref{prop:reconstruction-stability-corrected}.

Then the following holds.

\paragraph{Lower bound.}
Any protocol that learns arbitrary \(m\)-mode bosonic unitaries under the metric \(\|\cdot\|_\diamond^{mE}\) must use
\begin{equation}
\label{eq:compact-lb-corrected}
n=\Omega\!\left(\frac{d_E^2}{\epsilon}\right)
\end{equation}
channel uses.

\paragraph{Upper bound in the high-accuracy regime.}
If the following holds
\begin{equation}
\label{eq:summary-high-accuracy}
\epsilon_{\rm recon}\le 2c_{\rm rec}C_{\rm rec},
\end{equation}
then protocol using coherent probes and heterodyne measurement returns an estimator \(\widehat{\mathcal V}\) satisfying $\|\widehat{\mathcal V}-\mathcal U\|_\diamond^{mE}\le \epsilon$ with probability at least \(1-\delta\), using
\begin{equation}
\label{eq:compact-ub-high-accuracy}
n
=
O\!\left(
d_L^{\,2b+c+5}\,
\epsilon_{\rm recon}^{-2}\,
\log\!\frac{d_L}{\delta}
\right)
\end{equation}
channel uses.

\paragraph{Upper bound in the coarse-accuracy regime.}
If the following holds
\begin{equation}
\label{eq:summary-coarse-accuracy}
\epsilon_{\rm recon}\ge 2c_{\rm rec}C_{\rm rec},
\end{equation}
then the same protocol satisfies $\|\widehat{\mathcal V}-\mathcal U\|_\diamond^{mE}\le \epsilon$ with probability at least \(1-\delta\), using
\begin{equation}
\label{eq:compact-ub-coarse-accuracy}
n
=
O\!\left(
d_L^{\,2b+c+5}\,
\log\!\frac{d_L}{\delta}
\right)
\end{equation}
channel uses.

In particular, for asymptotic comparison with the lower bound as \(\epsilon\to 0\), the relevant regime is Eq.~\eqref{eq:summary-high-accuracy}. In that regime, if one assigns a fixed fraction of the total error budget to reconstruction, namely \(\epsilon_{\rm recon}=\Theta(\epsilon)\), then
\begin{equation}
\label{eq:compact-ub-asymptotic}
n
=
O\!\left(
d_L^{\,2b+c+5}\,
\epsilon^{-2}\,
\log\!\frac{d_L}{\delta}
\right).
\end{equation}
The hidden constants in Eqs.~\eqref{eq:compact-ub-high-accuracy}--\eqref{eq:compact-ub-asymptotic} depend only on
\(C_M,C_\mu,C_\nu,C_{\rm rob},c_{\rm rec},C_{\rm rec}\).
\end{theorem}

\begin{proof}
We first prove the lower bound.
Let \(q\coloneqq \lfloor N_{\rm in}\rfloor\).
Every state supported on \(\mathcal H_{\le q}\) has photon number at most \(q\le N_{\rm in}\), hence is admissible in the energy-constrained diamond norm.
Therefore, any protocol that learns arbitrary bosonic unitaries under \(\|\cdot\|_\diamond^{mE}\) must in particular learn the subclass of physical unitaries that act arbitrarily on \(\mathcal H_{\le q}\) and as identity on \(\mathcal H_{\le q}^{\perp}\).
This reduces the problem to learning an arbitrary \(d_E\)-dimensional unitary channel in the ordinary diamond norm.
The finite-dimensional minimax lower bound for unitary-channel learning in the general quantum query model then gives
\begin{align}
n=\Omega\!\left(\frac{d_E^2}{\epsilon}\right),
\end{align}
which proves Eq.~\eqref{eq:compact-lb-corrected}; see Ref.~\cite{haah2023query}.

We next prove the upper bound.
Choose a decomposition
\begin{equation}
\epsilon=\epsilon_{\rm model}+\epsilon_{\rm recon},
\qquad
\epsilon_{\rm model},\epsilon_{\rm recon}\in(0,1).
\end{equation}
By Theorem~\ref{thm:effective-dimension-corrected}, there exists an effective cutoff-space unitary \(V_L\) and a full-space extension \(V=V_L\oplus V_L^\perp\) such that
\begin{equation}
\|\mathcal U-\mathcal V\|_\diamond^{mE}\le \epsilon_{\rm model},
\qquad
\mathcal V(\rho)\coloneqq V\rho V^\dagger.
\end{equation}

Set
\begin{align}
N_{\rm in}\coloneqq m\left(E-\frac12\right),
\qquad
M\coloneqq
\left\lceil
\frac{256N_{\rm in}}{\epsilon_{\rm model}^2}
\right\rceil .
\end{align}
Then set
\begin{align}
N_{\rm out}
\coloneqq
\left\lceil
\frac{256E_U(M)}{\epsilon_{\rm model}^2}
\right\rceil,
\qquad
L\coloneqq \max\{M,N_{\rm out}\}.
\end{align}

Since \(L\in\mathbb N\), the cutoff dimension is exactly
\begin{equation}
D=\dim(\mathcal H_{\le L})=\binom{L+m}{m}=d_L.
\end{equation}

Now apply the reconstruction result derived in \ref{subsec:reconstruction}, \ref{subsec:probe-design-stability}, and Proposition~\ref{prop:reconstruction-stability-corrected}.
The exact sufficient query bound proved there is
\begin{equation}
\label{eq:proof-exact-upper}
n
=
O\!\left(
M_{\rm probe}\,
\mu_\infty(\Phi)^2\,
\nu_{\max}\,
\eta_E^{-2}\,
\log\!\frac{M_{\rm probe}D^2}{\delta}
\right),
\end{equation}
where
\begin{equation}
\eta_E
=
\min\!\left\{
c_{\rm rec}D^{-3/2},
\frac{\epsilon_{\rm recon}}{2C_{\rm rec}D^{3/2}}
\right\},
\end{equation}
and the associated model-bias constraint is
\begin{equation}
\label{eq:proof-bias-condition}
\mu_\infty(\Phi)e^{N_{\rm probe}}\epsilon_{\rm model}\le \frac{\eta_E}{2}.
\end{equation}

Using the assumptions in Eq.~\eqref{eq:summary-assumptions-probes}, we have
\begin{equation}
M_{\rm probe}\,\mu_\infty(\Phi)^2\,\nu_{\max}\,D^3
\le
C_M C_\mu^2 C_\nu\, d_L^{\,2+2b+c+3}
=
C_M C_\mu^2 C_\nu\, d_L^{\,2b+c+5}.
\end{equation}
Moreover,
\begin{equation}
\log\!\frac{M_{\rm probe}D^2}{\delta}
\le
\log\!\frac{C_M d_L^4}{\delta}
=
O\!\left(\log\!\frac{d_L}{\delta}\right).
\end{equation}
Substituting these bounds into Eq.~\eqref{eq:proof-exact-upper} gives
\begin{equation}
\label{eq:proof-after-substitution}
n
=
O\!\left(
d_L^{\,2b+c+2}\,
\eta_E^{-2}\,
\log\!\frac{d_L}{\delta}
\right).
\end{equation}

We now distinguish the two regimes.

If \(\epsilon_{\rm recon}\le 2c_{\rm rec}C_{\rm rec}\), then
\begin{equation}
\eta_E
=
\frac{\epsilon_{\rm recon}}{2C_{\rm rec}d_L^{3/2}},
\qquad
\eta_E^{-2}
=
4C_{\rm rec}^2\,d_L^3\,\epsilon_{\rm recon}^{-2}.
\end{equation}
Substituting this into Eq.~\eqref{eq:proof-after-substitution} yields
\begin{equation}
n
=
O\!\left(
d_L^{\,2b+c+5}\,
\epsilon_{\rm recon}^{-2}\,
\log\!\frac{d_L}{\delta}
\right),
\end{equation}
which proves Eq.~\eqref{eq:compact-ub-high-accuracy}.

If \(\epsilon_{\rm recon}\ge 2c_{\rm rec}C_{\rm rec}\), then
\begin{equation}
\eta_E
=
c_{\rm rec}d_L^{-3/2},
\qquad
\eta_E^{-2}
=
c_{\rm rec}^{-2}d_L^3.
\end{equation}
Substituting this into Eq.~\eqref{eq:proof-after-substitution} yields
\begin{equation}
n
=
O\!\left(
d_L^{\,2b+c+5}\,
\log\!\frac{d_L}{\delta}
\right),
\end{equation}
which proves Eq.~\eqref{eq:compact-ub-coarse-accuracy}.

It remains to verify the bias condition.
Using \(\mu_\infty(\Phi)\le C_\mu d_L^b\) and
\begin{align}
\eta_E
=
d_L^{-3/2}
\min\!\left\{
c_{\rm rec},
\frac{\epsilon_{\rm recon}}{2C_{\rm rec}}
\right\},
\end{align}
a sufficient condition for Eq.~\eqref{eq:proof-bias-condition} is
\begin{equation}
C_\mu d_L^b e^{N_{\rm probe}}\epsilon_{\rm model}
\le
\frac{1}{2}d_L^{-3/2}
\min\!\left\{
c_{\rm rec},
\frac{\epsilon_{\rm recon}}{2C_{\rm rec}}
\right\},
\end{equation}
which is exactly Eq.~\eqref{eq:summary-bias-condition}.

Under this condition, the reconstruction theorem gives
\begin{equation}
\|\widehat{\mathcal V}-\mathcal V\|_\diamond^{mE}\le \epsilon_{\rm recon},
\end{equation}
while the effective-dimension reduction gives
\begin{equation}
\|\mathcal U-\mathcal V\|_\diamond^{mE}\le \epsilon_{\rm model}.
\end{equation}
The triangle inequality then yields
\begin{equation}
\|\widehat{\mathcal V}-\mathcal U\|_\diamond^{mE}\le \epsilon_{\rm model}+\epsilon_{\rm recon}=\epsilon
\end{equation}
with probability at least \(1-\delta\).
This completes the proof.
\end{proof}

Theorem~\ref{thm:query-summary-corrected} expresses both bounds through the same dimension function \(d_x\).
The lower bound depends on the energy-accessible photon-number scale \(N_{\rm in}=m(E-\frac12)\), while the upper bound depends on the larger effective cutoff scale \(L\).
In the experimentally relevant high-accuracy regime, the present protocol therefore has query scaling
\begin{align}
n
=
O\!\left(
d_L^{\,2b+c+5}\,
\epsilon^{-2}\,
\log\!\frac{d_L}{\delta}
\right),
\end{align}
provided the probe family is realizable with Eq.~\eqref{eq:summary-assumptions-probes} and the truncation bias is made small enough through Eq.~\eqref{eq:summary-bias-condition}. This is the cleanest rigorous scaling statement obtainable from the present analysis.

\end{widetext}

\end{document}